\documentclass[11pt,twoside,openright]{report}
\usepackage{epsfig}
\usepackage[greek,german,british]{babel}
\usepackage[T1]{fontenc}
\usepackage{times}
\usepackage{mathptm}
\usepackage{amssymb}
\usepackage{amsmath}
\usepackage{amsthm}
\usepackage[mathscr]{eucal}
\usepackage{exscale}
\usepackage{fancyheadings}

\makeatletter
\renewcommand{\@endtheorem}{\endtrivlist}
\makeatother

\newcommand{\clearemptydoublepage}{\newpage{\pagestyle{empty}%
                                    \cleardoublepage}}

\theoremstyle{definition}
\newtheorem{Def}{Definition}[chapter]

\theoremstyle{plain}
\newtheorem{The}[Def]{Theorem}
\newtheorem{Pro}[Def]{Proposition}
\newtheorem{Cor}[Def]{Corollary}
\newtheorem{Lem}[Def]{Lemma}

\theoremstyle{remark}
\newtheorem*{Rem}{Remark}

\newtheoremstyle{assumption}{1.5ex}{1ex}{\normalfont}{}%
  {\normalfont\bfseries}{.}{\newline}{}

\theoremstyle{assumption}
\newtheorem*{Com}{Compactness Criterion}

\DeclareMathAlphabet{\mathib}{T1}{ptm}{b}{it}
\DeclareMathAlphabet{\mathcmmib}{OML}{cmm}{b}{it}
\DeclareMathAlphabet{\mathecm}{U}{eur}{m}{n}

\DeclareMathOperator{\Aut}{Aut}
\DeclareMathOperator{\id}{id}
\DeclareMathOperator{\linhull}{span}
\DeclareMathOperator{\nusup}{\nu-ess\:sup}
\DeclareMathOperator{\supp}{supp}
\DeclareMathOperator{\sw}{\sigma-weak}
\DeclareMathOperator{\str}{strong}
\DeclareMathOperator{\trace}{Tr \,}

\newcommand{\set}[1]{\{ #1 \}}
\newcommand{\bset}[1]{\bigl\{ #1 \bigr\}}
\newcommand{\Bset}[1]{\Bigl\{ #1 \Bigr\}}

\newcommand{\bcomm}[2]{\bigl[ #1 , #2 \bigr]}
\newcommand{\Bcomm}[2]{\Bigl[ #1 , #2 \Bigr]}

\newcommand{\abs}[1]{\lvert #1 \rvert}
\newcommand{\babs}[1]{\bigl\lvert #1 \bigr\rvert}
\newcommand{\Babs}[1]{\Bigl\lvert #1 \Bigr\rvert}

\newcommand{\norm}[1]{\lVert #1 \rVert}
\newcommand{\bnorm}[1]{\bigl\lVert #1 \bigr\rVert}
\newcommand{\Bnorm}[1]{\Bigl\lVert #1 \Bigr\rVert}

\newcommand{\Norm}[1]{\lvert \negthinspace \lvert \negthinspace \lvert
  #1 \rvert \negthinspace \rvert \negthinspace \rvert}

\newcommand{\FNorm}[1]{\lvert \negthinspace \lvert \negthinspace \lvert
  #1 \rvert \negthinspace \rvert \negthinspace \rvert_F}

\newcommand{\pD}{p_\Delta}
\newcommand{\pDx}[1]{p_\Delta ( #1 )}
\newcommand{\bpDx}[1]{p_\Delta \bigl( #1 \bigr)}
\newcommand{\BpDx}[1]{p_\Delta \Bigl( #1 \Bigr)}

\newcommand{\pprimeD}{p'_\Delta}
\newcommand{\pprimeDx}[1]{p'_\Delta ( #1 )}

\newcommand{\pDprime}{p_{\Delta'}}
\newcommand{\pDprimex}[1]{p_{\Delta'} ( #1 )}

\newcommand{\POG}{\mathcal{P}_{\Gamma_1 , \Gamma_2}^{\; \mathscr{O}}}
\newcommand{\POGx}[1]{\mathcal{P}_{\Gamma_1 , \Gamma_2}^{\;
    \mathscr{O}} ( #1 )}
\newcommand{\bPOGx}[1]{\mathcal{P}_{\Gamma_1 , \Gamma_2}^{\;
    \mathscr{O}} \bigl( #1 \bigr)}

\newcommand{\Pm}{\mathcal{P}_m}
\newcommand{\Pmx}[1]{\mathcal{P}_m ( #1 )}

\newcommand{\Qm}{\mathcal{Q}^m}
\newcommand{\Qmx}[1]{\mathcal{Q}^m ( #1 )}

\newcommand{\Qtwos}{\mathcal{Q}^{2 s}}
\newcommand{\Qtwosx}[1]{\mathcal{Q}^{2 s} ( #1 )}

\newcommand{\Rk}{\mathcal{R}^k}
\newcommand{\Rkx}[1]{\mathcal{R}^k ( #1 )}

\newcommand{\RMx}[1]{\mathcal{R}^M ( #1 )}

\newcommand{\qD}{q_\Delta}
\newcommand{\qDx}[1]{q_\Delta ( #1 )}
\newcommand{\bqDx}[1]{q_\Delta \bigl( #1 \bigr)}
\newcommand{\BqDx}[1]{q_\Delta \Bigl( #1 \Bigr)}

\newcommand{\qprimeD}{q'_\Delta}
\newcommand{\qprimeDx}[1]{q'_\Delta ( #1 )}
\newcommand{\bqprimeDx}[1]{q'_\Delta \bigl( #1 \bigr)}

\newcommand{\qDprimex}[1]{q_{\Delta'} ( #1 )}

\newcommand{\ql}{q_\lambda}

\newcommand{\bqlx}[1]{q_\lambda \bigl( #1 \bigr)}
\newcommand{\Bqlx}[1]{q_\lambda \Bigl( #1 \Bigr)}
\newcommand{\bgqlx}[1]{q_\lambda \biggl( #1 \biggr)}

\newcommand{\qprim}{q'_\mu}

\newcommand{\bqprimx}[1]{q'_\mu \bigl( #1 \bigr)}

\newcommand{\qs}{q_\varsigma}
\newcommand{\qsx}[1]{q_\varsigma ( #1 )}

\newcommand{\qw}{q_w}
\newcommand{\qwx}[1]{q_w ( #1 )}

\newcommand{\ket}[1]{\vert #1 \rangle}
\newcommand{\bket}[1]{\big\vert #1 \bigr\rangle}

\newcommand{\bullket}[1]{\vert #1 \rangle^\bullet}
\newcommand{\bbullket}[1]{\big\vert #1 \bigr\rangle^\bullet}

\newcommand{\xiket}[1]{\vert #1 \rangle_\xi}
\newcommand{\bxiket}[1]{\big\vert #1 \bigr\rangle_\xi}
\newcommand{\Bxiket}[1]{\Big\vert #1 \Bigr\rangle_\xi}

\newcommand{\scp}[2]{\langle #1 \vert #2 \rangle}
\newcommand{\bscp}[2]{\bigl\langle #1 \big\vert #2 \bigr\rangle}

\newcommand{\Wx}[2]{W ( #1 \vert #2 )}
\newcommand{\bWx}[2]{W \bigl( #1 \big\vert #2 \bigr)}

\newcommand{\bWxK}[2]{W_K \bigl( #1 \big\vert #2 \bigr)}

\newcommand{\scpx}[3]{\langle #1 \vert #2 \vert #3 \rangle}
\newcommand{\bscpx}[3]{\bigl\langle #1 \big\vert #2 \big\vert #3
  \bigr\rangle}

\newcommand{\xiscpx}[3]{\langle #1 \vert #2 \vert #3 \rangle_\xi}
\newcommand{\bxiscpx}[3]{\bigl\langle #1 \big\vert #2 \big\vert #3
  \bigr\rangle_\xi}

\newcommand{\Cbb}{\mathbb{C}}
\newcommand{\Kbb}{\mathbb{K}}
\newcommand{\Nbb}{\mathbb{N}}
\newcommand{\Qbb}{\mathbb{Q}}
\newcommand{\Rbb}{\mathbb{R}}

\newcommand{\Mbf}{\mathbf{M}}
\newcommand{\Nbf}{\mathbf{N}}

\newcommand{\Qcal}{\mathcal{Q}}
\newcommand{\Tcal}{\mathcal{T}}
\newcommand{\Zcal}{\mathcal{Z}}

\newcommand{\Becm}{\mathecm{B}}
\newcommand{\Kecm}{\mathecm{K}}
\newcommand{\Mecm}{\mathecm{M}}
\newcommand{\Secm}{\mathecm{S}}
\newcommand{\Wecm}{\mathecm{W}}
\newcommand{\Xecm}{\mathecm{X}}

\newcommand{\zerofrak}{\mathfrak{0}}
\newcommand{\Afrak}{\mathfrak{A}}
\newcommand{\Bfrak}{\mathfrak{B}}
\newcommand{\Cfrak}{\mathfrak{C}}
\newcommand{\Dfrak}{\mathfrak{D}}

\newcommand{\Gfrak}{\mathfrak{G}}
\newcommand{\Kfrak}{\mathfrak{K}}
\newcommand{\Lfrak}{\mathfrak{L}}
\newcommand{\Mfrak}{\mathfrak{M}}
\newcommand{\Nfrak}{\mathfrak{N}}

\newcommand{\Tfrak}{\mathfrak{T}}
\newcommand{\Ufrak}{\mathfrak{U}}
\newcommand{\Vfrak}{\mathfrak{V}}
\newcommand{\Wfrak}{\mathfrak{W}}
\newcommand{\Xfrak}{\mathfrak{X}}
\newcommand{\Yfrak}{\mathfrak{Y}}

\newcommand{\hfrak}{\mathfrak{h}}

\newcommand{\vfrak}{\mathfrak{v}}
\newcommand{\xfrak}{\mathfrak{x}}
\newcommand{\yfrak}{\mathfrak{y}}

\newcommand{\zeroib}{\mathib{0}}

\newcommand{\Kib}{\mathib{K}}
\newcommand{\Oib}{\mathib{O}}

\newcommand{\eib}{\mathib{e}}
\newcommand{\hib}{\mathib{h}}
\newcommand{\pib}{\mathib{p}}

\newcommand{\sib}{\mathib{s}}
\newcommand{\tib}{\mathib{t}}

\newcommand{\vib}{\mathib{v}}
\newcommand{\xib}{\mathib{x}}
\newcommand{\yib}{\mathib{y}}
\newcommand{\zib}{\mathib{z}}

\newcommand{\Dscr}{\mathscr{D}}
\newcommand{\Fscr}{\mathscr{F}}
\newcommand{\Gscr}{\mathscr{G}}
\newcommand{\Hscr}{\mathscr{H}}

\newcommand{\Mscr}{\mathscr{M}}
\newcommand{\Nscr}{\mathscr{N}}
\newcommand{\Oscr}{\mathscr{O}}
\newcommand{\Qscr}{\mathscr{Q}}
\newcommand{\Sscr}{\mathscr{S}}

\newcommand{\Uscr}{\mathscr{U}}
\newcommand{\Vscr}{\mathscr{V}}
\newcommand{\Xscr}{\mathscr{X}}

\newcommand{\Csf}{\mathsf{C}}
\newcommand{\Lsf}{\mathsf{L}}
\newcommand{\Psf}{\mathsf{P}}

\newcommand{\Ssf}{\mathsf{S}}

\newcommand{\Usf}{\mathsf{U}}
\newcommand{\Vsf}{\mathsf{V}}

\newcommand{\neutral}{\mathcmmib{1}}

\newcommand{\unit}{\mathbf{1}}

\newcommand{\Rs}{\mathbb{R}^s}
\newcommand{\Rsone}{\mathbb{R}^{s + 1}}

\newcommand{\AS}{\mathfrak{A}_{\mathscr{S}}}
\newcommand{\Aann}{\mathfrak{A}_{\text{\itshape ann}}}

\newcommand{\Xecmbar}{\overline{\mathecm{X}}}
\newcommand{\Afrakbar}{\overline{\mathfrak{A}}}
\newcommand{\Afrakubar}{\underline{\mathfrak{A}}}
\newcommand{\Cfrakbar}{\overline{\mathfrak{C}}}

\newcommand{\Lfrakbar}{\overline{\mathfrak{L}}}
\newcommand{\Tfrakbar}{\overline{\mathfrak{T}}}
\newcommand{\Vfrakbar}{\overline{\mathfrak{V}}}
\newcommand{\Hscrbar}{\overline{\mathscr{H}}}
\newcommand{\Hscrubar}{\underline{\mathscr{H}}}

\newcommand{\Ubarxi}{\overline{U}_\xi}
\newcommand{\Uubar}{\underline{U}}
\newcommand{\Wbar}{\overline{W}}

\newcommand{\alphaubar}{\underline{\alpha}}
\newcommand{\nubar}{\overline{\nu}}
\newcommand{\pibar}{\overline{\pi}}
\newcommand{\piubar}{\underline{\pi}}
\newcommand{\Deltabar}{\overline{\Delta}}
\newcommand{\Deltaprimebar}{\overline{\Delta'}}
\newcommand{\Deltaetabar}{\overline{\Delta}_\eta}

\newcommand{\AC}{\mathfrak{A}_\mathfrak{C}}
\newcommand{\AL}{\mathfrak{A}_\mathfrak{L}}
\newcommand{\BH}{\mathfrak{B} ( \mathscr{H} )}

\newcommand{\BHbullet}{\mathfrak{B} ( \mathscr{H}^\bullet )}

\newcommand{\BHw}{\mathfrak{B} ( \mathscr{H}_w )}
\newcommand{\BHxi}{\mathfrak{B} ( \mathscr{H}_\xi )}

\newcommand{\Lor}{\mathsf{L}_+^\uparrow}
\newcommand{\Poin}{\mathsf{P}_{\negthinspace +}^\uparrow}

\newcommand{\fwcone}{\overline{V}_{\negthinspace +}}

\newcommand{\ED}{E ( \Delta )}
\newcommand{\EDG}{E \bigl( \Delta' ( \Delta , \Gamma ) \bigr)}
\newcommand{\EDdG}{E \bigl( \Delta'' ( \Delta , \Gamma ) \bigr)}
\newcommand{\EDdGl}{E \bigl( \Delta'' ( \Delta , \Gamma_l ) \bigr)}
\newcommand{\EDdGone}{E \bigl( \Delta'' ( \Delta , \Gamma_1 ) \bigr)}
\newcommand{\EDdGtwo}{E \bigl( \Delta'' ( \Delta , \Gamma_2 ) \bigr)}
\newcommand{\EDprime}{E ( \Delta' )}
\newcommand{\EDGplus}{E ( \overline{\Delta} + \Gamma)}
\newcommand{\EDGoneplus}{E ( \overline{\Delta} + \Gamma_1)}
\newcommand{\EDGtwoplus}{E ( \overline{\Delta} + \Gamma_2)}
\newcommand{\EDGprimebarminus}{E ( \overline{\Delta'} - \Gamma)}
\newcommand{\EDGprimebarplus}{E ( \overline{\Delta'} + \Gamma)}
\newcommand{\EDdprime}{E ( \Delta'' )}
\newcommand{\EDbar}{E ( \overline{\Delta} )}
\newcommand{\EDbarG}{E ( \overline{\Delta}_0 + \Gamma)}
\newcommand{\EDetabar}{E ( \overline{\Delta}_\eta )}
\newcommand{\EDzero}{E ( \Delta_0 )}
\newcommand{\EDzerodprime}{E ( \Delta_0'' )}
\newcommand{\EbulletDelta}{E^\bullet ( \Delta )}
\newcommand{\EbulletDeltaprime}{E^\bullet ( \Delta' )}
\newcommand{\EbulletGammaN}{E^\bullet ( \Gamma_N )}
\newcommand{\EbulletGammaNzero}{E^\bullet ( \Gamma_{N_0} )}
\newcommand{\Ecupminus}{E \bigl( \Delta' \cup ( \overline{\Delta'} -
  \Gamma) \bigr)}
\newcommand{\Ecupplus}{E \bigl( \Delta' \cup ( \overline{\Delta'} +
  \Gamma) \bigr)}
\newcommand{\EsDprime}{E_\sigma ( \Delta' )}
\newcommand{\EsDprimeeta}{E_\sigma ( \Delta'_\eta )}
\newcommand{\EsGamma}{E_\sigma ( \Gamma )}
\newcommand{\EubarDelta}{\underline{E} ( \Delta )}
\newcommand{\EubarDeltaprime}{\underline{E} ( \Delta' )}
\newcommand{\EwDprime}{E_w ( \Delta' )}
\newcommand{\EwDelta}{E_w ( \Delta )}

\newcommand{\EwGamma}{E_w ( \Gamma )}
\newcommand{\ExiDelta}{E_\xi ( \Delta )}
\newcommand{\ExiDeltaprime}{E_\xi ( \Delta' )}
\newcommand{\ExiGammaN}{E_\xi ( \Gamma_N )}
\newcommand{\ExiGammaNzero}{E_\xi ( \Gamma_{N_0} )}

\newcommand{\ACstar}{{\mathfrak{A}_\mathfrak{C}}^{\negthickspace *}}
\newcommand{\Cstar}{\mathfrak{C}^*}
\newcommand{\Cstarplus}{\mathfrak{C}^{* \thinspace +}}
\newcommand{\Cbarstar}{\overline{\mathfrak{C}}^*}
\newcommand{\CDstar}{{\mathfrak{C}_\Delta}^{\negthickspace *}}
\newcommand{\CDstarplus}{{\mathfrak{C}_\Delta}^{\negthickspace *
    \thinspace +}}
\newcommand{\Lstar}{\mathfrak{L}^*}
\newcommand{\Lzerostar}{{\mathfrak{L}_0}^*}

\newcommand{\aLax}{\alpha_{( \Lambda , x )}}
\newcommand{\aLaxprime}{\alpha_{( \Lambda' , x' )}}
\newcommand{\aLaxdprime}{\alpha_{( \Lambda'' , x'' )}}
\newcommand{\aLaxzero}{\alpha_{( \Lambda_0 , x_0 )}}
\newcommand{\aLaxone}{\alpha_{( \Lambda_1 , x_1 )}}
\newcommand{\aLaxtwo}{\alpha_{( \Lambda_2 , x_2 )}}

\newcommand{\abulletLax}{\alpha^\bullet_{( \Lambda , x )}}
\newcommand{\abulletLaxzero}{\alpha^\bullet_{( \Lambda_0 , x_0 )}}
\newcommand{\abulletLaxone}{\alpha^\bullet_{( \Lambda_1 , x_1 )}}
\newcommand{\abulletLaxtwo}{\alpha^\bullet_{( \Lambda_2 , x_2 )}}
\newcommand{\abulletLaxn}{\alpha^\bullet_{( \Lambda_n , x_n )}}
\newcommand{\abulletx}{\alpha^\bullet_x}
\newcommand{\abulletxk}{\alpha^\bullet_{x_k}}
\newcommand{\abulletxl}{\alpha^\bullet_{x_l}}
\newcommand{\abulletxprime}{\alpha^\bullet_{x'}}
\newcommand{\abullety}{\alpha^\bullet_y}
\newcommand{\aubarLax}{\underline{\alpha}_{( \Lambda , x )}}
\newcommand{\aubarx}{\underline{\alpha}_x}
\newcommand{\agGamma}{\alpha_{g_\Gamma}}
\newcommand{\aibx}{\alpha_\mathib{x}}
\newcommand{\aibxprime}{\alpha_{\mathib{x}'}}
\newcommand{\aibtwox}{\alpha_{2\mathib{x}}}
\newcommand{\aiby}{\alpha_\mathib{y}}
\newcommand{\aminusibx}{\alpha_{(-\mathib{x})}}
\newcommand{\ax}{\alpha_x}

\newcommand{\Oi}{\mathscr{O}_i}
\newcommand{\Ok}{\mathscr{O}_k}

\newcommand{\AO}{\mathfrak{A} ( \mathscr{O} )}
\newcommand{\AOone}{\mathfrak{A} ( \mathscr{O}_1 )}
\newcommand{\AOtwo}{\mathfrak{A} ( \mathscr{O}_2 )}
\newcommand{\AoneO}{\mathfrak{A}_1 ( \mathscr{O} )}

\newcommand{\AOr}{\mathfrak{A} ( \mathscr{O}_r )}
\newcommand{\AOtwor}{\mathfrak{A} ( \mathscr{O}_{2 r} )}
\newcommand{\AOrplustwox}{\mathfrak{A} ( \mathscr{O}_r + 2 \mathib{x}
  )} 
\newcommand{\AOk}{\mathfrak{A} ( \mathscr{O}_k )}

\newcommand{\ArO}{\mathfrak{A}_r ( \mathscr{O} )}

\newcommand{\Acountbar}{\overline{\mathfrak{A}^c}}
\newcommand{\Lcount}{\mathsf{L}^{\negthinspace c}}
\newcommand{\Lbarcount}{\overline{\mathsf{L}}^{\negthinspace c}}
\newcommand{\Pcount}{\mathsf{P}^c}
\newcommand{\Pbarcount}{\overline{\mathsf{P}}^c}
\newcommand{\Pubarcount}{\underline{\mathsf{P}}^c}
\newcommand{\Rcount}{\mathscr{R}^c}
\newcommand{\Tcount}{\mathsf{T}^c}
\newcommand{\Tbarcount}{\overline{\mathsf{T}}^c}

\newcommand{\Acount}{\mathfrak{A}^c}
\newcommand{\Aubarcount}{\underline{\mathfrak{A}}^c}
\newcommand{\AcountO}{\mathfrak{A}^c ( \mathscr{O} )}
\newcommand{\AcountOi}{\mathfrak{A}^c ( \mathscr{O}_i )}
\newcommand{\AcountOk}{\mathfrak{A}^c ( \mathscr{O}_k )}

\newcommand{\detectcount}{\mathfrak{C}^c}
\newcommand{\idealcount}{\mathfrak{L}^c}
\newcommand{\vacbar}{\overline{\mathfrak{L}_0}}
\newcommand{\vaccount}{\mathfrak{L}_0^c}
\newcommand{\vaccountbar}{\overline{\mathfrak{L}_0^c}}

\newcommand{\Hbullet}{\mathscr{H}^\bullet}

\newcommand{\Abullet}{\mathfrak{A}^\bullet}

\newcommand{\AbulletOk}{\mathfrak{A}^\bullet ( \mathscr{O}_k )}

\newcommand{\ArbulletOk}{\mathfrak{A}_r^\bullet ( \mathscr{O}_k )}
\newcommand{\AsubbulletO}{\mathfrak{A}_\bullet ( \mathscr{O} )}

\newcommand{\piwDprime}{\pi_{w , \Delta'}}

\newcommand{\etaDprime}{\eta_{\Delta'}}

\newcommand{\SDG}{S_\Delta^{\, \Gamma}}

\newcommand{\TODprime}{T_{\Delta'}^\Oscr}
\newcommand{\TODbar}{T_{\overline{\Delta}}^\Oscr}
\newcommand{\TwODprime}{T_{w , \Delta'}^\Oscr}

\newcounter{defitem}

\newenvironment{deflist}{\begin{list}{(\alph{defitem})}%
  {\usecounter{defitem} \setlength{\topsep}{0ex}%
   \setlength{\parsep}{0.2ex} \setlength{\itemsep}{0.4ex}%
   \setlength{\leftmargin}{0em} \setlength{\itemindent}{0.5em}%
   }}{\end{list}}

\newcounter{proofitem}

\newenvironment{prooflist}{\begin{list}{(\roman{proofitem})}%
  {\usecounter{proofitem} \setlength{\topsep}{0ex}%
   \setlength{\parsep}{0.2ex} \setlength{\itemsep}{0.4ex}%
   \setlength{\leftmargin}{0em} \setlength{\itemindent}{0.5em}%
   \setlength{\listparindent}{1em}}}{\qed \end{list}}

\newcounter{Proofitem}

\newenvironment{Prooflist}{\begin{list}{Part (\Roman{Proofitem}):}%
  {\usecounter{Proofitem} \setlength{\topsep}{0ex}%
   \setlength{\parsep}{0.2ex} \setlength{\itemsep}{0.4ex}%
   \setlength{\leftmargin}{0em} \setlength{\itemindent}{0.5em}%
   \setlength{\listparindent}{1em}}}{\qed \end{list}}

\newcounter{Proofsubitem}

\newenvironment{Proofsublist}{\begin{list}{(\roman{Proofsubitem})}%
  {\usecounter{Proofsubitem} \setlength{\topsep}{0ex}%
   \setlength{\parsep}{0.2ex} \setlength{\itemsep}{0.4ex}%
   \setlength{\leftmargin}{0em} \setlength{\itemindent}{2em}%
   \setlength{\listparindent}{1em} \setlength{\labelsep}{1ex}%
   }}{\end{list}}

\newcounter{propitem}

\newenvironment{proplist}{\begin{list}{(\roman{propitem})}%
  {\usecounter{propitem} \setlength{\topsep}{0ex}%
   \setlength{\parsep}{0.2ex} \setlength{\itemsep}{0.4ex}%
   \setlength{\leftmargin}{0em} \setlength{\itemindent}{0.5em}%
   }}{\end{list}}

\newcounter{remitem}

\newenvironment{remlist}{\begin{list}{(\roman{remitem})}%
  {\usecounter{remitem} \setlength{\topsep}{0ex}%
   \setlength{\parsep}{0.2ex} \setlength{\itemsep}{0.4ex}%
   \setlength{\leftmargin}{0em} \setlength{\itemindent}{0.5em}%
   }}{\end{list}}

\newcounter{theoitem}

\newenvironment{theolist}{\begin{list}{(\Roman{theoitem})}%
  {\usecounter{theoitem} \setlength{\topsep}{0ex}%
   \setlength{\parsep}{0.2ex} \setlength{\itemsep}{0.4ex}%
   \setlength{\leftmargin}{0em} \setlength{\itemindent}{0.5em}%
   }}{\end{list}}

\newcounter{theosubitem}

\newenvironment{theosublist}{\begin{list}{(\roman{theosubitem})}%
  {\usecounter{theosubitem} \setlength{\topsep}{0ex}%
   \setlength{\parsep}{0.2ex} \setlength{\itemsep}{0.4ex}%
   \setlength{\leftmargin}{0em} \setlength{\itemindent}{2em}%
   }}{\end{list}}

\newcounter{abcitem}

\newenvironment{abclist}{\begin{list}{(\Alph{abcitem})}%
  {\usecounter{abcitem} \setlength{\topsep}{0ex}%
   \setlength{\parsep}{0.2ex} \setlength{\itemsep}{0.4ex}%
   \setlength{\leftmargin}{2em}}}{\end{list}}

\newcounter{stepitem}

\newenvironment{steplist}{\begin{list}{(\arabic{stepitem})}%
  {\usecounter{stepitem} \setlength{\topsep}{0ex}%
   \setlength{\parsep}{0.2ex} \setlength{\itemsep}{0.6ex}%
   \setlength{\leftmargin}{0em} \setlength{\itemindent}{1.7em}%
   }}{\end{list}}

\newcounter{latinitem}

\newenvironment{latinlist}{\begin{list}{(\Roman{latinitem})}%
  {\usecounter{latinitem} \setlength{\topsep}{0ex}%
   \setlength{\parsep}{0.2ex} \setlength{\itemsep}{0.6ex}%
   \setlength{\leftmargin}{0em} \setlength{\itemindent}{2.2em}%
   }}{\end{list}}

\newenvironment{trilist}{\begin{list}%
  {$\mspace{-50mu} \blacktriangleright$}%
  {\setlength{\topsep}{0ex} \setlength{\parsep}{0.2ex}%
   \setlength{\itemsep}{0.4ex} \setlength{\leftmargin}{1.3em}%
   }}{\end{list}}

\newenvironment{plaintrilist}{\begin{list}%
  {$\mspace{-50mu} \blacktriangleright$}%
  {\setlength{\topsep}{1ex} \setlength{\parsep}{0.2ex}%
   \setlength{\itemsep}{0.4ex} \setlength{\leftmargin}{0em}%
   \setlength{\itemindent}{1.25em}}}{\end{list}}

\newenvironment{bulletlist}{\begin{list}{$\mspace{-50mu} \bullet$}%
  {\setlength{\topsep}{0ex} \setlength{\parsep}{0.2ex}%
   \setlength{\itemsep}{0.4ex} \setlength{\leftmargin}{0em}%
   \setlength{\itemindent}{1em}}}{\end{list}}

\newenvironment{indentbulletlist}{\begin{list}%
  {$\mspace{-50mu} \bullet$}%
  {\setlength{\topsep}{0ex} \setlength{\parsep}{0.2ex}%
   \setlength{\itemsep}{0.4ex} \setlength{\leftmargin}{1.5em}%
   \setlength{\rightmargin}{1.5em}\setlength{\itemindent}{1em}}}%
   {\end{list}}

\newenvironment{lindentbulletlist}{\begin{list}%
  {$\mspace{-50mu} \bullet$}%
  {\setlength{\topsep}{0ex} \setlength{\parsep}{0.2ex}%
   \setlength{\itemsep}{0.4ex} \setlength{\leftmargin}{1.5em}%
   \setlength{\itemindent}{1em}}}{\end{list}}

\begin{document}

\begin{titlepage}
  \begin{center}
    \vspace*{\fill}
    {\Huge\bfseries The\\ Concept of Particle Weights\\ in\\
      Local Quantum Field Theory\\}
    \vfill
    \vfill
    \vfill
    Dissertation\\
    zur Erlangung des Doktorgrades\\
    der Mathematisch-Naturwissenschaftlichen Fakult\"{a}ten\\
    der Georg-August-Universit\"at zu G\"{o}ttingen\\
    \vspace{1cm}
    vorgelegt von\\
    \vspace{0.5cm}
    \textsc{Martin Porrmann}\\
    \vspace{0.5cm}
    aus\\
    Wolfenb\"{u}ttel\\
    \vspace{1cm}
    G\"{o}ttingen 1999\\
  \end{center}
\end{titlepage}

\thispagestyle{empty}
\vspace*{\fill}
\vfill
\noindent D 7\\
Referent: \textsc{Prof.~Dr.~Detlev Buchholz}\footnote{Institut
  f\"{u}r Theoretische Physik der Universit\"{a}t G\"{o}ttingen} \\
Korreferent: \textsc{Prof.~Dr.~Klaus
  Fredenhagen}\footnote{II.~Institut f\"{u}r Theoretische Physik
  der Universit\"{a}t Hamburg} \\
Tag der m\"{u}ndlichen Pr\"{u}fung: 26.~Januar~2000
\vfill

\clearemptydoublepage

\setcounter{page}{0}
\pagenumbering{roman}

\tableofcontents

\clearemptydoublepage

\pagenumbering{arabic}

\chapter{Introduction}
  \label{chap-introduction}

Physical phenomena occurring in high energy physics are analysed in
terms of `particles', arising as asymptotic configurations of
elementary entities in scattering experiments. These particles are
characterized by certain specific intrinsic properties, which are
expressed by quantum numbers whose integration in the framework of a
consistent and complete theoretical description is an aim of quantum
field theory. The usual theoretical description of particles goes back
to the famous analysis by Wigner of the irreducible representations of
the Poincar\'{e} group \cite{wigner:1939}. He gives a complete
classification of all these representations, which are labelled by two
parameters $m$ and $s$. It is assumed that a particle pertains to a
specific representation of this group, in which case the parameters
$m$ and $s$ are interpreted as its intrinsic mass and spin,
respectively. However, this approach to a theoretical description of
mass and spin is not universally applicable. There are quantum field
theories in which particles coupled to particles of zero rest mass
cannot be described in terms of eigenstates of the mass operator. An
example is quantum electrodynamics where charged particles are
inevitably accompanied by soft photons. It is an open question, known
as the infraparticle problem \cite{schroer:1963}, how mass and spin of
a particle are to be described in the framework of quantum field
theory. Moreover, standard collision theory does not work in these
cases.

A closer analysis of quantum electrodynamics shows that the
infraparticle problem is connected with Gauss' law
\cite{froehlich/morchio/strocchi:1979,buchholz:1986a}. An outline of 
the underlying  mechanism, following arguments of Buchholz in
\cite{buchholz:1986a}, may be appropriate at this point. Due to Gauss'
law, the charge of a physical state can be determined by measuring the
electromagnetic field at asymptotic spacelike distances. These
measurements do not interfere with those performed within bounded
regions; therefore, being a $c$-number, the asymptotic field
configuration is a superselection rule of the theory. Its dependence
on the state of motion of the charged particle implies that there
exists a multitude of superselection sectors and that the Lorentz
symmetry is broken. Consequently, charged particles cannot be
described according to Wigner's theory.

The present thesis proposes a novel approach to the concept of
particles, elaborating some of the ideas of Buchholz' which he
introduced in \cite{buchholz:1986b}. In a model-independent framework,
especially without excluding massless states and without assuming
asymptotic completeness of the theory, an approach of Araki and Haag
\cite{araki/haag:1967} to scattering theory is reconsidered.
Chapter~\ref{chap-localizing-operators} introduces the concept of 
detectors to be used in this work and investigates the suitable
topologies that the corresponding algebraic structures are furnished
with. A basic ingredient here is the interplay between locality and
the spectrum condition. In Chapter~\ref{chap-particle-weights} we pass
to the dual point of view and analyse the resulting continuous
functionals. Then, on physical grounds, a certain subclass is
distinguished, arising as asymptotic limits of certain functionals
constructed from physical states of bounded energy. These limits
exhibit properties of singly localized systems (particles). The
limiting procedure to be presented here is able to directly reproduce
charged systems, in contrast to the LSZ-theory where charge-carrying
unobservable operators are necessary.

The representations induced by these asymptotic functionals (the
particle weights) are highly reducible, so the obvious task is to work
out a disintegration theory in terms of irreducible representations
(pure particle weights). This will be done in
Chapters~\ref{chap-disintegration} and \ref{chap-choquet}. The
approach of Chapter~\ref{chap-disintegration} makes use of the
standard decomposition theory for representations of
$C^*$-algebras. To be able to apply this theory, the mathematical
structures under consideration have to be adapted to its needs. Great
care is taking to ensure that the resulting irreducible
representations have all the properties allowing for their
interpretation as representatives of elementary particles. As
demonstrated by Buchholz \cite{buchholz/porrmann}, it is then possible
to classify the pure particle weights according to their spin and
mass even in the case of charged systems. This shows that the notion
of particle weights provides a promising approach to the
aforementioned infraparticle problem. In
Chapter~\ref{chap-local-normality} a compactness criterion due to
Fredenhagen and Hertel is used to impose certain restrictions on the
phase space of quantum field theory. The additional information is
used to demonstrate that the particle weight representations of
Chapter~\ref{chap-disintegration} are locally normal. This implies
that one does not lose essential information about the physical
systems in the course of the constructions needed to adapt the problem
at hand to the needs of spatial disintegration.
Chapter~\ref{chap-choquet}, again drawing on the mentioned compactness
criterion, presents the first steps in an alternative approach to
disintegration: Choquet theory. Chapter~\ref{chap-summary} gives a
brief summary.

\subsubsection{Assumptions of Local Quantum Physics}

We collect here the main structural postulates upon which Local
Quantum Physics is built in the abstract setting of the algebraic
approach \cite{haag/kastler:1964,araki:1961}, principally in order to
fix notation.
\begin{plaintrilist}
\item The basis of the present investigations is a net
  \begin{subequations}
    \begin{equation}
      \label{eq-local-net}
      \Oscr \mapsto \AO
    \end{equation}
    of $C^*$-algebras, which are indexed by the bounded regions
    $\Oscr$ in space-time $\Rsone$ and which are \emph{concrete} in
    the sense that they all belong to the algebra of bounded operators
    $\BH$ on a certain Hilbert space $\Hscr$. The so-called
    quasi-local algebra $\Afrak$ is the $C^*$-inductive limit of the
    net \eqref{eq-local-net}
    (cf.~\cite[Definition~2.63]{bratteli/robinson:1987}):
    \begin{equation}
      \label{eq-quasi-local}
      \Afrak \doteq \bigcup_\Oscr^{C^*} \AO \text{.}
    \end{equation}
  \end{subequations}
\item On the $C^*$-algebra $\Afrak$ the symmetry transformations in
  the inhomogeneous Lorentz group, the Poincar\'{e} group $\Poin =
  \Lor \ltimes \Rsone$, are implemented via a strongly continuous
  group of automorphisms:
  \begin{equation}
    \label{eq-Poin-auto}
    \Poin \ni ( \Lambda , x ) \mapsto \aLax \in \Aut \Afrak \text{.}
  \end{equation}
\item The net \eqref{eq-local-net} is subject to the following
  conditions:
  \begin{subequations}
    \begin{lindentbulletlist}
    \item Isotony: For any two bounded regions $\Oscr_1$ and $\Oscr_2$
      in $\Rsone$
      \begin{equation}
        \label{eq-isotony}
        \Oscr_1 \subseteq \Oscr_2 \Rightarrow \AOone \subseteq \AOtwo
        \text{.}
      \end{equation}
    \item Locality: If the bounded regions $\Oscr_1$ and $\Oscr_2$ are
      spacelike separated, i.\,e., $\Oscr_1$ belongs to the spacelike
      complement of $\Oscr_2$, formally $\Oscr_1 \subseteq \Oscr_2'$,
      then
      \begin{equation}
        \label{eq-locality}
        \AOone \subseteq \AOtwo' \text{,}
      \end{equation}
      where the prime in \eqref{eq-locality} denotes the commutant in
      $\BH$.
    \item Relativistic Covariance: For arbitrary bounded regions
      $\Oscr$ and arbitrary transformations $( \Lambda , x ) \in
      \Poin$ there hold the relations 
      \begin{equation}
        \label{eq-covariance}
        \Afrak ( \Lambda \Oscr + x ) = \aLax \bigl( \AO \bigr)
        \text{.}
      \end{equation}
    \end{lindentbulletlist}
  \end{subequations}
\item The subgroup $\Rsone$ of translations in $\Poin$ is implemented
  on $\Afrak$ by a strongly continuous unitary group, i.\,e., one
  which is continuous with respect to the strong-operator
  topology. These unitaries can be expressed through the (unbounded)
  generators $P^{\, \mu}$, $\mu = 1 \text{,} \dots \text{,} s + 1$, of
  space-time translations according to
  \begin{subequations}
    \begin{equation}
      \label{eq-unitaries-by-generators}
      U ( x ) = \exp ( i \: P^{\, \mu} x_\mu ) \text{,}
    \end{equation}
    and, by virtue of \eqref{eq-Poin-auto}, one has for any $x \in
    \Rsone$
    \begin{equation}
      \label{eq-unitary-auto}
      \alpha_x ( A ) = U ( x ) A {U ( x )}^* \text{,} \quad A \in
      \Afrak \text{.}
    \end{equation}
  \end{subequations}
  The joint spectrum of the generators $P^{\, \mu}$, expressed by the
  pertinent spectral resolution $E (~.~)$ in terms of projections in
  $\Afrak''$, is supposed to lie in the closed forward light cone
  \begin{equation*}
    \fwcone \doteq \bset{p \in \Rsone : p \cdot p = p^{\, \mu} p_\mu
    \geqslant 0} \text{.}
  \end{equation*}
  This assumption is known under the term `positive-energy
  representation.'
\item Physical states are represented by normalized positive linear
  functionals on the quasi-local algebra $\Afrak$, which are normal,
  i.\,e., continuous with respect to the $\sigma$-weak topology that
  $\Afrak$ inherits from $\BH$. The set of all physical states
  $\omega$ is denoted by $\Sscr$; it is in one-to-one correspondence
  to the entirety of all density matrices, the positive trace-class
  operators in $\BH$ with unit trace, via
  \begin{subequations}
    \begin{equation}
      \label{eq-state-trace}
      \omega ( A ) = \trace ( \rho_\omega A ) \text{,} \quad A \in
      \Afrak \text{,}
    \end{equation}
    where $\rho_\omega$ denotes the unique operator of the above
    kind. The fact that a physical state $\omega$ possesses
    energy-momentum in the Borel set $\Delta \subseteq \Rsone$ is
    expressed by the condition
    \begin{equation}
      \label{eq-energy-momentum-content}
      \omega \bigl( \ED \bigr) = \trace \bigl( \ED \rho_\omega \ED
      \bigr) = 1 \text{.}
    \end{equation}
  \end{subequations}
  The corresponding subset of $\Sscr$ is written $\Sscr ( \Delta )$.
\end{plaintrilist}

At this point, for the sake of clarity, a few remarks concerning
topological notions seem advisable. The norm topology on $\Afrak$ is
sometimes called the uniform topology and leaves no room for a
possible misunderstanding. The situation is more complicated in case
of the term `strong continuity:'
\begin{bulletlist}
\item An automorphism group $\bset{\alpha_g : g \in \Gscr} \subseteq
  \Aut \Afrak$ on the $C^*$-algebra $\Afrak$, $\Gscr$ a topological
  group, is called strongly continuous if the mapping
  \begin{equation*}
    \Gscr \ni g \mapsto \alpha_g ( A ) \in \Afrak
  \end{equation*}
  is continuous for arbitrary $A \in \Afrak$ with respect to the
  initial topology of the group $\Gscr$ and with respect to the
  uniform topology of $\Afrak$.
\item A unitary group $\bset{U ( g ) : g \in \Gscr} \subseteq \BH$,
  $\Gscr$ again a topological group, is called strongly continuous if
  the mapping
  \begin{equation*}
    \Gscr \ni g \mapsto U ( g ) \in \BH
  \end{equation*}
  is continuous with respect to the topology of $\Gscr$ and with
  respect to the strong-operator topology on $\BH$.
\end{bulletlist}
The term `$\sigma$-weak topology' is used to denote the locally convex
topology on the algebra $\BH$ that is defined through the family of
seminorms
\begin{equation*}
  \Qcal_{\thickspace \set{\phi_n , \psi_n}} : \BH \rightarrow \Rbb_0^+
  \qquad A \mapsto \Qcal_{\thickspace \set{\phi_n , \psi_n}} ( A )
  \doteq \Babs{\sum_{n = 1}^\infty ( \phi_n , A \psi_n )} \text{,}
\end{equation*}
where the sequences $\set{\phi_n}_{n \in \Nbb}$ and $\set{\psi_n}_{n
\in \Nbb}$ of vectors in the Hilbert space $\Hscr$ are subject to the
conditions $\sum_{n = 1}^\infty \norm{\phi_n}^2 < \infty$ and $\sum_{n
= 1}^\infty \norm{\psi_n}^2 < \infty$. This designation is synonymous
with `ultra-weak topology.' Mappings which are continuous with respect
to this topology are called normal.

\chapter{Localizing Operators and Spectral Seminorms}
  \label{chap-localizing-operators}

The results presented in Chapters~\ref{chap-localizing-operators} and
\ref{chap-particle-weights} have been worked out in close
collaboration with Detlev Buchholz, whose ideas, as set out in
\cite{buchholz:1986b}, constituted the foundation. Their somewhat
complicated presentation is the author's responsibility. The particle
concept to be set forth in the sequel is motivated by the experimental
situation encountered in high energy physics where certain physical
systems show up as `particles,' being traced by specific measuring
devices called `detectors.' The common characteristic of these
physical systems is that they are localized in the course of the
measuring process. Haag and Kastler stated in their fundamental
article \cite{haag/kastler:1964} on algebraic quantum field theory
that `\dots ultimately all physical processes are analyzed in terms of
geometric relations of unresolved phenomena,' emphasizing localization
as the very nature of all measurements. To represent the experimental
set-up in the framework of the algebraic approach to local quantum
physics elements of the quasi-local algebra $\Afrak$ have to be
singled out first that exhibit properties of particle detectors.

\section{The Algebra of Detectors}

As argued by Araki and Haag \cite{araki/haag:1967} a particle detector
$C \in \Afrak$ should be insensitive to the vacuum $\Omega$: $C \Omega
= 0$. In view of the actual experimental situation one can be more
specific, noting that a minimal energy, depending on the detector
used, has to be deposited to produce a signal. In the present thesis
we shall therefore work with a smaller class of operators: the
algebraic representatives corresponding to a particle counter are to
annihilate all physical states with bounded energy below a specific
threshold, to be precise. Now, on account of the
Reeh--Schlieder-Theorem, this feature is incompatible with locality
since an algebra pertaining to a region $\Oscr$ with non-void causal
complement $\Oscr'$ does not contain any operator annihilating states
of bounded energy (cf.~\cite{reeh/schlieder:1961,haag:1996}). As a
consequence, the operators which comply with the above annihilation
property cannot be strictly local; instead their localization has to
be weakened. This is done in a way that resembles the introduction of
rapidly decreasing functions on $\Rbb^n$: the operators in question
are not contained in a local algebra, but they are almost local in the
sense of the following definition (`quasilocal of infinite order' is
the designation used in \cite{araki/haag:1967}).
\begin{Def}[Almost Locality]
  \label{Def-almost-locality}
  Let $\Oscr_r \doteq \bset{( x^0 , \xib ) \in \Rsone : \abs{x^0} +
  \abs{\xib} < r}$, $r > 0$, denote the double cone (standard diamond)
  with basis $\Oib_r \doteq \bset{\xib \in \Rs : \abs{\xib} < r}$. An
  operator $A \in \Afrak$ is called almost local if there exists a net
  $\bset{A_r \in \AOr : r > 0}$ of local operators such that 
  \begin{equation}
    \label{eq-almost-locality}
    \lim_{r \rightarrow \infty} r^k \norm{A - A_r} = 0
  \end{equation}
  for any $k \in \Nbb_0$. The set of almost local operators is a
  $^*$-subalgebra of $\Afrak$ denoted by $\AS$.
\end{Def}
\begin{Rem}
  \begin{remlist}
  \item Let $A$ and $B$ be almost local operators with approximating
    nets of local operators $\bset{A_r \in \AOr : r > 0}$ and
    $\bset{B_r \in \AOr : r > 0}$, respectively. Then, since $\Oscr_r$
    and $\Oscr_r + 2 \xib$ are spacelike separated for $r \leqslant
    \abs{\xib}$ so that the associated algebras $\AOr$ and
    $\AOrplustwox$ commute, the following estimate holds for any $\xib
    \in \Rs \setminus \set{\zeroib}$
    \begin{subequations}
      \begin{equation}
        \label{eq-commutator-norm-estimate}
        \bnorm{\bcomm{\aibtwox ( A )}{B}} \leqslant 2 \, \bigl(
        \norm{A - A_{\abs{\xib}}} \, \norm{B} + \norm{A -
        A_{\abs{\xib}}} \, \norm{B - B_{\abs{\xib}}} + \norm{A} \,
        \norm{B - B_{\abs{\xib}}} \bigr)
      \end{equation}
      The right-hand side of this inequality is bounded and falls off
      more rapidly than any power of $\abs{\xib}^{-1}$, therefore the
      continuous mapping $\Rs \ni \xib \mapsto
      \bnorm{\bcomm{\alpha_{\xib} ( A )}{B}}$ turns out to be
      integrable:
      \begin{equation}
        \label{eq-commutator-integral}
        \int_{\Rs} d^s x \; \bnorm{\bcomm{\aibx ( A )}{B}} < \infty
        \text{.}
      \end{equation}
    \item The approximating net of local operators $\bset{A_r \in \AOr
      : r > 0}$ for $A \in \AS$ can be used to construct a second
      approximating net $\bset{A'_r \in \AOr : r > 0}$ with the
      additional property $\norm{A'_r} \leqslant \norm{A}$ for any $r
      > 0$, which at the same time is subject to the inequality
      $\norm{A - A'_r} \leqslant 2 \, \norm{A - A_r}$ and thus
      satisfies condition \eqref{eq-almost-locality} for almost
      locality. Estimates of this kind will later on turn out to be
      important in solving the problem of existence of uniform bounds
      for integrals of the form \eqref{eq-commutator-integral},
      evaluated for sequences or even nets of almost local
      operators. With approximating nets of local operators of this
      special kind the estimate \eqref{eq-commutator-norm-estimate}
      can be improved for arbitrary $A \text{,} B \in \AS$ to yield
      \begin{equation}
        \label{eq-special-commutator-norm-estimate}
        \bnorm{\bcomm{\aibtwox ( A )}{B}} \leqslant 2 \, \bigl(
        \norm{A - A_{\abs{\xib}}} \, \norm{B} + \norm{A} \, \norm{B -
        B_{\abs{\xib}}} \bigr) \text{,} \quad \xib \in \Rs \setminus
        \set{\zeroib} \text{.}
      \end{equation}
    \end{subequations}
  \end{remlist}
\end{Rem}

The feature of annihilating states of bounded energy below a certain
threshold is called vacuum annihilation property in the sequel and
finds its rigorous mathematical expression in the following definition.
\begin{Def}[Vacuum Annihilation Property]
  \label{Def-vacuum-annihilation}
  An operator $A \in \Afrak$ is said to have the vacuum annihilation
  property if, in the sense of operator-valued distributions, the
  mapping
  \begin{equation}
    \label{eq-vacuum-annihilation}
    \Rsone \ni x \mapsto \ax ( A ) \doteq U ( x ) \, A \, U ( x )^*
    \in \Afrak
  \end{equation}
  has a Fourier transform with compact support $\Gamma$ contained in
  the complement of the forward light cone $\fwcone$. The collection
  of all vacuum annihilation operators is a subspace of $\Afrak$
  denoted $\Aann$.
\end{Def}
\begin{Rem}
The support of the Fourier transform of \eqref{eq-vacuum-annihilation}
is precisely the energy-momen\-tum transfer of $A$, and the
energy-threshold for the annihilation of states depends on the
distance $d ( \Gamma , \fwcone )$ between $\Gamma$ and $\fwcone$. Let
$\Gamma_0$ be a closed subset of $\Rsone$ and let $\widetilde{\Afrak}
( \Gamma_0 )$ denote the set of all operators $A \in \Afrak$ having
energy-momentum transfer $\Gamma_A \subseteq \Gamma_0$. Then
$\widetilde{\Afrak} ( \Gamma_0 )$ is easily seen to be a uniformly
closed subspace of $\Afrak$, invariant under space-time translations.
\end{Rem}

The construction of a subalgebra $\Cfrak$ in $\Afrak$ containing
self-adjoint operators to be interpreted as representatives of
particle detectors is accomplished in three steps
(Definitions~\ref{Def-annihilator-space}--\ref{Def-counter-algebra}),
starting with a subspace $\Lfrak_0 \subseteq \Afrak$ consisting of
operators which, in addition to the properties mentioned above, are
infinitely often differentiable with respect to the automorphism
group $\bset{\aLax : ( \Lambda , x ) \in \Poin}$
(cf.~Definition~\ref{Def-partial-derivations} in
Appendix~\ref{chap-differentiability}).
\begin{Def}
  \label{Def-annihilator-space}
  The almost local vacuum annihilation operators $L_0 \in \Afrak$
  which are infinitely often differentiable with respect to the
  group $\bset{\aLax : ( \Lambda , x ) \in \Poin}$ constitute a
  subspace $\AS \cap \Aann \cap \Dscr^{( \infty )} ( \Afrak )$ of
  $\Afrak$. The intersection of this set with all the pre-images of
  $\AS$ under arbitrary products of partial derivations $\delta^{k_1}
  \dotsm \delta^{k_N}$ for any $N \in \Nbb$ and any $1 \leqslant k_i
  \leqslant d_\Psf$, $d_\Psf$ the dimension of $\Poin$, is again a
  linear space denoted $\Lfrak_0$. Explicitly, $\Lfrak_0$ consists of
  all almost local vacuum annihilation operators which are infinitely
  often differentiable, having \emph{almost local} partial derivatives
  of any order.
\end{Def}
\begin{Rem}
  \begin{remlist}
  \item The space $\Lfrak_0$ is stable under the action of the
    Poincar\'{e} group. This means that $\aLax ( \Lfrak_0 ) =
    \Lfrak_0$ for any $( \Lambda , x) \in \Poin$. Due to the
    properties of Fourier transformation, $\aLax ( L_0 )$ has
    energy-momentum transfer in $\Lambda \Gamma$ if $L_0 \in \Lfrak_0
    ( \Gamma ) \doteq \Lfrak_0 \cap \widetilde{\Afrak} ( \Gamma )$;
    the adjoint ${L_0}^*$ of this $L_0$ belongs to $\widetilde{\Afrak}
    ( -\Gamma )$.
  \item Furthermore $\Lfrak_0$ is invariant under differentiation: The
    partial derivatives are almost local and infinitely often
    differentiable  operators by definition, and, as uniform limits of
    vacuum annihilation operators, they inherit the energy-momentum
    transfer of these so that they belong to $\Aann$, too.
  \end{remlist}
\end{Rem}
A huge number of elements of $\Lfrak_0$ can be constructed by
regularizing almost local operators with respect to rapidly decreasing
functions on the Poincar\'e group. The semi-direct product Lie group
$\Poin = \Lor \ltimes \Rsone$ is unimodular by \cite[Proposition~II.29
and Corollary]{nachbin:1965} since $\Lor$ is a simple thus semisimple
Lie group \cite[Proposition~I.1.6]{helgason:1984}. So let $\mu$ be the
Haar measure on $\Poin$ and $A \in \AS$, then the operator
\begin{equation}
  \label{eq-regularized-annihilator}
  A ( F ) = \int d \mu ( \Lambda , x ) \; F ( \Lambda , x ) \, \aLax (
  A )
\end{equation}
belongs to $\Lfrak_0 ( \Gamma )$ if the infinitely differentiable
function $F$ is rapidly decreasing on the subgroup $\Rsone$ and
compactly supported on $\Lor$, i.\,e.~$F \in \Sscr_0 \bigl( \Poin
\bigr) = \Sscr_0 \bigl( \Lor \ltimes \Rsone \bigr)$ in the notation
introduced in \cite{baumgaertel/wollenberg:1992}, and has the
additional property that the Fourier transforms of the partial
functions $F_\Lambda (~.~) \doteq F ( \Lambda ,~.~)$ have common
support in the compact set $\Gamma \subseteq \complement \fwcone$ for
any $\Lambda \in \Lor$.

The following definition specifies a left ideal $\Lfrak$ of the
algebra $\Afrak$. 
\begin{Def}
  \label{Def-localizer-ideal}
  Let $\Lfrak$ denote the linear span of all operators $L \in \Afrak$
  of the form $L = A \, L_0$ where $A \in \Afrak$ and $L_0 \in
  \Lfrak_0$; i.\,e.
  \begin{equation*}
    \Lfrak \doteq \Afrak \, \Lfrak_0 = \linhull \bset{A \, L_0 : A \in
      \Afrak , L_0 \in \Lfrak_0} \text{.}
  \end{equation*}
  Then $\Lfrak$ is a left ideal of $\Afrak$, called the `left ideal of
  localizing operators.' 
\end{Def}
By their very construction, the elements of $\Lfrak$ annihilate the
vacuum and all states of bounded energy below a certain threshold that
depends on the minimum of $d ( \Gamma_i , \fwcone )$, $i = 1 \text{,}
\dots \text{,} N$, with respect to all representations $L =
\sum_{i = 1}^N A_i \, L_i \in \Lfrak$, where $\Gamma_i$ is the
energy-momentum transfer of $L_i$. The algebra of operators whose
self-adjoint elements are to be interpreted as representatives of
particle detectors is laid down in the next definition.
\begin{Def}
  \label{Def-counter-algebra}
  Let $\Cfrak$ denote the linear span of all operators $C \in \Afrak$
  which can be represented in the form $C = {L_1}^* \, L_2$ with $L_1
  \text{,} L_2 \in \Lfrak$; i.\,e.
  \begin{equation*}
    \Cfrak \doteq \Lfrak^* \, \Lfrak = \linhull \bset{{L_1}^* \, L_2 :
     L_1 , L_2 \in \Lfrak} \text{.}
  \end{equation*}
  Then $\Cfrak$ is a $^*$-subalgebra of $\Afrak$, called the `algebra
  of detectors.'
\end{Def}
\begin{Rem}
  The algebra $\Cfrak$ is smaller than that used by Araki and Haag in
  \cite{araki/haag:1967}. It is not closed in the uniform topology of
  $\Afrak$ and does not contain a unit. 
\end{Rem}

\section{Spectral Seminorms on the Algebra of Detectors}

The analysis of physical states is performed by use of the algebra of
detectors $\Cfrak$. In a state $\omega$ of bounded energy $E$ we
expect to encounter a finite number of localization centres, since the
triggering of a detector $C \in \Cfrak$ requires a minimal energy
$\epsilon$ to be deposited, the number $N$ of localization centres
being equal to or less than $E/\epsilon$. Now, according to this
heuristic picture, placing the counter $C$ for given time $t$ at every
point $\xib \in \Rs$ and adding up the corresponding expectation
values $\omega \bigl( \alpha_{( t , \xib )} ( C ) \bigr)$ should
result in the finite integral
\begin{equation}
  \label{eq-heuristic-integral}
  \int_{\Rs} d^s x \; \babs{\omega \bigl( \alpha_{( t , \xib )} ( C )
    \bigr)} < \infty \text{.} 
\end{equation}
As a matter of fact, the operators $C \in \Cfrak$ turn out to have the
property \eqref{eq-heuristic-integral} as was shown by Buchholz in
\cite{buchholz:1990}. For the sake of completeness and to demonstrate
how phase-space properties of the theory (localization in space
combined with energy-bounds) enter the present investigation, we give
an elaborate proof.
\begin{Pro}
  \label{Pro-harmonic-analysis}
  Let $E (~.~)$ be the spectral resolution of the space-time
  translations $U ( x )$, $x \in \Rsone$, and let $L_0 \in \Lfrak_0$
  have energy-momentum transfer $\Gamma$ in a convex subset of
  $\complement \fwcone$. Then for any bounded Borel set $\Delta
  \subseteq \Rsone$ the net of operator-valued Bochner integrals
  indexed by compact $\Kib \subseteq \Rs$,
  \begin{equation*}
    \begin{split}
      Q_{\Delta , \Kib}^{( {L_0}^* L_0 )} & \doteq \ED \int_\Kib d^s x
      \; \aibx ( {L_0}^* L_0 ) \; \ED \\
      & = \int_\Kib d^s x \; \ED \aibx ( {L_0}^* L_0 ) \ED \text{,}
    \end{split}
  \end{equation*}
  is $\sigma$-strongly convergent as $\Kib \nearrow \Rs$ and the limit
  $Q_\Delta^{( {L_0}^* L_0 )} \in \BH^+$ satisfies the estimate
  \begin{equation}
    \label{eq-commutator-integral-estimate}
    \Bnorm{Q_\Delta^{( {L_0}^* L_0 )}} \leqslant N ( \Delta , \Gamma )
    \int_{\Rs} d^s x \; \bnorm{\bcomm{\aibx ( L_0 )}{{L_0}^*}}
  \end{equation}
  for suitable $N ( \Delta , \Gamma ) \in \Nbb$, depending on $\Delta$
  and $\Gamma$. Moreover the mapping
  \begin{equation*}
    \xib \mapsto \ED \aibx ( {L_0}^* L_0 ) \ED
  \end{equation*}
  is integrable with respect to the $\sigma$-weak topology on $\BH$
  and its integral coincides with the operator $Q_\Delta^{( {L_0}^*
  L_0 )}$:
  \begin{equation*}
    Q_\Delta^{( {L_0}^* L_0 )} = \sw \int_{\Rs} d^s x \; \ED \aibx (
    {L_0}^* L_0 ) \ED \text{.}
  \end{equation*}
\end{Pro}
\begin{proof}
  $\Delta$ being a bounded Borel set, the same is true of its closure
  $\overline{\Delta}$, so that, due to compactness and convexity of
  $\Gamma$, there exists a number $n \in \Nbb$ for which the relation
  \label{page-proof-harmonic-analysis} $( \overline{\Delta} + \Gamma_n
  ) \cap \fwcone = \emptyset$ is satisfied, where $\Gamma_n$ denotes
  the sum $\Gamma_n \doteq \Gamma + \dots + \Gamma$ with $n$
  terms. The spectrum condition then entails:
  \begin{equation}
    \label{eq-spectrum-condition-implication}
    E ( \overline{\Delta} + \Gamma_n ) = 0 \text{.}
  \end{equation}
  Note, that in the derivation of this result compactness of $\Gamma$
  is needed to ensure that the distance between $\Gamma$ and $\fwcone$
  is positive; other shapes of $\Gamma$ are possible as long as
  convexity and the condition $d ( \Gamma , \fwcone ) > 0$ are
  preserved, e.\,g.~wedges in $\complement \fwcone$. For arbitrary
  $\xib_1 \text{,} \dots \text{,} \xib_n \in \Rs$ all the operators
  $\alpha_{\xib_i} ( L_0 )$, $i = 1 \text{,} \dots \text{,} n$, belong
  to $\widetilde{\Afrak} ( \Gamma )$ whilst their product
  $\prod_{i = 1}^n \alpha_{\xib_i} ( L_0 )$ is an element of
  $\widetilde{\Afrak} ( \Gamma_n )$, hence by
  \eqref{eq-spectrum-condition-implication}
  \begin{equation}
    \label{eq-leaving-of-lightcone}
    \prod_{i = 1}^n \alpha_{\xib_i} ( L_0 ) \: \ED = E (
    \overline{\Delta} + \Gamma_n ) \, \prod_{i = 1}^n \alpha_{\xib_i}
    ( L_0 ) \: \ED = 0 \text{.}
  \end{equation}
  Now, \cite[Lemma~2.2]{buchholz:1990} states that for any $B \in \BH$
  and any $k \in \Nbb$
  \begin{equation}
    \label{eq-harmonic-analysis-integral}
    \Bnorm{P_k \int_\Kib d^s x \; \aibx ( B^* B ) \; P_k} \leqslant (
    k - 1 ) \, \sup_\Psi \Bigl( \int_{\Kib - \Kib} d^s x \;
    \bnorm{\bcomm{\aibx ( B )}{B^*} \Psi} \Bigr) \text{,} 
  \end{equation}
  where $P_k$ is the orthogonal projection onto the intersection of
  the kernels of $k$-fold products $\prod_{i = 1}^k \alpha_{\yib_i} (
  B )$ for arbitrary $\yib_1 \text{,} \dots \text{,} \yib_k \in \Rs$,
  $\Kib \subseteq \Rs$ is compact and the supremum extends over all
  unit vectors $\Psi \in P_{k - 1} \Hscr$. According to
  \eqref{eq-leaving-of-lightcone} $\ED \leqslant P_n$ if we take $B
  \doteq L_0$, so  that the following  estimate, uniform in $\Kib$, is
  a consequence of \eqref{eq-harmonic-analysis-integral} combined with
  almost locality of $L_0$ (cf.~\eqref{eq-commutator-integral}):
  \begin{equation}
    \label{eq-harmonic-analysis-commutator-estimate}
    \Bnorm{Q_{\Delta , \Kib}^{( {L_0}^* L_0 )}} = \Bnorm{\ED \int_\Kib
      d^s x \; \aibx ( {L_0}^* L_0 ) \; \ED} \leqslant ( n - 1 ) \,
    \int_{\Rs} d^s x \; \bnorm{\bcomm{\aibx ( L_0 )}{{L_0}^*}}
    \text{.}
  \end{equation}
  The positive operators $\Bset{Q_{\Delta , \Kib}^{( {L_0}^* L_0 )} :
  \Kib \subseteq \Rs \; \text{compact}}$ thus constitute an increasing
  net which is bounded in $\BH^+$. According to
  \cite[Lemma~2.4.19]{bratteli/robinson:1987} this net has a least
  upper bound in $\BH^+$, which is its $\sigma$-strong limit
  $Q_\Delta^{( {L_0}^* L_0 )}$ and satisfies
  \begin{equation}
    \label{eq-commutator-integral-bound}
    \Bnorm{Q_\Delta^{( {L_0}^* L_0 )}} \leqslant ( n - 1 ) \,
    \int_{\Rs} d^s x \; \bnorm{\bcomm{\aibx ( L_0 )}{{L_0}^*}}
    \text{.}
  \end{equation}
  For $N ( \Delta , \Gamma ) \doteq n - 1$ this is the desired
  estimate \eqref{eq-commutator-integral-estimate}.

  The $\sigma$-weak topology of $\BH$ is induced by the positive
  normal functionals of the space $\BH^+_*$, so that integrability of
  $\xib \mapsto \ED \aibx ( {L_0}^* L_0 ) \ED$ in the $\sigma$-weak
  topology is implied by integrability of the functions
  \begin{equation*}
    \xib \mapsto \babs{\psi \bigl( \ED \aibx ( {L_0}^* L_0 ) \ED
    \bigr)} = \psi \bigl( \ED \aibx ( {L_0}^* L_0 ) \ED \bigr)
  \end{equation*}
  for any $\psi \in \BH^+_*$. Now, given any compact subset $\Kib$ of
  $\Rs$, there holds the estimate  
  \begin{multline*}
    \int_\Kib d^s x \; \babs{\psi \bigl( \ED \aibx ( {L_0}^* L_0 ) \ED
    \bigr)} = \int_\Kib d^s x \; \psi \bigl( \ED \aibx ( {L_0}^* L_0 )
    \ED \bigr) \\
    = \psi \Bigl( \int_\Kib d^s x \; \ED \aibx ( {L_0}^* L_0 ) \ED
    \Bigr) \leqslant \norm{\psi} \Bnorm{Q_{\Delta , \Kib}^{( {L_0}^* L_0
    )}} \leqslant \norm{\psi} \Bnorm{Q_\Delta^{( {L_0}^* L_0 )}}
    \text{,}
  \end{multline*}
  and, as a consequence of the Monotone Convergence
  Theorem~\cite[II.2.7]{fell/doran:1988a}, the functions $\xib \mapsto
  \babs{\psi \bigl( \ED \aibx ( {L_0}^* L_0 ) \ED \bigr)}$ indeed turn
  out to be integrable for any $\psi \in \BH^+_*$. Thus the integral
  of the mapping $\xib \mapsto \ED \aibx ( {L_0}^* L_0 ) \ED$ with
  respect to the $\sigma$-weak topology exists
  (cf.~\cite[II.6.2]{fell/doran:1988a}) and, through an application of
  Lebesgue's Dominated Convergence
  Theorem~\cite[II.5.6]{fell/doran:1988a}, is seen to be the
  $\sigma$-weak limit of the net of operators $Q_{\Delta , \Kib}^{(
  {L_0}^* L_0 )}$ which coincides with the $\sigma$-strong limit
  $Q_\Delta^{( {L_0}^* L_0 )}$ established above. Formally
  \begin{equation*}
    Q_\Delta^{( {L_0}^* L_0 )} = \sw \int_{\Rs} d^s x \; \ED \aibx (
    {L_0}^* L_0 ) \ED \text{,}
  \end{equation*}
  which is the last of the above assertions.
\end{proof}
\begin{Pro}
  \label{Pro-counter-integrals}
  Suppose that $\Delta \subseteq \Rsone$ is a bounded Borel set.
  \begin{proplist}
  \item Let $L \in \Lfrak$ be arbitrary, then the net of operators for
    compact $\Kib \subseteq \Rs$
    \begin{equation*}
      \begin{split}
        Q_{\Delta , \Kib}^{( L^* L )} & \doteq \ED \int_\Kib d^s x \;
        \aibx ( L^* L ) \; \ED \\
        & = \int_\Kib d^s x \; \ED \aibx ( L^* L ) \ED \text{,}
      \end{split}
    \end{equation*}
    converges $\sigma$-strongly to $Q_\Delta^{( L^* L )} \in \BH^+$ in
    the limit $\Kib \nearrow \Rs$. Moreover the mapping $\xib \mapsto
    \ED \aibx ( L^* L ) \ED$ is integrable with respect to the
    $\sigma$-weak topology on $\BH$ and satisfies 
    \begin{equation*}
      Q_\Delta^{( L^* L )} = \sw \int_{\Rs} d^s x \; \ED \aibx ( L^* L
      ) \ED \text{.}
    \end{equation*}
  \item Let $C \in \Cfrak$ be arbitrary, then the net of operators
    indexed by compact $\Kib \subseteq \Rs$
    \begin{equation*}
      \begin{split}
        Q_{\Delta , \Kib}^{( C )} & \doteq \ED \int_\Kib d^s x \;
        \aibx ( C ) \; \ED \\
        & = \int_\Kib d^s x \; \ED \aibx ( C ) \ED 
      \end{split}
    \end{equation*}
    is $\sigma$-strongly convergent to $Q_\Delta^{( C )} \in \BH$ for
    $\Kib \nearrow \Rs$. In addition to this the mapping $\xib \mapsto
    \ED \aibx ( C ) \ED$ is integrable with respect to the
    $\sigma$-weak topology on $\BH$ and the integral is given by
    \begin{equation*}
      Q_\Delta^{( C )} = \sw \int_{\Rs} d^s x \; \ED \aibx ( C ) \ED
      \text{.} 
    \end{equation*}
    Furthermore
    \begin{equation}
      \label{eq-supremum-seminorm}
      \sup \Bset{\int_{\Rs} d^s x \; \babs{\phi \bigl( \ED \aibx ( C )
      \ED \bigr)} : \phi \in \BH_{*,1}} < \infty \text{.}
    \end{equation}
  \end{proplist}
\end{Pro}
\begin{Rem}
  Note, that relation \eqref{eq-supremum-seminorm} is a sharpened
  version of \eqref{eq-heuristic-integral} which, based on heuristic
  considerations, was the starting point of the present
  investigation. 
\end{Rem}
\begin{proof}
  \begin{prooflist}
  \item By partition of unity (cf.~\cite[Satz~8.1]{jantscher:1971}),
    applied to elements of $\Lfrak_0$ which have arbitrary
    energy-momentum transfer in $\complement \fwcone$, any $L \in
    \Lfrak$ can be written as a finite sum $L = \sum_{j = 1}^m A_j
    L_j$ where the $A_j$ belong to $\Afrak$ and the operators $L_j \in
    \Lfrak_0$ have energy-momentum transfer in compact and convex
    subsets $\Gamma_j$ of $\complement \fwcone$. Since
    \begin{equation*}
      L^* L \leqslant 2^{m - 1} \Bigl( \sup_{1 \leqslant j \leqslant
      m} \norm{A_j}^2 \Bigr) \sum_{j = 1}^m {L_j}^* L_j \text{,} 
    \end{equation*}
    we infer
    \begin{equation*}
      Q_{\Delta , \Kib}^{( L^* L )} \leqslant 2^{m - 1} \Bigl( \sup_{1
      \leqslant j \leqslant m} \norm{A_j}^2 \Bigr) \sum_{j = 1}^m
      Q_{\Delta , \Kib}^{( {L_j}^* L_j )} \text{,} 
    \end{equation*}
    so that by \eqref{eq-harmonic-analysis-commutator-estimate}
    the increasing net $\Bset{Q_{\Delta , \Kib}^{( L^* L )} : \Kib
    \subset \Rs \mathrm{compact}}$ turns out to be bounded, having a
    least upper bound in $\BH^+$ that is its $\sigma$-strong limit
    $Q_\Delta^{( L^* L )}$. Making again use of the above order
    relation for $L^* L$ one arrives at
    \begin{equation*}
      \psi \bigl( \ED \aibx ( L^* L ) \ED \bigr) \leqslant 2^{m - 1}
      \Bigl( \sup_{1 \leqslant j \leqslant m} \norm{A_j}^2 \Bigr)
      \sum_{j = 1}^m \psi \bigl( \ED \aibx ( {L_j}^* L_j ) \ED \bigr) 
    \end{equation*}
    for any $\psi \in \BH^+_*$ and any $\xib \in \Rs$, where the
    right-hand side of this relation is integrable as was shown in the
    proof of Proposition~\ref{Pro-harmonic-analysis}. Then the
    reasoning applied there establishes the $\sigma$-weak
    integrability of $\xib \mapsto \ED \aibx ( L^* L ) \ED$ together
    with the relation 
    \begin{equation*}
      Q_\Delta^{( L^* L )} = \sw \int_{\Rs} d^s x \; \ED \aibx ( L^* L
      ) \ED \text{.}
    \end{equation*}
  \item Consider $C_0 = {L_1}^* L_2 \in \Cfrak$ with $L_1 \text{,} L_2
    \in \Lfrak$. By polarization 
    \begin{equation*}
      C_0 = \frac{1}{4} \sum_{k = 0}^3 i^{-k} ( L_1 + i^k L_2 )^* (
      L_1 + i^k L_2 ) = \frac{1}{4} \sum_{k = 0}^3 i^{-k} \, {L^{( k
      )}}^* L^{( k )} \text{,} 
    \end{equation*}
    where $L^{( k )} \doteq L_1 + i^k L_2 \in \Lfrak$ for $k = 0
    \text{,} \dots \text{,} 3$, and according to (i)
    \begin{multline*}
      Q_{\Delta , \Kib}^{( C_0 )} = \ED \int_\Kib d^s x \; \aibx ( C_0
      ) \; \ED \\
      = \frac{1}{4} \sum_{k = 0}^3 i^{-k} \Bigl( \ED \int_\Kib d^s x
      \; \aibx \bigl( {L^{( k )}}^* L^{( k )} \bigr) \; \ED \Bigr) =
      \frac{1}{4} \sum_{k = 0}^3 i^{-k} Q_{\Delta , \Kib}^{\left(
      {L^{( k )}}^* L^{( k )} \right)}
    \end{multline*}
    converges $\sigma$-strongly to
    \begin{equation}
      \label{eq-polarization-limit}
      Q_\Delta^{( C_0 )} \doteq \frac{1}{4} \sum_{k = 0}^3 i^{-k}
      Q_\Delta^{\left( {L^{( k )}}^* L^{( k )} \right)} \text{.}
    \end{equation}
    Now, let $\phi$ be a normal functional on $\BH$. By polar
    decomposition (cf.~\cite[Theorem~III.4.2(i),
    Proposition~III.4.6]{takesaki:1979}) there exist a partial
    isometry $V \in \BH$ and a positive normal functional $\abs{\phi}$
    subject to the relation $\norm{\abs{\phi}} = \norm{\phi}$, such
    that $\phi(~.~) = \abs{\phi} (~.~V )$, allowing for the following
    estimate ($\xib \in \Rs$ arbitrary):
    \begin{multline*}
      2 \, \babs{\phi \bigl( \ED \aibx ( C_0 ) \ED \bigr)} = 2 \,
      \babs{\abs{\phi} \bigl( \ED \aibx ( {L_1}^* L_2 ) \ED V \bigr)}
      \\
      \mspace{-80mu} \leqslant 2 \, \sqrt{\abs{\phi} \bigl( \ED \aibx
      ( {L_1}^* L_1 ) \ED \bigr)} \sqrt{\abs{\phi} \bigl( V^* \ED
      \aibx ( {L_2}^* L_2 ) \ED V \bigr)} \\ 
      = \inf_{\lambda > 0} \Bigl( \lambda^{-1} \abs{\phi} \bigl( \ED
      \aibx ( {L_1}^* L_1 ) \ED \bigr) + \lambda \, \abs{\phi} \bigl(
      V^* \ED \aibx ( {L_2}^* L_2 ) \ED V \bigr) \Bigr) \text{,} 
    \end{multline*}
    where we made use of the fact that $2 \sqrt{a b} = \inf_{\lambda >
    0} ( \lambda^{-1} a + \lambda b )$ for any $a \text{,} b \geqslant
    0$. Now, from the first part of this Proposition we infer that it
    is possible to integrate the above expression over all of $\Rs$
    to get for any $\lambda > 0$ the estimate
    \begin{equation*}
      2 \int_{\Rs} d^s x \; \babs{\phi \bigl(\ED \aibx ( C_0 ) \ED
      \bigr)} \leqslant \lambda^{-1} \norm{\phi}
      \Bnorm{Q_\Delta^{({L_1}^* L_1 )}} + \lambda \norm{\phi}
      \Bnorm{Q_\Delta^{( {L_2}^* L_2 )}} \text{.} 
    \end{equation*}
    Note, that the normal functionals $\phi$ and the $\sigma$-weak
    integrals commute due to \cite[Proposition~II.5.7 adapted to
    integrals in locally convex spaces]{fell/doran:1988a}. Taking the
    infimum with respect to $\lambda$ one finally arrives at 
    \begin{equation}
      \label{eq-C-integrability}
      \int_{\Rs} d^s x \; \babs{\phi \bigl( \ED \aibx ( C_0 ) \ED
      \bigr)} \leqslant \norm{\phi} \Bnorm{Q_\Delta^{( {L_1}^* L_1
        )}}^{1/2} \Bnorm{Q_\Delta^{( {L_2}^* L_2 )}}^{1/2} \text{.} 
    \end{equation}
    This relation is valid for any normal functional in $\BH_*$, so
    that the $\sigma$-weak integrability of $\xib \mapsto \ED \aibx (
    C_0 ) \ED$ is established, the relation
    \begin{equation}
      \label{eq-C-limit}
      Q_\Delta^{( C_0 )} = \sw \int_{\Rs} d^s x \; \ED \aibx ( C_0 )
      \ED 
    \end{equation}
    being an immediate consequence (cf.~the proof of
    Proposition~\ref{Pro-harmonic-analysis}). Another fact implied by
    inequality \eqref{eq-C-integrability} is the estimate
    \begin{equation}
      \label{eq-supremum-estimate-prepared}
      \sup \Bset{\int_{\Rs} d^s x \; \babs{\phi \bigl( \ED \aibx ( C_0
      ) \ED \bigr)} : \phi \in \BH_{*,1}} 
      \leqslant \Bnorm{Q_\Delta^{( {L_1}^* L_1 )}}^{1/2}
      \Bnorm{Q_\Delta^{( {L_2}^* L_2 )}}^{1/2} \text{.}
    \end{equation}
    Since any $C \in \Cfrak$ is a linear combination of operators of
    the form $C_0$, the above relations \eqref{eq-polarization-limit}
    through \eqref{eq-supremum-estimate-prepared} are easily
    generalized to establish the second part of the Proposition.
  \end{prooflist}
  \renewcommand{\qed}{}
\end{proof}

The preceding result suggests the introduction of topologies on the
left ideal $\Lfrak$ and on the $^*$-algebra $\Cfrak$, respectively,
using specific seminorms indexed by bounded Borel subsets $\Delta$ of
$\Rsone$.
\begin{subequations}
  \begin{Def}
    \label{Def-seminorms}
    \begin{deflist}
    \item The left ideal $\Lfrak$ is equipped with a family of
      seminorms $\qD$ via
      \begin{equation}
        \label{eq-q-seminorm}
        \qDx{L} \doteq \Bnorm{Q_\Delta^{( L^* L )}}^{1/2} \text{,}
        \quad L \in \Lfrak \text{.}
      \end{equation}
    \item The $^*$-algebra $\Cfrak$ is furnished with seminorms $\pD$ by
      assigning
      \begin{equation}
        \label{eq-p-seminorm}
        \pDx{C} \doteq \sup \Bset{\int_{\Rs} d^s x \; \babs{\phi
        \bigl( \ED \aibx( C ) \ED \bigr)} : \phi \in \BH_{*,1}}
        \text{,} \quad C \in \Cfrak \text{.} 
      \end{equation}
    \item The completions of the locally convex (Hausdorff) spaces $(
      \Lfrak , \Tfrak_q )$ and $( \Cfrak , \Tfrak_p )$ arising from
      topologization by these seminorms are denoted $( \Lfrakbar ,
      \Tfrakbar_q )$ and $( \Cfrakbar , \Tfrakbar_p )$,
      respectively. Accordingly, the complete locally convex subspace
      of $\Lfrakbar$ generated by $\Lfrak_0$ is designated as $(
      \Lfrakbar_0 , \Tfrakbar_q )$.
    \item The completions of the locally convex spaces $( \Lfrak ,
      \Tfrak^u_q )$ and $( \Cfrak , \Tfrak^u_p )$ arising from
      topologization by all the seminorms $\qD$ and $\pD$,
      respectively, together with the initial uniform (norm) topology
      inherited from the quasi-local algebra $\Afrak$ are denoted $(
      \AL , \Tfrakbar^u_q )$ and $( \AC , \Tfrakbar^u_p )$.
    \end{deflist}
  \end{Def}
  \begin{Rem}
    \begin{remlist}
    \item Let $\BH^+_*$ denote the positive cone in $\BH_*$, then for
      any $L \in \Lfrak$
      \begin{equation}
        \label{eq-alternative-q-seminorm}
        \qDx{L}^2 = \sup \Bset{\int_{\Rs} d^s x \; \omega \bigl( \ED
        \aibx ( L^* L ) \ED \bigr) : \omega \in \BH^+_{*,1}} \text{,}
      \end{equation}
      a formulation that will frequently be used.
    \item The seminorm properties of $\qD$ and $\pD$ are easily
      checked. To establish the subadditivity of $\qD$ one has to
      observe that
      \begin{multline*}
        \qDx{L_1 + L_2}^2 \leqslant \qDx{L_1}^2 + \inf_{\lambda > 0}
        \bigl[ \lambda^{-1} \qDx{L_1}^2 + \lambda \, \qDx{L_2}^2
        \bigr] + \qDx{L_2}^2 \\
        = \qDx{L_1}^2 + 2 \, \qDx{L_1} \, \qDx{L_2} + \qDx{L_2}^2 =
        \bigl( \qDx{L_1} + \qDx{L_2} \bigr)^2 \text{,}
      \end{multline*}
      where we made use of the fact that ${L_1}^* L_2 + {L_2}^* L_1
      \leqslant \lambda^{-1} {L_1}^* L_1 +\lambda \, {L_2}^* L_2$ for
      any $\lambda > 0$ and $L_1 \text{,} L_2 \in \Lfrak$.
    \item The Hausdorff property of the locally convex spaces $(
      \Lfrak , \Tfrak_q )$ and $( \Cfrak , \Tfrak_p )$ can be
      established using the fact that vectors corresponding to states
      of bounded energy constitute a dense subspace of $\Hscr$. From
      the very definition of the seminorms $\qD$ and $\pD$ we infer
      that the conditions $\qDx{L} = 0$ and $\pDx{C} = 0$, $L \in
      \Lfrak$, $C \in \Cfrak$, imply $L \ED = 0$ and $\ED C \ED = 0$
      for any bounded Borel set $\Delta$, since the integrands
      occurring in \eqref{eq-alternative-q-seminorm} and
      \eqref{eq-p-seminorm} vanish identically on $\Rs$, and
      $\BH^+_{*,1}$ as well as $\BH_{*,1}$ are separating sets of
      functionals for $\BH$. By the density property just mentioned,
      it then follows that $L = 0$ and $C = 0$, and the nets of
      seminorms turn out to separate the elements of the left ideal
      $\Lfrak$ and the $^*$-algebra $\Cfrak$, respectively.
    \item The completions $( \Lfrakbar , \Tfrakbar_q )$ and $(
      \Cfrakbar , \Tfrakbar_p )$ as well as $( \AL , \Tfrakbar^u_q )$
      and $( \AC , \Tfrakbar^u_p )$ are again locally convex
      spaces with topologies defined by the unique extensions of 
      the seminorms $\qD$ and $\pD$ and of the norm $\norm{~.~}$ to
      $\Lfrakbar$, $\AL$ and $\Cfrakbar$, $\AC$, respectively
      \cite[Chapter~Four, \S\,18,\,4.]{koethe:1983}. Therefore, in the
      sequel, we shall apply these seminorms to elements of the
      completions without special mention. Depending on the relation
      between the underlying uniform structures as being finer or
      coarser, we infer that $\AL \subseteq \Lfrakbar$ and $\AC
      \subseteq \Cfrakbar$. Furthermore $\AL$ and $\AC$ are uniformly
      closed subspaces of the quasi-local algebra $\Afrak$.
    \end{remlist}
  \end{Rem}
\end{subequations}

\section{Characteristics of the Spectral Seminorms}

The investigations of the subsequent chapters very much depend on
special properties of the seminorms defined above, so these are
collected in this section. Interesting in their own right as they may
be, we are, in the present context, not aiming at utmost generality of
statements, but have future applications in mind.

\subsection{Basic Properties}

\begin{Pro}
  \label{Pro-seminorm-nets}
  The families of seminorms $\qD$ and $\pD$ on $\Lfrakbar$ and
  $\Cfrakbar$, respectively, where the symbols $\Delta$ denote bounded
  Borel sets, constitute nets with respect to the inclusion
  relation. For any $\Delta$ and $\Delta'$ we have
  \begin{align*}
    \Delta \subseteq \Delta' \qquad & \Rightarrow \qquad \qDx{L}
    \leqslant \qDprimex{L} \text{,} \quad L \in \Lfrakbar \text{,} \\
    \Delta \subseteq \Delta' \qquad & \Rightarrow \qquad \pDx{C}
    \leqslant \pDprimex{C} \text{,} \quad C \in \Cfrakbar \text{.} 
  \end{align*}
\end{Pro}
\begin{proof}
  For the $\qD$-seminorms on $\Lfrakbar$ the assertion follows from
  the order relation for operators in $\BH^+$. Let $L$ belong to the
  left ideal $\Lfrak$, then
  \begin{equation*}
    Q_\Delta^{( L^* L )} \leqslant Q_{\Delta'}^{( L^* L )} \text{,}
  \end{equation*}
  which by Definition~\ref{Def-seminorms} has the consequence
  \begin{equation*}
    \qDx{L}^2 = \Bnorm{Q_\Delta^{( L^* L )}} \leqslant
    \Bnorm{Q_{\Delta'}^{( L^* L )}} = \qDprimex{L}^2 \text{.}
  \end{equation*}
  This relation extends by continuity of the seminorms to all of
  $\Lfrakbar$. 

  In case of the $\pD$-topologies, note that for any Borel set
  $\Delta$ the functional $\phi^{\ED}$, defined through $\phi^{\ED}
  (~.~) \doteq \phi (\ED~.~\ED)$, belongs to $\BH_{*,1}$ if $\phi$
  does. From this we infer, since moreover $\Delta \subseteq \Delta'$
  implies $\ED = \ED \EDprime = \EDprime \ED$, that
  \begin{multline*}
    \Bset{\int_{\Rs} d^s x \; \babs{\phi \bigl( \ED \aibx( C ) \ED
    \bigr)} : \phi \in \BH_{*,1}} \\ 
    \subseteq \Bset{\int_{\Rs} d^s x \; \babs{\phi \bigl( \EDprime
    \aibx( C ) \EDprime \bigr)} : \phi \in \BH_{*,1}}  
  \end{multline*}
  for any $C \in \Cfrak$ and thus, by \eqref{eq-p-seminorm}, that
  $\pDx{C} \leqslant \pDprimex{C}$, a relation which by continuity of
  the seminorms is likewise valid for any operator in the completion
  $\Cfrakbar$.
\end{proof}

The continuous extensions of the seminorms $\qD$ and $\pD$ to
$\Lfrakbar$ and $\Cfrakbar$, respectively, can be explicitly computed
on the subspaces $\AL$ and $\AC$ of $\Afrak$.
\begin{Lem}
  \label{Lem-explicit-pq-extension}
  Let $\Delta$ denote an arbitrary bounded Borel subset of $\Rsone$.
  \begin{subequations}
    \begin{proplist}
    \item For any $L \in \AL$ we have
      \begin{equation}
        \label{eq-explicit-q-extension}
        \qDx{L} = \sup \Bset{\int_{\Rs} d^s x \; \omega \bigl( \ED
        \aibx ( L^* L ) \ED \bigr) : \omega \in \BH^+_{*,1}}^{1/2}
        \text{.}
      \end{equation}
    \item For any $C \in \AC$ there holds the relation
      \begin{equation}
        \label{eq-explicit-p-extension}
        \pDx{C} = \sup \Bset{\int_{\Rs} d^s x \; \babs{\phi \bigl( \ED
        \aibx( C ) \ED \bigr)} : \phi \in \BH_{*,1}} \text{.} 
      \end{equation}
    \end{proplist}
  \end{subequations}
\end{Lem}
\begin{proof}
  \begin{prooflist}
  \item Note, that we can define a linear subspace $\Afrak_{q'}$ of
    $\Afrak$ consisting of all those operators $L'$ which fulfill
    \begin{equation*}
      \qprimeDx{L'}^2 \doteq \sup \Bset{\int_{\Rs} d^s x \; \omega
      \bigl( \ED \aibx ( {L'}^* L' ) \ED \bigr) : \omega \in
      \BH^+_{*,1}} < \infty
    \end{equation*}
    for any bounded Borel set $\Delta$. On this space the mappings
    $\qprimeD$ act as seminorms whose restrictions to $\Lfrak$
    coincide with $\qD$ (cf.~the Remark following
    Definition~\ref{Def-seminorms}). Now let $L \in \AL$ be
    arbitrary. Given a bounded Borel set $\Delta$ we can then find a
    \emph{sequence} $\bset{L_n}_{n \in \Nbb}$ in $\Lfrak$ satisfying
    \begin{equation*}
      \lim_{n \rightarrow \infty} \qDx{L - L_n} = 0 \qquad \text{and}
      \qquad \lim_{n \rightarrow \infty} \norm{L - L_n} = 0 \text{.}
    \end{equation*}
    The second equation implies
    \begin{equation*}
      \lim_{n \rightarrow \infty} \norm{L \ED - L_n \ED} = 0 \text{,}
    \end{equation*}
    so that Lebesgue's Dominated Convergence Theorem can be applied to
    get for any functional $\omega \in \BH^+_{*,1}$ and any compact
    $\Kib \subseteq \Rs$
    \begin{equation*}
      \int_\Kib d^s x \; \omega \bigl( \ED \aibx ( L^* L ) \ED \bigr)
      = \lim _{n \rightarrow \infty} \int_\Kib d^s x \; \omega \bigl(
      \ED \aibx ( {L_n}^* L_n ) \ED \bigr) \text{.}
    \end{equation*}
    According to \eqref{eq-alternative-q-seminorm} each term in the
    sequence on the right-hand side is majorized by the corresponding
    $\qDx{L_n}^2$ and this sequence in turn converges to $\qDx{L}^2$
    by assumption, so that in passing from $\Kib$ to $\Rs$ and to the
    supremum over all $\omega \in \BH^+_{*,1}$ we get
    \begin{equation*}
      \sup \Bset{\int_{\Rs} d^s x \; \omega \bigl( \ED \aibx ( L^* L )
      \ED \bigr) : \omega \in \BH^+_{*,1}} \leqslant \qDx{L}^2
      \text{.}
    \end{equation*}
    This final estimate shows, by arbitrariness of $L \in \AL$ and the
    selected $\Delta$, that $\AL$ is a subspace of $\Afrak_{q'}$ and,
    from $\qprimeD \restriction \Lfrak = \qD$, it eventually follows
    that for all these $L$ and $\Delta$
    \begin{equation*}
      \qDx{L} = \sup \Bset{\int_{\Rs} d^s x \; \omega \bigl( \ED \aibx
      ( L^* L ) \ED \bigr) : \omega \in \BH^+_{*,1}}^{1/2} \text{.}
    \end{equation*}
  \item The proof of the second part follows the same lines of
    thought. We introduce the subspace $\Afrak_{p'} \subseteq \Afrak$
    consisting of operators $C'$ satisfying
    \begin{equation*}
      \pprimeDx{C'} \doteq \sup \Bset{\int_{\Rs} d^s x \; \babs{\phi
      \bigl( \ED \aibx( C' ) \ED \bigr)} : \phi \in \BH_{*,1}} <
      \infty 
    \end{equation*}
    for any bounded Borel set $\Delta$ and furnish it with the locally
    convex topology defined by the seminorms $\pprimeD$. An arbitrary
    $C \in \AC$ is, for given $\Delta$, approximated by a
    \emph{sequence} $\bset{C_n}_{n \in \Nbb}$ with respect to the norm
    and the $\pD$-topology. As above one has
    \begin{equation*}
      \lim_{n \rightarrow \infty} \norm{\ED C \ED - \ED C_n \ED} = 0
    \end{equation*}
    and infers
    \begin{equation*}
      \sup \Bset{\int_{\Rs} d^s x \; \babs{\phi \bigl( \ED \aibx( C' )
      \ED \bigr)} : \phi \in \BH_{*,1}} \leqslant \pDx{C} \text{.} 
    \end{equation*}
    This establishes, by arbitraryness of $C \in \AC$ and $\Delta$,
    that $\AC \subseteq \Afrak_{p'}$, and the equation $\pprimeD
    \restriction \Cfrak = \pD$ implies that for these $C$ and $\Delta$
    \begin{equation*}
      \pDx{C} = \sup \Bset{\int_{\Rs} d^s x \; \babs{\phi \bigl( \ED
      \aibx( C ) \ED \bigr)} : \phi \in \BH_{*,1}} \text{.}
      \tag*{\qed}
    \end{equation*}
    \renewcommand{\qed}{}
  \end{prooflist}
  \renewcommand{\qed}{}
\end{proof}
An immediate consequence of this result is the subsequent lemma, which
in some way reverts the arguments given in the concluding remark of
the last section in order to establish the Hausdorff property for $(
\Lfrak , \Tfrak_q )$ and $( \Cfrak , \Tfrak_p )$.
\begin{Lem}
  \label{Lem-delta-annihilation}
  Let $\Delta$ be a bounded Borel set.
  \begin{proplist}
  \item For $L \in \AL$ with $L \ED = 0$ there holds $\qDx{L} = 0$.
  \item If $C \in \AC$ satisfies $\ED C \ED = 0$ one has $\pDx{C} =
    0$.
  \end{proplist}
\end{Lem}

Next we deal with an implication of the fact, that $\Lfrak$ is an
ideal of the $C^*$-algebra $\Afrak$, and clarify the relationship
between the seminorms $\qD$ and $\pD$.
\begin{Lem}
  \label{Lem-basic-estimate}
  Let $\Delta$ denote bounded Borel subsets of $\Rsone$.               
  \begin{proplist}
  \item $\AL$ is a left ideal of the quasi-local algebra $\Afrak$ and
    satisfies 
    \begin{equation}
      \label{eq-ideal-estimate}
      \qDx{A L} \leqslant \norm{A} \, \qDx{L}
    \end{equation}
    for any $L \in \AL$ and $A \in \Afrak$.
  \item Let $L_i$, $i = 1 \text{,} 2$, be operators in $\AL$ and $A
    \in \Afrak$, then ${L_1}^* A L_2$ belongs to $\AC$. If in addition
    the operators $L_i$ have energy-momentum transfer in $\Gamma_i
    \subseteq \Rsone$ and $\Delta_i$ are Borel subsets of $\Rsone$
    containing $\Delta + \Gamma_i$, respectively, then
    \begin{equation}
      \label{eq-counter-estimate}
      \pDx{{L_1}^* A L_2} \leqslant \norm{E ( \Delta_1 ) A E (
      \Delta_2 )} \, \qDx{L_1} \, \qDx{L_2} \text{.} 
    \end{equation}
  \end{proplist}
\end{Lem}
\begin{proof}
  \begin{prooflist}
  \item For any $L \in \AL \subseteq \Afrak$ and arbitrary $A \in
    \Afrak$ the relation $L^* A^* A L \leqslant \norm{A}^2 L^* L$
    leads to the estimate
    \begin{equation*}
      \int_{\Rs} d^s x \; \omega \bigl( \ED \aibx ( L^* A^* A L ) \ED
      \bigr) \leqslant \norm{A}^2 \int_{\Rs} d^s x \; \omega \bigl(
      \ED \aibx ( L^* L ) \ED \bigr)
    \end{equation*}
    for any $\omega \in \BH^+_{*,1}$ and thus, by
    \eqref{eq-explicit-q-extension} and the notation of the proof of
    Lemma~\ref{Lem-explicit-pq-extension}, to
    \begin{multline*}
      \qprimeDx{A L} = \sup \Bset{\int_{\Rs} d^s x \; \omega \bigl(
      \ED \aibx ( L^* A^* A L ) \ED \bigr) : \omega \in
      \BH^+_{*,1}}^{1/2} \\
      \leqslant \norm{A} \sup \Bset{\int_{\Rs} d^s x \; \omega \bigl(
      \ED \aibx ( L^* L ) \ED \bigr) : \omega \in \BH^+_{*,1}}^{1/2} =
      \norm{A} \, \qDx{L} \text{.}
    \end{multline*}
    This shows that $A L$ belongs to $\AL$ and at the same time that
    the seminorm $\qprimeD$ (on $\Afrak_{q'}$) can be replaced by
    $\qD$ to yield \eqref{eq-ideal-estimate}.
  \item Let $\phi$ be a normal functional on $\BH$ with $\norm{\phi}
    \leqslant 1$. By polar decomposition there exist a partial
    isometry $V$ and a positive normal functional $\abs{\phi}$ with
    $\norm{\abs{\phi}} \leqslant 1$ such that $\phi (~.~) = \abs{\phi}
    (~.~V )$. Then
    \begin{multline*}
      \babs{\phi \bigl( \ED \aibx ( {L_1}^* A L_2 ) \ED \bigr)} \\
      \shoveleft{= \abs{\phi} \bigl( \ED \aibx ( {L_1}^* ) E (
      \Delta_1 ) \aibx ( A ) E ( \Delta_2 ) \aibx ( L_2 ) \ED V
      \bigr)} \\
      \leqslant \norm{E ( \Delta_1 ) \aibx ( A ) E ( \Delta_2 )}
      \sqrt{\abs{\phi} \bigl( \ED \aibx ( {L_1}^* L_1 ) \ED \bigr)}
      \sqrt{\abs{\phi} \bigl( V^* \ED \aibx ( {L_2}^* L_2 ) \ED V
      \bigr)} 
    \end{multline*}
    for any $\xib \in \Rs$ and the method used in the proof of
    Proposition~\ref{Pro-counter-integrals} can be applied to get, in
    analogy to \eqref{eq-supremum-estimate-prepared},
    \begin{multline*}
      \sup \Bset{\int_{\Rs} d^s x \; \babs{\phi \bigl( \ED \aibx (
      {L_1}^* A L_2 ) \ED \bigr)} : \phi \in \BH_{*,1}} \\
      \leqslant \norm{E ( \Delta_1 ) A E ( \Delta_2 )} \, \qDx{L_1} \,
      \qDx{L_2} \text{,}
    \end{multline*}
    where we made use of \eqref{eq-explicit-q-extension}. According to
    the notation introduced in the proof of
    Lemma~\ref{Lem-explicit-pq-extension} this result expressed in
    terms of the seminorm $\pprimeD$ on $\Afrak_{p'}$ reads
    \begin{equation*}
      \pprimeDx{{L_1}^* A L_2} \leqslant \norm{E ( \Delta_1 ) A E (
      \Delta_2 )} \, \qDx{L_1} \, \qDx{L_2} \text{,}
    \end{equation*}
    from which we infer, as in the first part of the present proof,
    not only that ${L_1}^* A L_2$ is an element of $\AC$ but also that
    $\pprimeD$ can be substituted by $\pD$ to give
    \eqref{eq-counter-estimate}.
  \end{prooflist}
  \renewcommand{\qed}{}
\end{proof}
The second part of the above lemma means that the product ${L_1}^*
L_2$, defined by two operators $L_1 \text{,} L_2 \in \AL$, is
continuous with respect to the locally convex spaces
(cf.~\cite[Chapter~Four, \S\,18,\,3.(5)]{koethe:1983}) $( \AL ,
\Tfrakbar^u_q ) \times ( \AL , \Tfrakbar^u_q )$ and $( \AC ,
\Tfrakbar^u_p )$.
\begin{Cor}
  \label{Cor-sesquilinear-product}
  The sesquilinear mapping on the topological product of the locally
  convex space $( \AL , \Tfrakbar^u_q )$ with itself, defined by 
  \begin{equation*}
    \AL \times \AL \ni ( L_1 , L_2 ) \mapsto {L_1}^* L_2 \in \AC
    \text{,}
  \end{equation*}
  is continuous with respect to the respective locally convex
  topologies.
\end{Cor}
In the special case of coincidence of both operators ($L_1 = L_2 =
L$) it turns out that $\pDx{L^* L}$ equals the square of
$\qDx{L}$. Another result involving  the operation of adjunction is
the fact, that this mapping leaves the $\pD$-seminorms invariant.
\begin{Lem}
  \label{Lem-involution-invariance}
  Let $\Delta$ denote the bounded Borel sets in $\Rsone$.
  \begin{proplist}
  \item For any operator $L \in \AL$ there hold the relations
    \begin{equation*}
      \pDx{L^* L} = \qDx{L}^2 \text{.}
    \end{equation*}
  \item Let $C$ be an element of $\AC$, then $C^*$ lies in $\AC$, too,
    and satisfies
    \begin{equation*}
      \pDx{C^*} = \pDx{C} \text{.}
    \end{equation*}
  \end{proplist}
\end{Lem}
\begin{proof}
  \begin{prooflist}
  \item According to Lemma~\ref{Lem-explicit-pq-extension}, we have
    for any $L \in \AL$ 
    \begin{multline*}
      \qDx{L}^2 = \sup \Bset{\int_{\Rs} d^s x \; \omega \bigl( \ED
      \aibx ( L^* L ) \ED \bigr) : \omega \in \BH^+_{*,1}} \\
      \leqslant \sup \Bset{\int_{\Rs} d^s x \; \babs{\phi \bigl( \ED
      \aibx ( L^* L ) \ED \bigr)} : \phi \in \BH_{*,1}} = \pDx{L^* L}
      \text{,}
    \end{multline*}
    whereas the reverse inequality is a consequence of
    Lemma~\ref{Lem-basic-estimate}. This proves the assertion.
  \item Note, that $\BH_{*,1}$ is invariant under the operation of
    taking adjoints defined by $\psi \mapsto \psi^*$ with $\psi^* ( A
    ) \doteq \overline{\psi ( A^* )}$, $A \in \BH$, for any linear
    functional $\psi$ on $\BH$. Thus 
    \begin{multline*}
      \pprimeDx{C^*} = \sup \Bset{\int_{\Rs} d^s x \; \babs{\phi
      \bigl( \ED \aibx ( C^* ) \ED \bigr)} : \phi \in \BH_{*,1}} \\
      = \sup \Bset{\int_{\Rs} d^s x \; \babs{\phi^* \bigl( \ED \aibx (
      C ) \ED \bigr)} : \phi \in \BH_{*,1}} = \pprimeDx{C}  
    \end{multline*}
    for any $C \in \Afrak_{p'}$ (cf.~the proof of
    Lemma~\ref{Lem-explicit-pq-extension}), which is sufficient to
    establish both of the assertions.
  \end{prooflist}
  \renewcommand{\qed}{}
\end{proof}

The last statement of this subsection on basic properties of the
spectral seminorms establishes their invariance under translations
in the $s +1$-dimensional configuration space.
\begin{Lem}
  \label{Lem-translation-invariance}
  The subspaces $\AL$ and $\AC$ of the quasi-local algebra $\Afrak$
  are invariant under translations. In particular, let $\Delta$ be a
  bounded Borel set in $\Rsone$ and let $x \in \Rsone$ be arbitrary,
  then
  \begin{proplist}
  \item \hfill $\bqDx{\ax ( L )} = \qDx{L}$, \quad $L \in \AL$; \hfill
    \text{~}
  \item \hfill $\bpDx{\ax ( C )} = \pDx{C}$, \quad $C \in \AC$. \hfill
    \text{~}
  \end{proplist}
\end{Lem}
\begin{proof}
  $\BH_{*,1}$ as well as its intersection $\BH^+_{*,1}$ with the
  positive cone $\BH^+_*$ are invariant under the mapping $\psi
  \mapsto \psi^U$ defined by $\psi^U (~.~) \doteq \psi ( U~.~U^* )$
  for any unitary operator $U \in \BH$ and any linear functional
  $\psi$ on $\BH$.
  \begin{prooflist}
  \item Now, $\ax ( L^* L ) = U_t \aibx ( L^* L ) {U_t}^*$ for any $x
    = ( t , \xib ) \in \Rsone$. This implies
    \begin{equation*}
      \omega \bigl( \ED \alpha_\yib \bigl( \aibx ( L^* L ) \bigr) \ED
      \bigr) = \omega \bigl( U_t \ED \alpha_{\xib + \yib} ( L^* L )
      \ED {U_t}^* \bigr) 
    \end{equation*}
    for any $\yib \in \Rs$ and any $\omega \in \BH_*$, henceforth
    \begin{multline*}
      \int_{\Rs} d^s y \; \omega \bigl( \ED \alpha_\yib \bigl(
      \ax ( L^* L ) \bigr) \ED \bigr) \\
      = \int_{\Rs} d^s y \; \omega \bigl( U_t \ED \alpha_{\xib + \yib}
      ( L^* L ) \ED {U_t}^* \bigr) = \int_{\Rs} d^s y \; \omega \bigl(
      U_t \ED \alpha_\yib ( L^* L ) \ED {U_t}^* \bigr) \text{.}
    \end{multline*}
    Therefore the introductory remark in combination with
    \eqref{eq-explicit-q-extension} yields for any $L \in \AL$: 
    \begin{multline*}
      \bqprimeDx{\ax ( L )}^2 = \sup \Bset{\int_{\Rs} d^s y \; \omega
      \bigl( \ED \alpha_\yib \bigl( \ax ( L^* L ) \bigr) \ED \bigr) :
      \omega \in \BH^+_{*,1}} \\
      \mspace{60mu} = \sup \Bset{\int_{\Rs} d^s y \; \omega \bigl( U_t
      \ED \alpha_\yib ( L^* L ) \ED {U_t}^* \bigr) : \omega \in
      \BH^+_{*,1}} \\
      = \sup \Bset{\int_{\Rs} d^s y \; \omega \bigl( \ED \alpha_\yib (
      L^* L ) \ED \bigr) : \omega \in \BH^+_{*,1}} = \qprimeDx{L}^2
      \text{,}
    \end{multline*}
    which, as in the proof of Lemma~\ref{Lem-basic-estimate},
    establishes the assertions. 
  \item The same argument applies to the seminorm $\pprimeD$, so that
    for $C \in \AC$
    \begin{multline*}
      \bpDx{\ax ( C )} = \sup \Bset{\int_{\Rs} d^s y \; \babs{\phi
      \bigl( \ED \alpha_\yib \bigl( \ax ( C ) \bigr) \ED \bigr)} :
      \phi \in \BH_{*,1}} \\
      = \sup \Bset{\int_{\Rs} d^s y \; \babs{\phi \bigl( U_t \ED
      \alpha_\yib ( C ) \ED {U_t}^* \bigr)} : \phi \in \BH_{*,1}} \\
      = \sup \Bset{\int_{\Rs} d^s y \; \babs{\phi \bigl( \ED
      \alpha_\yib ( C ) \ED \bigr)} : \phi \in \BH_{*,1}} = \pDx{C}
      \text{.} \tag*{\qed}  
    \end{multline*}
    \renewcommand{\qed}{}
  \end{prooflist}
  \renewcommand{\qed}{}
\end{proof}

\subsection{Continuity and Differentiability}

The assumed strong continuity of the automorphism group $\bset{\aLax :
( \Lambda , x ) \in \Poin}$ acting on the $C^*$-algebra $\Afrak$
carries over to the locally convex spaces $( \Lfrak , \Tfrak_q )$ and
$( \Cfrak , \Tfrak_p )$; and even the infinite differentiability of $(
\Lambda , x ) \mapsto \aLax ( L_0 )$ for $L_0 \in \Lfrak_0$ is
conserved in passing from the uniform topology on $\Lfrak_0$ to that
induced by the seminorms $\qD$.
\begin{Pro}
  \label{Pro-lcs-continuity-differentiability}
  \begin{proplist}
  \item For fixed $L \in \Lfrak$ the mapping
    \begin{equation*}
      \Xi_L : \Poin \rightarrow \Lfrak \quad ( \Lambda , x ) \mapsto
      \Xi_L ( \Lambda , x ) \doteq \aLax ( L )
    \end{equation*}
    is continuous with respect to the locally convex space $( \Lfrak ,
    \Tfrak_q )$.
  \item For given $C \in \Cfrak$ the mapping
    \begin{equation*}
      \Xi_C : \Poin \rightarrow \Cfrak \quad ( \Lambda , x ) \mapsto
      \Xi_C ( \Lambda , x ) \doteq \aLax ( C )
    \end{equation*}
    is continuous with respect to the locally convex space $( \Cfrak ,
    \Tfrak_p )$.
  \item Let $\id_{\Lfrak_0}$ denote the identity mapping
    \begin{equation*}
      \id_{\Lfrak_0} : ( \Lfrak_0 , \norm{~.~} ) \rightarrow (
      \Lfrak_0 , \Tfrak_q ) \quad L_0 \mapsto \id_{\Lfrak_0} ( L_0 )
      \doteq L_0
    \end{equation*}
    on the space $\Lfrak_0$ , once endowed with the norm topology and
    once with $\Tfrak_q$. Consider furthermore the family
    $\Xscr_{\Lfrak_0}$ of infinitely often differentiable mappings:
    \begin{gather*}
      \Xscr_{\Lfrak_0} \doteq \bset{\Xi_{L_0} : L_0 \in \Lfrak_0}
      \text{,} \\
      \Xi_{L_0} : \Poin \rightarrow \Lfrak_0 \quad ( \Lambda , x )
      \mapsto \Xi_{L_0} ( \Lambda , x ) \doteq \aLax ( L_0 ) \text{.}
    \end{gather*}
    Then the linear operator $\id_{\Lfrak_0}$ is
    $\Xscr_{\Lfrak_0}$-differentiable in the sense of
    Definition~\ref{Def-generalized-diff-lin-mappings}.
  \end{proplist}
\end{Pro}
\begin{Rem}
  The last assertion means, due to the invariance of
  $\Xscr_{\Lfrak_0}$ under differentiation of arbitrary order, that
  all the mappings $\Xi_{L_0}$, $L_0 \in \Lfrak_0$, are infinitely
  often differentiable in the locally convex space $( \Lfrak_0 ,
  \Tfrak_q )$ and, as $\id_{\Lfrak_0}$ and the operator of
  differentiation $\Dfrak$ commute, inherit the derivatives from the
  presupposed differentiability of the mappings $\Xi_{L_0}$ with
  respect to the uniform topology.
\end{Rem}
\begin{proof}
  \begin{prooflist}
  \item Note, that continuity of the mapping $( \Lambda , x ) \mapsto
    \aLax ( L )$ with respect to the locally convex space $( \Lfrak ,
    \Tfrak_q )$ is equivalent to its continuity with respect to each
    of the topologizing seminorms $\qD$.  

    Let the Borel subset $\Delta$ of $\Rsone$ be arbitrary but
    fixed. We shall first consider the special point $( \unit , 0 ) 
    \in \Poin$ and restrict attention to an operator $L' \in \Lfrak_0$
    having energy-momentum transfer $\Gamma$ which, under
    transformations from a sufficiently small neighbourhood $\Nscr'$
    of the neutral element $( \unit , 0 )$, stays bounded in a compact
    and convex subset $\widehat{\Gamma}$ of $\complement
    \fwcone$. This means that all operators $\aLax ( L' ) \in
    \Lfrak_0$, $( \Lambda , x ) \in \Nscr'$, have energy-momentum
    transfer in the common set $\widehat{\Gamma}$ and relation
    \eqref{eq-commutator-integral-estimate} of 
    Proposition~\ref{Pro-harmonic-analysis} applies to the differences
    $\aLax ( L' ) - L'$ yielding
    \begin{multline}
      \label{eq-applied-harmonic-analysis}
      \bqDx{\aLax ( L' ) - L'}^2 \\
      = \Bnorm{\ED \int_{\Rs} d^s y \; \alpha_\yib \bigl( \bigl( \aLax
      ( L' ) - L' \bigr)^* \bigl( \aLax ( L' ) - L' \bigr) \bigr) \;
      \ED} \\ 
      \leqslant N ( \Delta , \widehat{\Gamma} ) \int_{\Rs} d^s y \;
      \bnorm{\bcomm{\alpha_\yib \bigl( \aLax ( L' ) - L'
      \bigr)}{\bigl( \aLax ( L' ) - L' \bigr)^*}} \text{.}
    \end{multline}
    An estimate for the integrand on the right-hand side can be based
    on relation \eqref{eq-special-commutator-norm-estimate}, requiring
    suitable approximating nets of local operators for $\aLax ( L' ) -
    L'$. Given $R_0 > 0$ there exists a neighbourhood $\Nscr''$ of $(
    \unit ,x )$ such that $\Nscr'' \Oscr_r \subseteq \Oscr_{2r}$ for
    $r > R_0$, and if $\bset{L'_r \in \AOr : r > 0}$ is an
    approximating net of local operators for $L'$, then $\aLax ( L'_r 
    ) \in \AOtwor$ for any $r > R_0$ and $( \Lambda , x ) \in
    \Nscr''$. Now 
    \begin{equation}
      \label{eq-relation-almost-locality}
      r^k \bnorm{\aLax ( L' ) - \aLax ( L'_r )} = r^k \norm{L' - L'_r}
      \xrightarrow[r \rightarrow \infty]{} 0
    \end{equation}
    holds for any $k \in \Nbb$, so that the operators $\aLax ( L'_r )
    - L'_r \in \AOtwor$, $r > R_0$, constitute the large radius part
    of approximating nets for each of $\aLax ( L' ) - L'$, $( \Lambda
    , x ) \in \Nscr''$, subject to the bound
    \begin{equation*}
      \bnorm{\bigl( \aLax ( L' ) - L' \bigr) - \bigl( \aLax ( L'_r ) -
      L'_r \bigr)} \leqslant \norm{L' - L'_r} \text{,}
    \end{equation*}
    which is independent of $( \Lambda ,x ) \in \Nscr''$. Then,
    according to the remark following
    Definition~\ref{Def-almost-locality}, there exist approximating
    nets $\bset{L'( \Lambda , x )_r \in \AOr : r > 0}$ for the almost
    local operators $\aLax ( L' ) - L'$ that fulfill the estimates
    $\norm{L'( \Lambda , x )_r} \leqslant \bnorm{\aLax ( L' ) - L'}$
    and, for $r > R_0$, $\bnorm{\bigl( \aLax ( L' ) - L' \bigr) - L'(
    \Lambda , x )_{2r}} \leqslant 2 \, \norm{L' - L'_r}$, where in
    view of \eqref{eq-relation-almost-locality} the second inequality
    amounts to $\bnorm{\bigl( \aLax ( L' ) - L' \bigr) - L'( \Lambda ,
    x )_{2r}} \leqslant 2 \, C_k \, r^{-k}$ for suitable $C_k >
    0$. Making use of relation
    \eqref{eq-special-commutator-norm-estimate} in the same remark we
    arrive at
    \begin{multline}
      \label{eq-majorizing-function}
      \bnorm{\bcomm{\alpha_\yib \bigl( \aLax ( L' ) - L'
      \bigr)}{\bigl( \aLax ( L' ) - L' \bigr)^*}} \\
      \leqslant 2 \, \bnorm{\aLax ( L' ) - L'}^2 \chi_< ( \yib ) + 8
      \, \bnorm{\aLax ( L' ) - L'} \bnorm{L' - L'_{4^{-1} \abs{\yib}}}
      \, \chi_> ( \yib ) \\
      \leqslant 8 \, \norm{L'}^2 \chi_< (\yib) + 16 \, \norm{L'} \,
      C_k \, 4^k \abs{\yib}^{-k} \chi_> (\yib) 
    \end{multline}
    for any $\yib \in \Rs$, where $\chi_<$ and $\chi_>$ denote the
    characteristic functions pertaining to the compact ball of radius
    $4 \, R_0$ in $\Rs$ and its complement, respectively. The above
    relation \eqref{eq-majorizing-function} holds for any $( \Lambda ,
    x ) \in \Nscr \doteq \Nscr' \cap \Nscr''$, and its right-hand side
    turns out to be an integrable majorizing function for the mapping
    \begin{equation}
      \label{eq-norm-commutator-integrand}
      \yib \mapsto \bnorm{\bcomm{\alpha_\yib \bigl( \aLax ( L' ) - L'
      \bigr)}{\bigl( \aLax ( L' ) - L' \bigr)^*}} \text{,}
    \end{equation}
    irrespective of $( \Lambda , x ) \in \Nscr$, if $k \geqslant s +
    2$. Another consequence of \eqref{eq-majorizing-function} is that
    the function \eqref{eq-norm-commutator-integrand} converges
    pointwise to 0 on $\Rs$ in the limit $( \Lambda , x ) \rightarrow
    ( \unit , 0 )$ due to strong continuity of the automorphism group
    $\bset{\aLax : ( \Lambda , x ) \in \Poin}$. Therefore we can apply
    Lebesgue's Dominated Convergence Theorem to the integral on the
    right-hand side of \eqref{eq-applied-harmonic-analysis}, evaluated
    for any sequence $\bset{( \Lambda_n , x_n )}_{n \in \Nbb}
    \subseteq \Nscr$ approaching $( \unit , 0 )$ and infer
    \begin{equation*}
      \lim_{n \rightarrow \infty} \bqDx{\alpha_{( \Lambda_n , x_n )} (
      L' ) - L'} = 0 \text{.}
    \end{equation*}
    Since $\Poin$ as a topological space satisfies the first axiom of
    countability, this suffices to establish continuity of the mapping
    $( \Lambda ,x ) \mapsto \aLax ( L' )$ in $( \unit , 0 )$ with
    respect to the $\qD$-topology.

    An arbitrary operator $L \in \Lfrak$ can be represented as $L =
    \sum_{i = 1}^N  A_i L'_i$ where $L'_i \in \Lfrak_0$ comply with
    the above assumptions on $L'$ and the operators $A_i$ belong to
    the quasi-local algebra $\Afrak$ for any $i=1 \text{,} \dots
    \text{,} N$. According to Lemma~\ref{Lem-basic-estimate} we have
    \begin{multline*}
      0 \leqslant \bqDx{\aLax ( L ) - L} \\
      \leqslant \sum_{i = 1}^N \Bigl( \bqDx{\aLax ( A_i ) \bigl( \aLax
      ( L'_i ) - L'_i \bigr)} + \bqDx{\bigl( \aLax ( A_i ) - A_i
      \bigr) L'_i} \Bigr) \\
      \leqslant \sum_{i = 1}^N \Bigl( \norm{A_i} \, \bqDx{\aLax ( L'_i
      ) - L'_i} + \bnorm{\aLax  ( A_i ) - A_i} \, \qDx{L'_i} \Bigr)
      \text{,}
    \end{multline*}
    where the right-hand side vanishes in the limit $( \Lambda , x )
    \rightarrow ( \unit , 0 )$ due to the preceding result and strong
    continuity of the group $\bset{\aLax : ( \Lambda , x ) \in
    \Poin}$. Thus the mapping $( \Lambda , x ) \mapsto \aLax ( L )$ 
    turns out to be continuous in $( \unit , 0 )$ with respect to
    $\qD$ for arbitrary $L \in \Lfrak$. The restriction to the
    specific point $( \unit , 0 ) \in \Poin$ is inessential in the
    last step since for arbitrary $( \Lambda' , x' ) \text{,} (
    \Lambda_0 , x_0 ) \in \Poin$ one has 
    \begin{equation*}
      \bqDx{\aLaxprime ( L ) - \alpha_{( \Lambda_0 , x_0 )} ( L )} =
      \BqDx{\alpha_{( \Lambda' , x' ) {( \Lambda_0 , x_0 )}^{-1}}
      \bigl( \alpha_{( \Lambda_0 , x_0 )} ( L ) \bigr) - \alpha_{(
      \Lambda_0 , x_0 )} ( L )} \text{,}
    \end{equation*}
    explicitly showing that continuity of $( \Lambda , x ) \mapsto
    \aLax ( L )$ in $( \Lambda_0 , x_0 )$ is equivalent to continuity
    of $( \Lambda , x ) \mapsto \aLax \bigl( \alpha_{( \Lambda_0 , x_0
    )} ( L ) \bigr)$ in $( \unit , 0 )$ with respect to any of the
    seminorms $\qD$, where $\alpha_{( \Lambda_0 , x_0 )} ( L )$
    belongs to $\Lfrak$.
  \item Continuity of a mapping with values in the locally convex
    space $( \Cfrak , \Tfrak_p )$ is equivalent to its continuity with
    respect to all seminorms $\pD$. The problem at hand thus reduces
    to the one already solved in the first part, if one takes into
    account the shape of general elements of $\Cfrak$ according to
    Definition~\ref{Def-counter-algebra} and
    Corollary~\ref{Cor-sesquilinear-product}. 
  \item According to
    Definition~\ref{Def-generalized-diff-lin-mappings} we have to show
    that for any vacuum annihilation operator $L_0 \in \Lfrak_0$ the
    mapping $( \Lambda , x ) \mapsto \Xi_{L_0} ( \Lambda , x ) = \aLax
    ( L_0 )$ is differentiable in the locally convex space $( \Lfrak_0
    , \Tfrak_q )$ and has derivatives coinciding with those existing
    in the uniform topology by assumption.

    Let $L_0 \in \Lfrak_0$ be given and consider the local chart $(
    \Usf , \phi )$ around $( \Lambda_0 , x_0 ) = \phi^{-1} ( \tib_0
    )$. Due to the presupposed differentiability of the mapping
    $\Xi_{L_0}$ with respect to the uniform topology the corresponding
    residual term at $( \Lambda_0 , x_0 )$ with respect to $( \Usf ,
    \phi )$ is given by
    \begin{subequations}
      \begin{equation}
        \label{eq-Ldiff-residual}
        R \bigl[ \Xi_{L_0} \circ \phi^{-1} , \tib_0 \bigr] ( \hib ) =
        \alpha_{( \Lambda_\hib , x_\hib )} ( L_0 ) - \alpha_{(
        \Lambda_0 , x_0 )} ( L_0 ) - \Dfrak_\phi \Xi_{L_0} (
        \Lambda_0 , x_0 ) \hib \text{,}
      \end{equation}
      using the notation $( \Lambda_{\hib'} , x_{\hib'} ) \doteq
      \phi^{-1} ( \tib_0 + \hib' )$ for elements of $\Usf$, and
      satisfies the limit condition
      \begin{equation}
        \label{eq-Ldiff-residual-limit}
        \lim_{\hib \rightarrow \zeroib} \abs{\hib}^{-1} \bnorm{R
        \bigl[ \Xi_{L_0} \circ \phi^{-1} , \tib_0 \bigr] ( \hib )} = 0
        \text{,}
      \end{equation}
    \end{subequations}
    To prove the assertion it has to be shown that
    \eqref{eq-Ldiff-residual-limit} stays true when the norm is
    replaced by any of the seminorms $\qD$. Now, according to the Mean
    Value Theorem~\ref{The-mean-value}, we have for small $\hib$
    \begin{equation*}
      \alpha_{( \Lambda_\hib , x_\hib )} ( L_0 ) - \alpha_{( \Lambda_0
      , x_0 )} ( L_0 ) = \int_0^1 d \vartheta \; \Dfrak_\phi \Xi_{L_0}
      ( \Lambda_{\vartheta \hib} , x_{\vartheta \hib} ) \hib \text{,}
    \end{equation*}
    where the integral is to be understood with respect to the norm
    topology of $\Afrak$. Thus the residual term
    \eqref{eq-Ldiff-residual} can be re-written as
    \begin{multline*}
      R \bigl[ \Xi_{L_0} \circ \phi^{-1} , \tib_0 \bigr] ( \hib ) =
      \int_0^1 d \vartheta \; \Bigl( \Dfrak_\phi \Xi_{L_0} (
      \Lambda_{\vartheta \hib} , x_{\vartheta \hib} ) - \Dfrak_\phi
      \Xi_{L_0} ( \Lambda_0 , x_0 ) \Bigr) \hib \\
      = \sum_{i , j = 1}^{d_\Psf} \int_0^1 d \vartheta \; h_i \Bigl(
      C_{i j} ( \Lambda_{\vartheta \hib} , x_{\vartheta \hib} ) \,
      \alpha_{( \Lambda_{\vartheta \hib} , x_{\vartheta \hib} )}
      \bigl( \delta^j ( L_0 ) \bigr) - C_{i j} ( \Lambda_0 , x_0 ) \,
      \alpha_{( \Lambda_0 , x_0 )} \bigl( \delta^j ( L_0 ) \bigr)
      \Bigr) \text{,}
    \end{multline*}
    where in the last equation \eqref{eq-xi-delta} is used to
    represent the linear operator $\Dfrak_\phi \Xi_{L_0}$ in terms of
    partial derivatives of $\Xi_{L_0}$ which can be expressed by means
    of analytic functions $C_{i j}$ on $\Usf$ and Poincar\'{e}
    transformed derivatives $\delta^j ( L_0 )$ of $L_0$ ($d_\Psf$ is
    the dimension of the Poincar\'{e} group). As a consequence of the
    first statement of this proposition, the integrand on the
    right-hand side is continuous with respect to all seminorms $\qD$,
    so that the integral exists in the complete locally convex space
    $( \Lfrakbar , \Tfrakbar_q )$. By \cite[II.6.2 and
    5.4]{fell/doran:1988a} this leads to the following estimate for
    the residual term
    \begin{multline*}
      \abs{\hib}^{-1} \bqDx{R \bigl[ \Xi_{L_0} \circ \phi^{-1} ,
      \tib_0 \bigr] ( \hib )} \\
      \leqslant \sum_{i , j = 1}^{d_\Psf} \int_0^1 d \vartheta \;
      \frac{\abs{h_i}}{\abs{\hib}} \, \BqDx{C_{i j} (
      \Lambda_{\vartheta \hib} , x_{\vartheta \hib} ) \, \alpha_{(
      \Lambda_{\vartheta \hib} , x_{\vartheta \hib} )} \bigl( \delta^j
      ( L_0 ) \bigr) - C_{i j} ( \Lambda_0 , x_0 ) \, \alpha_{(
      \Lambda_0 , x_0 )} \bigl( \delta^j ( L_0 ) \bigr)} \\
      \leqslant \sum_{i , j = 1}^{d_\Psf} \max_{0 \leqslant \vartheta
      \leqslant 1} \BqDx{C_{i j} ( \Lambda_{\vartheta \hib} ,
      x_{\vartheta \hib} ) \, \alpha_{( \Lambda_{\vartheta \hib} ,
      x_{\vartheta \hib} )} \bigl( \delta^j ( L_0 ) \bigr) - C_{i j} (
      \Lambda_0 , x_0 ) \, \alpha_{( \Lambda_0 , x_0 )} \bigl(
      \delta^j ( L_0 ) \bigr)} \text{,}
    \end{multline*}
    where evidently the right-hand side vanishes in the limit $\hib
    \rightarrow \zeroib$. Thus condition \eqref{eq-diff-lqs-residual}
    for differentiability of mappings with values in a locally convex
    space is fulfilled, and according to the counterpart
    \eqref{eq-Ldiff-residual} of \eqref{eq-diff-lqs} the derivatives
    of $\Xi_{L_0}$ with respect to both the uniform and locally convex
    topologies on $\Lfrak_0$ coincide.
  \end{prooflist}
  \renewcommand{\qed}{}
\end{proof}

\subsection{Integrability}
\label{subsec-integrability}

Having established
Proposition~\ref{Pro-lcs-continuity-differentiability} on continuity
of the mappings $( \Lambda , x ) \mapsto \aLax ( L )$ and $( \Lambda ,
x ) \mapsto \aLax ( C )$ for given $L \in \Lfrak$ and $C \in \Cfrak$,
it turns out to be possible to construct new elements of $( \Lfrak_0 ,
\Tfrak_q )$, $( \AL , \Tfrakbar^u_q )$ and $( \AC , \Tfrakbar^u_p )$
through integration with respect to the Haar measure on $\Poin$.
\begin{Lem}
  \label{Lem-Poin-Bochner-integrals}
  Let the function $F \in L^1 \bigl( \Poin , d \mu ( \Lambda , x )
  \bigr)$ have compact support $\Ssf$.
  \begin{proplist}
  \item For any $L_0 \in \Lfrak_0$ the operator
    \begin{equation}
      \label{eq-L0-integral}
      \alpha_F ( L_0 ) \doteq \int d \mu ( \Lambda , x ) \; F (
      \Lambda , x ) \, \aLax ( L_0 )
    \end{equation}
    belongs to $\Lfrak_0$, too.
  \item If $L \in \Lfrak$ and $C \in \Cfrak$, then
    \begin{subequations}
      \label{eq-LC-integral}
      \begin{align}
        \label{eq-L-integral}
        \alpha_F ( L ) & \doteq \int d \mu ( \Lambda , x ) \; F (
        \Lambda , x ) \, \aLax ( L ) \text{,} \\
        \label{eq-C-integral}
        \alpha_F ( C ) & \doteq \int d \mu ( \Lambda , x ) \; F (
        \Lambda , x ) \, \aLax ( C )
      \end{align}
    \end{subequations}
    exist as integrals in the complete locally convex spaces $( \AL ,
    \Tfrakbar^u_q )$ and $( \AC , \Tfrakbar^u_p )$, respectively, and
    for any bounded Borel set $\Delta$ there hold the estimates
    \begin{subequations}
      \label{eq-LC-integral-estimate}
      \begin{align}
        \label{eq-L-integral-estimate}
        \bqDx{\alpha_F ( L )} & \leqslant \norm{F}_1 \sup_{( \Lambda ,
        x ) \in \Ssf} \bqDx{\aLax ( L )} \text{,} \\
        \label{eq-C-integral-estimate}
        \bpDx{\alpha_F ( C )} & \leqslant \norm{F}_1 \sup_{( \Lambda ,
        x ) \in \Ssf} \bpDx{\aLax ( C )} \text{.}
      \end{align}
    \end{subequations}
  \end{proplist}
\end{Lem}
\begin{proof}
  \begin{prooflist}
  \item By assumption $( \Lambda , x ) \mapsto \abs{F ( \Lambda , x
    )} \, \norm{\aLax ( L_0 )} = \abs{F ( \Lambda , x )} \,
    \norm{L_0}$ \cite[Corollary~I.5.4]{takesaki:1979} is an integrable
    majorizing function for the integrand of \eqref{eq-L0-integral},
    so $\alpha_F ( L_0 )$ exists as a Bochner integral in $\Afrak$. 
    The same holds true for the integrals constructed by use of an
    approximating net $\bset{L_{0 , r} \in \AOr : r > 0}$ for the
    almost local operator $L_0 \in \Lfrak_0$:
    \begin{equation*}
      \alpha_F ( L_{0,r} ) \doteq \int_\Ssf d \mu ( \Lambda , x ) \; F
      ( \Lambda , x ) \, \aLax ( L_{0,r} ) \text{.}
    \end{equation*}
    Due to compactness of $\Ssf$, these operators belong to the local
    algebras $\Afrak \bigl( \Oscr_{r ( \Ssf )} \bigr)$ pertaining to
    standard diamonds in $\Rsone$ which have each an $s$-dimensional
    basis of radius $r ( \Ssf ) \doteq a ( \Ssf ) r + b ( \Ssf )$
    where $a ( \Ssf )$ and $b ( \Ssf )$ are suitable positive
    constants. Now,
    \begin{equation*}
      \alpha_F ( L_0 ) - \alpha_F ( L_{0,r} ) = \int_\Ssf d \mu (
      \Lambda , x ) \; F ( \Lambda , x ) \bigl( \aLax ( L_0 ) - \aLax
      ( L_{0,r} ) \bigr) \text{,}
    \end{equation*}
    so that we arrive at the estimate ($\mu ( \Ssf )$ is the measure
    of the compact set $\Ssf$)
    \begin{equation*}
      r ( \Ssf )^k \bnorm{\alpha_F ( L_0 ) - \alpha_F ( L_{0,r} )}
      \leqslant \mu ( \Ssf ) \norm{F}_1 \bigl( a ( \Ssf ) r + b ( \Ssf
      ) \bigr)^k \norm{L_0 - L_{0,r}}
    \end{equation*}
    which holds for any $k \in \Nbb$. Due to almost locality of $L_0$,
    the right-hand side vanishes in the limit of large $r$, so that
    the operator $\alpha_F ( L_0 )$ itself turns out to be almost
    local: $\alpha_F ( L_0 ) \in \AS$ with approximating net
    $\Bset{\alpha_F ( L_{0,r} ) \in \Afrak \bigl( \Oscr_{r ( \Ssf )}
    \bigr) : r > 0}$.

    Let $\Gamma \subseteq \complement \fwcone$ denote the
    energy-momentum transfer of the vacuum annihilation operator
    $L_0$, then, by the Fubini Theorem
    \cite[II.16.3]{fell/doran:1988a}, the following equation is valid
    for any $g \in L^1 \bigl( \Rsone , d^{s + 1} y \bigr)$
    \begin{equation*}
      \int_{\Rsone} d^{s + 1} y \; g(y) \, \alpha_y \bigl( \alpha_F (
      L_0 ) \bigr) = \int_\Ssf d \mu ( \Lambda , x ) \; F ( \Lambda ,
      x ) \int_{\Rsone} d^{s + 1} y \; g(y) \, \alpha_y \bigl( \aLax (
      L_0 ) \bigr) \text{.}
    \end{equation*}
    In the special case $\supp \tilde{g} \subseteq \bigcap_{( \Lambda
    , x ) \in \Ssf} \complement( \Lambda \Gamma )$, $\tilde{g}$ the
    Fourier transform of $g$, the inner integrals on the right-hand
    side vanish for any $( \Lambda , x ) \in \Ssf$ so that we infer
    \begin{equation*}
      \int_{\Rsone} d^{s + 1} y \; g(y) \, \alpha_y \bigl( \alpha_F (
      L_0 ) \bigr) = 0 \text{,}
    \end{equation*}
    which shows that the energy-momentum transfer of $\alpha_F ( L_0
    )$ is contained in the compact subset $\bigcup_{( \Lambda , x )
    \in \Ssf} \Lambda \Gamma$ of $\complement \fwcone$. Therefore
    $\alpha_F ( L_0 )$ is indeed a vacuum annihilation operator from
    $\Aann$. 

    Finally, infinite differentiability with respect to the uniform
    topology of the mapping
    \begin{equation*}
      ( \Lambda , x ) \mapsto \Xi_{\alpha_F ( L_0 )} ( \Lambda , x ) =
      \aLax \bigl( \alpha_F ( L_0 ) \bigr)
    \end{equation*}
    has to be established. By assumption $L_0$ is infinitely often
    differentiable with respect to the Poincar\'{e} group, which
    implies that likewise all the operators $\aLax ( L_0 )$ belong to
    $\Dscr^{( \infty )} ( \Afrak )$ for any $( \Lambda , x ) \in
    \Poin$. Their residual terms at $( \unit ,  0 ) = \phi_0^{-1} (
    \zeroib )$ with respect to the canonical coordinates $( \Usf_0 ,
    \phi_0 )$ of the first kind, as introduced in
    \cite[Section~2.10]{varadarajan:1984}, can, using the notation $(
    \Lambda_{\hib'} , x_{\hib'} ) \doteq \phi_0^{-1} ( \hib' )$ for
    the transformations in $\Usf_0$, be expressed by
    \begin{multline}
      \label{eq-residual-integrand}
      R \bigl[ \Xi_{\aLax ( L_0 )} \circ \phi_0^{-1} , \zeroib \bigr]
      ( \hib ) \\
      = \alpha_{( \Lambda_\hib , x_\hib )} \bigl( \aLax ( L_0 ) \bigr)
      - \aLax ( L_0 ) - \Dfrak_{\phi_0} \Xi_{\aLax ( L_0 )} ( \unit ,
      0 ) \hib \\ 
      = \int_0^1 d \vartheta \; \Bigl( \Dfrak_{\phi_0} \Xi_{\aLax (
      L_0 )} ( \Lambda_{\vartheta \hib} , x_{\vartheta \hib} ) -
      \Dfrak_{\phi_0} \Xi_{\aLax ( L_0 )} ( \unit , 0 ) \Bigr) \hib
      \text{,}
    \end{multline}
    where the last equation stems from an application of the Mean
    Value Theorem~\ref{The-mean-value}, which holds true for small
    $\hib$. By Proposition~\ref{Pro-G-differentiability} the term
    $\Dfrak_{\phi_0} \Xi_{\aLax ( L_0 )} ( \unit , 0 ) \hib$ on the
    second line is continuous in $( \Lambda , x )$, so that it is
    possible to multiply \eqref{eq-residual-integrand} with the
    function $F ( \Lambda , x )$ and subsequently integrate over its
    compact support $\Ssf$.  Taking into account that each of the
    automorphisms $\alpha_{( \Lambda_\hib , x_\hib )}$ is uniformly
    continuous, thus commuting with Bochner integrals, this yields
    \begin{multline}
      \label{eq-residual-integral}
      \int_\Ssf d \mu ( \Lambda , x ) \; F ( \Lambda , x ) \, R \bigl[
      \Xi_{\aLax ( L_0 )} \circ \phi_0^{-1} , \zeroib \bigr] ( \hib )
      \\
      = \alpha_{( \Lambda_\hib , x_\hib )} \bigl( \alpha_F ( L_0 )
      \bigr) - \alpha_F ( L_0 ) - \int_\Ssf d \mu ( \Lambda , x ) \; F
      ( \Lambda , x ) \, \Dfrak_{\phi_0} \Xi_{\aLax ( L_0 )} ( \unit ,
      0 ) \hib \text{,}
    \end{multline}
    which has the shape of a residual term for $\Xi_{\alpha_F ( L_0
    )}$ at $( \unit , 0 )$. Now, the operator-norm of $\Dfrak_{\phi_0}
    \Xi_{\aLax ( L_0 )} ( \Lambda_{\vartheta \hib} , x_{\vartheta
    \hib} )$ can be estimated according to 
    \eqref{eq-generalized-transformed-G-derivatives} by 
    \begin{equation*}
      \bnorm{\Dfrak_{\phi_0} \Xi_{\aLax ( L_0 )} ( \Lambda_{\vartheta
      \hib} , x_{\vartheta \hib} )} \leqslant \bnorm{\Dfrak_{\phi_0} 
      \Xi_{L_0} ( \neutral )} \bnorm{\Nbf ( \Lambda , x )}
      \bnorm{\Mbf^{\phi_0} ( \Lambda_{\vartheta \hib} , x_{\vartheta
      \hib} )} \text{,}
    \end{equation*}
    which, due to continuity of $( \Lambda , x ) \mapsto \Nbf (
    \Lambda , x )$ and $\vartheta \mapsto \Mbf^{\phi_0} (
    \Lambda_{\vartheta \hib} , x_{\vartheta \hib} )$ with respect to 
    the operator-norm topology, is majorized on the compact set $\Ssf
    \times [ 0 , 1 ]$ by a constant $K ( \Ssf )$. As a consequence of
    the last equation in \eqref{eq-residual-integrand} we then get for
    any $( \Lambda , x ) \in \Ssf$ and small $\hib$ the bound
    \begin{equation}
      \label{eq-residual-integrand-estimate}
      \abs{\hib}^{-1} \bnorm{F ( \Lambda , x ) \, R \bigl[ \Xi_{\aLax
      ( L_0 )} \circ \phi_0^{-1} , \zeroib \bigr] ( \hib )} \leqslant
      2 \, K ( \Ssf ) \, \abs{F ( \Lambda , x )} \text{,}
    \end{equation}
    which is integrable over $\Ssf$ by assumption; restricting
    furthermore attention to sequences $\bset{\hib_n}_{n \in \Nbb}$
    converging to $\zeroib$, we see that the left-hand side of
    \eqref{eq-residual-integrand-estimate} converges pointwise to $0$.
    With this information at hand it is possible to apply Lebesgue's
    Dominated Convergence Theorem \cite[II.5.6]{fell/doran:1988a} to
    the left-hand side of \eqref{eq-residual-integral} to get
    \begin{equation}
      \label{eq-residual-integral-condition}
      \lim_{n \rightarrow \infty} \abs{\hib_n}^{-1} \int_\Ssf d \mu (
      \Lambda , x ) \; F ( \Lambda , x ) \, R \bigl[ \Xi_{\aLax ( L_0
      )} \circ \phi_0^{-1} , \zeroib \bigr] ( \hib_n ) = 0 \text{,}
    \end{equation}
    which is sufficient to establish condition
    \eqref{eq-diff-lqs-residual} for differentiability of the mapping
    $\Xi_{\alpha_F ( L_0 )}$ at $( \unit , 0 )$. The linear operator
    defining the corresponding derivative is according to the
    right-hand side of \eqref{eq-residual-integral} in connection with
    \eqref{eq-generalized-transformed-G-derivatives} given by 
    \begin{multline*}
      \Dfrak_{\phi_0} \Xi_{\alpha_F ( L_0 )} ( \unit , 0 ) \hib =
      \int_\Ssf d \mu ( \Lambda , x ) \; F ( \Lambda , x ) \,
      \Dfrak_{\phi_0} \Xi_{\aLax ( L_0 )} ( \unit , 0 ) \hib \\ 
      = \int_\Ssf d \mu ( \Lambda , x ) \; F ( \Lambda , x ) \, \aLax
      \circ \Dfrak_{\phi_0} \Xi_{L_0} ( \unit , 0 ) \circ \Nbf (
      \Lambda , x ) \hib \\
      = \sum_{i , j = 1}^{d_\Psf} h_i \int_\Ssf d \mu ( \Lambda , x )
      \; F_{j i} ( \Lambda , x ) \, \aLax \bigl( \delta^j ( L_0 )
      \bigr) = \sum_{i , j = 1}^{d_\Psf} h_i \, \alpha_{F_{j i}}
      \bigl( \delta^j ( L_0 ) \bigr) \text{,}
    \end{multline*}
    where $F_{j i} ( \Lambda , x ) \doteq F ( \Lambda , x ) \Nbf_{j i}
    ( \Lambda , x )$ are functions from $L^1 \bigl( \Poin , d \mu (
    \Lambda , x ) \bigr)$ with compact support $\Ssf$. Since
    $\Lfrak_0$ is invariant under differentiation we conclude from the
    first two paragraphs of the present proof and the above
    considerations that the partial derivatives
    \begin{equation*}
      \delta^i \bigl( \alpha_F ( L_0 ) \bigr) = \sum_{j = 1}^{d_\Psf}
      \alpha_{F_{j i}} \bigl( \delta^j ( L_0 ) \bigr)
    \end{equation*}
    are again almost local vacuum annihilation operators which belong
    to $\Dscr^{( 1 )} ( \Afrak )$. Thus by induction, repeatedly using
    these methods, $\alpha_F ( L_0 )$ is seen to be an element of
    $\Dscr^{( \infty )} ( \Afrak )$ with almost local derivatives of
    any order, i.\,e. $\alpha_F ( L_0 ) \in \Lfrak_0$.
  \item By Proposition~\ref{Pro-lcs-continuity-differentiability} the
    mappings $( \Lambda , x ) \mapsto \aLax ( L )$ and $( \Lambda , x
    ) \mapsto \aLax ( C )$ are continuous with respect to the uniform
    topology and all the $\qD$- and $\pD$-topologies, respectively,
    staying bounded on the compact set $\Ssf$. This implies their
    measurability in the locally convex spaces $( \AL , \Tfrakbar^u_q
    )$ and $( \AC , \Tfrakbar^u_p )$ together with the fact that their
    product with the integrable function $f$ is majorized in each of
    the norm and seminorm topologies by a multiple of $\abs{F}$. As a
    consequence the integrals $\alpha_F ( L )$ and $\alpha_F ( C )$
    exist in the complete locally convex spaces $( \AL ,
    \Tfrakbar^u_q )$ and $( \AC , \Tfrakbar^u_p )$, respectively, and
    \eqref{eq-LC-integral-estimate} is an immediate upshot
    \cite[II.6.2 and 5.4]{fell/doran:1988a}.
  \end{prooflist}
  \renewcommand{\qed}{}
\end{proof}

There exists a version of the second part of the above lemma for
functions on $\Rsone$ that are Lebesgue-integrable but no longer have
to be compactly supported.
\begin{Lem}
  \label{Lem-Lebesgue-Bochner-integrals}
  Let $L \in \Lfrak$ and let $g \in L^1 \bigl( \Rsone , d^{s + 1} x
  \bigr)$. Then
  \begin{equation}
    \label{eq-alpha-g-def}
    \alpha_g ( L ) \doteq \int_{\Rsone} d^{s + 1} x \; g ( x ) \,
    \ax ( L )
  \end{equation}
  is an operator in $( \AL , \Tfrakbar^u_q )$, satisfying the
  estimates 
  \begin{equation}
    \label{eq-alpha-g-estimates}
    \bqDx{\alpha_g ( L )} \leqslant \norm{g}_1 \qDx{L}
  \end{equation}
  for any bounded Borel set $\Delta$. The energy-momentum transfer of
  $\alpha_g (L)$ is contained in $\supp \, \tilde{g}$, the support of
  the Fourier transform $\tilde{g}$ of $g$.
\end{Lem}
\begin{proof}
  By translation invariance of the norm $\norm{~.~}$ as well as of the
  seminorms $\qD$ (cf.~Lemma~\ref{Lem-translation-invariance}) the
  (measurable) integrand on the right-hand side of
  \eqref{eq-alpha-g-def} is majorized by the functions $x \mapsto
  \abs{g ( x )} \, \norm{L}$ and $x \mapsto \abs{g ( x )} \, \qDx{L}$
  for any bounded Borel set $\Delta$. These are Lebesgue-integrable
  and therefore $\alpha_g ( L )$ exists as a unique element of $( \AL
  , \Tfrakbar^u_q )$, satisfying the claimed estimates
  \eqref{eq-alpha-g-estimates}.

  Next, we consider an arbitrary function $h \in L^1 \bigl( \Rsone ,
  d^{s + 1} x \bigr)$. By Fubini's Theorem
  \cite[II.16.3]{fell/doran:1988a} and translation invariance of
  Lebesgue measure
  \begin{multline*}
    \int_{\Rsone} d^{s + 1} y \; h ( y ) \, \alpha_y \bigl( \alpha_g (
    L ) \bigr) = \int_{\Rsone} d^{s + 1} y \; h ( y ) \, \alpha_y
    \Bigl( \int_{\Rsone} d^{s + 1} x \; g ( x ) \, \ax ( L ) \Bigr) \\
    \mspace{-40mu} = \int_{\Rsone} d^{s + 1} y \int_{\Rsone} d^{s + 1}
    x \; h ( y ) \, g ( x ) \; \alpha_{x + y} ( L ) \\
    = \int_{\Rsone} d^{s + 1} x \; \Bigl( \int_{\Rsone} d^{s + 1} y \;
    h ( y ) \, g ( x - y ) \Bigr) \ax ( L ) \text{,}
  \end{multline*}
  where the term in brackets on the right-hand side of the last
  equation is the convolution product $h \ast g$ of $h$ and $g$. Its
  Fourier transform $\widetilde{h \ast g}$ is given by $\widetilde{h
  \ast g} ( p ) = ( 2 \pi )^{( s + 1 ) / 2} \tilde{h} ( p ) \tilde{g}
  ( p )$ (cf.~\cite[Theorem~VI.(21.41)]{hewitt/stromberg:1969}), so
  that this function vanishes if $\tilde{h}$ and $\tilde{g}$ have
  disjoint supports. Therefore $\supp \tilde{h} \cap \supp \tilde{g} =
  \emptyset$ entails
  \begin{equation*}
    \int_{\Rsone} d^{s + 1} y \; h ( y ) \, \alpha_y \bigl( \alpha_g (
    L ) \bigr) = 0 \text{,}
  \end{equation*}
  and this shows that the Fourier transform of $y \mapsto \alpha_y
  \bigl( \alpha_g ( L ) \bigr)$ has support in $\supp \tilde{g}$,
  which henceforth contains the energy-momentum transfer of $\alpha_g
  ( L )$.
\end{proof}

\subsection{Decay Property}
\label{subsec-decay-property}

Eventually we are able to establish a property of rapid decay with
respect to the seminorms $\qD$ for commutators of elements of $\Lfrak$
which are almost local.
\begin{Lem}
  \label{Lem-commutator-qd-decay}
  Let $L_1$ and $L_2$ belong to $\Lfrak_0$ and let $A_1 \text{,} A_2
  \in \Afrak$ be almost local. Then for any bounded Borel subset
  $\Delta$ of $\Rsone$
  \begin{equation*}
    \Rs \ni \xib \mapsto \bqDx{\bcomm{\aibx ( A_1 L_1 )}{A_2 L_2}}
  \end{equation*}
  decreases with $\abs{\xib} \rightarrow \infty$ faster than any power
  of $\abs{\xib}^{-1}$.
\end{Lem}
\begin{proof}
  First we consider the special case of two elements $L_a$ and $L_b$
  in $\Lfrak_0$ having energy-momentum transfer in compact and convex
  subsets $\Gamma_a$ and $\Gamma_b$ of $\complement \fwcone$,
  respectively, with the additional property that $\Gamma_{a , b}
  \doteq  ( \Gamma_a + \Gamma_b ) - \Gamma_a$ and $\Gamma_{b , a}
  \doteq ( \Gamma_a + \Gamma_b ) - \Gamma_b$ lie in the complement of
  $\fwcone$, too. According to the
  Lemmas~\ref{Lem-involution-invariance} and~\ref{Lem-basic-estimate}
  \begin{multline}
    \label{eq-specialized-Ep-transfer-estimate}
    \bqDx{\bcomm{\aibx ( L_a )}{L_b}}^2 = \bpDx{\bcomm{\aibx ( L_a
    )}{L_b}^* \bcomm{\aibx ( L_a )}{L_b}} \\
    \leqslant \qDx{L_b} \, \bqDx{\aibx ( L_a )^* \bcomm{\aibx ( L_a
    )}{L_b}} + \qDx{L_a} \, \bqDx{{L_b}^* \bcomm{\aibx ( L_a )}{L_b}}
    \text{,}
  \end{multline}
  and we are left with the task to investigate for large $\abs{\xib}$
  the behaviour of the functions $\bqDx{\aibx ( L_a )^* \bcomm{\aibx (
  L_a )}{L_b}}$ and $\bqDx{{L_b}^* \bcomm{\aibx ( L_a )}{L_b}}$. Since
  the arguments of both terms belong to $\Lfrak_0$, having
  energy-momentum transfer in the compact and convex subsets
  $\Gamma_{a , b}$ and $\Gamma_{ b , a}$ of $\complement \fwcone$, we
  can apply \eqref{eq-commutator-integral-estimate} of
  Proposition~\ref{Pro-harmonic-analysis} in connection with
  \eqref{eq-q-seminorm} to get the estimate (for the second term)
  \begin{multline}
    \label{eq-L0-commutator-estimate}
    \abs{\xib}^{2k} \bqDx{{L_b}^* \bcomm{\aibx ( L_a )}{L_b}}^2 \\
    \leqslant N ( \Delta , \Gamma_{b , a} ) \int_{\Rs} d^s y \;
    \abs{\xib}^{2k} \Bnorm{\Bcomm{\aiby \bigl({L_b}^* \bcomm{\aibx (
    L_a )}{L_b} \bigr)}{\bigl({L_b}^* \bcomm{\aibx ( L_a )}{L_b}
    \bigr)^*}} \text{.}
  \end{multline}
  Let $\bset{ L_{a , r} \in \AOr : r > 0}$ and $\bset{ L_{b , r} \in
  \AOr : r > 0}$ be approximating nets for $L_a$ and $L_b$,
  respectively, satisfying $\norm{L_{a , r}} \leqslant \norm{L_a}$
  and $\norm{L_{b , r}} \leqslant \norm{L_b}$. Then the elements
  \begin{equation*}
    {L_{b , r}}^* \bcomm{\aibx ( L_{a , r} )}{L_{b , r}} \in \Afrak (
    \Oscr_r + \xib ) \subseteq \Afrak ( \Oscr_{r + \abs{\xib}} )
  \end{equation*}
  constitute the large radius part of approximating nets for the
  almost local operators ${L_b}^* \bcomm{\aibx ( L_a )}{L_b}$, $\xib
  \in \Rs$, subject to the estimate
  \begin{multline}
    \label{eq-commutator-approximant}
      \bnorm{{L_b}^* \bcomm{\aibx ( L_a )}{L_b} - {L_{b , r}}^*
      \bcomm{\aibx ( L_{a , r} )}{L_{b , r}}} \\
      \leqslant 4 \, \norm{L_a} \norm{L_b} \norm{L_b - L_{b , r}} + 2
      \, \norm{L_b}^2 \norm{L_a - L_{a , r}} \leqslant C_l \, r^{-l}
  \end{multline}
  for any $l \in \Nbb$ with suitable $C_l > 0$. Now, as suggested by
  the remark following Definition~\ref{Def-almost-locality}, there
  exist approximating nets $\bset{L ( a , b ; \xib )_r \in \AOr : r >
  0}$, $\xib \in \Rs$, with $\norm{L ( a , b ; \xib )_r} \leqslant
  \bnorm{{L_b}^* \bcomm{\aibx ( L_a )}{L_b}}$ and $\bnorm{{L_b}^*
  \bcomm{\aibx ( L_a )}{L_b}  - L ( a , b ; \xib )_{r + \abs{\xib}}}
  \leqslant 2 \, C_l \,  r^{-l}$, so that, according to
  \eqref{eq-special-commutator-norm-estimate}, the integrand of
  \eqref{eq-L0-commutator-estimate} is bounded by
  \begin{multline}
    \label{eq-La-Lb-commutator-estimate}
    \abs{\xib}^{2 k} \Bnorm{\Bcomm{\aiby \bigl( {L_b}^* \bcomm{\aibx (
    L_a )}{L_b} \bigr)}{\bigl( {L_b}^* \bcomm{\aibx ( L_a )}{L_b}
    \bigr)^*}} \\
    \mspace{-90mu} \leqslant \abs{\xib}^{2 k} 4 \, \bnorm{{L_b}^*
    \bcomm{\aibx ( L_a )}{L_b}} \bnorm{{L_b}^* \bcomm{\aibx ( L_a
    )}{L_b} - L ( a , b ; \xib )_{2^{-1} \abs{\yib}}} \\
    \leqslant
    \begin{cases}
      8 \, \abs{\xib}^{2 k} \norm{L_b}^2 \bnorm{\bcomm{\aibx ( L_a
      )}{L_b}}^2 & \text{,} \abs{\yib} \leqslant 2 ( \abs{\xib} + 1 )
      \text{,} \\
      8 \, \norm{L_b} \abs{\xib}^{2 k} \bnorm{\bcomm{\aibx ( L_a
      )}{L_b}} C_l ( 2^{-1} \abs{\yib} - \abs{\xib} )^{-l} & \text{,}
      \abs{\yib} >  2 ( \abs{\xib} + 1 ) \text{,}
      \end{cases}
  \end{multline}
  which implies
  \begin{multline}
    \label{eq-La-Lb-integral-estimate}
    \abs{\xib}^{2 k} \bqDx{{L_b}^* \bcomm{\aibx ( L_a )}{L_b}}^2 \\
    \shoveleft{\leqslant N ( \Delta , \Gamma_{b , a} ) \biggl[ 8 \,
    \norm{L_b}^2 \abs{\xib}^{2k} \bnorm{\bcomm{\aibx ( L_a )}{L_b}}^2
    \int\limits_{\abs{\yib} \leqslant 2 ( \abs{\xib} + 1 )} d^s y} \\
    + 8 \, C_l \norm{L_b} \abs{\xib}^{2 k} \bnorm{\bcomm{\aibx ( L_a
    )}{L_b}} \int\limits_{\abs{\yib} > 2 ( \abs{\xib} + 1 )} d^s y \;
    ( 2^{-1} \abs{\yib} - \abs{\xib} )^{-l} \biggr] \text{.}
  \end{multline}
  Evaluation of the integrals on the right-hand side yields (for $l
  \geqslant s + 2$) polynomials of degree $s$ in $\abs{\xib}$, so
  that, due to the decay properties of $\xib \mapsto
  \bnorm{\bcomm{\aibx ( L_a )}{L_b}}$, there exists a uniform bound
  \begin{equation}
    \label{eq-qD-decay}
    \abs{\xib}^k \bqDx{{L_b}^* \bcomm{\aibx ( L_a )}{L_b}}^2 \leqslant
    M \text{,} \quad \xib \in \Rs \text{.}
  \end{equation}
  The same reasoning applies to the term $\bqDx{\aibx ( L_a )^*
  \bcomm{\aibx ( L_a )}{L_b}}$, thus establishing the asserted rapid
  decrease for the mapping $\xib \mapsto \bqDx{\bcomm{\aibx ( L_a
  )}{L_b}}$, according to relation
  \eqref{eq-specialized-Ep-transfer-estimate}.

  In the general case of almost local elements $A_1 \text{,} A_2 \in
  \Afrak$ and $L_1 \text{,} L_2 \in \Lfrak_0$ one has, by
  Lemma~\ref{Lem-basic-estimate}
  \begin{multline*}
    \bqDx{\bcomm{\aibx ( A_1 L_1 )}{A_2 L_2}} \\
    \shoveleft{\leqslant \norm{A_1} \bnorm{\bcomm{\aibx ( L_1 )}{A_2}}
    \, \qDx{L_2} + \norm{A_1} \norm{A_2} \, \bqDx{\bcomm{\aibx ( L_1
    )}{L_2}}} \\
    + \bnorm{\bcomm{\aibx ( A_1 )}{A_2}} \norm{L_2} \, \qDx{L_1} +
    \norm{A_2} \bnorm{\bcomm{\aibx ( A_1 )}{L_2}} \, \qDx{L_1}
    \text{,}
  \end{multline*}
  and rapid decay is an immediate consequence of almost locality for
  all terms but the second one on the right-hand side of this
  inequality. Using suitable decompositions of $L_1$ and $L_2$ in
  terms of elements of $\Lfrak_0$ complying pairwise with the special
  properties exploited in the previous paragraph, the remaining
  problem of decrease of the mapping $\xib \mapsto \bqDx{\bcomm{\aibx
  ( L_1 )}{L_2}}$ reduces to the case that has already been solved
  above, thus completing the proof. 
\end{proof}

\chapter{Particle Weights as Asymptotic Plane Waves}
  \label{chap-particle-weights}

Having analysed in great detail the nets of seminorms $\qD$ and $\pD$,
indexed by the bounded Borel sets $\Delta \subseteq \Rsone$, on
$\Lfrak$ and $\Cfrak$, respectively, we now turn to the investigation
of the topological dual spaces:
\begin{Def}
  \label{Def-varsigma-continuity}
  \begin{deflist}
  \item The linear functionals on $\Cfrak$ which are continuous with
    respect to the seminorm $\pD$ constitute a vector space $\CDstar$,
    which is a normed space via
    \begin{equation*}
      \norm{\varsigma}_\Delta \doteq \sup \bset{\abs{\varsigma ( C )}
      : C \in \Cfrak, \pDx{C} \leqslant 1} \text{,} \quad \varsigma
      \in \CDstar \text{.}
    \end{equation*}
  \item The topological duals of the locally convex spaces $( \Lfrak_0
    , \Tfrak_q )$, $( \Lfrak , \Tfrak_q )$ and $( \Cfrak , \Tfrak_p)$
    are denoted $\Lzerostar$, $\Lstar$ and $\Cstar$, respectively.
  \end{deflist}
\end{Def}
\begin{Rem}
  Due to the net property (Proposition~\ref{Pro-seminorm-nets}) of the
  family of seminorms $\pD$, a linear functional belongs to the
  topological dual $\Cstar$ of $( \Cfrak , \Tfrak_p )$ if and only if
  it is continuous with respect to one specific seminorm $\pDprime$,
  $\Delta'$ a bounded Borel subset of $\Rsone$
  \cite[Proposition~1.2.8]{kadison/ringrose:1983}. Hence
  \begin{equation}
    \label{eq-Cstar-CDstar-relation}
    \Cstar = \bigcup \bset{\CDstar : \Delta \subseteq \Rsone \text{~a
    bounded Borel set}} \text{.}
  \end{equation}
  By continuous linear extension \cite[Chapter~One,
  \S\,5,\,4.(4)]{koethe:1983}, the functionals from $\Cstar$ are
  moreover in one-to-one correspondence with the elements of the
  topological dual $\Cbarstar$ of the complete locally convex space $(
  \Cfrakbar , \Tfrakbar_p )$. By the same argument, they are
  furthermore embedded in the topological dual $\ACstar$ of $( \AC ,
  \Tfrakbar^u_p )$. We shall make use of these properties without
  special mention.
\end{Rem}

\section{General Properties}
  \label{sec-gen-prop}

Before proceeding to extract certain elements from $\Cstar$ to be
interpreted, on the grounds of their specific properties, as
representing asymptotic mixtures of particle-like quantities, we are
first going to collect a number of important properties common to
\emph{all} functionals from the topological dual of $\Cfrak$ whose
proof does not depend on special assumptions. First of all, continuity
as established in
Proposition~\ref{Pro-lcs-continuity-differentiability} directly
carries over to functionals in $\Cstar$.
\begin{Lem}
  \label{Lem-varsigma-continuity}
  Continuous linear functionals $\varsigma \in \Cstar$ have the
  following properties.
  \begin{proplist}
  \item The mapping $\Poin \ni ( \Lambda , x ) \mapsto \varsigma
    \bigl( {L_1}^* \aLax ( L_2 ) \bigr)$ is continuous for arbitrary
    but fixed $L_1 \text{,} L_2 \in \Lfrak$.
  \item The mapping $\Poin \ni ( \Lambda , x ) \mapsto \varsigma
    \bigl( \aLax ( C ) \bigr)$ is continuous for given $C \in \Cfrak$.
  \end{proplist}
\end{Lem}
\begin{proof}
  Due to the assumed continuity of $\varsigma$, the assertions follow
  from Proposition~\ref{Pro-lcs-continuity-differentiability} in
  connection with Corollary~\ref{Cor-sesquilinear-product}.
\end{proof}

Every \emph{positive} functional $\varsigma$ on the $^*$-algebra
$\Cfrak = \Lfrak^* \, \Lfrak$ defines a non-negative sesquilinear
form on $\Lfrak$ through
\begin{subequations}
  \label{eq-varsigma-quotient-constructions}
  \begin{equation}
    \label{eq-varsigma-sesquilinear-form}
    \scp{~.~}{~.~}_\varsigma : \Lfrak \times \Lfrak \rightarrow \Cbb
    \quad ( L_1 , L_2) \mapsto \scp{L_1}{L_2}_\varsigma \doteq
    \varsigma ( {L_1}^* L_2 ) \text{,} 
  \end{equation}
  and thus induces a seminorm $\qs$ on $\Lfrak$ via
  \begin{equation}
    \label{eq-varsigma-seminorm}
    \qs : \Lfrak \rightarrow \Rbb_+ \quad L \mapsto \qsx{L} \doteq
    \scp{L}{L}_\varsigma^{1/2} \text{.} 
  \end{equation}
  Denoting by $\Nfrak_\varsigma$ the null space of $\qs$, one can
  construct the quotient $\Lfrak_\varsigma \doteq \Lfrak /
  \Nfrak_\varsigma$, which is a normed space through the definition
  \begin{equation}
    \label{eq-varsigma-norm}
    \norm{~.~}_\varsigma : \Lfrak / \Nfrak_\varsigma \rightarrow
    \Rbb_+ \quad [ L ]_\varsigma \mapsto \norm{[ L
    ]_\varsigma}_\varsigma \doteq \qsx{L} \text{,}
  \end{equation}
\end{subequations}
where we used square brackets to designate the cosets in $\Lfrak /
\Nfrak_\varsigma$. These concepts can be applied to formulate,
parallel to Proposition~\ref{Pro-lcs-continuity-differentiability},
differentiability of the Poincar\'{e} automorphisms with respect to
continuous positive functionals on $\Cfrak$.
\begin{Lem}
  \label{Lem-varsigma-differentiability}
  Let $\varsigma$ be a continuous positive functional on the
  $^*$-algebra $\Cfrak$, i.\,e.~$\varsigma \in \Cfrak^*_+$. Then the
  restriction of the canonical homomorphism
  \begin{equation*}
    \Qscr_\varsigma : \Lfrak \rightarrow \Lfrak / \Nfrak_\varsigma
    \quad L \mapsto \Qscr_\varsigma ( L ) \doteq [ L ]_\varsigma
  \end{equation*}
  to the subspace $\Lfrak_0$ is $\Xscr_{\Lfrak_0}$-differentiable in
  the sense of Definition~\ref{Def-generalized-diff-lin-mappings},
  where
  \begin{equation*}
    \Xscr_{\Lfrak_0} = \bset{\Xi_{L_0} : L_0 \in \Lfrak_0}
  \end{equation*}
  is the family of infinitely often differentiable mappings defined in
  Proposition~\ref{Pro-lcs-continuity-differentiability}.
\end{Lem}
\begin{proof}
  Due to the assumed continuity of the functional $\varsigma$, there
  exists a bounded Borel set $\Delta$ such that, according to
  \eqref{eq-varsigma-quotient-constructions} in connection with
  Definition~\ref{Def-varsigma-continuity} and
  Lemma~\ref{Lem-involution-invariance}, for any $L \in \Lfrak$ there
  holds the inequality
  \begin{equation*}
    \norm{[ L ]_\varsigma}_\varsigma^2 = \qsx{L}^2 = \varsigma ( L^* L
    ) \leqslant \norm{\varsigma}_\Delta \pDx{L^* L} =
    \norm{\varsigma}_\Delta \qDx{L}^2 \text{.}
  \end{equation*}
  Therefore the linear operator
  \begin{equation*}
    \Qscr_\varsigma \restriction \Lfrak_0 : ( \Lfrak_0 , \Tfrak_q)
    \rightarrow ( \Lfrak / \Nfrak_\varsigma , \norm{~.~}_\varsigma )
  \end{equation*}
  turns out to be continuous, so that the assertion follows by an
  application of Corollary~\ref{Cor-differentiable-M-lin-mappings}
  from the result of
  Proposition~\ref{Pro-lcs-continuity-differentiability}, stating that
  the mappings
  \begin{equation*}
    \Xi_{L_0} : \Poin \rightarrow ( \Lfrak_0 , \Tfrak_q ) \quad (
    \Lambda , x ) \mapsto \Xi_{L_0} ( \Lambda , x ) \doteq \aLax ( L_0
    ) 
  \end{equation*}
  are differentiable for any $L_0 \in \Lfrak_0$ (cf.~the remark of
  that place).
\end{proof}

The next lemmas are concerned with integrability properties of
functionals $\varsigma \in \Cstar$, parallel to those established in
Subsection~\ref{subsec-integrability}. The first one,
Lemma~\ref{Lem-varsigma-integrals}, is an immediate consequence of
Lemmas~\ref{Lem-Poin-Bochner-integrals}
and~\ref{Lem-Lebesgue-Bochner-integrals}, whereas the second one,
Lemma~\ref{Lem-cluster-prep}, prepares the proof of a kind of Cluster
Property for \emph{positive} functionals in $\Cstar$, formulated in
the subsequent Proposition~\ref{Pro-cluster}.
\begin{Lem}
  \label{Lem-varsigma-integrals}
  Let $\varsigma \in \Cstar$, $L_1 \text{,} L_2 \in \Lfrak$ and $C \in
  \Cfrak$.
  \begin{proplist}
  \item Let $F \in L^1 \bigl( \Poin , d \mu ( \Lambda , x ) \bigr)$
    have compact support $\Ssf$, then
    \begin{subequations}
      \begin{align}
        \varsigma \bigl( {L_1}^* \alpha_F ( L_2 ) \bigr) & = \int d
        \mu ( \Lambda , x ) \; F ( \Lambda , x ) \, \varsigma \bigl(
        {L_1}^* \aLax ( L_2 ) \bigr) \text{,} \\
        \varsigma \bigl( \alpha_F ( C ) \bigr) & = \int d \mu (
        \Lambda , x ) \; F ( \Lambda , x ) \, \varsigma \bigl( \aLax (
        C ) \bigr) \text{,}
      \end{align}
    \end{subequations}
    and there hold the estimates
    \begin{subequations}
      \begin{align}
        \babs{\varsigma \bigl( {L_1}^* \alpha_F ( L_2 ) \bigr)} &
        \leqslant \norm{F}_1 \norm{\varsigma}_\Delta \, \qDx{L_1}
        \sup_{( \Lambda , x ) \in \Ssf} \bqDx{\aLax ( L_2 )} \text{,}
        \\
        \babs{\varsigma \bigl( \alpha_F ( C ) \bigr)} & \leqslant
        \norm{F}_1 \norm{\varsigma}_\Delta \sup_{( \Lambda , x ) \in
        \Ssf} \bpDx{\aLax(C)}
      \end{align}
    \end{subequations}
    for any $\Delta$ such that $\varsigma \in \CDstar$.
  \item For any function $g \in L^1 \bigl( \Rsone , d^{s + 1} x
    \bigr)$
    \begin{equation}
      \varsigma \bigl( {L_1}^* \alpha_g ( L_2 ) \bigr) = \int_{\Rsone}
      d^{s + 1} x \; g ( x ) \, \varsigma \bigl( {L_1}^* \ax ( L_2 )
      \bigr) \text{,}
    \end{equation}
    and a bound is given by
    \begin{equation}
      \babs{\varsigma \bigl( {L_1}^* \alpha_g ( L_2 ) \bigr)}
      \leqslant \norm{g}_1 \norm{\varsigma}_\Delta \, \qDx{L_1} \,
      \qDx{L_2}
    \end{equation}
    for any $\Delta$ satisfying $\varsigma \in \CDstar$.
  \end{proplist}
\end{Lem}
\begin{proof}
  Lemmas~\ref{Lem-Poin-Bochner-integrals}
  and~\ref{Lem-Lebesgue-Bochner-integrals} state that
  \begin{align*}
    \alpha_F ( L_2 ) & = \int d \mu ( \Lambda , x ) \; F ( \Lambda , x
    ) \, \aLax ( L_2 ) \text{,} \\
    \alpha_F ( C ) & = \int d \mu ( \Lambda , x ) \; F ( \Lambda , x )
    \, \aLax ( C ) \text{,} \\
    \alpha_g ( L_2 ) & = \int_{\Rsone} d^{s + 1} x \; g ( x ) \,
    \ax ( L_2 )
  \end{align*}
  exist in the complete locally convex spaces $( \AL , \Tfrakbar^u_q
  )$ and $( \AC , \Tfrakbar^u_p )$, respectively. Now, the functional
  $\varsigma$, which lies in $\ACstar$ according to the remark
  following Definition~\ref{Def-varsigma-continuity}, is linear and
  continuous with respect to $\alpha_F ( C ) \in \AC$ and, by
  Corollary~\ref{Cor-sesquilinear-product}, also with respect to both
  $\alpha_F ( L_2 ) \text{,} \alpha_g ( L_2 ) \in \AL$. Therefore it
  commutes with the locally convex integrals \cite[Proposition~II.5.7
  adapted to integrals in locally convex spaces]{fell/doran:1988a},
  which proves the assertion. The annexed estimates are a further
  simple application of the results contained in
  Lemmas~\ref{Lem-Poin-Bochner-integrals}
  and~\ref{Lem-Lebesgue-Bochner-integrals}.
\end{proof}
\begin{Lem}
  \label{Lem-cluster-prep}
  Let $L' \in \Lfrak$ and let $L \in \Lfrak ( \Gamma ) = \Lfrak \cap
  \widetilde{\Afrak} ( \Gamma )$, $\Gamma \subseteq \Rsone$ compact,
  i.\,e.~$L$ has energy-momentum transfer in $\Gamma$. If $\varsigma
  \in \Cstarplus$ is a positive functional which belongs to $\CDstar$
  and $\Delta'$ denotes any bounded Borel set containing $\Delta +
  \Gamma$, then
  \begin{equation}
    \label{eq-varsigmaintegral-prep}
    \int_{\Rs} d^s x \; \varsigma \bigl( L^* \aibx ( {L'}^* L' ) \, L
    \bigr) \leqslant \norm{\varsigma}_\Delta \, \qDx{L}^2
    \qDprimex{L'}^2 \text{.}
  \end{equation}
\end{Lem}
\begin{proof}
  Let $\Kib$ be an arbitrary compact subset of $\Rs$ and note that
  \begin{equation*}
    \int_\Kib d^s x \; \aibx ( {L'}^* L' ) \in \Afrak \text{.}
  \end{equation*}
  Thus, according to the construction of $\Cfrak$,
  \begin{equation*}
    \int_\Kib d^s x \; L^* \aibx ( {L'}^* L' ) \, L = L^* \int_\Kib
    d^s x \; \aibx ( {L'}^* L' ) \; L
  \end{equation*}
  belongs to the algebra of counters and exists furthermore as an
  integral in the locally convex space $( \AC , \Tfrakbar^u_p
  )$. Therefore the functional $\varsigma \in \ACstar$ can be
  interchanged with the integral
  \cite[Proposition~II.5.7]{fell/doran:1988a} to give
  \begin{equation*}
    \int_\Kib d^s x \; \varsigma \bigl( L^* \aibx ( {L'}^* L' ) \, L
    \bigr) = \varsigma \Bigl( L^* \int_\Kib d^s x \; \aibx ( {L'}^* L'
    ) \; L \Bigr) \text{.}
  \end{equation*}
  Application of Lemma~\ref{Lem-basic-estimate} then leads to the
  estimate
  \begin{multline*}
    0 \leqslant \int_\Kib d^s x \; \varsigma \bigl( L^* \aibx ( {L'}^*
    L' ) \, L \bigr) \leqslant \norm{\varsigma}_\Delta \, \BpDx{L^*
    \int_\Kib d^s x \; \aibx ( {L'}^* L' ) \; L} \\
    \leqslant \norm{\varsigma}_\Delta \, \qDx{L}^2 \Bnorm{\EDprime
    \int_\Kib d^s x \; \aibx ( {L'}^* L' ) \; \EDprime} =
    \norm{\varsigma}_\Delta \, \qDx{L}^2 \Bnorm{Q_{\Delta' , \Kib}^{(
    {L'}^* L' )}} \text{,}
  \end{multline*}
  where we made use of the positivity of $\varsigma$. The above
  inequality survives in the limit $\Kib \nearrow \Rs$ and the
  convergence of the right-hand side to a finite real number
  establishes the integrability of the function
  \begin{equation*}
    \Rs \ni \xib \mapsto \varsigma \bigl( L^* \aibx ( {L'}^* L' ) \, L
    \bigr)
  \end{equation*}
  as a consequence of the Monotone Convergence
  Theorem~\cite[II.2.7]{fell/doran:1988a}. In view of
  \eqref{eq-q-seminorm}, one finally arrives at the asserted bound
  \begin{equation*}
    \int_{\Rs} d^s x \; \varsigma \bigl( L^* \aibx ( {L'}^* L' ) \, L
    \bigr) \leqslant \norm{\varsigma}_\Delta \, \qDx{L}^2
    \Bnorm{Q_{\Delta'}^{( {L'}^* L' )}} = \norm{\varsigma}_\Delta \,
    \qDx{L}^2 \qDprimex{L'}^2 \text{.} \tag*{\qed}
  \end{equation*}
  \renewcommand{\qed}{}
\end{proof}

After these preparations we are in a position to prove the announced
Cluster Property for positive functionals in $\Cstar$.
\begin{Pro}[Cluster Property]
  \label{Pro-cluster}
  Let $L_i$ and $L'_i$ be elements of $\Lfrak_0$ and let $A_i \in
  \Afrak$, $i = 1 \text{,} 2$, be almost local operators, then the
  function
  \begin{equation*}
    \Rs \ni \xib \mapsto \varsigma \bigl( ( {L_1}^* A_1 L'_1 ) \aibx (
    {L_2}^* A_2 L'_2 ) \bigr) \in \Cbb
  \end{equation*}
  is an element of $L^1 \bigl( \Rs , d^s x \bigr)$ for any $\varsigma
  \in \Cstarplus$ and satisfies
  \begin{equation}
    \label{eq-cluster}
    \int_{\Rs} d^s x \; \babs{\varsigma \bigl( ( {L_1}^* A_1 L'_1 )
    \aibx ( {L_2}^* A_2 L'_2 ) \bigr)} \leqslant
    \norm{\varsigma}_\Delta \, M_\Delta
  \end{equation}
  for any bounded Borel set $\Delta$ for which $\varsigma$ belongs to
  $\CDstar$, where the constant $M_\Delta$ depends on $\Delta$ and the
  operators involved.
\end{Pro}
\begin{proof}
  First, we re-write the argument $( {L_1}^* A_1 L'_1 ) \aibx (
  {L_2}^* A_2 L'_2 )$, commuting the operators $A_1 L'_1$ and $\aibx (
  {L_2}^* A_2 )$, to get
  \begin{multline}
    \label{eq-commuted-argument}
    ( {L_1}^* A_1 L'_1 ) \aibx ( {L_2}^* A_2 L'_2 ) \\
    = {L_1}^* \bcomm{A_1 L'_1}{\aibx ( {L_2}^* A_2 )} \aibx ( L'_2 ) +
    {L_1}^* \aibx ( {L_2}^* A_2 ) A_1 L'_1 \aibx ( L'_2 ) \text{.}
  \end{multline}
  This implies
  \begin{multline}
    \label{eq-varsigma-sum-estimate}
    \babs{\varsigma \bigl( ( {L_1}^* A_1 L'_1 ) \aibx ( {L_2}^* A_2
    L'_2 ) \bigr)} \\
    \leqslant \babs{\varsigma \bigl( {L_1}^* \bcomm{A_1 L'_1}{\aibx (
    {L_2}^* A_2 )} \aibx ( L'_2 ) \bigr)} + \babs{\varsigma \bigl(
    {L_1}^* \aibx ( {L_2}^* A_2 ) A_1 L'_1 \aibx ( L'_2 ) \bigr)}
    \text{,}
  \end{multline}
  where the first term on the right-hand side is evidently integrable
  over $\Rs$, due to almost locality of the operators encompassed by
  the commutator. For $\varsigma \in \CDstar$ we have the estimate
  \begin{multline}
    \label{eq-first-varsigmaintegral-estimate}
    \int_{\Rs} d^s x \; \babs{\varsigma \bigl( {L_1}^* \bcomm{A_1
    L'_1}{\aibx ( {L_2}^* A_2 )} \aibx ( L'_2 ) \bigr)} \\
    \leqslant \norm{\varsigma}_\Delta \, \qDx{L_1} \, \qDx{L'_2}
    \int_{\Rs} d^s x \; \bnorm{\bcomm{A_1 L'_1}{\aibx ( {L_2}^* A_2
    )}} \text{.}
  \end{multline}
  The second term can be estimated by use of the Cauchy-Schwarz
  inequality applied to the positive functional $\varsigma$:
  \begin{multline}
    \label{eq-varsigma-cauchy-schwarz}
    2 \, \babs{\varsigma \bigl( {L_1}^* \aibx ( {L_2}^* A_2 ) A_1 L'_1
    \aibx ( L'_2 ) \bigr)} \\
    \mspace{-115mu} \leqslant 2 \, \varsigma \bigl( {L_1}^* \aibx (
    {L_2}^* A_2 {A_2}^* L_2 ) L_1 \bigr)^{1/2} \, \varsigma \bigl(
    \aibx ( {L'_2}^* ) {L'_1}^* {A_1}^* A_1 L'_1 \aibx ( L'_2 )
    \bigr)^{1/2} \\
    = \inf_{\lambda > 0} \Bigl( \lambda^{-1} \varsigma \bigl( {L_1}^*
    \aibx ( {L_2}^* A_2 {A_2}^* L_2 ) L_1 \bigr) + \lambda \,
    \varsigma \bigl( \aibx ( {L'_2}^* ) {L'_1}^* {A_1}^* A_1 L'_1
    \aibx ( L'_2 ) \bigr) \Bigr) \text{.}
  \end{multline}
  Integration of the first term on the right-hand side is possible
  according to the previous Lemma~\ref{Lem-cluster-prep} and gives
  \begin{equation}
    \label{eq-second-varsigmaintegral-estimate}
    \int_{\Rs} d^s x \; \varsigma \bigl( {L_1}^* \aibx ( {L_2}^* A_2
    {A_2}^* L_2 ) L_1 \bigr) \leqslant \norm{\varsigma}_\Delta \,
    \qDx{L_1}^2 q_{\Delta_1} ( {A_2}^* L_2 )^2 \text{,}
  \end{equation}
  where $\Delta_1$ is any bounded Borel set containing the sum of
  $\Delta$ and the energy-momentum transfer $\Gamma_1$ of $L_1$. 
  Concerning the second term on the right of
  \eqref{eq-varsigma-cauchy-schwarz}, we get, upon commuting $\aibx (
  {L'_2}^* )$ and $\aibx ( L'_2 )$ to the interior,
  \begin{multline}
    \label{eq-second-varsigma-comm-estimate}
    \varsigma \bigl( \aibx ( {L'_2}^* ) {L'_1}^* {A_1}^* A_1 L'_1
    \aibx ( L'_2 ) \bigr) \\
    \leqslant \babs{\varsigma \bigl( \bcomm{\aibx ( {L'_2}^*
    )}{{L'_1}^*} {A_1}^* A_1 L'_1 \aibx ( L'_2 ) \bigr)} +
    \babs{\varsigma \bigl( {L'_1}^* \aibx ( {L'_2}^* ) {A_1}^* A_1
    \bcomm{L'_1}{\aibx ( L'_2 )} \bigr)} \\
    + \norm{A_1}^2 \babs{\varsigma \bigl( {L'_1}^* \aibx ( {L'_2}^*
    L'_2 ) L'_1 \bigr)} \text{,}
  \end{multline}
  where again use was made of the positivity of $\varsigma$. The rapid
  decay of commutators of almost local operators with respect to the
  $\qD$-seminorm established in Lemma~\ref{Lem-commutator-qd-decay} of
  Subsection~\ref{subsec-decay-property} can be combined with
  Lemma~\ref{Lem-cluster-prep} to show integrability over $\Rs$:
  \begin{multline}
    \label{eq-third-varsigmaintegral-estimate}
    \int_{\Rs} d^s x \; \varsigma \bigl( \aibx ( {L'_2}^* ) {L'_1}^*
    {A_1}^* A_1 L'_1 \aibx ( L'_2 ) \bigr) \\
    \mspace{-70mu} \leqslant \norm{\varsigma}_\Delta \norm{A_1}^2
    \Bigl( \qDx{L'_1}^2 q_{\Delta'_1} ( L'_2 )^2 + \bigl( \norm{L'_1}
    \, \qDx{L'_2} + \norm{L'_2} \, \qDx{L'_1} \bigr) \cdot \\
    \cdot \int_{\Rs} d^s x \; \bqDx{\bcomm{L'_1}{\aibx ( L'_2 )}}
    \Bigr) \text{,}
  \end{multline}
  which holds for any bounded Borel set $\Delta'_1 \supseteq \Delta +
  \Gamma'_1$, where $\Gamma'_1$ denotes the energy-momentum transfer
  of $L'_1$. By \eqref{eq-second-varsigma-comm-estimate} and
  \eqref{eq-third-varsigmaintegral-estimate}, the left-hand side of
  \eqref{eq-varsigma-cauchy-schwarz} turns out to be integrable, and a
  bound for this integral is proportional to
  $\norm{\varsigma}_\Delta$. In connection with
  \eqref{eq-first-varsigmaintegral-estimate} this establishes the
  assertion for a suitable constant $M_\Delta$ that can be deduced
  from relations
  \eqref{eq-first-varsigmaintegral-estimate},
  \eqref{eq-varsigma-cauchy-schwarz},
  \eqref{eq-second-varsigma-comm-estimate}
  and \eqref{eq-third-varsigmaintegral-estimate}.
\end{proof}

The Cluster Property has been proved above under the fairly general
assumption of almost locality of the operators involved. If for given
$L_1 \text{,} L_2 \in \Lfrak$ the mapping
\begin{equation*}
  \Rs \ni \xib \mapsto \bpDx{{L_1}^* \aibx ( L_2 )}
\end{equation*}
happens to belong to the space $L^1 \bigl( \Rs , d^s x \bigr)$ for the
bounded Borel set $\Delta$, \eqref{eq-cluster} is obviously fulfilled
in case that $\varsigma \in \Cstar$ belongs to $\CDstar$. As an
example consider almost local operators $L'_1 \text{,} L'_2 \in
\Lfrak$ having energy-momentum transfer $\Gamma_1$ and $\Gamma_2$,
respectively, such that $( \Delta + \Gamma_1 + \Gamma_2 ) \cap \fwcone
= \emptyset$. This implies $L'_1 \aibx ( L'_2 ) \ED = 0$ for any $\xib
\in \Rs$ and, by Lemmas~\ref{Lem-delta-annihilation}
and~\ref{Lem-basic-estimate}, $\bpDx{{L'_1}^* \aibx ( {L'_2}^* ) L'_1
\aibx ( L'_2 )} = 0$. An application of Lemma~\ref{Lem-basic-estimate}
in connection with translation invariance of $\qD$
(Lemma~\ref{Lem-translation-invariance}) then yields for the counters
$C'_i \doteq {L'_i}^* L'_i$, $i = 1 \text{,} 2$,
\begin{multline*}
  \bpDx{{C'_1}^* \aibx ( C'_2 )} = \bpDx{{L'_1}^* L'_1 \aibx (
  {L'_2}^* L'_2 )} \\
  = \bpDx{{L'_1}^* \bcomm{L'_1}{\aibx ( {L'_2}^* )} \aibx ( L'_2 )}
  \leqslant \bnorm{\bcomm{L'_1}{\aibx ( {L'_2}^* )}} \, \qDx{L'_1} \,
  \qDx{L'_2} \text{,}
\end{multline*}
where, due to the assumed almost locality of $L'_1$ and $L'_2$, the
right-hand side is seen to belong to $L^1 \bigl( \Rs , d^s x \bigr)$.
The integrability of a mapping $\xib \mapsto \bpDx{{L_1}^* \aibx ( L_2
)}$, $L_1 \text{,} L_2 \in \Lfrak$, has another consequence concerning
weakly convergent nets $\bset{\varsigma_\iota : \iota \in J}$ of
functionals from $\Cstar$, which are contained in bounded subsets of
$\CDstar$ with respect to the norm $\norm{~.~}_\Delta$: a kind of
Dominated Convergence Theorem.
\begin{Lem}
  \label{Lem-varsigmanet-convergence}
  Let $L_1 \text{,} L_2 \in \Lfrak$ be such that $\xib \mapsto
  \bpDx{{L_1}^* \aibx ( L_2 )}$ is integrable and consider the weakly
  convergent net $\bset{\varsigma_\iota : \iota \in J}$ in the
  $D$-ball of $\CDstar$ with limit $\varsigma$.This means that for any
  $C \in \Cfrak$
  \begin{equation*}
    \lim_\iota \varsigma_\iota ( C ) = \varsigma ( C )
  \end{equation*}
  and for any $\iota \in J$
  \begin{subequations}
    \label{eq-varsigma-balls}
    \begin{align}
      \label{eq-varsigma-balls-1}
      \abs{\varsigma_\iota ( C )} & \leqslant D \cdot \pDx{C} \text{,}
      \\
      \label{eq-varsigma-balls-2}
      \abs{\varsigma ( C )} & \leqslant D \cdot \pDx{C} \text{,}
    \end{align}
  \end{subequations}
  the latter relation being implied by the former. Then
  \begin{equation}
    \label{eq-integral-net-limits}
    \int_{\Rs} d^s x \; \varsigma \bigl( {L_1}^* \aibx ( L_2 ) \bigr)
    = \lim_\iota \int_{\Rs} d^s x \; \varsigma_\iota \bigl( {L_1}^*
    \aibx ( L_2 ) \bigr) \text{.}
  \end{equation}
\end{Lem}
\begin{proof}
  As implied by Proposition~\ref{Pro-lcs-continuity-differentiability}
  and Corollary~\ref{Cor-sesquilinear-product}, $\xib \mapsto {L_1}^*
  \aibx ( L_2 )$ is a continuous mapping on $\Rs$ with respect to the
  $\pD$-topology, hence it is uniformly continuous on any compact set
  $\Kib$. This means that to $\epsilon > 0$ there exists $\delta > 0$
  such that $\xib \text{,} \xib' \in \Kib$ and $\abs{\xib - \xib'} <
  \delta$ imply
  \begin{equation*}
    \bpDx{{L_1}^* \aibx ( L_2 ) - {L_1}^* \aibxprime ( L_2 )} <
    \frac{\epsilon}{6 D \abs{\Kib}} \text{,}
  \end{equation*}
  where $\abs{\Kib}$ denotes the $s$-dimensional volume of $\Kib$.
  Consequently, under the above assumption on $\xib$ and $\xib'$, we
  infer from \eqref{eq-varsigma-balls}
  \begin{align*}
    \babs{\varsigma_\iota \bigl( {L_1}^* \aibx ( L_2 ) \bigr) -
    \varsigma_\iota \bigl( {L_1}^* \aibxprime ( L_2 ) \bigr)} & =
    \babs{\varsigma_\iota \bigl( {L_1}^* \aibx ( L_2 ) - {L_1}^*
    \aibxprime ( L_2 ) \bigr)} < \frac{\epsilon}{6 \abs{\Kib}}
    \text{,} \\ 
    \babs{\varsigma \bigl( {L_1}^* \aibx ( L_2 ) \bigr) - \varsigma
    \bigl( {L_1}^* \aibxprime ( L_2 ) \bigr)} & = \babs{\varsigma
    \bigl( {L_1}^* \aibx ( L_2 ) - {L_1}^* \aibxprime ( L_2 ) \bigr)}
    < \frac{\epsilon}{6 \abs{\Kib}} \text{.}
  \end{align*}
  By compactness of $\Kib$, there exist finitely many elements $\xib_1
  \text{,} \dots \text{,} \xib_N \in \Kib$ such that the
  $\delta$-balls around these points cover all of $\Kib$; moreover,
  since $\varsigma$ is the weak limit of the net
  $\bset{\varsigma_\iota : \iota \in J}$, we can find $\iota_0 \in J$
  such that $\iota \succ \iota_0$ implies
  \begin{equation*}
    \babs{\varsigma \bigl( {L_1}^* \alpha_{\xib_i} ( L_2 ) \bigr) -
    \varsigma_\iota \bigl( {L_1}^* \alpha_{\xib_i} ( L_2 ) \bigr)} <
    \frac{\epsilon}{6 \abs{\Kib}}
  \end{equation*}
  for any $i=1 \text{,} \dots \text{,} N$. Now, for $\iota \in J$ and
  $k \in \set{1 , \dots , N}$,
  \begin{multline*}
    \babs{\varsigma \bigl( {L_1}^* \aibx ( L_2 ) \bigr) -
    \varsigma_\iota \bigl( {L_1}^* \aibx ( L_2 ) \bigr)} \\
    \shoveleft{\leqslant \babs{\varsigma \bigl( {L_1}^* \aibx ( L_2 )
    \bigr) - \varsigma \bigl( {L_1}^* \alpha_{\xib_i} ( L_2 ) \bigr)}
    + \babs{\varsigma \bigl( {L_1}^* \alpha_{\xib_i} ( L_2 ) \bigr) -
    \varsigma_\iota \bigl( {L_1}^* \alpha_{\xib_i} ( L_2 ) \bigr)}} \\
    + \babs{\varsigma_\iota \bigl( {L_1}^* \alpha_{\xib_i} ( L_2 )
    \bigr) - \varsigma_\iota \bigl( {L_1}^* \aibx ( L_2 ) \bigr)}
    \text{,}
  \end{multline*} 
  and, selecting for $\xib \in \Kib$ an appropriate $\xib_k$ in a
  distance less than $\delta$, we can put the above results together
  to get the estimate
  \begin{equation*}
    \babs{\varsigma \bigl( {L_1}^* \aibx ( L_2 ) \bigr) -
    \varsigma_\iota \bigl( {L_1}^* \aibx ( L_2 ) \bigr)} <
    \frac{\epsilon}{2 \abs{\Kib}} \text{,}
  \end{equation*}
  which holds for any $\xib \in \Kib$ and $\iota \succ \iota_0$. Thus
  weak (i.\,e.~pointwise) convergence of the net
  $\bset{\varsigma_\iota : \iota \in J}$ is indeed uniform convergence on
  compact subsets of $\Rs$. Upon integration over $\Kib$ we arrive at
  \begin{multline}
    \label{eq-varsigma-K-integral}
    \Babs{\int_\Kib d^s x \; \Bigl( \varsigma \bigl( {L_1}^* \aibx (
    L_2 ) \bigr) - \varsigma_\iota \bigl( {L_1}^* \aibx ( L_2 ) \bigr)
    \Bigr)} \\
    \leqslant \int_\Kib d^s x \; \Babs{\varsigma \bigl( {L_1}^* \aibx
    ( L_2 ) \bigr) - \varsigma_\iota \bigl( {L_1}^* \aibx ( L_2 )
    \bigr)} < \frac{\epsilon}{2} \text{.}
  \end{multline}
  Now, by assumption
  \begin{equation*}
    \int_{\Rs} d^s x \; \bpDx{{L_1}^* \aibx ( L_2 )} < \infty \text{,}
  \end{equation*}
  so that to $\epsilon > 0$ there exists a compact subset
  $\Kib_\epsilon$ satisfying
  \begin{equation*}
    \int_{\complement \Kib_\epsilon} d^s x \; \bpDx{{L_1}^* \aibx (
    L_2 )} < \frac{\epsilon}{4D} \text{.}
  \end{equation*}
  Then, as a consequence of \eqref{eq-varsigma-balls-1} and
  \eqref{eq-varsigma-balls-2}, we get for any $\iota \in J$
  \begin{subequations}
    \label{eq-varsigma-integrals}
    \begin{align}
      \label{eq-varsigma-int-1}
      \int_{\complement \Kib_\epsilon} d^s x \; \babs{\varsigma_\iota
      \bigl( {L_1}^* \aibx ( L_2 ) \bigr)} & < \frac{\epsilon}{4}
      \text{,} \\
      \label{eq-varsigma-int-2}
      \int_{\complement \Kib_\epsilon} d^s x \; \babs{\varsigma \bigl(
      {L_1}^* \aibx ( L_2 ) \bigr)} & < \frac{\epsilon}{4} \text{.}
    \end{align}
  \end{subequations}
  Combining \eqref{eq-varsigma-integrals} with
  \eqref{eq-varsigma-K-integral} for the compact set $\Kib_\epsilon$
  yields for $\iota \succ \iota_0$ (note, that $\iota_0$ only depends
  on $\epsilon$)
  \begin{multline*}
    \Babs{\int_{\Rs} d^s x \; \varsigma \bigl( {L_1}^* \aibx ( L_2 )
    \bigr) - \int_{\Rs} d^s x \; \varsigma_\iota \bigl( {L_1}^* \aibx
    ( L_2 ) \bigr)} \\
    \shoveleft{\leqslant \int_{\complement \Kib_\epsilon} d^s x \;
    \babs{\varsigma \bigl( {L_1}^* \aibx ( L_2 ) \bigr)} + \int_\Kib
    d^s x \; \babs{\varsigma \bigl( {L_1}^* \aibx ( L_2 ) \bigr) -
    \varsigma_\iota \bigl( {L_1}^* \aibx ( L_2 ) \bigr)}} \\
    + \int_{\complement \Kib_\epsilon} d^s x \; \babs{\varsigma_\iota
    \bigl( {L_1}^* \aibx ( L_2 ) \bigr)} < \frac{\epsilon}{4} +
    \frac{\epsilon}{2} + \frac{\epsilon}{4} = \epsilon \text{.}
  \end{multline*}
  By arbitrariness of $\epsilon$ this proves the possibility to
  interchange integration and the limit with respect to $\iota$ as
  asserted in \eqref{eq-integral-net-limits}.
\end{proof}

The spectral support of not necessarily positive functionals
$\varsigma \in \Cstar$ (considered as distributions) depends, as
expressed in the subsequent proposition, on the bounded Borel sets
$\Delta$ for which $\varsigma \in \CDstar$. This property will prove
to be of importance when it comes to defining the energy-momentum of
particle weights.
\begin{Pro}[Spectral Property]
  \label{Pro-spectral-property}
  Let $L_1 \text{,} L_2 \in \Lfrak$ and $\varsigma \in \Cstar$. Then
  the support of the Fourier transform of the distribution
  \begin{equation*}
    \Rsone \ni x \mapsto \varsigma \bigl( {L_1}^* \ax ( L_2 ) \bigr)
    \in \Cbb
  \end{equation*}
  is contained in the shifted light cone $\fwcone - q$ for some
  $q \in \fwcone$. More specifically, $q$ is such that a bounded Borel
  set $\Delta$, satisfying $\varsigma \in \CDstar$, is contained in $q
  - \fwcone$.
\end{Pro}
\begin{proof}
  If a function $g$ belongs to the space $L^1 \bigl( \Rsone , d^{s +
  1} x \bigr)$, then the operator
  \begin{equation*}
    \alpha_g ( L_2 ) = \int_{\Rsone} d^{s + 1} x \; g ( x ) \,
    \ax ( L_2 )
  \end{equation*}
  lies in $\AL$, according to
  Lemma~\ref{Lem-Lebesgue-Bochner-integrals}, and has energy-momentum
  transfer in $\supp \tilde{g}$, the support of the Fourier transform
  of $g$. If this happens to satisfy $\supp \tilde{g} \subseteq
  \complement ( \fwcone - \Delta )$, we infer $\alpha_g ( L_2 ) \ED =
  0$ and henceforth, by Lemma~\ref{Lem-delta-annihilation},
  $\bqDx{\alpha_g ( L_2 )} = 0$. Since $\varsigma$ is assumed to
  belong to $\CDstar$, Lemma~\ref{Lem-varsigma-integrals} results in
  \begin{equation*}
    \Babs{\int_{\Rsone} d^{s + 1} x \; g ( x ) \, \varsigma \bigl(
    {L_1}^* \ax ( L_2 ) \bigr)} =  \babs{\varsigma \bigl( {L_1}^*
    \alpha_g ( L_2 ) \bigr)} \leqslant \norm{\varsigma}_\Delta \,
    \qDx{L_1} \, \bqDx{\alpha_g ( L_2 )} \text{,}
  \end{equation*}
  which, according to the preceding considerations, entails
  \begin{equation}
    \label{eq-spectral-support-integral}
    \int_{\Rsone} d^{s + 1} x \; g ( x ) \, \varsigma \bigl( {L_1}^*
    \ax ( L_2 ) \bigr) = 0 \text{.}
  \end{equation}
  Now, let $g'$ be an arbitrary function from $L^1 \bigl( \Rsone ,
  d^{s + 1} x \bigr)$ with $\supp \tilde{g'} \subseteq \complement (
  \fwcone - q )$, $\Delta \subseteq q -\fwcone$, then $\supp
  \tilde{g'} \subseteq \complement ( \fwcone - \Delta )$, so that
  \eqref{eq-spectral-support-integral} is fulfilled for any function
  of this kind, proving the assertion.
\end{proof}

\section{Asymptotic Functionals}

Now we turn to functionals in $\Cstar$ that carry additional
properties, reflecting the fact that the present investigation is
concerned with the structure of the totality of physical states at
asymptotic times (scattering states). The temporal development of such
a state of bounded energy, $\omega \in \Sscr ( \Delta )$, $\Delta$ a
bounded Borel set, can be explored by considering an integral of the
following shape:
\begin{equation}
  \label{eq-heuristic-velocity-integral}
  \int_{\Rs} d^s v \; h ( \vib ) \, \omega \bigl( \alpha_{( \tau ,
  \tau \vib )} ( C ) \bigr) \text{,}
\end{equation}
where $h$ denotes a bounded measurable function on the unit ball of
$\Rs$, where the elements $\vib$ represent velocities. Apart from this
function, \eqref{eq-heuristic-velocity-integral} coincides with the
integral \eqref{eq-heuristic-integral} encountered on page
\pageref{eq-heuristic-integral} in the heuristic considerations of
Chapter~\ref{chap-localizing-operators}. The investigations carried
through in that part (cf.~Proposition~\ref{Pro-counter-integrals})
imply that \eqref{eq-heuristic-velocity-integral} takes on a finite
value for any counter $C \in \Cfrak$ at any time $\tau$ and, according
to Lemma~\ref{Lem-explicit-pq-extension}, the integral
\eqref{eq-heuristic-velocity-integral} even exists for all $C \in
\AC$.

The physical interpretation is as follows: Consider a function $h$ of
bounded support $\Vsf \subseteq \Rs \setminus \set{0}$ in velocity
space, then the integral \eqref{eq-heuristic-velocity-integral}
corresponds to summing up, for given time $\tau$, the expectation
values of measurements of $C$ in the state $\omega$, where these
measurements extend over the bounded section $\tau \cdot \Vsf$ of
configuration space. For growing $\tau$ the distance of this portion
from the origin increases together with its total extension. More
exactly, the measurements take place in a cone with apex at the point
$0$ of space-time, its direction is determined by the support of $h$,
and for different times $\tau$ the counter $C$ is set up in specific
parts of that cone, their extension growing as $\abs{\tau}^{s}$
(compensating for the quantum mechanical spreading of wave packets)
while their distance from the origin increases proportional to
$\abs{\tau}$. If the physical state $\omega$ has, in the limit of
large (positive or negative) times, evolved into a configuration
containing a particle (incoming or outgoing) travelling with velocity
$\vib_0 \in \Vsf$, then a counter $C_0$, sensitive for that specific
particle, is expected to asymptotically produce a stable signal under
the above experimental conditions.

The mathematical equivalent of this situation is the existence of
limits of the above integral at asymptotic times, evaluated for the
counter $C_0$ and a function $h_0$ with support containing
$\vib_0$. Thus the problem has to be settled in which (topological)
sense such limits can be established, if they happen to exist at
all. To tackle this assignment we turn to a slightly modified version
of \eqref{eq-heuristic-velocity-integral} in
Definition~\ref{Def-rho-definition}, involving, for technical reasons,
a certain time average.
\begin{Def}
  \label{Def-rho-definition}
  Let $\Delta$ be a bounded Borel subset of $\Rsone$, let $\omega \in
  \Sscr ( \Delta )$ denote a physical state of bounded energy and let
  $\vib \mapsto h ( \vib )$ be a bounded measurable function on the
  unit ball of $\Rs$. Furthermore suppose that $t \mapsto T ( t )$ is
  a continuous real-valued function, approaching $+ \infty$ or $-
  \infty$ for asymptotic positive or negative times, respectively, not
  as fast as $\abs{t}$. Then we define a net $\bset{ \rho_{h , t} : t
  \in \Rbb}$ of linear functionals on $\Cfrak$ by setting
  \begin{equation}
    \label{eq-finite-times-functionals}
    \begin{split}
      \rho_{h , t} ( C ) & \doteq T ( t )^{-1} \int_t^{t + T ( t )} d
      \tau \; \tau^s \int_{\Rs} d^s v \, h ( \vib ) \, \omega \bigl(
      \alpha_{( \tau , \tau \vib )} ( C ) \bigr) \\
      & = T ( t )^{-1} \int_t^{t + T ( t )} d \tau \int_{\Rs} d^s x \;
      h ( \tau^{-1} \xib ) \, \omega \bigl( \alpha_{( \tau ,
      \xib )} ( C ) \bigr) \text{,} \quad C \in \Cfrak \text{.}
    \end{split}
  \end{equation}
\end{Def}

Under the above assumptions the functionals $\rho_{h , t}$ turn out to
be continuous with respect to the seminorm $\pD$ pertaining to the
energy-momentum support of the physical state $\omega (~.~)= \omega
\bigl( \ED~.~\ED \bigr)$, i.\,e.~$\rho_{h , t} \in \CDstar$. This can
be seen as follows: First, note that the operators $U ( \tau )$
implementing time translations commute with $\ED$, so that
\begin{equation*}
  \omega \bigl( \ED \alpha_{( \tau , \xib )} ( C ) \ED \bigr) = \omega
  \bigl( U ( \tau ) \ED \aibx ( C ) \ED U ( \tau )^* \bigr) \text{,}
\end{equation*}
which allows \eqref{eq-finite-times-functionals} to be re-written as
\begin{equation}
  \label{eq-rho-alt-form}
  \rho_{h , t} ( C ) = T ( t )^{-1} \int_t^{t + T ( t )} d \tau
  \int_{\Rs} d^s x \; h ( \tau^{-1} \xib ) \, \omega \bigl( U ( \tau )
  \ED \aibx ( C ) \ED U ( \tau )^* \bigr) \text{.}
\end{equation}
Now, all the functionals $\omega \bigl( U ( \tau )~.~U ( \tau )^*
\bigr)$, $\tau \in \Rbb$, belong to $\BH_{*,1}$, so that the absolute
value of $\rho_{h , t} ( C )$ can be estimated, making use of $\pD$ as
defined in \eqref{eq-p-seminorm}. Abbreviating the interval of
$\tau$-integration depending on $t$ as $I_t$, this gives
\begin{multline}
  \label{eq-rho-estimate}
  \babs{\rho_{h , t} ( C )} \leqslant \sup_{\tau \in I_t} \Babs{
  \int_{\Rs} d^s x \; h ( \tau^{-1} \xib ) \, \omega \bigl( U ( \tau )
  \ED \aibx ( C ) \ED U ( \tau )^* \bigr)} \\
  \leqslant \norm{h}_\infty \sup_{\phi \in \BH_{*,1}} \int_{\Rs} d^s x
  \babs{\phi \bigl( \ED \aibx ( C ) \ED \bigr)} = \norm{h}_\infty
  \pDx{C} \text{.}
\end{multline}
The above inequality implies that  the functionals $\rho_{h , t}$
belong to the dual space $\Cstar$ of $( \Cfrak , \Tfrak_p
)$. Moreover, the estimate \eqref{eq-rho-estimate} is uniform in $t$,
so that the net $\bset{\rho_{h , t} : t \in \Rbb}$ is even an
equicontinuous subset of $\Cstar$. The Theorem of Alao\u{g}lu-Bourbaki
\cite[Theorem~8.5.2]{jarchow:1981} then tells us, that this net is
relatively compact with respect to the weak topology, leading to the
following fundamental result.
\begin{The}[Existence of Limits]
  \label{The-singular-limits}
  Under the assumptions of Definition~\ref{Def-rho-definition} the net
  $\bset{\rho_{h , t} : t \in \Rbb} \subseteq \CDstar$ possesses weak
  limit points in $\Cstar$ at asymptotic times. This means that there
  exist functionals $\sigma_{h , \omega}^{( + )}$ and $\sigma_{h ,
  \omega}^{( - )}$ on $\Cfrak$ together with corresponding subnets
  $\bset{\rho_{h , t_\iota} : \iota \in J}$ and $\bset{\rho_{h ,
  t_\kappa} : \kappa \in K}$, i.\,e.~$\lim_\iota t_\iota = + \infty$
  and $\lim_\kappa t_\kappa = - \infty$, such that for arbitrary $C
  \in \Cfrak$
  \begin{subequations}
    \label{eq-singular-limits}
    \begin{align}
      \rho_{h , t_\iota} ( C ) & \underset{\iota}{\longrightarrow}
      \sigma_{h , \omega}^{( + )} ( C ) \text{,} \\
      \rho_{h , t_\kappa} ( C ) & \underset{\kappa}{\longrightarrow}
      \sigma_{h , \omega}^{( - )} ( C ) \text{.}
    \end{align}
  \end{subequations}
\end{The}

The heuristic picture laid open above suggests, that in theories which
are reasonable from a physicist's point of view the net $\bset{\rho_{h
, t} : t \in \Rbb}$ actually converges, but as yet we have not been
able to give rigorously formulated conditions under which to prove
this conjecture. This question seems to be connected with the problem
of asymptotic completeness of quantum field theoretic models; one has
to assure that in the limit of large times multiple scattering does no
longer withhold the measurement results $\rho_{h , t} ( C )$ from
growing stable. Another possibility is the disappearance of the limit
functionals $\sigma_{h , \omega}^{( + )}$ and $\sigma_{h , \omega}^{(
- )}$ on all of the algebra of counters $\Cfrak$, a phenomenon that we
anticipate to encounter in theories without a particle interpretation
(e.\,g.~generalized free field). The denomination of the asymptotic
functionals `$\sigma$' is chosen to reflect their \emph{singular}
nature: the values that the functionals $\rho_{h , t}$ return for
finite times $t$ when applied to the identity operator $\unit$ (which
is not contained in $\Cfrak$) are divergent as $\abs{t}^s$ at
asymptotic times.

The convergence problem as yet only partially solved in the sense of
Theorem~\ref{The-singular-limits}, one can nevertheless establish a
number of distinctive properties of the limit functionals $\sigma$
(from now on we will skip sub- and superscripts not to overburden the
notation), that allow for their interpretation in terms of asymptotic
configurations of particles. An immediate first consequence of the
above construction is the following proposition.
\begin{Pro}[Positivity and Continuity of Limits]
  \label{Pro-positivity-of-limits}
  Suppose that $\Delta$ is a bounded Borel subset of $\Rsone$, $\omega
  \in \Sscr ( \Delta )$ a physical state of bounded energy and $h \in
  L^\infty ( \Rs , d^s x )$ a non-negative function. Then the limit
  functionals $\sigma$ for the net $\bset{\rho_{h , t} : t \in \Rbb}$
  are positive elements of $\CDstar$:
  \begin{subequations}
    \begin{align}
      \babs{\sigma ( C )} & \leqslant \norm{h}_\infty \, \pDx{C}
      \text{,} \quad C \in \Cfrak \text{;} \\
      0 & \leqslant \sigma ( C ) \text{,} \quad C \in \Cfrak^+
      \text{.}
    \end{align}
  \end{subequations}
\end{Pro}
\begin{Rem}
  Due to the continuity of $\rho_{h , t}$ and $\sigma$ with respect to
  the $\pD$-topology, these functionals can be continuously extended
  to $\Cfrakbar$ as well as $\AC$, where $\rho_{h , t}$ are explicitly
  given on $\AC$ by the formula \eqref{eq-finite-times-functionals}
  with $C \in \AC$. It is then easily established, by use of elements
  $C'$ from $\Cfrak$ lying in suitable $\pD$-neighbourhoods of $C$,
  that the relations \eqref{eq-singular-limits} remain valid on this
  larger subspace of the quasi-local algebra $\Afrak$.
\end{Rem}
The next result deals  with the effect that space-time translations
exert on these limit functionals. A further assumption on the velocity
implementation $h \in L^\infty ( \Rs , d^s v )$ turns out to be
indispensible in their proof: $h$ has to be continuous, approximating
a constant value in the limit $\abs{\vib} \rightarrow \infty$,
i.\,e.~$h - M_h \in C_0 ( \Rs )$ for a suitable constant $M_h$; these
functions constitute a subspace of $C ( \Rs )$ that will be denoted
$C_{0 , c} ( \Rs )$ in the sequel.
\begin{Pro}[Translation Invariance]
  \label{Pro-translation-invariance}
  Let $\Delta \subseteq \Rsone$ be a bounded Borel set, let $\omega
  \in \Sscr ( \Delta )$ and $h \in C_{0 , c} ( \Rs )$. Then the limit
  functionals $\sigma$ of $\bset{\rho_{h , t_\iota} : \iota \in J}$
  are invariant under space-time translations:
  \begin{equation}
  \label{eq-sigma-invariance}
    \sigma \bigl( \ax ( C ) \bigr) = \sigma ( C )
  \end{equation}
  for any $C \in \AC$ and any $x \in \Rsone$.
\end{Pro}
\begin{proof}
  Taking into account the fact that the Lebesgue measure on $\Rsone$
  is invariant under translations, one can express $\rho_{h , t}
  \bigl( \alpha_{( x^0 , \xib )} ( C ) \bigr)$ for any finite time $t$
  and any given $x = ( x^0 , \xib ) \in \Rsone$ by the following
  integral
  \begin{equation*}
    \rho_{h , t} \bigl( \alpha_{( x^0 , \xib )} ( C ) \bigr) = T ( t
    )^{-1} \int_{t + x^0}^{t + x^0 + T ( t )} d \tau \int_{\Rs} d^s y
    \; h \bigl( ( \tau - x^0 )^{-1} ( \yib - \xib ) \bigr) \, \omega
    \bigl( \alpha_{( \tau , \yib )} ( C ) \bigr) \text{.}
  \end{equation*}
  Next, we want to evaluate $\babs{\rho_{h , t} ( C ) - \rho_{h , t}
  \bigl( \alpha_{( x^0 , \xib )} ( C ) \bigr)}$ which, according to
  the respective limits of $\tau$-integration, can be split into a sum
  of three integrals to be estimated separately:
  \begin{align*}
    \Babs{T ( t )^{-1} \int_t^{t + x^0} d \tau \int_{\Rs} d^s y \; h (
    \tau^{-1} \yib ) \, \omega \bigl( \alpha_{( \tau , \yib )} ( C )
    \bigr)} & \leqslant \abs{T ( t )}^{-1} \abs{x^0} \,
    \norm{h}_\infty \, \pDx{C} \text{,} \\
    \Babs{T ( t )^{-1} \int_{t + x^0 + T ( t )}^{t + T ( t )} d \tau
    \int_{\Rs} d^s y \; h ( \tau^{-1} \yib ) \, \omega \bigl(
    \alpha_{( \tau , \yib )} ( C ) \bigr)} & \leqslant \abs{T ( t
    )}^{-1} \abs{x^0} \, \norm{h}_\infty \, \pDx{C} \text{;}
  \end{align*}
  both $\rho_{h , t} ( C )$ and $\rho_{h , t} \bigl( \alpha_{( x^0 ,
  \xib )} ( C ) \bigr)$ contribute to the third integral
  \begin{multline*}
    \Babs{T ( t )^{-1} \int_{t + x^0}^{t + x^0 + T ( t )} d \tau
    \int_{\Rs} d^s y \; \bigl[ h ( \tau^{-1} \yib ) - h \bigl( ( \tau
    - x^0 )^{-1} ( \yib - \xib ) \bigr) \bigr] \, \omega \bigl(
    \alpha_{( \tau , \yib )} ( C ) \bigr)} \\
    \leqslant \sup_{\tau \in I_{t , x^0}} \, \sup_{\yib \in \Rs}
    \babs{h ( \tau^{-1} \yib ) - h \bigl( ( \tau - x^0 )^{-1} ( \yib -
    \xib ) \bigr)} \, \pDx{C} \text{,}
  \end{multline*}
  where we used the abbreviation $I_{t , x^0}$ for the interval of
  $\tau$-integration. Setting (for $\abs{\tau}$ large enough)
  \begin{equation*}
    \zib_\tau \doteq \zib + ( \tau - x^0 )^{-1} ( x^0 \zib - \xib )
  \end{equation*}
  we finally arrive at the estimate
  \begin{equation}
    \label{eq-translation-invariance-estimate}
    \babs{\rho_{h , t} ( C ) - \rho_{h , t} \bigl( \alpha_{( x^0 ,\xib
    )} ( C ) \bigr)} 
    \leqslant \Bigl( 2 \, \abs{T ( t )}^{-1} \abs{x^0} \,
    \norm{h}_\infty + \sup_{\tau \in I_{t , x^0}} \, \sup_{\zib \in
    \Rs} \abs{h ( \zib ) - h ( \zib_\tau )} \Bigr) \, \pDx{C} \text{.}
  \end{equation}
  The net $\bset{\zib_\tau : \tau \in \Rbb}$ approximates $\zib$
  uniformly on compact subsets of $\Rs$ in the limit of large
  $\abs{\tau}$, i.\,e.~given $\epsilon > 0$ and $R' > 0$ there exists
  a positive number $T'$ such that $\abs{\tau} > T'$ implies
  $\abs{\zib - \zib_\tau} < \epsilon$ for any $\zib \in \Rs$ with
  $\abs{\zib} \leqslant R'$. On the other hand, given $R'' > 0$ there
  exists $T'' > 0$ such that $\abs{\zib_\tau} > \frac{1}{2} R''$ for
  any $\abs{\zib} > R'' $ and any $\abs{\tau} > T''$. Combining these
  results with the special properties of $h \in C_{0 , c} ( \Rs )$,
  i.\,e.~uniform continuity on compact balls in $\Rs$ and approximate
  constancy at infinity, we infer that for large $\abs{\tau}$ the term
  $\sup_{\zib \in \Rs} \babs{h ( \zib ) - h ( \zib_\tau )}$ falls
  below any given positive bound. Therefore the right-hand side of
  \eqref{eq-translation-invariance-estimate} vanishes with $\abs{t}
  \rightarrow \infty$ since $\abs{T ( t )}$ exceeds any positive value
  in this limit.

  Now, let $\sigma$ be the weak limit of the subnet $\bset{\rho_{h ,
  t_\iota} : \iota \in J}$, i.\,e. 
  \begin{equation*}
    \rho_{h , t_\iota} ( C ) \underset{\iota}{\longrightarrow} \sigma
    ( C ) \text{,}
  \end{equation*}
  then there holds for any $\iota \in J$, any $C \in \AC$ and any $x
  \in \Rsone$ the subsequent inequality
  \begin{multline*}
    0 \leqslant \babs{\sigma \bigl( \ax ( C ) \bigr) - \sigma ( C )}
    \\
    \leqslant \babs{\sigma \bigl( \ax ( C ) \bigr) - \rho_{h ,
    t_\iota} \bigl( \ax ( C ) \bigr)} + \babs{\rho_{h , t_\iota}
    \bigl( \ax ( C ) \bigr) - \rho_{h , t_\iota} ( C )} +
    \babs{\rho_{h , t_\iota} ( C ) - \sigma ( C )} \text{.}
  \end{multline*}
  By the reasoning of the preceding paragraph and the above condition
  for subnet convergence, all three terms on the right-hand side
  vanish with respect to the directed set $J$, since in this limit
  $\abs{t_\iota} \rightarrow \infty$. As a result the intermediate
  term has to be equal to $0$, thereby establishing translation
  invariance of $\sigma$.
\end{proof}

The last property that we are going to demonstrate in this section for
those special elements $\sigma \in \CDstarplus$, that arise as limits
of nets of functionals $\bset{\rho_{h , t_\iota} : \iota \in J}$,
complements the Cluster Property~\ref{Pro-cluster}. It asserts, given
certain specific operators $C \in \Cfrak$, the existence of
\emph{lower} bounds for integrals of the functions $\xib \mapsto
\sigma \bigl( C^* \aibx ( C ) \bigr)$.
\begin{Pro}[Existence of Lower Bounds]
  \label{Pro-lower-bounds}
  Let $C \in \Cfrak$ be a counter which has the property that the
  function $\xib \mapsto \bpDx{C^* \aibx ( C )}$ is integrable
  (cf.~Lemma~\ref{Lem-varsigmanet-convergence}). Let furthermore
  $\sigma \in \CDstarplus$ be the limit of a net of functionals
  $\bset{\rho_{h , t_\iota} : \iota \in J}$, each defined by
  \eqref{eq-finite-times-functionals}, where the velocity function $h$
  is non-negative and belongs to $C_{0 , c} ( \Rs )$. Under these
  assumptions
  \begin{equation}
    \label{eq-lower-bounds}
    \babs{\sigma ( C )}^2 \leqslant \norm{h}_\infty \int_{\Rs} d^s x
    \; \sigma \bigl( C^* \aibx ( C ) \bigr) \text{.}
  \end{equation}
\end{Pro}
\begin{proof}
  Consider the functional $\rho_{h , t}$ at finite time $t$. Applying
  to the absolute value of its defining equation
  \eqref{eq-finite-times-functionals} the Cauchy-Schwarz inequality
  with respect to the inner product ($\abs{t}$ large enough)
  \begin{equation*}
    ( f , g )_t \doteq T ( t )^{-1} \int_t^{t + T ( t )} d \tau \;
    \overline{f ( \tau )} \, g ( \tau )
  \end{equation*}
  of square-integrable functions $f$ and $g$ depending on the time
  variable $\tau \in I_t$, one gets in the special case of
  \begin{equation*}
    f ( \tau ) \equiv 1 \quad \text{and} \quad g ( \tau ) = \int_{\Rs}
    d^s x \; h ( \tau^{-1} \xib ) \, \omega \bigl( \alpha_{( \tau ,
    \xib )} ( C ) \bigr)
  \end{equation*}
  the estimate
  \begin{multline}
    \label{eq-csi-estimate}
    \babs{\rho_{h , t} ( C )}^2 = \Babs{T ( t )^{-1} \int_t^{t + T ( t
    )} d \tau \int_{\Rs} d^s x \; h ( \tau^{-1} \xib ) \, \omega
    \bigl( \alpha_{( \tau , \xib )} ( C ) \bigr)}^2 \\
    \leqslant T ( t )^{-1} \int_t^{t + T ( t )} d \tau \;
    \Babs{\int_{\Rs} d^s x \; h ( \tau^{-1} \xib ) \, \omega \bigl(
    \alpha_\tau \bigl( \aibx ( C ) \bigr) \bigr)}^2 \text{.}
  \end{multline}
  Now, let $\Kib$ be a compact subset of $\Rs$; then, by positivity of
  the functional $\omega \in \Sscr ( \Delta )$,
  \cite[Proposition~2.3.11(b)]{bratteli/robinson:1987} together with
  the Fubini Theorem \cite[II.16.3]{fell/doran:1988a} leads for
  arbitrary $\tau \in \Rbb$ to
  \begin{multline*}
    \Babs{\omega \Bigl( \alpha_\tau \Bigl( \int_\Kib d^s x \; h (
    \tau^{-1} \xib ) \, \aibx ( C ) \Bigr) \Bigr)}^2 \\
    \leqslant \omega \Bigl( \alpha_\tau \Bigl( \int_\Kib d^s x
    \int_\Kib d^s y \; h ( \tau^{-1} \yib ) \, h ( \tau^{-1} \xib ) \,
    \aiby ( C^* ) \, \aibx ( C ) \Bigr) \Bigr) \text{,}
  \end{multline*}
  which is preserved in the limit $\Kib \nearrow \Rs$, which exists
  on account of the assumed integrability of the mapping $\xib \mapsto
  \bpDx{C^* \aibx ( C )}$. On commuting $\omega \circ \alpha_\tau$ and
  the integrals one arrives at
  \begin{multline}
    \label{eq-pos-omega-estimate}
      \Babs{\int_{\Rs} d^s x \; h ( \tau^{-1} \xib ) \, \omega \bigl(
      \alpha_\tau \bigl( \aibx ( C ) \bigr) \bigr)}^2 \\
      \leqslant \int_{\Rs} d^s x \int_{\Rs} d^s y \; h ( \tau^{-1}
      \yib ) \, h ( \tau^{-1} \xib ) \, \omega \bigl( \alpha_\tau
      \bigl( \aiby ( C^* ) \aibx ( C ) \bigr) \bigr) \leqslant
      \norm{h}_\infty^2 \int_{\Rs} d^s x \; \bpDx{C^* \aibx ( C )}
  \end{multline}
  and the combination of \eqref{eq-csi-estimate} and
  \eqref{eq-pos-omega-estimate} gives
  \begin{equation}
    \label{eq-square-rho-estimate}
    \babs{\rho_{h , t} ( C )}^2 \leqslant T ( t )^{-1} \int_t^{t + T (
    t )} d \tau \int_{\Rs} d^s x \int_{\Rs} d^s y \; h ( \tau^{-1}
    \yib ) \, h ( \tau^{-1} \xib ) \, \omega \bigl( \alpha_\tau \bigl(
    \aiby ( C^* ) \aibx ( C ) \bigr) \bigr) \text{.}
  \end{equation}
  We want to replace the term $h ( \tau^{-1} \xib )$ by the norm
  $\norm{h}_\infty$ and, to do so, define the function $h_+ \doteq
  ( \norm{h}_\infty h - h^2 )^{1/2}$, which is a non-negative element
  of $C_{0 , c} ( \Rs )$ as is $h$ itself. Then for any $\zib \text{,}
  \zib' \in \Rs$ there holds the equation
  \begin{equation}
    \label{eq-hh-decomposition}
    \norm{h}_\infty h ( \zib ) = h ( \zib ) \, h ( \zib' ) + h_+ (
    \zib ) \, h_+ ( \zib' ) +  h_+ ( \zib ) \bigl( h_+ ( \zib ) - h_+
    ( \zib' ) \bigr) + h ( \zib ) \bigl( h ( \zib ) - h ( \zib' )
    \bigr) \text{.}
  \end{equation}
  Next, consider for an arbitrary function $g \in C_{0 , c} ( \Rs )$
  the following inequality, based on an application of Fubini's
  Theorem and the reasoning of \eqref{eq-rho-estimate},
  \begin{multline}
    \label{eq-intermediate-integral-estimate}
    \Babs{T ( t )^{-1} \int_t^{t + T ( t )} d \tau \int_{\Rs} d^s x
      \int_{\Rs} d^s y \; g ( \tau^{-1} \yib ) \bigl( g ( \tau^{-1}
      \yib ) - g ( \tau^{-1} \xib ) \bigr) \, \omega \bigl(
      \alpha_\tau \bigl( \aiby ( C^* ) \aibx ( C ) \bigr) \bigr)} \\
      = \Babs{\int_{\Rs} d^s x \; T ( t )^{-1} \int_t^{t + T ( t )} d
      \tau \int_{\Rs} d^s z \; \tau^s g ( \zib ) \bigl( g ( \zib ) - g
      \bigl( \zib_\tau ( \xib ) \bigr) \bigr) \, \omega \bigl(
      \alpha_{( \tau , \tau \zib )} \bigl( C^* \aibx ( C ) \bigr)
      \bigr)} \\
      \leqslant \norm{g}_\infty \int_{\Rs} d^s x \; \sup_{\tau \in
      I_t} \, \sup_{\zib \in \Rs} \babs{g ( \zib ) - g \bigl(
      \zib_\tau ( \xib ) \bigr)} \, \bpDx{C^* \aibx ( C )} \text{,}
  \end{multline}
  where we made use of the coordinate transformation $\xib
  \rightsquigarrow \xib + \yib$ followed by the transformation $\yib
  \rightsquigarrow \zib \doteq \tau^{-1} \yib$ and introduced the
  abbreviations $\zib_\tau ( \xib ) \doteq \tau^{-1} \xib + \zib$ as
  well as $I_t$ for the interval of $\tau$-integration. Similar to the
  proof of Proposition~\ref{Pro-translation-invariance}, the
  expression $\sup_{\tau \in I_t} \sup_{\zib \in \Rs} \abs{g ( \zib )
  - g \bigl( \zib_\tau ( \xib ) \bigr)}$ is seen to vanish for all
  $\xib \in \Rs$ in the limit of large $\abs{t}$, so that by
  Lebesgue's Dominated Convergence Theorem the left-hand side of
  \eqref{eq-intermediate-integral-estimate} converges to 0. This
  reasoning in particular applies to the functions $h$ as well as
  $h_+$ and thus to the third and fourth term on the right of equation
  \eqref{eq-hh-decomposition}. On the other hand, substitution of $h$
  by $h_+$ in the integral of \eqref{eq-square-rho-estimate} likewise
  gives a non-negative result for all times $t$. Combining all these
  informations and specializing to a subnet $\bset{t_\iota : \iota \in
  J}$ approximating $+ \infty$ or $- \infty$, one arrives at the
  following version of \eqref{eq-square-rho-estimate}, valid for
  asymptotic times:
  \begin{multline*}
    \lim_\iota \babs{\rho_{h , t_\iota} ( C )}^2 \\
    \leqslant \lim_\iota \norm{h}_\infty T ( t_\iota)^{-1}
    \int_{t_\iota}^{t_\iota + T( t_\iota )} d \tau \int_{\Rs} d^s x
    \int_{\Rs} d^s y \; h ( \tau^{-1} \yib ) \, \omega \bigl(
    \alpha_\tau \bigl( \aiby ( C^* ) \aibx ( C ) \bigr) \bigr) \\
    \leqslant \norm{h}_\infty \lim_\iota \int_{\Rs} d^s x \; T (
    t_\iota )^{-1} \int_{t_\iota}^{t_\iota + T ( t_\iota )} d \tau
    \int_{\Rs} d^s y \; h ( \tau^{-1} \yib ) \, \omega \bigl(
    \alpha_{( \tau , \yib )} \bigl( C^* \aibx ( C ) \bigr) \bigr) \\
    = \norm{h}_\infty \lim_\iota \int_{\Rs} d^s x \; \rho_{h ,
    t_\iota} \bigl( C^* \aibx ( C ) \bigr) \text{.}
  \end{multline*}
  Making use of Lemma~\ref{Lem-varsigmanet-convergence}, this result
  can be expressed in terms of the functional $\sigma = \lim_\iota
  \rho_{h , t_\iota}$ to yield
  \begin{equation*}
    \babs{\sigma ( C )}^2 \leqslant \norm{h}_\infty \int_{\Rs} d^s x
    \; \sigma \bigl( C^* \aibx ( C ) \bigr) \text{.} \tag*{\qed}
  \end{equation*}
  \renewcommand{\qed}{}
\end{proof}

\section{Particle Weights}
  \label{sec-particle-weights}

The features of limit functionals $\sigma \in \CDstarplus$ collected
thus far, point to their interpretation as representatives of
mixtures of particle-like quantities with sharp energy-momentum: being
translationally invariant according to
Proposition~\ref{Pro-translation-invariance}, they appear as plane
waves, i.\,e.~energy-momentum eigenstates, on the other hand they are
singly localized at all times by Proposition~\ref{Pro-cluster},
thereby exhibiting properties of particle-like systems, their
energy-momentum spectrum being determined by
Proposition~\ref{Pro-spectral-property}. We shall summarize systems of
the above kind under the concept of \emph{particle weights}, a term
chosen to reflect the connection to the notion of `weights' or
`extended positive functionals' in the theory of $C^*$-algebras, going
back to Dixmier \cite[Section~I.4.2]{dixmier:1981} (cf.~also
\cite[Section~5.1]{pedersen:1979} and \cite{pedersen:1966}). These
designate functions on the positive cone $\Afrak^+$ of a $C^*$-algebra
$\Afrak$ which can attain infinite values, a property they share with
the singular functionals constructed in
Theorem~\ref{The-singular-limits}: it was seen to be of importance
that their domain $\Cfrak$ does not comprise the element $\unit$ of
the quasi-local algebra, for the defining approximation would then
lead to the value $\sigma ( \unit ) = + \infty$.

As already mentioned in Section~\ref{sec-gen-prop}, every positive
functional $\sigma$ on $\Cfrak = \Lfrak^* \, \Lfrak$ defines a
non-negative sesquilinear form $\scp{~.~}{~.~}_\sigma$ on $\Lfrak
\times \Lfrak$ via
\begin{equation}
  \label{eq-sesquilinear-form}
  \scp{L_1}{L_2}_\sigma \doteq \sigma ( {L_1}^* L_2 )
\end{equation}
for any $L_1 \text{,} L_2 \in \Lfrak$, which induces a seminorm
$q_\sigma$ on $\Lfrak$ and a norm $\norm{~.~}_\sigma$ on the
corresponding quotient of $\Lfrak$ by the null space $\Nfrak_\sigma$
of $q_\sigma$. Taking advantage of these constructions, we shall
depart from functionals and proceed to sesquilinear forms, a step
which is necessitated by the special demands of the subsequent
analysis. The following definition consists of a r\'{e}sum\'{e} of the
essence of our knowledge on asymptotic functionals acquired in the
above sequence of propositions.
\begin{Def}
  \label{Def-particle-weight}
  A particle weight is a non-trivial, non-negative sesquilinear form
  on $\Lfrak$, written $\scp{~.~}{~.~}$, which induces by
  \eqref{eq-varsigma-quotient-constructions} on the ideal $\Lfrak$ a
  seminorm $q_w$ with null space $\Nfrak_w$ as well as a norm
  $\norm{~.~}_w$ on the quotient $\Lfrak / \Nfrak_w$, and which
  complies with the following assumptions:
  \begin{proplist}
  \item for any $L_1 \text{,} L_2 \in \Lfrak$ and $A \in \Afrak$ there
    holds the relation 
    \begin{equation*}
      \scp{L_1}{A \, L_2} = \scp{A^* L_1}{L_2} \text{;}
    \end{equation*}
  \item for given $L \in \Lfrak$ the following mapping is continuous
    with respect to $\qw$:
    \begin{equation*}
      \Xi_L : \Poin \rightarrow \Lfrak \qquad ( \Lambda , x ) \mapsto
      \Xi_L ( \Lambda , x ) = \aLax ( L ) \text{;}
    \end{equation*}
  \item the restriction to the subspace $\Lfrak_0$ of the canonical
    homomorphism
    \begin{equation*}
      \Qscr_w : \Lfrak \rightarrow \Lfrak / \Nfrak_w \quad L \mapsto
      \Qscr_w ( L ) \doteq [ L ]_w
    \end{equation*}
    is $\Xscr_{\Lfrak_0}$-differentiable in the sense of
    Definition~\ref{Def-generalized-diff-lin-mappings};
  \item the sesquilinear form is invariant with respect to space-time
    translations $x \in \Rsone$, i.\,e.
    \begin{equation*}
      \bscp{\ax ( L_1 )}{\ax ( L_2 )} = \scp{L_1}{L_2} \text{,} \quad
      L_1 \text{,} L_2 \in \Lfrak \text{,}
    \end{equation*}
    and the $(s + 1)$-dimensional Fourier transforms of the
    distributions
    \begin{equation*}
      x \mapsto \bscp{L_1}{\ax ( L_2 )}
    \end{equation*}
    have support in a shifted forward light cone $\fwcone - q$, where
    $q \in \fwcone$.
  \end{proplist}
\end{Def}
\begin{Rem}
  \begin{remlist}
  \item Note, that we did not impose on $\scp{~.~}{~.~}$ any
    restrictions concerning continuity with respect to the
    $\qD$-topology of $\Lfrak$, for in general such conditions will
    get lost in the disintegration of particle weights to be expounded
    in Chapter~\ref{chap-disintegration}. The continuity property,
    which actually depends on the topology of $\Lfrak$, is formulated
    in terms of the seminorm $q_w$ induced by the sesquilinear form
    under consideration. The constituent properties of the above
    definition are preserved under the operations of addition and of
    multiplication by positive numbers, so that the totality of
    particle weights supplemented by the trivial form proves to be a
    positive (proper convex) cone
    (cf.~\cite{peressini:1967,asimow/ellis:1980}), denoted $\Wecm$, in
    the linear space of all sesquilinear forms on $\Lfrak$. This
    ascertainment is the foundation for the constructions of
    Chapter~\ref{chap-choquet}.
  \item One could be tempted to go the way back from a sesquilinear
    form of the above type to a positive linear functional on
    $\Cfrak$, but this is by no means self-evident. It is only
    possible under restrictive assumptions on the structure of the
    algebra $\Cfrak$ to make the definition of the associated
    functional unambiguous.
  \end{remlist}
\end{Rem}

A completely equivalent characterization of particle weights can be
given in terms of representations $( \pi_w , \Hscr_w )$ of the
quasi-local algebra $\Afrak$, obtained by means of a GNS-construction
(cf.~\cite[Theorem~3.2]{pedersen:1966} and
\cite[Proposition~5.1.3]{pedersen:1979}).
\begin{The}
  \label{The-particle-weight}
  \begin{theolist}
  \item To any particle weight $\scp{~.~}{~.~}$ there corresponds a
    non-zero, non-degenerate representation $( \pi_w , \Hscr_w )$ of
    the quasi-local $C^*$-algebra $\Afrak$ with the following
    properties:
    \begin{theosublist}
    \item there exists a linear mapping $\ket{~.~}$ from $\Lfrak$ onto
      a dense subspace of $\Hscr_w$
      \begin{equation*}
        \ket{~.~} : \Lfrak \rightarrow \Hscr_w \qquad L \mapsto
        \ket{L} \text{,}
      \end{equation*}
      such that the representation $\pi_w$ is given by
      \begin{equation*}
        \pi_w ( A ) \ket{L} = \ket{A L} \text{,} \quad A \in \Afrak
        \text{,} \quad L \in \Lfrak \text{;}
      \end{equation*}
    \item the following mapping is continuous for given $L \in
      \Lfrak$:
      \begin{equation*}
        \ket{\Xi_L (~.~)} : \Poin \rightarrow \Hscr_w \quad ( \Lambda
        , x ) \mapsto \bket{\Xi_L ( \Lambda , x )} = \bket{\aLax ( L
        )} \text{;}
      \end{equation*}
    \item the restriction of the linear mapping $\ket{~.~}$ to
      $\Lfrak_0$  with range in the subspace of $\Hscr_w$ spanned by
      all vectors $\ket{L_0}$, $L_0 \in \Lfrak_0$, is
      $\Xscr_{\Lfrak_0}$-differentiable;
    \item there exists a strongly continuous unitary representation $x
      \mapsto U_w ( x )$ of space-time translations $x \in \Rsone$ on
      $\Hscr_w$ defined by
      \begin{equation*}
        U_w ( x ) \ket{L} \doteq \bket{\ax ( L )} \text{,} \quad L \in
        \Lfrak \text{,}
      \end{equation*}
      with spectrum in a displaced forward light cone $\fwcone - q$,
      $q \in \fwcone$.
    \end{theosublist}
  \item Any representation $( \pi_w , \Hscr_w )$ which has the above
    characteristics defines a particle weight through the scalar
    product on $\Hscr_w$.
  \end{theolist}
\end{The}
\begin{Rem}
  By their very definition, the unitaries $U_w ( x )$ implement the
  automorphism group $\bset{\ax : x \in \Rsone} \subseteq \Aut \Afrak$
  through
  \begin{equation}
    \label{eq-auto-implement}
    U_w ( x ) \pi_w ( A ) {U_w ( x )}^* = \pi_w \bigl( \ax ( A )
    \bigr) \text{,} \quad A \in \Afrak \text{,~} x \in \Rsone \text{,}
  \end{equation}
  in the representation $( \Hscr_w , \pi_w )$.
\end{Rem}
\begin{proof}
  \begin{Prooflist}
  \item The proof of the various properties stated in the Theorem is
    readily carried out, once the GNS-construction has been realized.
    \begin{Proofsublist}
    \item Since a particle weight satisfies the Cauchy-Schwarz
      inequality its null space
      \begin{equation*}
        \Nfrak_w \doteq \bset{N \in \Lfrak : \scp{N}{N} = 0}
      \end{equation*}
      turns out to be a left ideal in $\Lfrak$ (and hence in
      $\Afrak$). The defining sesquilinear form endows the quotient
      space of $\Lfrak$ by $\Nfrak_w$ with a pre-Hilbert space
      structure; its completion $\Hscr_w$ contains by construction the
      range of the canonical homomorphism
      \begin{equation*}
        \ket{~.~} : \Lfrak \rightarrow \Lfrak / \Nfrak_w \quad L
        \mapsto \ket{L} \doteq [ L ]_w
      \end{equation*}
      as a dense subspace. $\Lfrak$ and $\Nfrak_w$ being left ideals
      in $\Afrak$, the definition
      \begin{equation*}
        \pi_w ( A ) \ket{L} \doteq \ket{A L} \text{,} \quad A \in
        \Afrak \text{,}
      \end{equation*}
      makes sense on the range of $\ket{~.~}$ and can be extended to
      all of $\Hscr_w$ due to the estimate
      \begin{equation}
        \label{eq-positivity-continuity}
        \bnorm{\pi_w ( A ) \ket{L}}^2 = \scp{A L}{A L} = \scp{L}{A^* A
        L} \leqslant \norm{A}^2 \scp{L}{L} = \norm{A}^2
        \norm{\ket{L}}^2 \text{,}
       \end{equation}
       which is founded on the fact that the particle weight is a
       non-negative sesquilinear form and the operator $\norm{A}^2 \,
       \unit - A^* A$ is positive. Since $\Afrak$ is unital, this
       yields a non-zero, non-degenerate representation of the
       quasi-local algebra on the Hilbert space $\Hscr_w$.
    \item The norm on $\Hscr_w$ induces a seminorm on $\Lfrak$ via the
      linear mapping $\ket{~.~}$ and this coincides with $q_w$ as
      defined for particle weights. Therefore the asserted continuity
      of the mapping $( \Lambda , x ) \mapsto \bket{\aLax ( L )}$ is
      an immediate consequence of the respective property in
      Definition~\ref{Def-particle-weight}.
    \item By construction, the canonical homomorphisms $\ket{~.~}$ and
      $\Qscr_w$ coincide and furthermore $\norm{\ket{L}} = \norm{[ L
      ]_w}_w$, so that the assumption of
      $\Xscr_{\Lfrak_0}$-differentiability is self-evident.
    \item The existence of a strongly continuous unitary
      representation of space-time translations in $( \pi_w , \Hscr_w
      )$ is a direct consequence of translation invariance of the
      particle weight $\scp{~.~}{~.~}$ and its continuity under
      Poincar\'{e} transformations with respect to $q_w$. Stone's
      Theorem (cf.~\cite[Chapter~6, \S\,2]{barut/raczka:1980} and
      \cite[Theorem~VIII.(33.8)]{hewitt/ross:1970}) connects the
      spectrum of its generator $P_w = ( P_w^{\, \mu} )$ to the
      support of the Fourier transform of $x \mapsto
      \bscp{L_1}{\ax ( L_2 )}$ in Definition~\ref{Def-particle-weight}
      by virtue of the relation
      \begin{multline}
        \label{eq-stones-theorem}
        \int_{\Rsone} d^{s + 1} x \; g ( x ) \, \bscp{L_1}{\ax ( L_2
        )} = \int_{\Rsone} d^{s + 1} x \; g ( x ) \, \bscpx{L_1}{U_w (
        x )}{L_2} \\
        = ( 2 \pi )^{( s + 1 )/2} \bscpx{L_1}{\tilde{g} ( P_w )}{L_2}
        \text{,}
      \end{multline}
      which holds for any $L_1 \text{,} L_2 \in \Lfrak$ and any $g \in
      L^1 \bigl( \Rsone , d^{s + 1} x \bigr)$. To clarify this fact,
      note, that the projection-valued measure $E_w (~.~)$
      corresponding to $P_w$ is regular, i.\,e.~$\EwDprime$ is for any
      Borel set $\Delta'$ the strong limit of the net $\bset{E_w (
      \Gamma' ) : \Gamma' \subseteq \Delta' \; \text{compact}}$. For
      each compact $\Gamma \subseteq \complement ( \fwcone - q )$
      consider an infinitely often differentiable function
      $\tilde{g}_\Gamma$ with support in $\complement ( \fwcone - q )$
      that envelops the characteristic function for $\Gamma$
      (cf.~\cite[Satz~7.7]{jantscher:1971}): $0 \leqslant \chi_\Gamma
      \leqslant \tilde{g}_\Gamma$. According to the assumption of
      Definition~\ref{Def-particle-weight} the left-hand side of
      \eqref{eq-stones-theorem} vanishes for any $g_\Gamma$ of the
      above kind, and this means that all the bounded operators
      $\tilde{g}_\Gamma ( P_w )$ equal $0$ not only on the dense
      subspace spanned by vectors $\ket{L}$, $L \in \Lfrak$, but on
      all of $\Hscr_w$. Due to the fact that $\tilde{g}_\Gamma$
      majorizes $\chi_\Gamma$, this in turn implies $\chi_\Gamma ( P_w
      ) = \EwGamma = 0$ and thus, by arbitrariness of $\Gamma
      \subseteq \complement ( \fwcone - q )$ in connection with
      regularity, the desired relation $E_w \bigl( \complement (
      \fwcone - q ) \bigr)= 0$.
  \end{Proofsublist}
  \item The reversion of the above arguments in order to establish
    that the scalar product on $\Hscr_w$ possesses the characteristics
    of a particle weight is self-evident.
  \end{Prooflist}
  \renewcommand{\qed}{}
\end{proof}

The following analogue of Lemmas~\ref{Lem-Poin-Bochner-integrals} and
\ref{Lem-Lebesgue-Bochner-integrals} in terms of the $q_w$-topology
induced on $\Lfrak$ by a particle weight is of importance not only for
the remaining results of this chapter, but plays an important role in
the constructions that underlie the theory of disintegration to be
expounded in Chapter~\ref{chap-disintegration}.
\begin{Lem}
\label{Lem-alpha-ket-integral}
  Let $L \in \Lfrak$ and let $\scp{~.~}{~.~}$ be a particle weight.
  \begin{proplist}
  \item Let $F \in L^1 \bigl( \Poin , d \mu ( \Lambda , x ) \bigr)$
    have compact support $\Ssf$, then the Bochner integral
    \begin{subequations}
      \begin{equation}
        \label{eq-alpha-F-weight-integral}
        \alpha_F ( L ) = \int d \mu ( \Lambda , x ) \; F ( \Lambda , x
        ) \, \aLax ( L ) 
      \end{equation}
      lies in the completion of $\Lfrak$ with respect to the locally
      convex topology induced on it by the initial norm $\norm{~.~}$
      and the $q_w$-seminorm defined by the particle weight. Moreover
      $\bket{\alpha_F ( L )}$ is a vector in the corresponding Hilbert
      space $\Hscr_w$ and can be written
      \begin{equation}
        \label{eq-F-ket-integral}
        \bket{\alpha_F ( L )} = \int d \mu ( \Lambda , x ) \; F (
        \Lambda , x ) \, \bket{\aLax ( L )} \text{,}
      \end{equation}
      satisfying the inequality
      \begin{equation}
        \label{eq-F-ket-integral-estimate}
        \bnorm{\bket{\alpha_F ( L )}} \leqslant \norm{F}_1 \sup_{(
        \Lambda , x ) \in \Ssf} \bnorm{\bket{\aLax ( L )}} \text{.}
      \end{equation}
    \end{subequations}
  \item For any function $g \in L^1 \bigl( \Rsone , d^{s + 1} x
    \bigr)$ the Bochner integral
    \begin{subequations}
      \begin{equation}
        \label{eq-alpha-g-weight-integral}
        \alpha_g ( L ) = \int_{\Rsone} d^{s + 1} x \; g ( x ) \,
        \ax ( L ) 
      \end{equation}
      likewise lies in the completion of $\Lfrak$ with respect to the
      locally convex topology mentioned above. $\bket{\alpha_g ( L )}$
      is a vector in the Hilbert space $\Hscr_w$ subject to the
      relation
      \begin{equation}
        \label{eq-g-ket-integral}
        \bket{\alpha_g ( L )} = \int_{\Rsone} d^{s + 1} x \; g ( x )
        \, \bket{\ax ( L )} = ( 2 \pi )^{( s + 1 )/2} \tilde{g} ( P_w
        ) \ket{L} \text{,}
      \end{equation}
      so that
      \begin{equation}
        \label{eq-g-ket-integral-estimate}
        \bnorm{\bket{\alpha_g ( L )}} \leqslant \norm{g}_1 \,
        \norm{\ket{L}} \text{.}
      \end{equation}
    \end{subequations}
  \end{proplist}
\end{Lem}
\begin{proof}
  \begin{prooflist}
  \item Due to continuity of the particle weight $\scp{~.~}{~.~}$ with
    respect to Poincar\'{e} transformations as claimed in
    Definition~\ref{Def-particle-weight}, the integrand of
    \eqref{eq-alpha-F-weight-integral} can be estimated with respect
    to the seminorm $\qw$ induced on $\Lfrak$, which gives the
    Lebesgue-integrable function $( \Lambda , x ) \mapsto \abs{F (
    \Lambda , x )} \cdot \sup_{( \Lambda , x ) \in \Ssf} \qwx{\aLax (
    L )}$. Therefore the integral in question indeed exists in the
    completion of the locally convex space $\Lfrak$ not only with
    respect to the norm topology but also with respect to the seminorm
    $\qw$. Furthermore the corresponding GNS-construction of $( \pi_w
    , \Hscr_w )$ implies that $\norm{\ket{L}}$ coincides with
    $\qwx{L}$ for any $L \in \Lfrak$, a relation which extends to the
    respective completions (cf.~\cite[Chapter~One,
    \S\,5\,4.(4)]{koethe:1983}) thus resulting in
    \eqref{eq-F-ket-integral}. \eqref{eq-F-ket-integral-estimate} is
    then an immediate consequence, again on grounds of continuity
    under Poincar\'{e} transformations.
  \item According to Definition~\ref{Def-particle-weight}, the
    particle weight $\scp{~.~}{~.~}$ is invariant under space-time
    translations and so is the seminorm $\qw$. Therefore the integrand
    of \eqref{eq-alpha-g-weight-integral} is majorized by the
    Lebesgue-integrable function $x \mapsto \abs{g ( x )} \, \qwx{L}$,
    so that the respective integral exists in the completion of
    $\Lfrak$. The first equation of \eqref{eq-g-ket-integral} arises
    from the same arguments that were already applied above, whereas
    the second one is then a consequence of Stone's Theorem
    (cf.~\eqref{eq-stones-theorem}). Again on the ground of
    translation invariance, the estimate
    \eqref{eq-g-ket-integral-estimate} is an immediate conclusion from
    \eqref{eq-g-ket-integral}.
  \end{prooflist}
  \renewcommand{\qed}{}
\end{proof}
Having this preparatory result at our disposal, we are in the position
to prove a statement on spectral subspaces of $\Hscr_w$, that will be
significant in the next chapter as well as for the subsequent proof of
the Cluster Property for particle weights.
\begin{Pro}[Spectral Subspaces]
  \label{Pro-spectral-subspace}
  Let $L$ be an element of $\Lfrak ( \Delta' ) = \Lfrak \cap
  \widetilde{\Afrak} ( \Delta' )$, which means that $L \in \Lfrak$ has
  energy-momentum transfer in the Borel subset $\Delta'$ of $\Rsone$. 
  Then, in the representation $( \pi_w , \Hscr_w )$ corresponding to
  the particle weight $\scp{~.~}{~.~}$, the vector $\ket{L}$ belongs
  to the spectral subspace which pertains to $\Delta'$ with respect to
  the intrinsic unitary representation $x \mapsto U_w ( x )$ of
  space-time translations:
  \begin{equation}
    \label{eq-energy-momentum-subspace}
    \ket{L} = \EwDprime \ket{L} \text{.}
  \end{equation}
\end{Pro}
\begin{proof}
  The energy-momentum transfer of an operator $A \in \Afrak$ can be
  stated in terms of the support properties of the Fourier transform
  of the mapping $x \mapsto \ax ( A )$ considered as an
  operator-valued distribution (cf.~the remark following
  Definition~\ref{Def-vacuum-annihilation}). For the operator $L \in
  \Lfrak ( \Delta' )$ this has the consequence that $\alpha_g ( L ) =
  0$ if $g$ is any Lebesgue-integrable function with $\supp \tilde{g}
  \cap \Delta' = \emptyset$. In this case we have, by an application
  of Lemma~\ref{Lem-alpha-ket-integral},
  \begin{equation}
    \label{eq-g-ket-integral-vanished}
    \int_{\Rsone} d^{s + 1} x \; g ( x ) \, \bket{\ax ( L )} =
    \bket{\alpha_g ( L )} = 0 \text{.}
  \end{equation}
  Upon insertion of \eqref{eq-g-ket-integral-vanished} into the
  formulation \eqref{eq-stones-theorem} of Stone's Theorem, the
  reasoning applied in the proof of Theorem~\ref{The-particle-weight}
  yields the assertion.
\end{proof}
The particle weights enjoy a Cluster Property parallel to that
established in Proposition~\ref{Pro-cluster} for functionals in
$\CDstarplus$. This characteristic, shared by the asymptotic
functionals $\sigma$, could have been included in
Definition~\ref{Def-particle-weight}, but it turns out, that it is
already enforced by the other features.
\begin{Pro}[Cluster Property for Particle Weights]
  \label{Pro-weights-cluster}
  Let $L_i$ and $L'_i$ be elements of $\Lfrak_0$ with energy-momentum
  transfer $\Gamma_i$ respectively $\Gamma'_i$, and let $A_i \in
  \Afrak$, $i = 1 \text{,} 2$, be almost local operators. Suppose
  furthermore that $\scp{~.~}{~.~}$ is a particle weight with
  associated GNS-representation $( \pi_w , \Hscr_w )$, then
  \begin{equation*}
    \Rs \ni \xib \mapsto \bscp{{L_1}^* A_1 L'_1}{\aibx ( {L_2}^* A_2
    L'_2 )} = \bscpx{{L_1}^* A_1 L'_1}{U_w ( \xib )}{{L_2}^* A_2 L'_2}
    \in \Cbb
  \end{equation*}
  is a function in $L^1 \bigl( \Rs , d^s x \bigr)$.
\end{Pro}
\begin{proof}
  To establish this result we follow in the main the strategy of the
  proof of Proposition~\ref{Pro-cluster}. Applied to the problem at
  hand in terms of $( \pi_w , \Hscr_w )$, this yields initially the
  estimate
  \begin{multline}
    \label{eq-weight-cluster-estimate1}
    \babs{\bscpx{{L_1}^* A_1 L'_1}{U_w ( \xib )}{{L_2}^* A_2 L'_2}} \\
    \leqslant \babs{\bscpx{L'_1}{\pi_w \bigl( \bcomm{{A_1}^*
    L_1}{\aibx ( {L_2}^* A_2 )} \bigr) U_w ( \xib )}{L'_2}} +
    \babs{\bscpx{L'_1}{\pi_w \bigl( \aibx ( {L_2}^* A_2 ) {A_1}^* L_1
    \bigr) U_w ( \xib )}{L'_2}}
  \end{multline}
  for any $\xib \in \Rs$. The first term on the right-hand side turns
  out to be majorized by $\bnorm{\bcomm{{A_1}^* L_1}{\aibx ( {L_2}^*
  A_2 )}} \, \norm{\ket{L'_1}} \, \norm{\ket{L'_2}}$ in view of the
  fact that the particle weight is invariant under translations and
  that the representation $\pi_w$ is continuous. As the operators
  involved are almost local without exception, the norm of the
  commutator taking part in this expression decreases rapidly, thus
  rendering it integrable. The second term requires a closer
  inspection. One has
  \begin{multline}
    \label{eq-weight-cluster-estimate2}
    2 \, \babs{\bscpx{L'_1}{\pi_w \bigl( \aibx ( {L_2}^* A_2 ) {A_1}^*
    L_1 \bigr) U_w ( \xib )}{L'_2}} \\
    \mspace{-90mu} \leqslant 2 \, \bnorm{\pi_w \bigl( \aibx (  {A_2}^*
    L_2 ) \bigr) \ket{L'_1}} \, \bnorm{\pi_w \bigl( {A_1}^* L_1 \bigr)
    U_w ( \xib ) \ket{L'_2}} \\
    \leqslant \bnorm{\pi_w \bigl( \aibx ( {A_2}^* L_2 ) \bigr)
    \ket{L'_1}}^2 + \bnorm{\pi_w \bigl( \aminusibx ( {A_1}^* L_1 )
    \bigr) \ket{L'_2}}^2 \text{,}
  \end{multline}
  again by translation invariance of the particle weight in the last
  estimate. Now, a substitute of Lemma~\ref{Lem-cluster-prep} has to
  be sought for, which was applied in the proof of
  Proposition~\ref{Pro-cluster} to get an estimate for
  \eqref{eq-varsigma-cauchy-schwarz}, corresponding to the right-hand
  side of \eqref{eq-weight-cluster-estimate2}. Note, that $\pi_w ( A'
  )$ has the same energy-momentum transfer with respect to the unitary
  group $\set{U_w ( x ) : x \in \Rsone}$ as the operator $A' \in
  \Afrak$ has regarding the underlying positive energy representation,
  and that, according to Proposition~\ref{Pro-spectral-subspace},
  $\ket{L'_1} = E_w ( \Gamma'_1 ) \ket{L'_1}$ and $\ket{L'_2} =
  E_w ( \Gamma'_2 ) \ket{L'_2}$ belong to the spectral subspaces
  pertaining to the compact sets $\Gamma'_1$ and
  $\Gamma'_2$. As in addition the spectrum of $\set{U_w ( x ) : x \in
  \Rsone}$ is restricted to a displaced forward light cone, all of the
  arguments given in the proofs of
  Propositions~\ref{Pro-harmonic-analysis} and
  \ref{Pro-counter-integrals} also apply to the representation $(\pi_w
  , \Hscr_w )$, so that e.\,g.
  \begin{equation*}
    \int_{\Rs} d^s x \; E_w ( \Gamma'_1 ) \pi_w \bigl( \aibx ( {L_2}^*
    A_2 \, {A_2}^* L_2 ) \bigr) E_w ( \Gamma'_1 )
  \end{equation*}
  is seen to exist in the $\sw$-topology on $\BHw$. For this term we
  thus have
  \begin{multline}
    \label{eq-weight-cluster-estimate3}
    \int_{\Rs} d^s x \; \bnorm{\pi_w \bigl( \aibx ( {A_2}^* L_2 )
    \bigr) E_w ( \Gamma'_1 ) \ket{L'_1}}^2 \\
    = \int_{\Rs} d^s x \; \bscpx{L'_1}{E_w ( \Gamma'_1 ) \pi_w \bigl(
    \aibx ( {L_2}^* A_2 \, {A_2}^* L_2 ) \bigr)
    E_w ( \Gamma'_1 )}{L'_1} < \infty \text{.}
  \end{multline}
  The same holds true for the other expression on the right-hand side
  of \eqref{eq-weight-cluster-estimate2}, which shows that $\xib
  \mapsto \babs{\bscpx{L'_1}{\pi_w \bigl( \aibx ( {L_2}^* A_2 )
  {A_1}^* L_1 \bigr) U_w ( \xib )}{L'_2}}$ is an integrable function,
  too. Altogether, we have thus established the Cluster Property for
  particle weights.
\end{proof}
\begin{Rem}
  Note, that the above result is independent of the differentiability
  properties of a particle weight (item (iii) in both
  Definition~\ref{Def-particle-weight} and
  Theorem~\ref{The-particle-weight}), since these did not enter into
  its proof.
\end{Rem}

At this point a brief comment on the notation chosen seems
appropriate (cf.~\cite{buchholz/porrmann/stein:1991}). We deliberately
utilize the typographical token $\ket{~.~}$ introduced by Dirac
\cite[\S\,23]{dirac:1958} for ket vectors describing improper momentum
eigenstates $\ket{\pib}$, $\pib \in \Rs$. These act as distributions
on the space of momentum wave functions with values in the physical
Hilbert space $\Hscr$, thereby presupposing a superposition principle
to hold without limitations. This assumption collapses in an
infraparticle situation as described in the Introduction. In contrast
to this, the \emph{pure} particle weights, that will shortly have
their appearance in connection with elementary physical systems, are
seen to be associated with sharp momentum and yet capable of
describing infraparticles. Here the operators $L \in \Lfrak$ take on
the role of the previously mentioned momentum space wave functions in
that they localize the particle weight in order to produce a
normalizable vector $\ket{L}$ in the pertaining Hilbert space
$\Hscr_w$. This in turn substantiates the terminology introduced in
Definition~\ref{Def-localizer-ideal}. As they describe elementary
physical systems, pure particle weights should give rise to
irreducible representations of the quasi-local algebra, thus
motivating the subsequent definition. It is supplemented by a certain
specific regularity condition of technical importance, which we
anticipate to hold in physically relevant situations, and by a notion
of boundedness which is in particular shared by the positive
asymptotic functionals $\sigma$, as shown in
Lemma~\ref{Lem-Delta-bound}.
\begin{Def}
  \label{Def-weight-classification}
  A particle weight is said to be
  \begin{deflist}
    \item \emph{pure}, if the corresponding representation $( \pi_w ,
      \Hscr_w )$ is irreducible;
    \item \emph{regular}, if for any $L \in \Lfrak$ the following
      implication is valid:
      \begin{equation*}
        \scp{L^* L}{L^* L} = 0 \quad \Longrightarrow \quad \scp{L}{L}
        = 0 \text{;}
      \end{equation*}
    \item \emph{$\Delta$-bounded}, if to any bounded Borel subset
      $\Delta'$ of $\Rsone$ there exists another such set $\Deltabar
      \supseteq \Delta + \Delta'$, such that the GNS-representation $(
      \pi_w , \Hscr_w )$ of the particle weight and the defining 
      representation are connected by the inequality
      \begin{equation}
        \label{eq-Delta-boundedness}
        \norm{\EwDprime \pi_w ( A ) \EwDprime} \leqslant c \cdot
        \norm{\EDbar A \EDbar}
      \end{equation}
      for any $A \in \Afrak$ with a suitable positive constant $c$ 
      (independent of the Borel sets). Evidently, $\Delta$ ought to be
      a bounded Borel set as well.
  \end{deflist}
\end{Def}
\begin{Lem}
  \label{Lem-Delta-bound}
  Any positive asymptotic functional $\sigma \in \CDstarplus$,
  constructed according to Theorem~\ref{The-singular-limits} under the
  assumptions of Proposition~\ref{Pro-positivity-of-limits}, gives
  rise to a $\Delta$-bounded particle weight $\scp{~.~}{~.~}_\sigma$.
\end{Lem}
\begin{proof}
  Let $( \pi_\sigma , \Hscr_\sigma )$ denote the GNS-representation of
  the particle weight $\sigma$ with associated spectral measure
  $E_\sigma (~.~)$ for the generator $P_\sigma = ( P_\sigma^\mu )$ of
  the intrinsic space-time translations. For the time being, suppose
  that $\Delta'$ is an \emph{open} bounded Borel set in $\Rsone$. Let
  furthermore $L$ be an arbitrary element of $\Lfrak$ and $A \in
  \Afrak$. We are interested in an estimate of the term
  $\scpx{L}{\EsDprime \pi_\sigma ( A ) \EsDprime}{L}_\sigma$. Note,
  that the spectral measure is regular, so that $\EsDprime$ is the
  strong limit of the net $\set{\EsGamma : \Gamma \subset \Delta'
  \thickspace \text{compact}}$. As $\Delta'$ is assumed to be open,
  there exists for each compact subset $\Gamma$ of $\Delta'$ an
  infinitely often differentiable function $\tilde{g}_\Gamma$ with
  $\supp \tilde{g}_\Gamma \subset \Delta'$ that fits between the
  corresponding characteristic functions
  \cite[Satz~7.7]{jantscher:1971}: $\chi_\Gamma \leqslant
  \tilde{g}_\Gamma \leqslant \chi_{\Delta'}$. Thus the respective
  operators are subject to the relation
  \begin{equation*}
    0 \leqslant \bigl( \EsDprime - \tilde{g}_\Gamma ( P_\sigma )
    \bigr)^2 \leqslant \bigl( \EsDprime - \EsGamma \bigr)^2 \text{,}
  \end{equation*}
  from which we infer that for arbitrary $L' \in \Lfrak$
  \begin{equation}
    \label{eq-regular-Esigma-approx}
    0 \leqslant \bnorm{\bigl( \EsDprime - \tilde{g}_\Gamma ( P_\sigma
    ) \bigr) \ket{L'}}^2 \leqslant \bnorm{\bigl( \EsDprime - \EsGamma
    \bigr) \ket{L'}}^2 \xrightarrow[\Gamma \nearrow \Delta']{} 0
    \text{.}
  \end{equation}
  By density of all the vectors $\ket{L'}$ in $\Hscr_\sigma$, it is
  thereby established that
  \begin{equation}
    \label{eq-regular-Esigma-strong-approx}
    \EsDprime = \str-\lim_{\Gamma \nearrow \Delta'} \tilde{g}_\Gamma (
    P_\sigma ) \text{,}
  \end{equation}
  which implies for the scalar product in to be considered here
  \begin{equation}
    \label{eq-regular-approximation}
    \scpx{L}{\EsDprime \pi_\sigma ( A ) \EsDprime}{L}_\sigma =
    \lim_{\Gamma \nearrow \Delta'} \scpx{L}{\tilde{g}_\Gamma (
    P_\sigma ) \pi_\sigma ( A ) \tilde{g}_\Gamma ( P_\sigma )}{L}
    \text{.}
  \end{equation}
  Since $\tilde{g}_\Gamma$ is the Fourier transform of a rapidly
  decreasing function $g_\Gamma$, which therefore belongs to the space
  $L^1 \bigl( \Rsone , d^{s + 1} x \bigr)$,
  Lemma~\ref{Lem-alpha-ket-integral} can be applied to yield for the
  right-hand side of \eqref{eq-regular-approximation}
  \begin{multline}
    \label{eq-alpha-g-reformulation}
    \scpx{L}{\tilde{g}_\Gamma ( P_\sigma ) \pi_\sigma ( A )
    \tilde{g}_\Gamma ( P_\sigma )}{L} = ( 2 \pi )^{- ( s + 1 )}
    \scpx{\agGamma ( L )}{\pi_\sigma ( A )}{\agGamma ( L )}_\sigma \\
    = ( 2 \pi )^{- ( s + 1 )} \sigma \bigl( \agGamma ( L )^* A
    \agGamma ( L ) \bigr) \text{,}
  \end{multline}
  where, following the remark pertaining to
  Proposition~\ref{Pro-positivity-of-limits}, the ultimate expression
  is based on the fact that $\agGamma ( L )^* A \agGamma ( L ) \in
  \AC$ as a consequence of Lemmas~\ref{Lem-Lebesgue-Bochner-integrals}
  and \ref{Lem-basic-estimate} in connection with
  Corollary~\ref{Cor-sesquilinear-product}. The approximating
  functionals $\rho_{h , t}$ for $\sigma$ in the form
  \eqref{eq-rho-alt-form} with a non-negative function $h \in
  L^\infty ( \Rs , d^s x )$ allow, through an application of
  \cite[Proposition~2.3.11]{bratteli/robinson:1987}, for the following
  estimate of their integrand:
  \begin{multline*}
    \babs{h ( \tau^{-1} \xib ) \, \omega \bigl( U ( \tau ) \ED \aibx (
    \agGamma ( L )^* A \agGamma ( L ) ) \ED U ( \tau )^* \bigr)} \\
    =  h ( \tau^{-1} \xib ) \, \babs{\omega \bigl( U ( \tau ) \ED
    \aibx (\agGamma ( L )^* ) \EDbar \aibx ( A ) \EDbar \aibx (
    \agGamma ( L ) ) \ED U ( \tau )^* \bigr)} \\
    \leqslant \norm{\EDbar A \EDbar} \; h ( \tau^{-1} \xib ) \, \omega
    \bigl( U ( \tau ) \ED \aibx ( \agGamma ( L )^* \agGamma ( L ) )
    \ED U ( \tau )^* \bigr) \text{.}
  \end{multline*}
  Here the spectral projections $\EDbar$ pertaining to the Borel set
  $\Deltabar = \Delta + \Delta'$, which is both bounded and open,
  could be introduced, since, according to
  Lemma~\ref{Lem-Lebesgue-Bochner-integrals}, the energy-momentum
  transfer of $\agGamma ( L )$ is contained in $\Delta'$ by
  construction. An immediate consequence of the above relation is
  \begin{equation*}
    \babs{\rho_{h , t} \bigl( \agGamma ( L )^* A \agGamma ( L )
    \bigr)} \leqslant \norm{\EDbar A \EDbar} \; \rho_{h , t} \bigl(
    \agGamma ( L )^* \agGamma ( L ) \bigr) \text{,}
  \end{equation*}
  which extends to the limit functional $\sigma$:
  \begin{equation}
    \label{eq-Delta-bound-estimate1}
    \babs{\sigma \bigl( \agGamma ( L )^* A \agGamma ( L ) \bigr)}
    \leqslant \norm{\EDbar A \EDbar} \; \sigma \bigl( \agGamma ( L )^*
    \agGamma ( L ) \bigr) \text{.}
  \end{equation}
  Insertion of this result into \eqref{eq-alpha-g-reformulation}
  yields
  \begin{equation}
    \label{eq-Delta-bound-estimate2}
    \babs{\scpx{L}{\tilde{g}_\Gamma ( P_\sigma ) \pi_\sigma ( A )
    \tilde{g}_\Gamma ( P_\sigma )}{L}} \leqslant \norm{\EDbar A
    \EDbar} \; \scpx{L}{\tilde{g}_\Gamma ( P_\sigma )^2}{L}
  \end{equation}
  and in the limit $\Gamma \nearrow \Delta'$, in compliance with
  \eqref{eq-regular-approximation},
  \begin{equation}
    \label{eq-Delta-bound-estimate3}
    \babs{\scpx{L}{\EsDprime \pi_\sigma ( A ) \EsDprime}{L}_\sigma}
    \leqslant \norm{\EDbar A \EDbar} \, \scpx{L}{\EsDprime}{L}_\sigma
    \leqslant \norm{\EDbar A \EDbar} \, \scp{L}{L}_\sigma \text{.}
  \end{equation}
  Passing to the supremum with respect to all $L \in \Lfrak$ such that
  $\norm{\ket{L}_\sigma} \leqslant 1$ (these constitute a dense subset
  of the unit ball in $\Hscr_\sigma$), we get through an application
  of \cite[Satz~4.4]{weidmann:1976}
  \begin{equation}
    \label{eq-Delta-final-bound1}
    \norm{\EsDprime \pi_\sigma ( A ) \EsDprime} \leqslant 2 \cdot
    \norm{\EDbar A \EDbar} \text{.}
  \end{equation}
  This establishes the defining condition \eqref{eq-Delta-boundedness}
  for $\Delta$-boundedness with $c = 2$ in the case of an \emph{open}
  bounded Borel set $\Delta'$. But this is not an essential
  restriction, since an arbitrary bounded Borel set $\Delta'$ is
  contained in the open set $\Delta'_\eta$, $\eta > 0$, consisting of
  all those points $p \in \Rsone$ for which $\inf_{p' \in \Delta'}
  \abs{p - p'} < \eta$. Since $\Delta'_\eta$ is likewise a bounded
  Borel set, we get
  \begin{equation}
    \label{eq-Delta-final-bound2}
    \norm{\EsDprime \pi_\sigma ( A ) \EsDprime} \leqslant
    \norm{\EsDprimeeta \pi_\sigma ( A ) \EsDprimeeta} \leqslant 2
    \cdot \norm{\EDetabar A \EDetabar}
  \end{equation}
  as an immediate consequence of \eqref{eq-Delta-final-bound1}, where
  $\Deltaetabar \doteq \Delta + \Delta'_\eta$. This covers the general
  case and thereby proves $\Delta$-boundedness for the asymptotic
  functionals $\sigma \in \CDstarplus$.
\end{proof}

\chapter{Disintegration of Particle Weights}
  \label{chap-disintegration}

\label{cit-prokrustes}
\renewcommand{\thefootnote}{\fnsymbol{footnote}}
\hspace*{\fill}
\epsfig{file=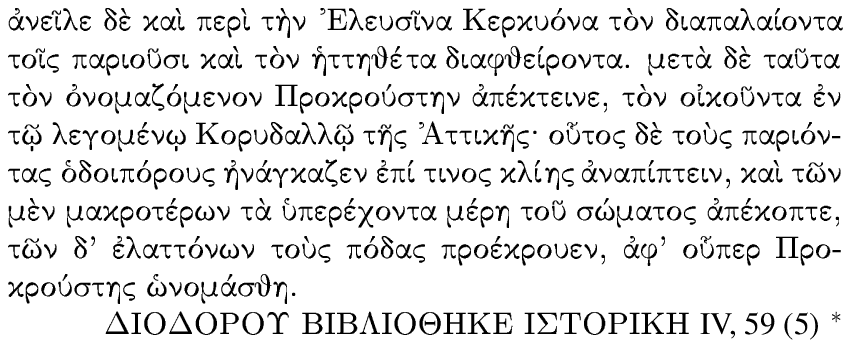}
\footnotetext[1]{A german translation can be found on page
  \pageref{cit-prokrustes-transl}.}
\vspace{1cm}

\noindent
In Section III of their treatment of collision cross sections for
massive theories within the framework of local quantum physics, Araki
and Haag got to the following asymptotic relation which holds true for
the counters $C$ they had selected, for arbitrary vectors $\Phi$ and
certain specific vectors $\Psi$ representing outgoing particle
configurations \cite[Theorem~4]{araki/haag:1967}:
\begin{equation}
  \label{eq-araki/haag-result}
  \lim_{t \rightarrow \infty} \bscpx{\Phi}{t^3 C ( h , t )}{\Psi} =
  \sideset{}{'} \sum_{i , j} \int d^3 p \; \Gamma_{i j} ( \pib ) \,
  \bscpx{\Phi}{a_j^{\dagger \text{out}} ( \pib ) \, a_i^{\text{out}} (
  \pib )}{\Psi} \, h ( \vib_i ) \text{,}
\end{equation}
where
\begin{align*}
  \Gamma_{i j} ( \pib ) & \doteq 8 \pi^3 \bscpx{\pib j}{C ( 0 )}{\pib
  \, i} \text{,} \\
  \vib_i & \doteq ( \pib^2 + m_i^2 )^{-1/2} \pib \text{.}
\end{align*}
The indices $i$ and $j$ in the above formula denote the particle types
including spin, and summation runs over pairs of particles with equal
mass: $m_i = m_j$. The structure of the right-hand side of this
equation is based on the \emph{a priori} knowledge of the particle
content of the theory they considered. Comparing this result with the
concepts developed in the preceding chapter
(cf.~Theorem~\ref{The-singular-limits}), one has an asymptotic
functional $\sigma_h^{( + )}$ standing on the left-hand side of
equation \eqref{eq-araki/haag-result} that is decomposed with respect
to momentum eigenstates $\ket{\pib \, i}$, hidden in the definition of
$\Gamma_{i j}$. If we accept such an interpretation of this theorem of
Araki and Haag, it is possible to re-write it in the form
\begin{equation*}
  \sigma_h^{( + )} ( C ) = \sideset{}{'} \sum_{i , j} \int d \mu_{i ,
  j} ( \pib ) \; \bscpx{\pib j}{C ( 0 )}{\pib \, i} \text{,}
\end{equation*}
where all expressions occurring in \eqref{eq-araki/haag-result} apart 
from $\Gamma_{i j}$ are absorbed into the measures $\mu_{i , j}$. This
presents the asymptotic functional as a mixture of linear forms on
$\Cfrak$ (an algebra which is part of that selected in
\cite{araki/haag:1967}) defined by Dirac kets representing improper
momentum eigenstates; thus we happen to meet exactly those constructs
that we already hinted at in the remarks concerning our notation
that led to Definition~\ref{Def-weight-classification}. The aim of the
present chapter is to establish a corresponding formula in the general
setting, i.\,e.~without any previous knowledge of the particle
content.

As indicated by \eqref{eq-araki/haag-result}, representations
resulting from the construction of asymptotic functionals as expounded
in Chapter~\ref{chap-particle-weights} will be highly reducible,
whereas elementary physical systems are expected to be connected with
pure particle weights, giving rise to irreducible representations of
the quasi-local $C^*$-algebra $\Afrak$. In view of the preceding
paragraph the obvious problem to be tackled now is to develop a theory
for the decomposition or rather \emph{disintegration} of generic
particle weights into pure ones. Two approaches to this problem will
be presented in this work:
\begin{abclist}
\item Decomposition of the GNS-representation pertaining to a particle
  weight into a direct integral of representations (spatial
  disintegration):
  \begin{equation*}
    ( \pi_w , \Hscr_w ) \simeq \int_\Xecm^\oplus d \nu ( \xi ) \; (
    \pi_\xi , \Hscr_\xi ) \text{.}
  \end{equation*}
\item Barycentric decomposition of a given particle weight with
  respect to a base $\Becm_\Wecm$ of the positive cone $\Wecm$ of all
  particle weights in the space of sesquilinear forms on $\Lfrak$
  (Choquet theory):
  \begin{equation*}
    \scp{~.~}{~.~} = \int_{\Becm_\Wecm} d \upsilon ( \zeta ) \;
    \scp{~.~}{~.~}_\zeta \text{.}
  \end{equation*}
\end{abclist}
Although the technical problems to come to grips with in these two
constructions are quite different, we anticipate equivalence of their
results: the separability assumptions essential in the first one are
substituted by compactness conditions in the second. So evidently both
of them require certain restrictions in the number of degrees of
freedom, which seem to be complementary in one way or another. While
the partial results achieved so far in connection with the barycentric
decomposition will be discussed in Chapter~\ref{chap-choquet}, the
spatial disintegration of the GNS-representation of a particle weight
is the subject we will elaborate on first.

\section[Separable Reformulation]{Separable Reformulation of Local
  Quantum Physics and its Associated Algebra of Detectors} 
  \label{sec-separable-reformulation}

The theory of spatial disintegration of representations $( \pibar ,
\Hscrbar )$ of a $C^*$-algebra $\Afrakbar$ is a common theme of the
pertinent textbooks (cf.~\cite{dixmier:1981,dixmier:1982,%
takesaki:1979,pedersen:1979,bratteli/robinson:1987}), an indispensable
presupposition being that of separability of the algebra $\Afrakbar$
as well as of the Hilbert space $\Hscrbar$ in their respective uniform
topologies. Note, that in this way the statements of
\cite[Section~4.4]{bratteli/robinson:1987} are incorrect (cf.~also
\cite[Corrigenda]{bratteli/robinson:1997}). These separability
assumptions are too restrictive to be encountered in physically
reasonable theories from the outset, so first of all a
\emph{countable} version of the fundamental assumptions of local
quantum field theory in terms of the net $\Oscr \mapsto \AO$ and of
the symmetry group $\Poin$ has to be formulated before one can benefit
from the extensive theory made available in the literature. This
construction will be accomplished in a sequence of steps:
\begin{steplist}
\item With respect to its initial topology, the Poincar\'{e} group
  $\Poin$ contains a numerable dense subgroup that we signify by
  $\Pcount$. It is itself the semi-direct product of countable dense
  subgroups of Lorentz transformations $\Lcount$ in $\Lor$ and of
  space-time translations $\Tcount$ in $\Rsone$: $\Pcount = \Lcount
  \ltimes \Tcount$.
\item Consider the standard diamonds with \emph{rational} radii,
  centred around the origin. Subjecting these regions to all of the
  transformations in $\Pcount$ yields a countable family $\Rcount$ of
  open bounded regions, which is invariant with respect to the
  selected Poincar\'{e} transformations and constitutes a covering of
  $\Rsone$. Note, that arbitrarily small regions belong to $\Rcount$
  in the sense, that any region in Minkowski space contains an element
  of this numerable collection as a subset.
\item As shown in Appendix~\ref{chap-separable-algebras}, any unital
  $C^*$-algebra of operators on a separable Hilbert space $\Hscr$
  contains a strongly dense (i.\,e.~dense with respect to the
  strong-operator topology), norm-separable $C^*$-subalgebra, that
  includes the identity. Applied to the local $C^*$-algebras $\AO$ of
  the defining positive-energy representation, this result has the
  consequence that to each open bounded region $\Oscr$ in Minkowski
  space one can associate a norm-separable, unital $C^*$-algebra
  $\AsubbulletO$, that lies strongly dense in $\AO$. This means, that
  the algebra $\AsubbulletO$ in turn contains a countable
  $^*$-subalgebra $\AcountO$ over the field $\Qbb + i \Qbb$, which is
  uniformly dense in $\AsubbulletO$, strongly dense in $\AO$ and can
  likewise be chosen to comprise the unit.

  Let $\Ok$, $k \in \Nbb$, be a denumeration of the countable family
  $\Rcount$ of open bounded regions in Minkowski space constructed
  above. We define $\AbulletOk$ as the $C^*$-algebra (over $\Cbb$)
  which is generated by the union of all $\aLax \bigl( \AcountOi
  \bigr)$, where $( \Lambda , x ) \in \Pcount$ and $\Oi \in \Rcount$
  run through all combinations for which $\Lambda \Oi + x \subseteq
  \Ok$. By construction this algebra is norm-separable and satisfies
  \begin{subequations}
    \begin{equation}
      \label{eq-netsubsetseq}
      \AcountOk \subseteq \AbulletOk \subseteq \AOk \text{,}
    \end{equation}
    so that $\AbulletOk$ turns out to be strongly dense in $\AOk$.

    The net of local $C^*$-algebras $\bset{\AbulletOk : k \in \Nbb}$
    fulfills the conditions of isotony, locality and covariance with
    respect to $\Rcount$ and $\Pcount$. Isotony is an immediate
    consequence of the construction whereas locality follows from
    \eqref{eq-netsubsetseq} in connection with locality of the
    defining net $\Oscr \mapsto \AO$. To establish covariance one has
    to observe that, given any $( \Lambda , x ) \in \Pcount$, the
    algebra $\aLax \bigl( \AbulletOk \bigr)$ is generated by all
    $\aLax \bigl( \aLaxprime \bigl( \AcountOi \bigr) \bigr)$, where $(
    \Lambda' , x' ) \in \Pcount$ and $\Oi \in \Rcount$ run through
    those combinations which satisfy the relation $\Lambda' \Oi + x'
    \subseteq \Ok$. This can equivalently be expressed by saying that
    the algebra in question is generated by all $\aLaxdprime \bigl(
    \AcountOi \bigr)$, for which $( \Lambda'' , x'' ) \in \Pcount$ and
    $\Oi \in \Rcount$ have the property $\Lambda'' \Oi + x'' \subseteq
    \Lambda \Ok + x$. In this formulation $\aLax \bigl( \AbulletOk
    \bigr)$ turns out to be equal to the algebra $\Abullet ( \Lambda
    \Ok + x )$. The somewhat intricate construction of $\AbulletOk$ is
    necessitated by the requisite to have the standard properties of a
    net of local algebras at our disposition.

    By construction, the countable $^*$-algebra $\Acount$ over $\Qbb +
    i \Qbb$, which is generated by the union of all the algebras
    $\AcountOk$, $\Ok \in \Rcount$, and thus invariant under
    transformations from $\Pcount$, lies uniformly dense in the
    $C^*$-inductive limit $\Abullet$ of the net $\Ok \mapsto
    \AbulletOk$, and is, on account of \eqref{eq-netsubsetseq}, even
    strongly dense in the quasi-local algebra $\Afrak$ itself. We thus
    have the inclusions
    \begin{equation}
      \label{eq-algebrasubsetseq}
      \Acount \subseteq \Abullet \subseteq \Afrak \text{,}
    \end{equation}
  \end{subequations}
  with a norm-separable $C^*$-algebra $\Abullet$, which lies strongly
  dense in $\Afrak$ and contains $\Acount$ as a numerable uniformly
  dense subalgebra (over $\Qbb + i \Qbb$).
\end{steplist}

Into this restricted setting of Local Quantum Physics defined above,
we now introduce countable counterparts of the left ideal of
localizing operators $\Lfrak$, of the algebra of detectors $\Cfrak$
and, most important of all, of the subspace $\Lfrak_0 \subseteq
\Lfrak$ of almost local vacuum annihilation operators.

First of all note, that it is possible to select a \emph{numerable}
subspace over $\Qbb + i \Qbb$ in $\Lfrak_0$, which consists of almost
local vacuum annihilation operators with energy-momentum transfer in
arbitrarily small regions. E.\,g.~let $\set{\Gamma_n}_{n \in \Nbb}$ be
a countable cover of $\complement \fwcone$, constituted by compact and
convex subsets of the complement of the forward light cone, with the
additional property that any bounded region in $\complement \fwcone$
contains one of these compacta. Let, for instance, $\set{p_i}_{i \in
\Nbb}$ be a dense sequence in $\complement \fwcone$ and associate to
each $p_i$ the compact balls of rational radius $r \in \Qbb$ that
satisfy $\overline{\mathscr{B}_r ( p_i )} \subseteq \complement
\fwcone$ in addition. The Lorentz group $\Lor$, being locally compact,
can be covered by a countable family of arbitrarily small compact sets
$\set{\Theta_m}_{m \in \Nbb}$ as well. Now, the spaces
$\Dscr_{\Gamma_n}$ and $\Dscr_{\Theta_m}$ of test functions with
support in $\Gamma_n$ or else $\Theta_m$
(cf.~\cite[\S\,12]{jantscher:1971}) are separable as subspaces of the
respective Banach spaces $L^p \bigl( \Rsone , d^{s+1} x \bigr)$ and
$L^p \bigl( \Rbb^{d_\Lsf} , d^{d_\Lsf} t \bigr)$ ($d_\Lsf \doteq
2^{-1} s ( s + 1 )$ is the dimension of $\Lor$), which in turn are
separable due to an application of
\cite[Theorem~IV.(13.20)]{hewitt/stromberg:1969} using elements of the
numerable set of simple functions with rational values on intervals
with rational end points. Thus there exist dense sequences
$\tilde{g}_n^l$ and $h_m^k$ in the spaces $\Dscr_{\Gamma_n}$ and
$\Dscr_{\Theta_m}$, respectively. Consider the countable family of
operators in $\Lfrak_0$, which are defined through
\begin{equation}
  \label{eq-countable-vacann}
  \alpha_{h_m^k \otimes g_n^l} ( A_j ) \doteq \int_{\Poin} d \mu (
  \Lambda , x ) \; h_m^k ( \Lambda ) \, g_n^l ( x ) \, \aLax ( A_j )
  \text{,}
\end{equation}
for any $A_j \in \Acount$ in the uniform topology of $\Afrak$, and
supplement this selection by all orders of partial derivatives with
respect to the canonical coordinates around $( \unit , 0 )$
(cf.~Appendix~\ref{chap-differentiability}):
\begin{equation*}
  \delta^{\iota_M} \bigl( \alpha_{h_m^k \otimes g_n^l} ( A_j ) \bigr)
  = \delta^{i_M} \dotsb \delta^{i_1} \bigl( \alpha_{h_m^k \otimes
  g_n^l} ( A_j ) \bigr) \in \Lfrak_0 
\end{equation*}
for any $M$-tuple $\iota_M = ( i_1 , \dots , i_M )$ with integer
entries from the set $\set{1 , \dots , d_\Psf}$, where $d_\Psf =
d_\Lsf + ( s + 1 )$. Upon application of all transformations from
$\Pcount$ to these constructs, we get a sequence of vacuum
annihilation operators, comprising elements with energy-momentum
transfer in arbitrarily small regions, which generates a countable
subspace $\vaccount$ over the field $\Qbb + i \Qbb$ in $\Lfrak_0$,
invariant under transformations from $\Pcount$ and under arbitrary
partial derivations. When this construct is to be used in connection
with a given particle weight $\scp{~.~}{~.~}$ that is non-negative by
definition, it does not cause any problems to supplement the set of
operators defined in \eqref{eq-countable-vacann} by a countable number
of other elements from $\Lfrak_0$, on which the particle weight
attains non-vanishing values. In this way the imminent restriction of
$\scp{~.~}{~.~}$ to a subset of $\Lfrak$ can be protected from getting
trivial.

The above selection of vacuum annihilation operators does not yet meet
the requirements for the disintegration. For it to be feasible we
compactly regularize these operators: Take a \emph{countable} set of
compactly supported test functions $F$ on $\Poin$ with a support
$\Ssf_F \doteq \supp F$ which contains the unit $( \unit , 0 )$ of
$\Poin$. Then all the Bochner integrals
\begin{equation}
  \label{eq-def-vaccountbar}
  \alpha_F ( L_0 ) = \int_{\Ssf_F} d \mu ( \Lambda , x ) \; F (
  \Lambda , x ) \, \aLax ( L_0 ) \text{,} \quad L_0 \in \vaccount
  \text{,}
\end{equation}
are elements of the $C^*$-algebra $\Abullet$ and of $\Lfrak_0$
according to Lemma~\ref{Lem-Poin-Bochner-integrals} with
energy-momentum transfer contained in $\bigcup_{( \Lambda , x ) \in
\Ssf_F} \Lambda \Gamma$ given $L_0 \in \Lfrak_0 ( \Gamma )$ (cf.~the
proof of the quoted Lemma). The specific property of operators of type
\eqref{eq-def-vaccountbar} in contrast to those from $\vaccount$ is,
that their differentiability with respect to the Poincar\'{e} group
can be expressed in terms of derivatives of the infinitely
differentiable test function $F \in \Dscr_{\Ssf_F}$, a feature that
will be of great significance later on. By choosing the support of the
functions $F$ small enough, one can impose an energy-momentum transfer
in arbitrarily small regions on the operators $\alpha_F ( L_0 )$ as
was the case for the elements of $\vaccount$ itself. Furthermore, a
particle weight that did not vanish on the set $\vaccount$ is also
non-zero when restricted to all of the operators $\alpha_F ( L_0 )$
constructed in \eqref{eq-def-vaccountbar}. This fact is easily
established with relation \eqref{eq-F-ket-integral} of
Lemma~\ref{Lem-alpha-ket-integral} and the continuity of the particle
weight under Poincar\'{e} transformations in mind. The numerable set
of vacuum annihilation operators that consists of those explicitly
presented in \eqref{eq-def-vaccountbar} together with all their
partial derivations of arbitrary order (that share this specific style
of construction) will be denoted $\vaccountbar$ in the sequel. It
might happen that two of these elements of $\Lfrak_0$ are connected by
a Poincar\'{e} transformation not yet included in $\Pcount$. For
technical reasons, which are motivated by the exigencies for the proof
of the central Theorem~\ref{The-spatial-disintegration} of this
chapter, we supplement $\Pcount$ by all of the (countably many)
transformations arising in this way and consider henceforth the
countable subgroup $\Pbarcount = \Lbarcount \ltimes \Tbarcount
\subseteq \Poin$ generated by them. The set $\vaccountbar$ is then
invariant under the operation of taking derivatives as well as under
all transformations from the numerable dense subgroup $\Pbarcount$.

Here is a list of the countable substitutes for the algebraic concepts
used thus far:
\begin{subequations}
  \begin{latinlist}
  \item We have defined an isotonous, local and $\Pcount$-covariant
    net $\Ok \mapsto \AbulletOk$, $\Ok \in \Rcount$, which has
    $\Abullet$ as $C^*$-inductive limit. This is a norm-separable
    $C^*$-algebra (over the field $\Cbb$) with unit $\unit$,
    containing $\Acount$, which is generated by the countable local
    algebras $\AcountOk \subseteq \AbulletOk$, as a likewise unital,
    numerable, uniformly dense $*$-subalgebra over $\Qbb + i
    \Qbb$. $\Abullet$ itself lies strongly dense in the quasi-local
    algebra $\Afrak$ and, due to uniform continuity of the mappings $(
    \Lambda , x ) \mapsto \aLax ( A )$, $A \in \Afrak$, it is
    invariant with respect to the whole Poincar\'{e} group. In
    contrast to this, note, that the invariance property for $\Acount$
    is restricted to $\Pcount$.
  \item $\vaccountbar \subseteq \Lfrak_0 \cap \Abullet$ is a countable
    set of vacuum annihilation operators of the special construction
    \eqref{eq-def-vaccountbar}, which is invariant under
    transformations from $\Pbarcount$ and under the operation of
    taking partial derivations. Depending on a given particle weight,
    it can be chosen in such a way, that the particle weight
    restricted to $\vaccountbar$ remains non-trivial.
  \item The image of $\vaccountbar$ under all Poincar\'{e}
    transformations is denoted $\vacbar$:
      \begin{equation}
        \label{eq-def-vacbar}
        \vacbar \doteq \bset{\aLax ( L_0 ) : L_0 \in \vaccountbar , (
        \Lambda , x ) \in \Poin} \text{.}
      \end{equation} 
  \item $\Acountbar \subseteq \Abullet$ in turn denotes the numerable,
    unital $^*$-algebra over $\Qbb + i \Qbb$ which is generated by
    $\Acount \cup \vaccountbar$. It is thus stable with respect to
    $\Pcount$ and uniformly dense in $\Abullet$.
  \item The countable counterpart $\idealcount$ of the left ideal
    $\Lfrak$ in $\Afrak$ is defined as the linear span with respect to
    the field $\Qbb + i \Qbb$ of operators of the form $L = A \, L_0$
    with $A \in \Acountbar$ and $L_0 \in \vaccountbar$:
    \begin{equation}
      \label{eq-def-idealcount}
      \idealcount \doteq \Acountbar \; \vaccountbar = \linhull_{\Qbb +
      i \Qbb} \bset{A \, L_0 : A \in \Acountbar , L_0 \in
      \vaccountbar} \text{.}
    \end{equation}
    This constitutes a left ideal of the algebra $\Acountbar$,
    likewise invariant under transformations from $\Pcount$.
  \item Finally, one can introduce the countable $^*$-subalgebra
    $\detectcount \subseteq \Cfrak$ via 
    \begin{equation}
      \label{eq-def-detectcount}
      \detectcount \doteq {\idealcount}^* \, \idealcount =
      \linhull_{\Qbb + i \Qbb} \bset{{L_1}^* \, L_2 : L_1 , L_2 \in
      \idealcount} \text{.}
    \end{equation}
  \end{latinlist}
\end{subequations}

\section[Restricted $\Kfrak_0^c$-Particle Weights]{Restricted
  $\boldsymbol{\Kfrak_0^c}$-Particle Weights}
  \label{sec-restricted-particle-weights}

The subsequent developments in this chapter have to be founded on a
mitigated version for the concept of particle weights as it was
introduced in Definition~\ref{Def-particle-weight}. The reason is that
the sesquilinear forms occurring in the decomposition theory of
Section~\ref{sec-spatial-disintegration} do not share all the desired
properties. Therefore we insert the present section which deals with
the necessary restrictions that have to be imposed on the concepts of
Chapter~\ref{chap-particle-weights}. The essential cuts are indicated
by the work previously accomplished.
\begin{Def}[Restricted $\boldsymbol{\Kfrak_0^c}$-Particle Weights]
  \label{Def-restr-particle-weight}
  Suppose that we are given a sextuple $( \piubar , \Hscrubar ,
  \Afrakubar , \alphaubar , \Pubarcount , \Kfrak_0^c )$ with entries
  of the following sense:
  \begin{subequations}
    \begin{trilist}
    \item $\Afrakubar$ is a norm separable $C^*$-subalgebra of the
      quasi-local algebra $\Afrak$, which arises as the
      $C^*$-inductive limit of a countable net of local
      $C^*$-algebras.
    \item $\Pubarcount$ is a numerable dense subgroup of the
      Poincar\'{e} group. $\Poin$ as a whole is implemented in
      $\Afrakubar$ by the strongly continuous group of automorphisms
      \begin{equation*}
        \bset{\aubarLax : ( \Lambda , x ) \in \Poin} \subseteq \Aut
        \Afrakubar \text{.}
      \end{equation*}
    \item $\Kfrak_0^c$ designates a countable set of almost local
      vacuum annihilation operators in $\Afrakubar$, stable with
      respect to transformations from $\Pubarcount$. The image of
      $\Kfrak_0^c$ under all Poincar\'{e} transformations is denoted
      $\Kfrak_0$:
      \begin{equation}
        \label{eq-def-Kzero}
        \Kfrak_0 \doteq \bset{\aubarLax ( K_0 ) : K_0 \in \Kfrak_0^c ,
        ( \Lambda , x ) \in \Poin} \text{.}
      \end{equation}
    \item Together with the numerable uniformly dense $^*$-subalgebra
      of $\Afrakubar$, which exists by construction, $\Kfrak_0^c$
      generates a countable $^*$-algebra over $\Qbb + i \Qbb$, denoted
      $\Aubarcount$ and likewise invariant under $\Pubarcount$.
    \item A countable left ideal in $\Aubarcount$ is then defined by 
      \begin{equation}
        \label{eq-def-Kcount}
        \Kfrak^c \doteq \Aubarcount \, \Kfrak_0^c = \linhull_{\Qbb + i
        \Qbb} \bset{A \, K_0 : A \in \Aubarcount , K_0 \in \Kfrak_0^c}
        \text{.}
      \end{equation}
      It is invariant under the automorphism group $\bset{\aubarLax :
      ( \Lambda , x ) \in \Pubarcount}$ as well. 
    \item Finally, $( \piubar , \Hscrubar )$ is a non-zero,
      non-degenerate representation of the $C^*$-algebra
      $\Afrakubar$.
    \end{trilist}
  \end{subequations}
  The sextuple $( \piubar , \Hscrubar , \Afrakubar , \alphaubar ,
  \Pubarcount , \Kfrak_0^c )$ is called a restricted
  $\Kfrak_0^c$-particle weight, in case that it complies with the
  following list of features:
  \begin{proplist}
  \item There exists a $( \Qbb + i \Qbb )$-linear mapping $\ket{~.~}$
    from $\Kfrak^c$ onto a dense subset $\Hscrubar^c \subseteq
    \Hscrubar$:
    \begin{subequations}
      \begin{equation}
        \label{eq-count-lin-map}
        \ket{~.~} : \Kfrak^c \rightarrow \Hscrubar^c \qquad K \mapsto
        \ket{K} \text{,}
      \end{equation}
      such that the representation $\piubar$ acts on this space
      according to
      \begin{equation}
        \label{eq-rep-count-lin-map}
        \piubar ( A ) \ket{K} = \ket{A K} \text{,} \quad A \in
        \Aubarcount \text{,} \quad K \in \Kfrak^c \text{.}
      \end{equation}
    \end{subequations}
  \item The above linear mapping allows for an extension to any
    operator in $\Kfrak_0$, such that (in the notation of
    Theorem~\ref{The-particle-weight})
    \begin{equation}
      \label{eq-count-lin-map-ext}
      \ket{\Xi_{K'} (~.~)} : \Poin \rightarrow \Hscrubar \qquad (
      \Lambda , x ) \mapsto \bket{\Xi_{K'} ( \Lambda , x )} \doteq
      \bket{\aubarLax ( K' )} \text{,} \quad K' \in \Kfrak_0 \text{,}
    \end{equation}
    is a continuous mapping.
  \item There exists a strongly continuous unitary representation $x
    \mapsto \Uubar ( x )$ of space-time translations $x \in \Rsone$
    with spectral measure $\Delta \mapsto \EubarDelta$, supported by a
    displaced forward light cone $\fwcone - q$, $q \in \fwcone$, which
    implements these transformations in the representation $( \piubar
    , \Hscrubar )$ via
    \begin{subequations}
      \label{eq-transl-relations}
      \begin{equation}
        \label{eq-transl-implement}
        \Uubar ( x ) \piubar ( A ) {\Uubar ( x )}^* = \piubar (
        \aubarx ( A ) ) \text{,} \quad A \in \Afrakubar \text{,} \quad
        x \in \Rsone \text{.}
      \end{equation}
      On the subset $\bset{\ket{K'} : K' \in \Kfrak_0}$ of $\Hscrubar$
      this unitary group acts according to 
      \begin{equation}
        \label{eq-transl-Kzero-action}
        \Uubar ( x ) \ket{K'} = \bket{\aubarx ( K' )} \text{,} \quad
        K' \in \Kfrak_0 \text{,}
      \end{equation}
      and there holds the relation
      \begin{equation}
        \label{eq-spectral-subspace}
        \EubarDeltaprime \ket{K} = \ket{K} \text{,} \quad K \in \Kfrak
        ( \Delta' ) \text{,}
      \end{equation}
      where $\Kfrak ( \Delta' )$ denotes the set of operators from
      $\Kfrak^c \cup \Kfrak_0$ with energy-momentum transfer in the
      Borel set $\Delta' \subseteq \Rsone$.
     \end{subequations}
  \end{proplist}
\end{Def}
Through \eqref{eq-spectral-subspace} we have explicitly installed into
the definition of restricted $\Kfrak_0^c$-particle weights the result
of Proposition~\ref{Pro-spectral-subspace} for generic particle
weights. A spectral assumption of this kind is of great importance
since it constitutes the basis for the proof of the Cluster Property
of Proposition~\ref{Pro-weights-cluster}, and the arguments presented
there can be adopted literally, on condition that the obvious
substitutions are observed, to implement it in the present reduced
setting as well.
\begin{Pro}
  \label{Pro-restr-weights-cluster}
  A restricted $\Kfrak_0^c$-particle weight $( \piubar , \Hscrubar ,
  \Afrakubar , \alphaubar , \Pubarcount , \Kfrak_0^c )$ has the
  Cluster Property presented in Proposition~\ref{Pro-weights-cluster},
  with the reservation that the replacements $\Lfrak_0
  \rightarrowtail \Kfrak_0^c$ and $\Afrak \rightarrowtail \Aubarcount$
  have to be carried out.
\end{Pro}
\begin{Rem}
  The rather intricate Definition~\ref{Def-restr-particle-weight} will
  find its justification in the subsequent section, where it turns
  out, that the characteristics listed above are exactly those which
  survive in the process of spatial disintegration---at least, it did
  regrettably not lie within our reach to establish a more complete
  list of features to be preserved. Nevertheless, it should be noted,
  that those characteristics motivating the interpretation of particle
  weights as asymptotic plane waves are perpetuated (cf.~the first
  paragraph of Section~\ref{sec-particle-weights}).
\end{Rem}

Now, it does not come as a surprise that, with respect to the countable
and separable notions introduced in
Section~\ref{sec-separable-reformulation}, a particle weight of the
general type gives rise to a restricted $\vaccountbar$-particle
weight.
\begin{The}
  \label{The-bullet-weight}
  Let $( \pi_w , \Hscr_w )$ be the GNS-representation corresponding to
  a given particle weight $\scp{~.~}{~.~}$ according to
  Theorem~\ref{The-particle-weight}. Then $( \pi^\bullet , \Hbullet ,
  \Abullet , \alpha^\bullet , \Pcount , \vaccountbar)$ is a restricted
  $\vaccountbar$-particle weight, where the individual entries (if not
  already fixed by Section~\ref{sec-separable-reformulation}) are
  defined as follows:
  \begin{trilist}
  \item $\Hbullet$ designates the Hilbert subspace of $\Hscr_w$, which
    is the closed $\Cbb$-linear span of the assortment of vectors
    $\bset{\ket{L} \in \Hscr_w: L \in \idealcount = \Acountbar \;
    \vaccountbar }$ and thus separable;
  \item $\pi^\bullet \doteq \pi_w \restriction \Abullet$ denotes the
    restriction of the initial representation to the algebra
    $\Abullet$, where the representatives have their limited domain
    as well as range on $\Hbullet$;
  \item $\bset{\abulletLax \doteq \aLax \restriction \Abullet : (
    \Lambda , x ) \in \Poin}$ is the restriction of the initial
    automorphism group to $\Abullet$.
  \end{trilist}
\end{The}
\begin{proof}
  With the definitions $\bullket{~.~} \doteq \ket{~.~} \restriction
  \idealcount$ and $U^\bullet ( x ) \doteq U_w ( x ) \restriction
  \Hbullet$, $x \in \Rsone$, where the latter obviously leaves
  invariant $\Hbullet$ and is such that the corresponding spectral
  measure turns out to be $\EbulletDelta \doteq \EwDelta \restriction
  \Hbullet$ for any Borel set $\Delta$, all features of the restricted
  $\vaccountbar$-particle weight are readily checked on the grounds of
  Theorem~\ref{The-particle-weight} and
  Proposition~\ref{Pro-spectral-subspace}.
\end{proof}

\section[Spatial Disintegration]{Spatial Disintegration of Particle
  Weights}
  \label{sec-spatial-disintegration}

We now get to the central result of this chapter: the construction of
the spatial disintegration of a particle weight in terms of pure ones,
or rather of the corresponding restricted $\vaccountbar$-particle
weight into a direct integral of pure representations, which again
are associated with restricted $\vaccountbar$-particle weights. In
Theorem~\ref{The-bullet-weight} the representation $( \pi^\bullet ,
\Hbullet )$ of the norm-separable $C^*$-algebra $\Abullet$ on the
separable Hilbert space $\Hbullet$ was derived from the given particle
weight $\scp{~.~}{~.~}$. This places the method of spatial
disintegration expounded in the relevant literature at our disposal to
apply it to the problem at hand. In order to express $\pi^\bullet$ in
terms of an integral of irreducible representations, a last
preparatory step has to be taken: a \emph{maximal abelian} von Neumann
algebra $\Mfrak$ in the commutant of $\pi^\bullet ( \Abullet )$ has to
be selected in view of \cite[Theorem~8.5.2]{dixmier:1982}. The choice
of such an algebra is restricted by our further objective to arrive at
a disintegration in terms of restricted $\vaccountbar$-particle
weights, which means that one has to provide for the possibility to
establish the relations \eqref{eq-transl-relations}.

The unitary group $\bset{U^\bullet ( x ) : x \in \Rsone}$ has
generators with joint spectrum in a displaced forward light
cone. Through multiplication by suitably chosen exponential factors
$\exp ( i \: q \, x )$ with fixed $q \in \fwcone$ we can pass to
another unitary group which likewise implements the space-time
translations but has spectrum contained in $\fwcone$. This places
\cite[Theorem~IV.5]{borchers:1984} at our disposal, implying that one
can find a strongly continuous unitary group of this kind with
elements belonging to $\pi^\bullet ( \Abullet )''$, the weak closure
of $\pi^\bullet ( \Abullet )$
(cf.~\cite[Corollary~2.4.15]{bratteli/robinson:1987}). This result can
again be tightened up by use of
\cite[Theorem~3.3]{borchers/buchholz:1985} in the sense that among all
the unitary groups complying with the above features there exists
exactly one which is characterized by the further requirement that the
lower boundary of the joint spectrum of its generators be Lorentz
invariant. It is denoted as
\begin{subequations}
  \begin{equation}
    \label{eq-can-unitary-group}
    \bset{U_c^\bullet ( x ) \in \pi^\bullet ( \Abullet )'' : x \in
    \Rsone} \text{.}
  \end{equation}
  At this point it turns out to be significant that the $C^*$-algebra
  $\Abullet$ has been constructed in
  Section~\ref{sec-separable-reformulation} by using local operators
  so that the reasoning given in \cite{borchers/buchholz:1985} applies
  to the present situation. Another unitary group can be defined
  through
  \begin{equation}
    \label{eq-unitary-renorm-group}
    \bset{V^\bullet ( x ) \doteq U_c^\bullet ( x ) {U^\bullet ( x
     )}^{-1} : x \in \Rsone} \text{.}
  \end{equation}
\end{subequations}
By their very construction, all the operators $V^\bullet ( x )$, $x \in
\Rsone$, are elements of $\pi^\bullet ( \Abullet )'$. The maximal
commutative von Neumann algebra $\Mfrak$ that we are going to work with
in the sequel is now selected in compliance with the condition
\begin{equation}
  \label{eq-condition-max-comm-alg}
  \bset{V^\bullet ( x ) : x \in \Rsone}'' \subseteq \Mfrak \subseteq
  \Bigl( \pi^\bullet ( \Abullet ) \cup \bset{U^\bullet ( x ) : x \in
  \Rsone} \Bigr)' \text{.}
\end{equation}
The main result to be acquired in the present chapter can then be
summarized in the subsequent theorem.
\begin{The}
  \label{The-spatial-disintegration}
  Let $\scp{~.~}{~.~}$ be a generic particle weight with
  representation $( \pi_w , \Hscr_w )$ inducing, by
  Theorem~\ref{The-bullet-weight}, the restricted
  $\vaccountbar$-particle weight $( \pi^\bullet , \Hbullet , \Abullet
  , \alpha^\bullet , \Pcount , \vaccountbar)$. With respect to the
  representation $( \pi^\bullet , \Hbullet )$ of the separable
  $C^*$-algebra $\Abullet$ on the separable Hilbert space $\Hbullet$,
  we select a maximal abelian von Neumann algebra $\Mfrak$ such that
  \eqref{eq-condition-max-comm-alg} is fulfilled. Then there exist a
  standard Borel space $\Xecm$, a bounded positive measure $\nu$ on
  $\Xecm$, and a field of restricted $\vaccountbar$-particle weights
  \begin{equation}
    \label{eq-def-pure-weight-field}
    \Xecm \ni \xi \mapsto ( \pi_\xi , \Hscr_\xi , \Abullet ,
    \alpha^\bullet , \Pcount , \vaccountbar) \text{,}
  \end{equation}
  such that the following assertions hold true:
  \begin{subequations}
    \label{eq-spatial-disintegration}
    \begin{proplist}
    \item The field $\xi \mapsto ( \pi_\xi , \Hscr_\xi )$, as part of
      \eqref{eq-def-pure-weight-field}, is a $\nu$-measurable field of
      irreducible representations of $\Abullet$.
    \item The non-zero representation $( \pi^\bullet , \Hbullet )$ is
      unitarily equivalent to the direct integral of this field of
      irreducible representations:
      \begin{equation}
        \label{eq-rep-disintegration}
        ( \pi^\bullet , \Hbullet ) \simeq \int_\Xecm^\oplus d \nu (
        \xi ) \; ( \pi_\xi , \Hscr_\xi ) \text{,}
      \end{equation}
      and, when $W$ denotes the unitary operator connecting both sides
      of \eqref{eq-rep-disintegration}, the vectors in both spaces are
      linked up by the relation
      \begin{equation}
        \label{eq-spatial-disintegration-of-vectors}
        W \, \bullket{L} = \bset{\xiket{L} : \xi \in \Xecm} \doteq
        \int_\Xecm^\oplus d \nu ( \xi ) \; \xiket{L} \text{,} \quad L
        \in \idealcount \cup \vacbar \text{,}
      \end{equation}
      where $\xiket{~.~}$ denotes the linear mapping characteristic
      for the restricted $\vaccountbar$-particle weight $( \pi_\xi ,
      \Hscr_\xi , \Abullet , \alpha^\bullet , \Pcount , \vaccountbar)$,
      according to \eqref{eq-count-lin-map} in
      Definition~\ref{Def-restr-particle-weight}.
    \item The von Neumann algebra $\Mfrak$ coincides with the algebra
      of those operators which are diagonalisable with respect to the
      above disintegration of $( \pi^\bullet , \Hbullet )$: any
      operator $T \in \Mfrak$ corresponds to an essentially bounded
      measurable complex-valued function $g_T$ according to
      \begin{equation}
        \label{eq-diagonalised-operators}
        W \, T \, W^* = \int_\Xecm^\oplus d \nu ( \xi ) \; g_T ( \xi )
        \, \unit_\xi \text{,}
      \end{equation}
      where $\unit_\xi$, $\xi \in \Xecm$, are the unit operators of
      the algebras $\BHxi$, respectively.
    \item Let $\bset{U_\xi ( x ) : x \in \Rsone} \subseteq \BHxi$
      denote the unitary group, which implements the space-time
      translations in the restricted $\vaccountbar$-particle weight
      pertaining to $\xi \in \Xecm$ according to
      \eqref{eq-transl-implement}, and let $E_\xi ( \Delta ) \in
      \BHxi$ designate the corresponding spectral measure belonging to
      the Borel set $\Delta \subseteq \Rsone$. Then the fields of
      operators
      \begin{equation*}
        \xi \mapsto U_\xi ( x ) \qquad \text{and} \qquad \xi \mapsto
        E_\xi ( \Delta )
      \end{equation*}
      are measurable and satisfy for any $x$ and any Borel set
      $\Delta$ the following equations:
      \begin{align}
        \label{eq-unitary-group-disint}
        W \, U^\bullet ( x ) \, W^* & = \int_\Xecm^\oplus d \nu ( \xi
        ) \; U_\xi ( x ) \text{,} \\
        \label{eq-spectral-measure-disint}
        W \, E^\bullet ( \Delta ) \, W^* & = \int_\Xecm^\oplus d \nu (
        \xi ) \; E_\xi ( \Delta ) \text{.}
      \end{align}
    \item There exists a canonical choice of a strongly continuous
      unitary group in each Hilbert space $\Hscr_\xi$
      \begin{equation}
        \label{eq-canonical-unitary-group}
        \bset{U_\xi^c ( x ) \in \pi_\xi ( \Abullet )'' = \BHxi : x \in
        \Rsone} \text{,}
      \end{equation}
      which is measurable with respect to $\xi$, implements the
      space-time translations in the representation $( \pi_\xi ,
      \Hscr_\xi )$ and has generators $P_\xi^c$ whose joint spectrum
      lies in the closed forward light cone $\fwcone$. It is defined
      by
      \begin{equation}
        \label{eq-unitary-group-connection}
        U_\xi^c ( x ) \doteq \exp ( i \: p_\xi x ) \, U_\xi ( x )
        \text{,} \quad x \in \Rsone \text{,}
      \end{equation}
      where $p_\xi$ is the unequivocal vector in $\Rsone$ that is to
      be interpreted as the sharp energy-momentum corresponding to the
      respective particle weight.
    \end{proplist}
  \end{subequations}
\end{The}
\begin{Rem}
  The concepts occurring in the theory of direct integrals of Hilbert
  spaces (standard Borel space, decomposable and diagonalisable
  operators, and the like) are expounded in
  \cite[Chapter~3]{arveson:1976}, \cite[Part~II]{dixmier:1981} and
  likewise \cite[Section~IV.8 and Appendix]{takesaki:1979}. 
\end{Rem}
\begin{proof}
  The presuppositions of this theorem meet the requirements for an
  application of \cite[Theorem~8.5.2]{dixmier:1982}. This supplies us
  with
  \begin{bulletlist}
  \item a standard Borel space $\Xecmbar$,
  \item a bounded positive measure $\nubar$ on $\Xecmbar$,
  \item a $\nubar$-measurable field $\xi \mapsto ( \pi_\xi ,
    \Hscr_\xi )$ on $\Xecmbar$ consisting of irreducible
    representations $\pi_\xi$ of the $C^*$-algebra $\Abullet$ on the
    Hilbert spaces $\Hscr_\xi$,
  \item and an isomorphism (a linear isometry) $\Wbar$ from $\Hbullet$
    onto the direct integral of these Hilbert spaces, such that
    \begin{subequations}
      \label{eq-direct-integral}
      \begin{equation}
        \label{eq-direct-integral-Hilbert-spaces}
        \Wbar : \Hbullet \rightarrow \int_{\Xecmbar}^\oplus d \nubar (
        \xi ) \; \Hscr_\xi \text{,}
      \end{equation}
      transforms $\pi^\bullet$ into the direct integral of the
      representations $\pi_\xi$ according to
      \begin{equation}
        \label{eq-direct-integral-representations}
        \Wbar \pi^\bullet ( A ) \Wbar^* = \int_{\Xecmbar}^\oplus d
        \nubar ( \xi ) \; \pi_\xi ( A ) \text{,} \quad A \in \Abullet
        \text{,}
      \end{equation}
      and the maximal abelian von Neumann algebra $\Mfrak$ can be
      identified with the algebra of diagonalisable operators via
      \begin{equation}
        \label{eq-direct-integral-M}
        \Wbar T \, \Wbar^* = \int_{\Xecmbar}^\oplus d \nubar ( \xi )
        \; g_T ( \xi ) \, \unit_\xi \text{,} \quad T \in \Mfrak
        \text{,}
      \end{equation}
      with an appropriate function $g_T \in L^\infty \bigl( \Xecmbar ,
      d \nubar ( \xi ) \bigr)$.
    \end{subequations}
  \end{bulletlist}

  At first sight, the different statements of
  \cite[Theorem~8.5.2]{dixmier:1982} listed above seem to cover almost
  all of the assertions of the present
  Theorem~\ref{The-spatial-disintegration}, but one must not forget
  that the disintegration is to be expressed in terms of a field of
  restricted $\vaccountbar$-particle weights. So we are left with the
  task to establish their defining properties in the representations
  $( \pi_\xi , \Hscr_\xi )$ supplied by the standard disintegration
  theory. In accomplishing this assignment, one has to see to it that
  simultaneously relation \eqref{eq-spatial-disintegration-of-vectors}
  is to be satisfied, which means that one is faced with the following
  problem: In general the isomorphism $\Wbar$ connects a given vector
  $\Psi \in \Hbullet$ not with a unique vector field $\bset{\Psi_\xi :
  \xi \in \Xecm}$ but rather with an equivalence class of such fields,
  characterized by the fact that its elements differ pairwise at most
  on $\nubar$-null sets. In contrast to this,
  \eqref{eq-spatial-disintegration-of-vectors} associates the vector
  field $\bset{\xiket{L} : \xi \in \Xecm}$ with $\bullket{L}$ for any
  $L \in \idealcount \cup \vacbar$, leaving no room for any
  ambiguity. In particular, the algebraic relations prevailing in the
  set $\idealcount \cup \vacbar$ which carry over to $\ket{~.~}$ have
  to be observed in defining each of the mappings $\xiket{~.~}$ which
  are characteristic of a restricted $\vaccountbar$-particle
  weight. The contents of the theorem quoted above, important as they
  are, can therefore only serve as the starting point for the
  constructions carried out below, in the course of which again and
  again $\nubar$-null sets have to be removed from $\Xecmbar$ to
  secure definiteness of the remaining components in the
  disintegration of a given vector. In doing so, one has to be
  cautious not to apply this procedure uncountably many times; for,
  otherwise, by accident the standard Borel space $\Xecm \subseteq
  \Xecmbar$ arising in the end could happen to be itself a
  $\nubar$-null set. Then, if $\nu$ denotes the restriction of
  $\nubar$ to this set, one would have $\nubar ( \Xecm ) = \nu ( \Xecm
  ) = 0$, in contradiction to the disintegration
  \eqref{eq-rep-disintegration} of the \emph{non-zero} representation
  $( \pi^\bullet , \Hbullet )$.
  \begin{prooflist}
  \item As indicated above, our first task in view of
    \eqref{eq-count-lin-map} and \eqref{eq-rep-count-lin-map} of
    Definition~\ref{Def-restr-particle-weight} will be to establish
    the existence of $(\Qbb + i \Qbb)$-linear mappings
    \begin{subequations}
      \begin{equation}
        \label{eq-def-xiket}
        \xiket{~.~} : \idealcount \rightarrow \Hscr_\xi^c \qquad L
        \mapsto \xiket{L} \text{,}
      \end{equation}
      from $\idealcount$ onto a dense subset $\Hscr_\xi^c$ of each of
      the component Hilbert spaces supplied by
      \cite[Theorem~8.5.2]{dixmier:1982} with the property
      \begin{equation}
        \label{eq-rep-xiket}
        \pi_\xi ( A ) \xiket{L} = \xiket{A L} \text{,} \quad A \in
        \Acountbar \text{,} \quad L \in \idealcount \text{.}
      \end{equation}
    \end{subequations}
    Now, by relation \eqref{eq-direct-integral-Hilbert-spaces}, there
    exists to each $L \in \idealcount$ an equivalence class of vector
    fields on $\Xecm$ which corresponds to the element $\bullket{L}$
    in $\Hbullet$. The assumed $(\Qbb + i \Qbb)$-linearity of the
    mapping $\bullket{~.~} : \idealcount \rightarrow \Hbullet$ carries
    first of all over to these equivalence classes, but, upon
    selection of a single representative from each class, it turns out
    that every algebraic relation in question is fulfilled in all
    components of the representatives involved, possibly apart from
    those pertaining to a $\nubar$-null set. So, if we pick out one
    representative of the vector $\bullket{L}$ for every $L$ in the
    numerable set $\idealcount$ and designate it as $\bset{\xiket{L}:
    \xi \in \Xecm}$, all of the countably many relations that
    constitute $(\Qbb + i \Qbb)$-linearity are satisfied for
    $\nubar$-almost all of the components of these representatives.
    They can thus be taken to define the linear mappings of the form
    \eqref{eq-def-xiket} for all $\xi$ in a Borel subset $\Xecmbar_1$
    of $\Xecmbar$, which is left by the procedure of dismissing an
    appropriate $\nubar$-null set for each algebraic relation to be
    satisfied.

    The same reasoning can be applied to the disintegration of vectors
    of the form $\bullket{A L} = \pi^\bullet ( A ) \bullket{L}$ with
    $A \in \Acountbar$ and $L \in \idealcount$. Again with
    \eqref{eq-spatial-disintegration-of-vectors} in mind, the number
    of relations \eqref{eq-rep-xiket} to be satisfied is countable, so
    that in view of relation
    \eqref{eq-direct-integral-representations} the mere removal of an 
    appropriate $\nubar$-null set from $\Xecmbar_1$ leaves only those
    indices $\xi$ behind, for which the mappings $\xiket{~.~}$ indeed
    have the desired property \eqref{eq-rep-xiket}.

    In this way we have implemented by hand the first defining
    property of restricted $\vaccountbar$-particle weights in the
    representations $( \pi_\xi , \Hscr_\xi)$ for $\nubar$-almost all
    indices $\xi$. The only thing that remains to be done in this
    connection is to show that $\bset{\xiket{L} : L \in \idealcount}$
    is a dense subset $\Hscr_\xi^c$ in $\Hscr_\xi$. But, according to
    \cite[Section~II.1.6, Proposition~8]{dixmier:1981}, the fact that
    the set $\bset{\bullket{L} : L \in \idealcount}$ is total in
    $\Hbullet$ by assumption implies that the corresponding property
    holds for $\nubar$-almost all $\xi$ in the disintegration. Thus
    there exists a non-null Borel set $\Xecmbar_2 \subseteq
    \Xecmbar_1$, such that the corresponding mappings $\xiket{~.~}$,
    $\xi \in \Xecmbar_2$, have this property, too. In this way all of
    the characteristics presented in the first item of
    Definition~\ref{Def-restr-particle-weight} are fulfilled for $\xi
    \in \Xecmbar_2$ by the mappings \eqref{eq-def-xiket} constructed
    above, and additionally we have
    \begin{equation}
      \label{eq-spatial-disintegration-of-idealcount-vectors}
      \Wbar \, \bullket{L} = \int_{\Xecmbar_2}^\oplus d \nubar ( \xi )
      \; \xiket{L} \text{,} \quad L \in \idealcount \text{.}
    \end{equation}
  \item In the next step, the mappings $\xiket{~.~}$ have to be
    extended to the set $\vacbar$ of all Poincar\'{e} transforms of
    operators from $\vaccountbar$ in such a way that the counterpart
    of \eqref{eq-count-lin-map-ext} in
    Definition~\ref{Def-restr-particle-weight} is continuous. In the
    present notation this is the mapping
    \begin{equation}
      \label{eq-xiket-ext}
      \Poin \ni ( \Lambda , x ) \mapsto \bxiket{\abulletLax ( L' )}
      \in \Hscr_\xi \text{,} \qquad L' \in \vacbar \text{.}
    \end{equation}
    At this point the special selection of $\vaccountbar$ as
    consisting of compactly regularized vacuum annihilation operators
    comes into play, and also the invariance of this set under
    transformations $( \Lambda , x ) \in  \Pbarcount$ will be of
    importance. Great care has to be taken in these investigations
    based on the differentiability properties of the operators in
    question, that not uncountably many conditions are imposed on the
    mappings $\xiket{~.~}$, since anew not all of them will share the
    claimed extension property, but only a $\nubar$-null subset of
    $\Xecmbar_2$ shall get lost on the way.

    To start with, note that the Poincar\'{e} group $\Poin$ can be
    covered by a sequence of open sets $\Vsf_i$ with compact closures
    $\Csf_i$, $i \in \Nbb$, contained in corresponding open charts $(
    \Usf_i , \phi_i )$ with the additional property that the sets
    $\phi_i ( \Csf_i ) \subseteq \Rbb^{d_\Psf}$ are convex
    (e.\,g.~consider the translates of the canonical coordinates $(
    \Usf_0 , \phi_0 )$ around $( \unit , 0 )$ to all elements of
    $\Pcount$ and take suitable open subsets thereof). Select one of
    these compacta, say $\Csf_k$, and fix an element $\Hat{L}_0 \in
    \vaccountbar$, which by assumption is given as a compactly
    supported regularization of an element $L_0 \in \vaccount$:
    \begin{subequations}
      \begin{equation}
        \label{eq-def-restr-vac-ann}
        \Hat{L}_0 = \alpha_F ( L_0 ) \doteq \int_{\Ssf_F} d \mu (
        \Lambda , x ) \; F ( \Lambda , x ) \, \aLax ( L_0 ) \text{,}
      \end{equation}
      where $F$ is an infinitely often differentiable function on
      $\Poin$ with compact support $\Ssf_F$ in the Poincar\'{e} group
      $\Poin$. According to Lemma~\ref{Lem-alpha-ket-integral} the
      mapping $\ket{~.~}$ commutes with this integral so that the
      vector $\ket{\Hat{L}_0}$ in $\Hscr_w$ takes on the shape
      \begin{equation}
        \label{eq-F-restr-ket-integral}
        \ket{\Hat{L}_0} = \int_{\Ssf_F} d \mu ( \Lambda , x ) \; F (
        \Lambda , x ) \, \bket{\aLax ( L_0 )} \text{.} 
      \end{equation}
      The same equation holds for the Poincar\'{e} transforms of the
      operator $\Hat{L}_0$ as well, so that invariance of the Haar
      measure on $\Poin$ implies for any $( \Lambda_0 , x_0 ) \in
      \Csf_k$ the equations
      \begin{multline}
        \label{eq-transl-F-restr-ket-integral}
        \bket{\aLaxzero ( \Hat{L}_0 )} = \int_{\Ssf_F} d \mu ( \Lambda
        , x ) \; F ( \Lambda , x ) \, \bket{\alpha_{( \Lambda_0 , x_0
        ) ( \Lambda , x )} ( L_0 )} \\
        \mspace{55mu} = \int_{( \Lambda_0 , x_0 )
        \cdot \Ssf_F} d \mu ( \Lambda , x ) \; F \bigl( ( \Lambda_0 ,
        x_0 )^{-1} ( \Lambda , x ) \bigr) \, \bket{\aLax ( L_0 )} \\
        = \int_{\Csf_k \cdot \Ssf_F} d \mu ( \Lambda , x ) \; F \bigl(
        ( \Lambda_0 , x_0 )^{-1} ( \Lambda , x ) \bigr) \, \bket{\aLax
        ( L_0 )} \text{.}
      \end{multline}
      The derivatives of the mapping $( \Lambda_0 , x_0 ) \mapsto
      \bket{\aLaxzero ( \Hat{L}_0 )}$, the domain of $( \Lambda_0 ,
      x_0 )$ restricted to the neighbourhood $\Vsf_k$ in $\Csf_k$, are
      thus explicitly seen to be expressible in terms of derivatives
      of the functions
      \begin{equation*}
        F^{( \Lambda , x )} : \Vsf_k \rightarrow \Cbb \qquad (
        \Lambda_0 , x_0 ) \mapsto F^{( \Lambda , x )} ( \Lambda_0 ,
        x_0 ) \doteq F \bigl( ( \Lambda_0 , x_0 )^{-1} ( \Lambda , x )
        \bigr) \text{.}
      \end{equation*}
      So, let $( \Lambda_1 , x_1 )$ and $( \Lambda_2 , x_2 )$ be a
      pair of Poincar\'{e} transformations lying in the common
      neighbourhood $\Vsf_k$; then the following equation results from
      an application of the Mean Value Theorem~\ref{The-mean-value} to
      the $\Xscr_{\Lfrak_0}$-differentiable mapping $\ket{~.~}$
      (cf.~Theorem~\ref{The-particle-weight}):
      \begin{multline}
        \label{eq-vacbar-extension-mean-value}
        \bket{\aLaxone ( \Hat{L}_0 ) - \aLaxtwo ( \Hat{L}_0 )} \\
        \mspace{-60mu} = \bket{\alpha_{\phi_k^{-1} ( \sib )} (
        \Hat{L}_0 ) - \alpha_{\phi_k^{-1} ( \tib )} ( \Hat{L}_0 )} =
        \int_0^1 d \vartheta \; \bket{\Dfrak ( \Xi_{\Hat{L}_0} \circ
        \phi_k^{-1} ) ( \tib + \vartheta ( \sib - \tib ) ) ( \sib -
        \tib )} \\
        = \int_0^1 d \vartheta \int_{\Csf_k \cdot \Ssf_F} d \mu (
        \Lambda , x ) \; \Dfrak ( F^{( \Lambda , x )} \circ
        \phi_k^{-1} ) ( \tib + \vartheta ( \sib - \tib ) ) ( \sib -
        \tib ) \, \bket{\aLax ( L_0 )} \text{,}
      \end{multline}
    \end{subequations}
    where $\sib \doteq \phi_k ( \Lambda_1 , x_1 )$ and $\tib \doteq
    \phi_k ( \Lambda_2 , x_2 )$ belong to the compact and
    \emph{convex} set $\phi_k ( \Csf_k )$.

    Now, the vector $\bket{\aLaxone ( \Hat{L}_0 ) - \aLaxtwo (
    \Hat{L}_0 )}$ defines a positive functional on the algebra $\BHw$,
    and we want to show that this vector functional can be majorized
    by a positive normal functional in $\BH_*$. To establish this
    fact, note, that the integrals in
    \eqref{eq-vacbar-extension-mean-value} exist in the uniform
    topology of $\Hscr_w$, so that they commute with every bounded
    linear operator $B \in \BHw$. Hence
    \begin{subequations}
      \begin{multline}
        \label{eq-vacbar-extension-matrix-element}
        \bscpx{\aLaxone ( \Hat{L}_0 ) - \aLaxtwo ( \Hat{L}_0
        )}{B}{\aLaxone ( \Hat{L}_0 ) - \aLaxtwo ( \Hat{L}_0 )} \\
        = \iint \limits_{[ 0 , 1 ]^2} d \vartheta \; d \vartheta' \;
        \bscpx{\Dfrak ( \Xi_{\Hat{L}_0} \circ \phi_k^{-1} ) ( \tib +
        \vartheta' ( \sib - \tib ) ) ( \sib - \tib )}{B}{\Dfrak (
        \Xi_{\Hat{L}_0} \circ \phi_k^{-1} ) ( \tib + \vartheta ( \sib
        - \tib ) ) ( \sib - \tib )} \text{.}
      \end{multline}
      This equation is invariant with respect to an exchange of
      $\vartheta$ and $\vartheta'$. In the case of a \emph{positive}
      operator $B$ the following relation holds for arbitrary vectors
      $\Psi$ and $\Phi$ in $\Hscr_w$:
      \begin{equation*}
        \scpx{\Psi}{B}{\Phi} + \scpx{\Phi}{B}{\Psi} \leqslant
        \scpx{\Psi}{B}{\Psi} + \scpx{\Phi}{B}{\Phi} \text{,}
      \end{equation*}
      which, applied to the integrand of
      \eqref{eq-vacbar-extension-matrix-element} and to that resulting
      from an interchange of $\vartheta$ and $\vartheta'$, yields
      \begin{multline}
        \label{eq-vacbar-extension-estimate1}
        \bscpx{\aLaxone ( \Hat{L}_0 ) - \aLaxtwo ( \Hat{L}_0
        )}{B}{\aLaxone ( \Hat{L}_0 ) - \aLaxtwo ( \Hat{L}_0 )} \\
        \leqslant \int_0^1 d \vartheta \; \bscpx{\Dfrak (
        \Xi_{\Hat{L}_0} \circ \phi_k^{-1} ) ( \tib + \vartheta ( \sib
        - \tib ) ) ( \sib - \tib )}{B}{\Dfrak ( \Xi_{\Hat{L}_0} \circ
        \phi_k^{-1} ) ( \tib + \vartheta ( \sib - \tib ) ) ( \sib -
        \tib )}
      \end{multline}
      upon execution of a trivial integration over $\vartheta$ and
      $\vartheta'$, respectively. As in
      \eqref{eq-vacbar-extension-mean-value} we can pass to the
      following representation for the integrand on the right-hand
      side of \eqref{eq-vacbar-extension-estimate1}:
      \begin{multline}
        \label{eq-vacbar-extension-estimate2}
        \bscpx{\Dfrak ( \Xi_{\Hat{L}_0} \circ \phi_k^{-1} ) ( \tib +
        \vartheta ( \sib - \tib ) ) ( \sib - \tib )}{B}{\Dfrak (
        \Xi_{\Hat{L}_0} \circ \phi_k^{-1} ) ( \tib + \vartheta ( \sib
        - \tib ) ) ( \sib - \tib )} \\
        \mspace{-20mu} = \int_{\Csf_k \cdot \Ssf_F} d \mu ( \Lambda ,
        x ) \int_{\Csf_k \cdot \Ssf_F} d \mu ( \Lambda' , x' ) \;
        \overline{\Dfrak ( F^{( \Lambda' , x' )} \circ \phi_k^{-1} ) (
        \tib + \vartheta ( \sib - \tib ) ) ( \sib - \tib )} \cdot \\
        \cdot \Dfrak ( F^{( \Lambda , x )} \circ \phi_k^{-1} ) ( \tib
        + \vartheta ( \sib - \tib ) ) ( \sib - \tib ) \,
        \bscpx{\aLaxprime ( L_0 )}{B}{\aLax ( L_0 )} \text{.}
      \end{multline}
      The derivatives which show up in
      \eqref{eq-vacbar-extension-estimate2} depend by construction
      continuously on the parameters $\sib$ and $\tib$, $\vartheta$
      and $\vartheta'$ as well as $( \Lambda , x )$ and $( \Lambda' ,
      x' )$, so that their absolute values, taken on the compact
      domains $\phi_k ( \Csf_k )$, $[ 0 , 1 ]$ and $\Csf_k \cdot
      \Ssf_F$, respectively, are bounded by
      \begin{equation*}
        \babs{\Dfrak ( F^{( \Lambda , x )} \circ \phi_k^{-1} ) ( \tib
        + \vartheta ( \sib - \tib ) ) ( \sib - \tib )} \leqslant D ( F
        ; \Csf_k ) \: \abs{\sib - \tib} < \infty 
      \end{equation*}
      for all $( \Lambda , x ) \in \Csf_k \cdot \Ssf_F$ with a
      suitable non-negative constant $D ( F ; \Csf_k )$. Hence the
      non-negative matrix element in
      \eqref{eq-vacbar-extension-estimate2} can be estimated by
      \begin{multline}
        \label{eq-vacbar-extension-estimate3}
        \bscpx{\Dfrak ( \Xi_{\Hat{L}_0} \circ \phi_k^{-1} ) ( \tib +
        \vartheta ( \sib - \tib ) ) ( \sib - \tib )}{B}{\Dfrak (
        \Xi_{\Hat{L}_0} \circ \phi_k^{-1} ) ( \tib + \vartheta ( \sib
        - \tib ) ) ( \sib - \tib )} \\
        \leqslant D ( F ; \Csf_k )^2 \, \abs{\sib - \tib}^2
        \int_{\Csf_k \cdot \Ssf_F} d \mu ( \Lambda , x ) \int_{\Csf_k
        \cdot \Ssf_F} d \mu ( \Lambda' , x' ) \;
        \babs{\bscpx{\aLaxprime ( L_0 )}{B}{\aLax ( L_0 )}} \text{,}
      \end{multline}
      which is independent of $\vartheta$, so that insertion into
      \eqref{eq-vacbar-extension-estimate1} yields
      \begin{multline}
        \label{eq-vacbar-extension-estimate4}
        \bscpx{\aLaxone ( \Hat{L}_0 ) - \aLaxtwo ( \Hat{L}_0
        )}{B}{\aLaxone ( \Hat{L}_0 ) - \aLaxtwo ( \Hat{L}_0 )} \\
        \leqslant D ( F ; \Csf_k )^2 \, \abs{\sib - \tib}^2
        \int_{\Csf_k \cdot \Ssf_F} d \mu ( \Lambda , x ) \int_{\Csf_k
        \cdot \Ssf_F} d \mu ( \Lambda' , x' ) \;
        \babs{\bscpx{\aLaxprime ( L_0 )}{B}{\aLax ( L_0 )}} \text{.}
      \end{multline}
      Since the positive operator $B$ can be written as $C^* C$ for
      suitable $C \in \BHw$, the integrand on the right-hand side
      allows for the following estimate, making use of the relation
      between the geometric and the arithmetic mean of two
      non-negative numbers:
      \begin{multline*}
        \babs{\bscpx{\aLaxprime ( L_0 )}{B}{\aLax ( L_0 )}} \leqslant
        \bnorm{C \bket{\aLaxprime ( L_0 )}} \, \bnorm{C \bket{\aLax (
        L_0 )}} \\
        \leqslant 2^{-1} \Bigl( \bscpx{\aLaxprime ( L_0
        )}{B}{\aLaxprime ( L_0 )} + \bscpx{\aLax ( L_0 )}{B}{\aLax (
        L_0 )} \Bigr) \text{.}
      \end{multline*}
      As a consequence of this inequality entered into
      \eqref{eq-vacbar-extension-estimate4}, one integration over
      $\Csf_k \cdot \Ssf_F$ can be carried out on its right-hand side
      for each resulting term of the sum, so that finally
      \begin{multline}
        \label{eq-vacbar-extension-estimate5}
        \bscpx{\aLaxone ( \Hat{L}_0 ) - \aLaxtwo ( \Hat{L}_0
        )}{B}{\aLaxone ( \Hat{L}_0 ) - \aLaxtwo ( \Hat{L}_0 )} \\
        \leqslant D ( F ; \Csf_k )^2 \, \abs{\sib - \tib}^2 \, \mu (
        \Csf_k \cdot \Ssf_F ) \int_{\Csf_k \cdot \Ssf_F} d \mu (
        \Lambda , x ) \; \bscpx{\aLax ( L_0 )}{B}{\aLax ( L_0 )}
        \text{,}
      \end{multline}
    \end{subequations}
    where the last integral can be viewed as a positive normal
    functional on $\BHw$ in the variable $B$, as announced at the
    beginning of this paragraph.

    Now, let $\Mecm$ be a measurable subset of $\Xecmbar$ then,
    according to \eqref{eq-direct-integral-M}, it corresponds via the
    associated characteristic function $\chi_\Mecm$ to a projection
    $P_\Mecm$ in the selected maximal abelian von Neumann algebra
    $\Mfrak$. If $P^\bullet$ in turn denotes the orthogonal projection
    from $\Hscr_w$ onto the Hilbert space $\Hbullet$, we can define
    $B_\Mecm \doteq P^\bullet P_\Mecm P^\bullet$ as a positive
    operator in $\BHw$, which is therefore subject to
    \eqref{eq-vacbar-extension-estimate5}. This relation can then be
    re-written for $B = B_\Mecm$ in terms of the restricted
    $\vaccountbar$-particle weight $( \pi^\bullet , \Hbullet ,
    \Abullet , \alpha^\bullet , \Pcount , \vaccountbar)$:
    \begin{subequations}
      \begin{multline}
        \label{eq-vacbar-bullet-extension-estimate}
        \bnorm{P_\Mecm \bbullket{\abulletLaxone ( \Hat{L}_0 ) -
        \abulletLaxtwo ( \Hat{L}_0 )}}^2 = \bnorm{P_\Mecm P^\bullet
        \bket{\aLaxone ( \Hat{L}_0 ) - \aLaxtwo ( \Hat{L}_0 )}}^2 \\
        \leqslant D ( F ; \Csf_k )^2 \, \abs{\sib - \tib}^2 \, \mu (
        \Csf_k \cdot \Ssf_F ) \int_{\Csf_k \cdot \Ssf_F} d \mu (
        \Lambda , x ) \; \bscpx{\aLax ( L_0 )}{P^\bullet P_\Mecm
        P^\bullet}{\aLax ( L_0 )} \text{,}
      \end{multline}
      where now the integral on the right-hand side defines a positive
      normal functional on the von Neumann algebra $\Mfrak$ through
      \begin{equation}
        \label{eq-def-normal-T-functional}
        \varphi [ \Hat{L}_0 ; \Csf_k ] ( T ) \doteq \int_{\Csf_k \cdot
        \Ssf_F} d \mu ( \Lambda , x ) \; \bscpx{\aLax ( L_0
        )}{P^\bullet T P^\bullet}{\aLax ( L_0 )} \text{,} \quad T \in
        \Mfrak \text{.} 
      \end{equation}        
      Specializing to Poincar\'{e} transformations $( \Lambda_1 , x_1
      )$ and $( \Lambda_2 , x_2 )$ from the countable subgroup
      $\Pbarcount$, the unique disintegration of the vector
      $\bbullket{\abulletLaxone ( \Hat{L}_0 ) - \abulletLaxtwo (
      \Hat{L}_0 )}$ occurring on the left-hand side of
      \eqref{eq-vacbar-bullet-extension-estimate} is already
      explicitly given by \eqref{eq-def-xiket} for all $\xi \in
      \Xecmbar_2$ so that
      \begin{equation}
        \label{eq-vacbar-bullet-disint}
        \Wbar \, \bbullket{\abulletLaxone ( \Hat{L}_0 ) -
        \abulletLaxtwo ( \Hat{L}_0 )} = \int_{\Xecmbar_2}^\oplus d \nu
        ( \xi ) \; \bxiket{\abulletLaxone ( \Hat{L}_0 ) -
        \abulletLaxtwo ( \Hat{L}_0 )} \text{.}
      \end{equation}
      On the other hand, the positive normal functional $\varphi [
      \Hat{L}_0 , \Csf_k ] \in \Mfrak_*$ of
      \eqref{eq-def-normal-T-functional} is easily seen by
      \cite[Proposition~IV.8.34]{takesaki:1979} in connection with
      \eqref{eq-direct-integral-M} to correspond to a unique
      integrable field $\bset{\varphi [ \Hat{L}_0 , \Csf_k ]_\xi : \xi
      \in \Xecmbar}$ of positive normal functionals on the von Neumann
      algebras $\Cbb \cdot \unit_\xi$ in the direct integral
      decomposition of $\Mfrak$. Explicitly,
      \begin{equation}
        \label{eq-normal-T-functional-disint}
        \varphi [ \Hat{L}_0 ; \Csf_k ] ( T ) = \int_{\Xecmbar} d
        \nubar ( \xi ) \; g_T ( \xi ) \: \varphi [ \Hat{L}_0 , \Csf_k
        ]_\xi ( \unit_\xi )
      \end{equation}
      for any $T \in \Mfrak$ with an appropriate function $g_T \in
      L^\infty \bigl( \Xecmbar , d \nubar ( \xi ) \bigr)$. The above
      relation stays true, if we replace $\Xecmbar$ by $\Xecmbar_2$,
      since both differ at most by a $\nubar$-null set. So, in view of
      relations \eqref{eq-def-normal-T-functional} through
      \eqref{eq-normal-T-functional-disint},
      \eqref{eq-vacbar-bullet-extension-estimate} can for any
      measurable subset $\Mecm$ of $\Xecmbar_2$ corresponding to the
      orthogonal projection $P_\Mecm \in \Mfrak$ be expressed in terms
      of integrals according to
      \begin{multline}
        \label{eq-vacbar-bullet-extension-integral-estimate}
        \int_{\Mecm} d \nubar ( \xi ) \; \bnorm{\bxiket{\abulletLaxone
        ( \Hat{L}_0 ) - \abulletLaxtwo ( \Hat{L}_0 )}}^2 \\
        \leqslant D ( F ; \Csf_k )^2 \, \abs{\sib - \tib}^2 \, \mu (
        \Csf_k \cdot \Ssf_F ) \int_{\Mecm} d \nubar ( \xi ) \; \varphi
        [ \Hat{L}_0 , \Csf_k ]_\xi ( \unit_\xi ) \text{.}
      \end{multline}
      Due to arbitrariness of $\Mecm \subseteq \Xecmbar_2$, we then
      infer, making use of elementary results of integration theory
      \cite[Chapter~V, viz.~\S\,25, Theorem~D]{halmos:1968}, that for
      $\nubar$-almost all $\xi \in \Xecmbar_2$ there holds the
      estimate
      \begin{multline}
        \label{eq-vacbar-bullet-extension-final-estimate}
        \bnorm{\bxiket{\abulletLaxone ( \Hat{L}_0 ) - \abulletLaxtwo (
        \Hat{L}_0 )}} \\ 
        \leqslant \babs{\phi_k ( \Lambda_1 , x_1 ) - \phi_k (
        \Lambda_2 , x_2 )} \cdot D ( F ; \Csf_k ) \cdot \mu ( \Csf_k
        \cdot \Ssf_F ) \cdot \varphi [ \Hat{L}_0 , \Csf_k ]_\xi (
        \unit_\xi ) \text{,}
      \end{multline}
    \end{subequations}
    where we replaced the points $\sib$ and $\tib$ from the space
    $\Rbb^{d_\Psf}$ of coordinates for $\Poin$ by their pre-images $(
    \Lambda_1 , x_1 )$ and $( \Lambda_2 , x_2 )$ from $\Vsf_k \cap
    \Pbarcount$. The important thing to notice at this point is that,
    apart from the factor $\babs{\phi_k ( \Lambda_1 , x_1 ) - \phi_k (
    \Lambda_2 , x_2 )}$, the terms on the right-hand side of
    \eqref{eq-vacbar-bullet-extension-final-estimate} hinge upon the
    operator $\Hat{L}_0$ (determining the function $F$ as well as its
    support $\Ssf_F$) and on the neighbourhood $\Vsf_k$ with compact
    closure $\Csf_k$ containing $( \Lambda_1 , x_1 ) \text{,} (
    \Lambda_2 , x_2 ) \in \Pbarcount$. Therefore this estimate also
    holds for any other pair of Lorentz transformations in $\Vsf_k
    \cap \Pbarcount$ with the same $( \Hat{L}_0 , \Vsf_k )$-dependent
    factor; of course, in each of the resulting countably many
    relations one possibly loses a further $\nubar$-null subset of
    $\Xecmbar_2$. The reasoning leading up to this point can then be
    applied to any combination of an operator in the numerable
    selection $\vaccountbar$ with an open set from the countable cover
    of $\Poin$ to produce in each case a relation of the form of
    \eqref{eq-vacbar-bullet-extension-final-estimate} for the
    respective Poincar\'{e} transformations in
    $\Pbarcount$. Simultaneously, the domain of indices $\xi$, for
    which \emph{all} of these inequalities are valid, shrinks to an
    appropriate $\nubar$-measurable non-null subset $\Xecmbar_3$ of
    $\Xecmbar_2$.

    Consider now an arbitrary element $( \Lambda_0 , x_0 ) \in \Poin$,
    which belongs to at least one of the open sets $\Vsf_j$ from the
    covering of the Poincar\'{e} group already used above. By density
    of $\Pbarcount$ in $\Poin$, the transformation $( \Lambda_0 , x_0
    )$ can be approximated by a sequence $\bset{( \Lambda_n , x_n
    )}_{n \in \Nbb} \subseteq \Pbarcount \cap \Vsf_j$. This is a
    Cauchy sequence in the initial topology of $\Poin$, so that
    relation \eqref{eq-vacbar-bullet-extension-final-estimate} implies
    that for each $\xi \in \Xecmbar_3$ the corresponding sequences
    \begin{subequations}
      \label{eq-vacbar-extension-xi}
      \begin{equation}
        \label{eq-xi-cauchy-sequence}
        \bset{\bxiket{\abulletLaxn ( \Hat{L}_0 )}}_{n \in \Nbb}
        \subseteq \Hscr_\xi \text{,} \quad \Hat{L}_0 \in
        \vaccountbar\text{,}
      \end{equation}
      likewise have the Cauchy property with respect to the Hilbert
      space norms. Their limits in each of the spaces $\Hscr_\xi$,
      $\xi \in \Xecmbar_3$, thus exist and are obviously independent
      of the approximating sequence of Lorentz transformations from
      $\Pbarcount$. Therefore, we can write
      \begin{equation}
        \label{eq-vaccountbar-xi-limits}
        \bxiket{\Hat{L}_0 ; ( \Lambda_0 , x_0 )} \doteq \lim_{n
        \rightarrow \infty} \bxiket{\abulletLaxn ( \Hat{L}_0
        )} \text{,}
      \end{equation}
      a result that holds for arbitrary $\Hat{L}_0 \in \vaccountbar$
      as long as $\xi$ is taken from the non-null set $\Xecmbar_3$. 
      According to \cite[Definition~II.4.1]{fell/doran:1988a}, which
      lays down the notion of measurability for vector fields, the
      mapping
      \begin{equation}
        \label{eq-vaccountbar-limit-mapping}
        \Xecmbar_3 \ni \xi \mapsto \bxiket{\Hat{L}_0 ; ( \Lambda_0 ,
        x_0 )} \in \Hscr_\xi \text{,}
      \end{equation}
      that arises as the pointwise limit of measurable vector fields
      on $\Xecmbar_3$, is itself measurable with respect to the
      restriction of $\nubar$ to this subset of $\Xecmbar$ and turns
      out to be a representative of the vector
      $\bbullket{\abulletLaxzero ( \Hat{L}_0 )} \in \Hbullet$
      (cf.~\cite[Section~II.1.5, Proof of
      Proposition~5(ii)]{dixmier:1981}, and note that we can neglect
      the null set missing in $\Xecmbar_3$ compared to
      $\Xecmbar$).

      The question now is, if the limits $\bxiket{\Hat{L}_0 ; (
      \Lambda_0 , x_0 )}$, constructed by the above method for
      arbitrary operators $\Hat{L}_0 \in \vaccountbar$ and any
      transformation $( \Lambda_0 , x_0 ) \in \Poin$, can
      \emph{unambiguously} be identified for all $\xi$ in $\Xecmbar_3$
      with vectors $\bxiket{\abulletLaxzero ( \Hat{L}_0 )} \in
      \Hscr_\xi$, which satisfy a relation of the form
      \eqref{eq-spatial-disintegration-of-vectors}. One of the
      situations, in which an inconsistency possibly arises, is the
      appearance of two different representations for a single element
      $L' \in \vacbar$:
      \begin{equation}
        \label{eq-vacbar-coincidence}
        L' = \abulletLaxone ( \Hat{L}_1 ) = \abulletLaxtwo ( \Hat{L}_2
        ) \text{,}
      \end{equation}
      where $\Hat{L}_1 \text{,} \Hat{L}_2 \in \vaccountbar$, and $(
      \Lambda_1 , x_1 ) \text{,} ( \Lambda_2 , x_2 ) \in \Poin$. In
      this case the pair of operators is connected by the Poincar\'{e}
      transformation $( \Lambda_1 , x_1 )^{-1} ( \Lambda_2 , x_2 )$,
      which belongs to the subgroup $\Pbarcount$ of $\Poin$ according
      to the constructions of
      Section~\ref{sec-separable-reformulation}. Therefore
      \begin{equation*}
        \Hat{L}_1 = \alpha^\bullet_{( \Lambda_1 , x_1 )^{-1} (
        \Lambda_2 , x_2 )} ( \Hat{L}_2 ) \text{,}
      \end{equation*}
      which implies that
      \begin{equation*}
        \alpha^\bullet_{( \Lambda_{1,n} , x_{1,n} )} ( \Hat{L}_1 ) =
        \alpha^\bullet_{( \Lambda_{1,n} , x_{1,n} ) ( \Lambda_1 , x_1
        )^{-1} ( \Lambda_2 , x_2 ) } ( \Hat{L}_2 )
      \end{equation*}
      for any sequence $\bset{( \Lambda_{1,n} , x_{1,n} )}_{n \in
      \Nbb} \subseteq \Pbarcount$ approximating $( \Lambda_1 , x_1
      )$. But then the transformations on the right-hand side of the
      last equation constitute another sequence in $\Pbarcount$, which
      in this case tends to $( \Lambda_2 , x_2 )$ in the limit $n
      \rightarrow \infty$. As a consequence of the independence of the
      limits \eqref{eq-vaccountbar-xi-limits} from the selected
      sequence in $\Pbarcount$, we could define
      \begin{equation}
        \label{eq-def-vacbar-xiket}
        \xiket{L'} \doteq \bxiket{\Hat{L}_1 ; ( \Lambda_1 , x_1 )} =
        \bxiket{\Hat{L}_2 ; ( \Lambda_2 , x_2 )} \text{.}
      \end{equation}
      The only problem that is still left open with respect to an
      unequivocal definition of vectors of the form $\xiket{L'}$, $L'
      \in \vacbar$, occurs when the vacuum annihilation operator $L'$
      happens to be an element of $\idealcount$, so that its
      components in the Hilbert spaces $\Hscr_\xi$ have already been
      fixed in the initial step. But, as $\idealcount$ is a numerable
      set, such a coincidence will be encountered at most countably
      often, so that relation \eqref{eq-def-vacbar-xiket} indeed turns
      out to be the unique definition of $\xiket{L'}$ for all $\xi \in
      \Xecmbar_4$, such that the relation
      \begin{equation}
        \label{eq-spatial-disintegration-of-vacbar-vectors}
        \Wbar \, \bullket{L'} = \int_{\Xecmbar_4}^\oplus d \nubar (
        \xi ) \; \xiket{L'} \text{,} \quad L' \in \vacbar \text{,}
      \end{equation}
    \end{subequations}
    is satisfied, where again $\Xecmbar_4$ is a $\nubar$-measurable
    subset which differs from $\Xecmbar_3$ only by a null set.

    The $\Hscr_\xi$-vectors corresponding to elements of $\vacbar$
    that arise as Poincar\'{e} transforms of $L' = \abulletLaxzero (
    \Hat{L}_0) \in \vacbar$ are defined according to
    \eqref{eq-vacbar-extension-xi}, in particular by the relations
    \eqref{eq-def-vacbar-xiket} and \eqref{eq-vaccountbar-xi-limits}.
    As a result, when $( \Lambda_1 , x_1 )$ and $( \Lambda_2 , x_2 )$
    are closely neighbouring elements of $\Poin$ so that their
    products with $( \Lambda_0 , x_0 )$ lie in the common open
    neighbourhood $\Vsf_k$, we get the following estimate, which is a
    direct consequence of the above constructions inserted into
    relation \eqref{eq-vacbar-bullet-extension-final-estimate} and
    which holds for any $\xi \in \Xecmbar_4$:
    \begin{multline}
      \label{eq-vacbar-bullet-continuity}
      \bnorm{\bxiket{\abulletLaxone ( L' )} - \bxiket{\abulletLaxtwo (
      L' )}} \\ 
      \leqslant \babs{\phi_k \bigl( ( \Lambda_0 , x_0 ) ( \Lambda_1 ,
      x_1 ) \bigr) - \phi_k \bigl( ( \Lambda_0 , x_0 ) ( \Lambda_2 ,
      x_2 ) \bigr)} \cdot D ( F ; \Csf_k ) \cdot \mu ( \Csf_k \cdot
      \Ssf_F ) \cdot \varphi [ \Hat{L}_0 , \Csf_k ]_\xi ( \unit_\xi )
      \text{.}
    \end{multline}
    This shows that the continuity property with respect to generic
    Poincar\'{e} transformations as expressed in
    \eqref{eq-count-lin-map-ext} of
    Definition~\ref{Def-restr-particle-weight} is fulfilled by all the
    extended mappings $\xiket{~.~}$ introduced above for arbitrary $L'
    \in \vacbar$.
  \item The last property of restricted $\vaccountbar$-particle
    weights to be established is the existence of unitary groups
    $\bset{U_\xi ( x ) : x \in \Rsone}$ in the representations $(
    \pi_\xi , \Hscr_\xi)$ which satisfy the relations
    \eqref{eq-transl-relations}. To construct them we first consider
    one element $L$ of the countable space $\idealcount$ together with
    a single space-time translation $y$ in the numerable dense
    subgroup $\Tcount$ of $\Rsone$. By assumption
    \eqref{eq-condition-max-comm-alg}, the von Neumann algebra
    $\Mfrak$ is contained in the commutant of $\bset{U^\bullet ( x ) :
    x \in \Rsone}$, which means that for any measurable subset $\Mecm$
    of $\Xecmbar_4$ with associated orthogonal projection $P_\Mecm \in
    \Mfrak$ there holds the equation
    \begin{subequations}
      \begin{multline}
        \label{eq-transl-inv-disint}
        \int_{\Mecm} d \nubar ( \xi ) \; \bnorm{\bxiket{\abullety ( L
        )}}^2 = \bnorm{P_\Mecm \bbullket{\abullety ( L )}}^2 =
        \bnorm{P_\Mecm U^\bullet ( y ) \bullket{L}}^2 \\
        = \bnorm{U^\bullet ( y ) P_\Mecm \bullket{L}}^2 =
        \bnorm{P_\Mecm \bullket{L}}^2 = \int_{\Mecm} d \nubar ( \xi )
        \; \bnorm{\bxiket{L}}^2 \text{.}
      \end{multline}
      Since this result is valid for arbitrary measurable sets
      $\Mecm$, we infer by \cite[Chapter~V, \S\,25,
      Theorem~E]{halmos:1968} that for $\nubar$-almost all $\xi$ the
      vectors are subject to the relation
      \begin{equation}
        \label{eq-xiket-transl-inv}
        \bnorm{\bxiket{\abullety ( L )}} = \bnorm{\bxiket{L}} \text{.}
      \end{equation}
      A corresponding equation can be derived for any other of the
      countable number of combinations of elements in $\idealcount$
      and $\Tcount$, so that \eqref{eq-xiket-transl-inv} is true in
      all of these cases when the domain of $\xi$ is restricted to the
      $\nubar$-measurable subset $\Xecmbar_5$, which again differs
      from $\Xecmbar_4$ only by a null set. On $\Xecmbar_5$ we can
      then define for arbitrary $y \in \Tcount$ the mappings
      \begin{equation}
        \label{eq-def-Uxi-group}
        \Ubarxi ( y ) : \Hscr_\xi^c \rightarrow \Hscr_\xi^c \qquad
        \Ubarxi ( y ) \xiket{L} \doteq \bxiket{\abullety ( L )}
      \end{equation}
    \end{subequations}
    which are indeed determined unambiguously according to
    \eqref{eq-xiket-transl-inv}. By the same relation they are
    norm-preserving and, moreover, turn out to be $(\Qbb + i
    \Qbb)$-linear operators on the countable spaces $\Hscr_\xi^c
    \subseteq \Hscr_\xi$.

    We want to extend the definition given by \eqref{eq-def-Uxi-group}
    in two respects: All space-time translations $y \in \Rsone$ should
    be permissible, and all vectors of $\Hscr_\xi$ are to belong to
    the domain of the resulting operators. Now, let $L$ be an
    arbitrary element of $\idealcount$, i.\,e.
    \begin{subequations}
      \begin{equation}
        \label{eq-idealcount-rep}
        L = \sum_{i = 1}^N A_i \, L_i \quad \text{with} \quad A_i \in
        \Acountbar \thickspace \text{and} \thickspace L_i \in
        \vaccountbar \text{,}
      \end{equation}
      and consider $x \in \Rsone$ approximated by the sequence
      $\bset{x_n}_{n \in \Nbb} \subseteq \Tcount$. Then, by definition
      \eqref{eq-def-Uxi-group} in connection with property
      \eqref{eq-rep-xiket}, the translates by $x_k$ and $x_l$ of the
      vectors $\xiket{L}$ are for $\xi \in \Xecmbar_5$ subject to the
      following relation:
      \begin{multline}
        \label{eq-idealcount-translates}
        \Ubarxi ( x_k ) \xiket{L} - \Ubarxi ( x_l ) \xiket{L} =
        \Bxiket{\abulletxk \Bigl( \sum_{i = 1}^N A_i \, L_i \Bigr)} -
        \Bxiket{\abulletxl \Bigl( \sum_{i = 1}^N A_i \, L_i \Bigr)} \\
        \mspace{-190mu}= \sum_{i = 1}^N \pi_\xi \bigl( \abulletxk (
        A_i ) \bigr) \bxiket{\abulletxk ( L_i )} - \sum_{i = 1}^N
        \pi_\xi \bigl( \abulletxl ( A_i ) \bigr) \bxiket{\abulletxl (
          L_i )} \\
        = \sum_{i = 1}^N \pi_\xi \bigl( \abulletxk ( A_i ) -
        \abulletxl ( A_i ) \bigr) \bxiket{\abulletxk ( L_i )} +
        \sum_{i = 1}^N \pi_\xi \bigl( \abulletxl ( A_i ) \bigr) \bigl(
        \bxiket{\abulletxk ( L_i )} - \bxiket{\abulletxl ( L_i )}
        \bigr) \text{.}
      \end{multline}
      Since $\Xecmbar_3$ is a subset of $\Xecmbar_5$, we have relation
      \eqref{eq-vacbar-bullet-extension-final-estimate} at our
      disposal; moreover, by assumption, the group of automorphisms
      $\bset{\aLax : ( \Lambda , x ) \in \Poin}$ is strongly
      continuous. As a result, the sequences
      \begin{equation}
        \label{eq-vacbar-sum-cauchy-sequence}
        \bset{\bxiket{\abulletxk ( L )}}_{k \in \Nbb} \text{,} \quad L
        \in \vaccountbar \text{,} \qquad \qquad \bset{\pi_\xi \bigl(
        \abulletxk ( A ) \bigr)}_{k \in \Nbb} \text{,} \quad A \in
        \Acountbar \text{,}
      \end{equation}
      both have the Cauchy property and are thus convergent as well as
      bounded in the respective norm topologies. Applied to the
      elements of $\vaccountbar$ and $\Acountbar$ appearing in the
      representation \eqref{eq-idealcount-rep} of $L \in \idealcount$
      this has the consequence, that the right-hand side of
      \eqref{eq-idealcount-translates} can be made arbitrarily small
      for all pairs $k \text{,} l \in \Nbb$ exceeding a certain
      number. The terms on the left-hand side of this inequality thus
      turn out to be part of Cauchy sequences $\bset{\Ubarxi ( x_k )
      \xiket{L}}_{k \in \Nbb}$ which converge in the Hilbert spaces
      $\Hscr_\xi$. Since a renewed application of the above arguments
      shows that the arising limits are independent of the
      approximating sequence in $\Tcount$, the following relation
      unambiguously defines the mappings $\Ubarxi ( x )$ for arbitrary
      $x \in \Rsone$ and $L \in \idealcount$:
      \begin{gather}
        \label{eq-def-Uxi-group-transl-extension}
        \Ubarxi ( x ) \xiket{L} \doteq \lim_{k \rightarrow \infty}
        \Ubarxi ( x_k ) \xiket{L} = \lim_{k \rightarrow \infty}
        \bxiket{\alpha^\bullet_{x_k} ( L )}\text{,} \\
        \text{where} \quad \Tcount \ni x_k \xrightarrow[k \rightarrow
        \infty]{} x \in \Rsone \text{.} \notag
      \end{gather}
      Again these mappings act as $( \Qbb + i \Qbb)$-linear operators
      on the spaces $\Hscr_\xi^c$ and preserve the Hilbert space
      norm. As a consequence they can, by the standard procedure used
      for completions of uniform spaces, be continuously extended in a
      unique fashion to all of the Hilbert space on condition that
      their countable domain constitutes a dense subset of
      $\Hscr_\xi$; but this is the case as $\Xecmbar_5$ is contained
      in $\Xecmbar_2$. Changing the notation from $\Ubarxi$ to $U_\xi$
      for these extensions, their definition on arbitrary vectors
      $\Psi_\xi \in \Hscr_\xi$ then reads for any $x \in \Rsone$
      \begin{gather}
        \label{eq-def-Uxi-group-Hxi-extension}
        U_\xi ( x ) \Psi_\xi \doteq \lim_{l \rightarrow \infty}
        \Ubarxi ( x ) \bxiket{L^{( l )}} \text{,} \\
        \text{where} \quad \Hscr_\xi^c \ni \bxiket{L^{( l )}}
        \xrightarrow[l \rightarrow \infty]{} \Psi_\xi \in \Hscr_\xi
        \text{,} \notag
      \end{gather}
      and this definition is again independent of the selected
      sequence. For any $L \in \idealcount$ and any $x \in \Rsone$ the
      vector field $\bset{U_\xi ( x ) \xiket{L} : \xi \in
      \Xecmbar_5}$, which is the pointwise limit of a sequence of
      measurable vector fields by
      \eqref{eq-def-Uxi-group-transl-extension} and hence itself
      measurable \cite[Definition~II.4.1]{fell/doran:1988a},
      corresponds to $\bbullket{\abulletx ( L )}$, the equivalent
      limit in $\Hbullet$ (where we neglect the difference between
      $\Xecmbar$ and $\Xecmbar_5$ which is of measure $0$):
      \begin{equation}
        \label{eq-spatial-disintegration-of-transl-vectors}
        \Wbar \, U^\bullet ( x ) \bullket{L} = \Wbar \,
        \bbullket{\abulletx ( L )} = \int_{\Xecmbar_5}^\oplus d \nubar
        ( \xi ) \; U_\xi ( x ) \xiket{L} \text{.}
      \end{equation}
    \end{subequations}

    Having defined the family of mappings $\bset{U_\xi ( x ) : x \in
    \Rsone} \subseteq \BHxi$ for $\xi \in \Xecmbar_5$, we now have to
    check that they obey \eqref{eq-transl-relations}. First of all,
    note that, as an immediate consequence of the way in which they
    were introduced, these mappings are $\Cbb$-linear and
    norm-preserving. Another property that is readily checked by use
    of the relations \eqref{eq-def-Uxi-group-Hxi-extension} and
    \eqref{eq-def-Uxi-group-transl-extension} in connection with the
    estimates arising from \eqref{eq-idealcount-translates} with $L^{(
    l )}$ replacing $L$ is the fact that for arbitrary $x \text{,} y
    \in \Rsone$
    \begin{equation}
      \label{eq-Uxi-group-property}
      U_\xi ( x ) \cdot U_\xi ( y ) = U_\xi ( x + y ) \text{.}
    \end{equation}
    From this we infer that, as $U_\xi ( 0 ) = \unit_\xi$, each
    operator $U_\xi ( x )$ has the inverse $U_\xi ( - x )$ and thus
    proves to be an isometric isomorphism of $\Hscr_\xi$. Hence, in
    accordance with \eqref{eq-Uxi-group-property}, the set
    $\bset{U_\xi ( x ) : x \in \Rsone}$ indeed turns out to be a
    unitary group in $\BHxi$.

    The strong continuity of this group is easily seen: Consider the
    operator $L \in \idealcount$ as defined in
    \eqref{eq-idealcount-rep} and two sequences $\set{x_k}_{k \in
    \Nbb}$, $\set{y_l}_{l \in \Nbb}$ in $\Tcount$ converging to $x$
    and $y$, respectively. Then \eqref{eq-idealcount-translates} stays
    valid if we replace the translations $x_l$ by $y_l$ in each
    case. In compliance with \eqref{eq-def-Uxi-group-transl-extension}
    it is then possible to pass to the limit, which results in the
    obvious estimate
    \begin{subequations}
      \begin{multline}
        \label{eq-Uxi-group-strong-cont}
        \bnorm{\Ubarxi ( x ) \xiket{L} - \Ubarxi ( y ) \xiket{L}} \\
        \leqslant \sum_{i = 1}^N \bnorm{\abulletx ( A_i ) - \abullety
        ( A_i )} \bnorm{\bxiket{\abulletx ( L_i )}} + \sum_{i = 1}^N
        \norm{A_i} \bnorm{\bxiket{\abulletx ( L_i )} -
        \bxiket{\abullety ( L_i )}} \text{.}
      \end{multline}
      This explicit inequality shows that the right-hand side can be
      made arbitrarily small for all $y$ in an appropriate
      neighbourhood of $x$; for the first term this is brought about
      by the strong continuity of the automorphism group $\bset{\aLax
      : ( \Lambda , x ) \in \Poin}$, whereas for the second term it is
      a consequence of relation
      \eqref{eq-vacbar-bullet-continuity}. The defining condition for
      strong continuity is therefore satisfied for vectors in the
      dense subset $\Hscr_\xi^c$. If now an arbitrary vector $\Psi_\xi
      \in \Hscr_\xi$ is considered, we can expand the difference
      $U_\xi ( x ) \Psi_\xi - U_\xi ( y ) \Psi_\xi$ by introducing the
      corresponding translates of any element $\xiket{L} \in
      \Hscr_\xi^c$ and, making use of the property of
      norm-preservation of the unitaries, arrive at
      \begin{multline}
        \label{eq-Uxi-group-strong-cont-intermediate}
        \bnorm{U_\xi ( x ) \Psi_\xi - U_\xi ( y ) \Psi_\xi} \\
        \leqslant \bnorm{\Psi_\xi - \xiket{L}} + \bnorm{\Ubarxi ( x )
        \xiket{L} - \Ubarxi ( y ) \xiket{L}} +  \bnorm{\xiket{L} -
        \Psi_\xi} \text{.}
      \end{multline}
    \end{subequations}
    The right-hand side of this inequality can again be made smaller
    than any given bound by first choosing a suitable element
    $\xiket{L} \in \Hscr_\xi^c$ from a small neighbourhood of
    $\Psi_\xi$ and then, in dependence on this selected vector
    $\xiket{L}$ but irrelevant for the statement, selecting an
    appropriate neighbourhood of translations $y$ around $x$ as
    implied by \eqref{eq-Uxi-group-strong-cont}. Thereby we have
    established strong continuity of the unitary group $\bset{U_\xi (
    x ) : x \in \Rsone}$.

    Before considering the support of the spectral measure $E_\xi
    (~.~)$ associated with this strongly continuous unitary group, we
    mention a result on the interchange of integrations with respect
    to the Lebesgue measure on $\Rsone$ and the bounded positive
    measure $\nubar$ on $\Xecmbar_5$, which proves to be necessary as
    Fubini's Theorem does not apply to the situation in question. Let
    $g$ be a continuous bounded function in $L^1 \bigl( \Rsone , d^{s
    + 1} x \bigr)$, then
    \label{par-R-integral-interchange}
    \begin{equation*}
      \Rsone \ni x \mapsto g ( x ) \, \bxiscpx{L_1}{U_\xi ( x )}{L_2}
      \in \Cbb
    \end{equation*}
    is an integrable mapping for any $L_1 \text{,} L_2 \in
    \idealcount$ and $\xi \in \Xecmbar_5$. Moreover, it is Riemann
    integrable over any compact $( s + 1 )$-dimensional interval
    $\Kib$, and this integral is the limit of a Riemann sequence
    (cf.~\cite[Kapitel~XXIII, Abschnitt 197 and Lebesguesches
    Integrabilit\"{a}tskriterium~199.3]{heuser:1993b}):
    \begin{subequations}
      \begin{equation}
        \label{eq-compact-R-xiintegral-limit}
        \int_\Kib d^{s + 1} x \; g ( x ) \, \bxiscpx{L_1}{U_\xi ( x
        )}{L_2} = \lim_{i \rightarrow \infty} \sum_{m = 1}^{n_i}
        \babs{\Zcal_{\, m}^{( i )}} \: g \bigl( x_m^{( i )} \bigr) \,
        \bxiscpx{L_1}{U_\xi \bigl( x_m^{( i )} \bigr)}{L_2} \text{,}
      \end{equation}
      where $\bset{\Zcal_{\, m}^{( i )} : m = 1 , \dots , n_i}$
      denotes the $i$-th subdivision of $\Kib$, $\babs{\Zcal_{\, m}^{(
      i )}}$ are the Lebesgue measures of these sets, and $x_m^{( i )}
      \in \Zcal_{\, m}^{( i )}$ are corresponding intermediate
      points. The sums on the right-hand side of this equation turn
      out to be $\nubar$-measurable when their dependence on $\xi$ is
      taken into account, and so is the limit on the left-hand
      side. Moreover this property is preserved in passing to the
      limit $\Kib \nearrow \Rsone$, so that
      \begin{equation*}
        \Xecmbar_5 \ni \xi \mapsto \int_{\Rsone} d^{s + 1} x \; g ( x
        ) \, \bxiscpx{L_1}{U_\xi ( x )}{L_2} \in \Cbb
      \end{equation*}
      is $\nubar$-measurable and, in addition, integrable since
      \begin{multline}
        \label{eq-R-xiintegral}
        \int_{\Xecmbar_5} d \nubar ( \xi ) \: \Babs{\int_{\Rsone} d^{s +
        1} x \; g ( x ) \, \bxiscpx{L_1}{U_\xi ( x )}{L_2}} \\
        \mspace{-150mu} \leqslant \int_{\Xecmbar_5} d \nubar ( \xi )
        \int_{\Rsone} d^{s + 1} x \; \abs{g ( x )} \,
        \norm{\xiket{L_1}} \norm{\xiket{L_2}} \\
        = \norm{g}_1 \int_{\Xecmbar_5} d \nubar ( \xi ) \;
        \norm{\xiket{L_1}} \norm{\xiket{L_2}} \leqslant \norm{g}_1 \,
        \norm{\bullket{L_1}} \norm{\bullket{L_2}} \text{.}
      \end{multline}
      The counterpart of \eqref{eq-compact-R-xiintegral-limit} is
      valid in $\Hbullet$, too, and, if $\Mecm$ denotes a measurable
      subset of $\Xecmbar_5$ with associated orthogonal projection
      $P_\Mecm \in \Mfrak$, this equation reads
      \begin{equation}
        \label{eq-compact-R-bullintegral-limit}
        \int_\Kib d^{s + 1} x \; g ( x ) \, \bscpx{L_1}{P_\Mecm
        U^\bullet ( x )}{L_2} = \lim_{i \rightarrow \infty} \sum_{m = 
        1}^{n_i} \babs{\Zcal_{\, m}^{( i )}} \: g \bigl( x_m^{( i )}
        \bigr) \, \bscpx{L_1}{P_\Mecm U^\bullet \bigl( x_m^{( i )}
        \bigr)}{L_2} \text{.}
      \end{equation}
      Then \eqref{eq-spatial-disintegration-of-idealcount-vectors}
      and \eqref{eq-spatial-disintegration-of-transl-vectors} in
      connection with \eqref{eq-direct-integral-M} imply
      \begin{multline}
        \label{eq-compact-R-xibullintegral-limit}
        \int_\Kib d^{s + 1} x \; g ( x ) \, \bscpx{L_1}{P_\Mecm
        U^\bullet ( x )}{L_2} = \lim_{i \rightarrow \infty}
        \int_{\Mecm} d \nubar ( \xi ) \; \sum_{m = 1}^{n_i}
        \babs{\Zcal_{\, m}^{( i )}} \: g \bigl( x_m^{( i )} \bigr) \,
        \bxiscpx{L_1}{U_\xi \bigl( x_m^{( i )} \bigr)}{L_2} \\
        = \int_{\Mecm} d \nubar ( \xi ) \int_\Kib d^{s + 1} x \; g ( x
        ) \, \bxiscpx{L_1}{U_\xi ( x )}{L_2} \text{,}
      \end{multline}
      where, in the second equation, use was made of Lebesgue's
      Dominated Convergence Theorem in view of the fact that the
      integrable function
      \begin{equation*}
        \Xecmbar_5 \ni \xi \mapsto \norm{g}_1 \norm{\xiket{L_1}}
        \norm{\xiket{L_2}} \in \Cbb
      \end{equation*}
      majorizes both sides of \eqref{eq-compact-R-xiintegral-limit}.
      Relation \eqref{eq-compact-R-xibullintegral-limit} again stays
      true in passing to the limit $\Kib \nearrow \Rsone$:
      \begin{equation}
        \label{eq-R-integral-interchange}
        \int_{\Rsone} d^{s + 1} x \; g ( x ) \,
        \bscpx{L_1}{P_\Mecm U^\bullet ( x )}{L_2} = \int_{\Mecm} d
        \nubar ( \xi ) \int_{\Rsone} d^{s + 1} x \; g ( x ) \,
        \bxiscpx{L_1}{U_\xi ( x )}{L_2} \text{,}
      \end{equation}
    \end{subequations}
    which is the announced result on the commutability of integrations
    in the present context.

    The support of the spectral measure $E_\xi (~.~)$ associated with
    the generators $P_\xi$ of the unitary group $\bset{U_\xi ( x ) : x
    \in \Rsone}$ can now be investigated by use of the method  applied
    in the proof of the fourth item of the first part of
    Theorem~\ref{The-particle-weight}. Note, that the complement of
    the closed set $\fwcone - q \subseteq \Rsone$ can be covered by an
    increasing sequence $\bset{\Gamma_N}_{N \in \Nbb}$ of compact
    subsets (take e.\,g.~the intersection of the compact ball of
    radius $N$ with the complement of the open $N^{-1}$-neighbourhood
    of $\fwcone - q$). To each of these sets one can find an
    infinitely often differentiable function $\tilde{g}_N$ with
    support in $\complement ( \fwcone - q )$ that has the property $0
    \leqslant \chi_{\Gamma_N} \leqslant \tilde{g}_N$. As before, let
    $\Mecm$ be a measurable subset of $\Xecmbar_5$ with associated
    orthogonal projection $P_\Mecm \in \Mfrak$, then, by assumption on
    the spectral support of the unitary group implementing space-time
    translations in the underlying particle weight, we have
    \begin{subequations}
      \begin{equation}
        \label{eq-spectral-support}
        \int_{\Rsone} d^{s + 1} x \; g_N ( x ) \, \bscpx{L_1}{P_\Mecm
        U^\bullet ( x )}{L_2} = 0
      \end{equation}
      for any $N \in \Nbb$ and any pair of vectors $\ket{L_1}$ and
      $\ket{L_2}$ with $L_1 \text{,} L_2 \in \idealcount$, and this,
      according to \eqref{eq-R-integral-interchange}, implies
      \begin{equation}
        \label{eq-spectral-support-Mecm-disint}
        \int_\Mecm d \nubar ( \xi ) \int_{\Rsone} d^{s + 1} x \; g_N (
        x ) \, \bxiscpx{L_1}{U_\xi ( x )}{L_2} = 0 \text{.} 
      \end{equation}
      By arbitrariness of $\Mecm$ in the last expression, we conclude
      once more that for $\nubar$-almost all $\xi \in \Xecmbar_5$
      \begin{equation}
        \label{eq-spectral-support-xi}
        \int_{\Rsone} d^{s + 1} x \; g_N ( x ) \, \bxiscpx{L_1}{U_\xi
        ( x )}{L_2} = 0 \text{.}
      \end{equation}
      Eventually, if we want this equation to hold for any element of
      the countable set of triples $\bigl( g_N , \xiket{L_1} ,
      \xiket{L_2} \bigr)$, a non-null set $\Xecmbar_6 \subseteq
      \Xecmbar_5$ is left, and \eqref{eq-spectral-support-xi} stays
      valid for the remaining $\xi \in \Xecmbar_6$ even if the special
      vectors $\xiket{L_1}$ and $\xiket{L_2}$ are replaced by
      arbitrary ones. Stone's Theorem then implies
      (cf.~\eqref{eq-stones-theorem}) that $\tilde{g}_N ( P_\xi ) = 0$
      and therefore, by the order relation inherent in the definition
      of $\tilde{g}_N$, we have $E_\xi ( \Gamma_N ) = \chi_{\Gamma_N}
      ( P_\xi ) = 0$ for any $N \in \Nbb$. As the spectral measure
      $E_\xi (~.~)$ is regular, one can pass to the limit $N
      \rightarrow \infty$ and thereby arrives at the desired result
      \begin{equation}
        \label{eq-spectral-support-xi-final}
        E_\xi \bigl( \complement ( \fwcone - q ) \bigr) = 0 \text{,}
        \qquad \xi \in \Xecmbar_6 \text{.}
      \end{equation}
    \end{subequations}

    The defining equation \eqref{eq-def-Uxi-group} in connection with
    \eqref{eq-rep-xiket} implies that for arbitrary operators $A' \in
    \Acountbar$ and $L \in \idealcount$ and for any translation $x'
    \in \Tcount$ one can write
    \begin{subequations}
      \begin{multline}
        \label{eq-transl-xi-implement-prep}
        \pi_\xi \bigl( \abulletxprime ( A' ) \bigr) \xiket{L} =
        \bxiket{\abulletxprime ( A' ) L} = \bxiket{\abulletxprime
        \bigl( A' \alpha^\bullet_{- x'} ( L ) \bigr)} = \Ubarxi ( x' )
        \bxiket{A' \alpha^\bullet_{- x'} ( L )} \\
        = \Ubarxi ( x' ) \pi_\xi ( A' ) \Ubarxi ( - x' ) \xiket{L} =
        \Ubarxi ( x' ) \pi_\xi ( A' ) {\Ubarxi ( x' )}^* \xiket{L}
        \text{.}
      \end{multline}
      Since the vectors $\xiket{L} \in \Hscr_\xi^c$, $L \in
      \idealcount$, constitute a dense subset of $\Hscr_\xi$ for $\xi
      \in \Xecmbar_2 \subseteq \Xecmbar_5$ we infer from this equation
      that
      \begin{equation}
        \label{eq-transl-xi-implement}
        \pi_\xi \bigl( \abulletxprime ( A' ) \bigr) = \Ubarxi ( x' )
        \pi_\xi ( A' ) {\Ubarxi ( x' )}^* \text{,}
      \end{equation}
      an equation that readily extends to all translations $x$ in
      $\Rsone$ and, by uniform density of $\Acountbar$ in $\Abullet$,
      also to any operator $A$ in the $C^*$-algebra $\Abullet$:
      \begin{equation}
        \label{eq-transl-xi-implement-final}
        \pi_\xi \bigl( \abulletx ( A ) \bigr) = U_\xi ( x ) \pi_\xi (
        A ) {U_\xi ( x )}^* \text{,}\quad A \in \Abullet \text{,}
        \quad x \in \Rsone \text{.}
      \end{equation}
    \end{subequations}
    This proves the counterpart of \eqref{eq-transl-implement}. The
    action of the group $\bset{U_\xi ( x ) : x \in \Rsone}$ on
    $\bset{\xiket{L'} : L' \in \vacbar}$ according to
    \eqref{eq-transl-Kzero-action} is an immediate consequence of the
    defining relations \eqref{eq-def-Uxi-group-transl-extension} and
    \eqref{eq-def-Uxi-group-Hxi-extension} in connection with
    \eqref{eq-def-vacbar-xiket} and the continuity property as
    expressed by \eqref{eq-vacbar-bullet-continuity}. In the present
    setting we thus have 
    \begin{equation}
      \label{eq-transl-vacbar-action}
      U_\xi ( x ) \xiket{L'} \doteq \bxiket{\abulletx ( L' )} \text{,}
      \quad L' \in \vacbar \text{.}
    \end{equation}
    Now, let $L$ be an arbitrary element of $\idealcount$ having
    energy-momentum transfer $\Gamma_L$. Defined as the support of the
    Fourier transform of an operator-valued distribution (cf.~the
    Remark following Definition~\ref{Def-vacuum-annihilation}),
    $\Gamma_L$ is a closed Borel set, so that the arguments given in
    the preceding paragraph can again be applied when $L$ replaces
    $L_1$ and $L_2$ and the functions $\tilde{g}_N$ now correspond to
    an increasing sequence of compact sets $\Gamma'_N$ which
    constitute a cover of $\complement \Gamma_L$. As a result we
    arrive at the equivalent of \eqref{eq-spectral-support-xi}, so
    that
    \begin{subequations}
      \begin{equation}
        \label{eq-GammaL-spectral-support-xi}
        \int_{\Rsone} d^{s + 1} x \; g_N ( x ) \, \bxiscpx{L}{U_\xi (
        x )}{L} = 0
      \end{equation}
      holds for $\nubar$-almost all $\xi \in \Xecmbar_6$ even if the
      index $N$ is allowed to run through all natural numbers. As in
      the preceding paragraph we then conclude that for all of these
      $\xi$ and all $N \in \Nbb$ one has $\bxiscpx{L}{\tilde{g}_N (
      P_\xi )}{L} = 0$ and hence $\bxiscpx{L}{E_\xi ( \Gamma'_N )}{L}
      = 0$. According to the regularity of the spectral measure $E_\xi
      (~.~)$, passage to the limit with respect to $N$ yields the
      equation $E_\xi \bigl( \complement \Gamma_L \bigr) \xiket{L} =
      0$. By countability, this last result is valid for
      arbitrary $L \in \idealcount$ if a $\nubar$-measurable non-null
      set $\Xecmbar_7 \subseteq \Xecmbar_6$ is appropriately selected,
      from which the indices $\xi$ are to be taken. The complementary
      statement thus constitutes a restricted version of the
      counterpart of \eqref{eq-spectral-subspace}:
      \begin{equation} 
        \label{eq-idealcount-spectral-support-xi}
        E_\xi ( \Gamma_L ) \xiket{L} = \xiket{L} \text{,} \quad L \in
        \idealcount \text{,} \quad \xi \in \Xecmbar_7 \text{.}
      \end{equation}
      Now, let $\Hat{L}_0$ be an arbitrary element of $\vaccountbar$,
      then the energy-momentum transfer of its Poincar\'{e} transform
      by $( \Lambda , x ) \in \Pcount$, i.\,e.~of the operator
      $\abulletLax ( \Hat{L}_0 ) \in \vaccountbar \subseteq
      \idealcount$, is given by $\Lambda \Gamma_{\Hat{L}_0}$, so that,
      according to \eqref{eq-idealcount-spectral-support-xi},
      \begin{equation}
        \label{eq-vaccountbar-transform-spectral-support-xi}
        E_\xi ( \Lambda \Gamma_{\Hat{L}_0} ) \bxiket{\abulletLax (
        \Hat{L}_0 )} = \bxiket{\abulletLax ( \Hat{L}_0 )} \text{,}
        \quad \xi \in \Xecmbar_7 \text{.}
      \end{equation}
      This result can be applied to investigate the case of a generic
      element from $\vacbar$. For arbitrary $( \Lambda_0 , x_0 ) \in
      \Poin$ approximated by the sequence $\bset{( \Lambda_n , x_n
      )}_{n \in \Nbb} \subseteq \Pcount$ we have, by virtue of the
      relevant parts of \eqref{eq-vacbar-extension-xi},
      \begin{equation*}
        \bxiket{\abulletLaxzero ( \Hat{L}_0 )} = \lim_{n \rightarrow
        \infty} \bxiket{\abulletLaxn ( \Hat{L}_0 )} \text{,}
      \end{equation*}
      and Lebesgue's Dominated Convergence Theorem in connection with
      Stone's Theorem yields for any function $g \in L^1 \bigl( \Rsone
      , d^{s + 1} x \bigr)$ and any index $\xi \in \Xecmbar_7$
      \begin{multline}
        \label{eq-stone-approximation}
        \int_{\Rsone} d^{s + 1} x \; g ( x ) \bxiscpx{\abulletLaxzero (
        \Hat{L}_0 )}{U_\xi ( x )}{\abulletLaxzero ( \Hat{L}_0 )} \\
        \mspace{-50mu}= \lim_{n \rightarrow \infty} \int_{\Rsone} d^{s
        + 1} x \; g ( x ) \bxiscpx{\abulletLaxn ( \Hat{L}_0 )}{U_\xi (
        x )}{\abulletLaxn ( \Hat{L}_0 )} \\
        = ( 2 \pi )^{( s + 1 )/2} \lim_{n \rightarrow \infty}
        \bxiscpx{\abulletLaxn ( \Hat{L}_0 )}{\tilde{g} ( P_\xi
        )}{\abulletLaxn ( \Hat{L}_0 )} \text{.}
      \end{multline}
      In the limit of large $n$ one finds the energy-momentum transfer
      $\Lambda_n \Gamma_{\Hat{L}_0}$ of $\abulletLaxn ( \Hat{L}_0 )$
      in a small $\varepsilon$-neighbourhood of $\Lambda_0
      \Gamma_{\Hat{L}_0}$. Therefore, in view of
      \eqref{eq-vaccountbar-transform-spectral-support-xi}, the
      right-hand side of \eqref{eq-stone-approximation} vanishes for
      all $n$ exceeding a certain bound $N$ if $g$ is chosen in such a
      way that $\supp \tilde{g} \subseteq \complement ( \Lambda_0
      \Gamma_{\Hat{L}_0} )$. The Fourier transform of the distribution
      \begin{equation*}
        \Rsone \ni x \mapsto \bxiscpx{\abulletLaxzero ( \Hat{L}_0
        )}{U_\xi ( x )}{\abulletLaxzero ( \Hat{L}_0 )} \in \Cbb
      \end{equation*}
      is thus seen to be supported by $\Lambda_0 \Gamma_{\Hat{L}_0}$
      from which we infer
      \begin{equation}
        \label{eq-vacbar-spectral-support-xi}
        E_\xi ( \Lambda_0 \Gamma_{\Hat{L}_0} ) \bxiket{\abulletLaxzero
        ( \Hat{L}_0 )} = \bxiket{\abulletLaxzero ( \Hat{L}_0 )}
        \text{,} \quad \xi \in \Xecmbar_7 \text{,}
      \end{equation}
      which is the equivalent of
      \eqref{eq-vaccountbar-transform-spectral-support-xi} for
      arbitrary operators in $\vacbar$. Equations
      \eqref{eq-idealcount-spectral-support-xi} and
      \eqref{eq-vacbar-spectral-support-xi} are readily generalized,
      making use of the order structure of spectral projections
      reflecting the inclusion relation of Borel subsets of
      $\Rsone$. If $\overline{\Lfrak} ( \Delta' )$ denotes the set of
      operators from $\idealcount \cup \vaccountbar$ having
      energy-momentum transfer in the Borel set $\Delta'$, then
      \begin{equation}
        \label{eq-vacbar-idealcount-spectral-support-xi}
        E_\xi ( \Delta') \xiket{L} = \xiket{L} \text{,} \quad L \in
        \overline{\Lfrak} ( \Delta' ) \text{,}
      \end{equation}
    \end{subequations}
    and thus the counterpart of \eqref{eq-spectral-subspace} is
    established for the remaining indices $\xi$ from the non-null
    subset $\Xecmbar_7$ of $\Xecmbar$.
    \renewcommand{\qed}{}
  \end{prooflist}

  The above construction has supplied us with a measurable subset
  $\Xecm \doteq \Xecmbar_7$ of the standard Borel space $\Xecmbar$,
  that was introduced at the outset, emerging from an application of
  \cite[Theorem~8.5.2]{dixmier:1982}. $\Xecm$ is a non-null set,
  differing from $\Xecmbar$ only by a $\nubar$-null set. Moreover it
  is itself a standard Borel space (cf.~the definition in
  \cite[Section~3.3]{arveson:1976}), and we shall denote the
  restriction of the measure $\nubar$ to it by $\nu$; $\nu \doteq
  \nubar \restriction \Xecm$ is again a bounded positive
  measure. Moreover, and this has been the central aim of the previous
  investigations, the field
  \begin{equation*}
    \Xecm \ni \xi \mapsto ( \pi_\xi , \Hscr_\xi , \Abullet ,
    \alpha^\bullet , \Pcount , \vaccountbar) \text{,}
  \end{equation*}
  indeed consists of $\vaccountbar$-particle weights. What remains to
  be done now is a verification of the properties listed in
  \eqref{eq-spatial-disintegration}.
  \begin{prooflist}
  \item Arising as the restriction to a measurable subset in
    $\Xecmbar$ of a field of irreducible representations, the family
    of representations
    \begin{equation*}
      \Xecm \ni \xi \mapsto ( \pi_\xi , \Hscr_\xi )
    \end{equation*}
    is obviously $\nu$-measurable and its components inherit the
    feature of irreducibility.
  \item As $\Xecm$ and $\Xecmbar$ only differ by a $\nubar$-null set,
    one has
    \begin{equation}
      \label{eq-direct-integral-isomorphy}
      \int_{\Xecmbar}^\oplus d \nubar ( \xi ) \; \Hscr_\xi \simeq
      \int_\Xecm^\oplus d \nu \; \Hscr_\xi \text{,}
    \end{equation}
    and the relations \eqref{eq-direct-integral} can be reformulated,
    using the right-hand side of \eqref{eq-direct-integral-isomorphy}
    and an isomorphism $W$, which is the composition of $\Wbar$ with
    the isometry that implements the above unitary equivalence. As an
    immediate consequence of \eqref{eq-direct-integral-Hilbert-spaces}
    and \eqref{eq-direct-integral-representations} we then get the
    equivalence assertion of \eqref{eq-rep-disintegration}. Moreover,
    by \eqref{eq-spatial-disintegration-of-idealcount-vectors} and
    \eqref{eq-spatial-disintegration-of-vacbar-vectors}, the operator
    $W$ connects the vector fields $\bset{\xiket{L} : \xi \in \Xecm}$
    with vectors $\bullket{L}$ for $L \in \idealcount \cup \vacbar$ in
    the desired way as expressed in
    \eqref{eq-spatial-disintegration-of-vectors}.
  \item \eqref{eq-diagonalised-operators} is the mere reformulation of
    \eqref{eq-direct-integral-M} in terms of $\Xecm$ and $W$.
  \item According to the argument preceding
    \eqref{eq-spatial-disintegration-of-transl-vectors}, the mappings
    $\xi \mapsto \xiscpx{L_1}{U_\xi ( x )}{L_2}$, with $\xi$
    restricted to $\Xecm$ and $L_1$ as well as $L_2$ taken from
    $\idealcount$, are measurable for all vectors $\xiket{L_1}$ and
    $\xiket{L_2}$ in the dense subsets $\Hscr_\xi^c$, and this
    suffices, by \cite[Section~II.2.1, Proposition~1]{dixmier:1981},
    to establish measurability of the field $\xi \mapsto U_\xi ( x )$
    for arbitrary $x \in \Rsone$. Moreover, this is a bounded field of
    operators, so that it defines a bounded operator on $\Hbullet$
    which is given by \eqref{eq-unitary-group-disint} as an immediate
    consequence of
    \eqref{eq-spatial-disintegration-of-transl-vectors}, bearing in
    mind that $\Xecm$ and $\Xecmbar_5$ only differ by a $\nubar$-null
    set. The demonstration of \eqref{eq-spectral-measure-disint} on
    the other hand is less straightforward. Assume first of all that
    the Borel set $\Delta$ in question is open. Then we can make use
    of the regularity of spectral measures and construct, according to
    \cite[Definition~II.8.2]{fell/doran:1988a}, a sequence of compact
    subsets $\bset{\Gamma_N}_{N \in \Nbb}$ as well as of infinitely
    often differentiable functions $\bset{\tilde{g}_N}_{N \in \Nbb}$
    with support in $\Delta$ such that $0 \leqslant \chi_{\Gamma_N}
    \leqslant \tilde{g}_N \leqslant \chi_\Delta$ and furthermore
    \begin{subequations}
      \begin{align}
        \label{eq-Exi-approximation}
        \bxiscpx{L}{\ExiDelta}{L} & = \lim_{N \rightarrow \infty}
        \bxiscpx{L}{\ExiGammaN}{L} = \lim_{N \rightarrow \infty}
        \bxiscpx{L}{\tilde{g}_N ( P_\xi )}{L} \text{,} \\
        \label{eq-Ebullet-approximation}
        \bscpx{L}{\EbulletDelta}{L} & = \lim_{N \rightarrow \infty}
        \bscpx{L}{\EbulletGammaN}{L} = \lim_{N \rightarrow \infty}
        \bscpx{L}{\tilde{g}_N ( P^\bullet )}{L}
      \end{align}
      for any $L \in \idealcount$. The discussion on page
      \pageref{par-R-integral-interchange}\,f.---with $\Xecm$
      replacing $\Xecmbar_5$ and $\nu$ instead of
      $\nubar$---demonstrates, making use of Stone's Theorem, that the
      sequence appearing on the right-hand side consists of
      $\nu$-measurable functions of $\xi$, so that we infer that its
      limit
      \begin{equation*}
        \Xecm \ni \xi \mapsto \bxiscpx{L}{\ExiDelta}{L} \in \Cbb
      \end{equation*}
      is $\nu$-measurable, too. Another application of Stone's Theorem
      in connection with \eqref{eq-R-integral-interchange} in terms of
      $\Xecm$ then shows that
      \begin{multline}
        \label{eq-open-Delta-gN-disint}
        ( 2 \pi )^{( s + 1 )/2} \bscpx{L}{\tilde{g}_N ( P^\bullet
        )}{L} = \int_{\Rsone} d^{s + 1} x \; g_N ( x ) \,
        \bscpx{L}{U^\bullet ( x )}{L} \\
        = \int_{\Xecm} d \nubar ( \xi ) \int_{\Rsone} d^{s + 1} x \;
        g_N ( x ) \, \bxiscpx{L}{U_\xi ( x )}{L} = ( 2 \pi )^{( s + 1
        )/2} \int_{\Xecm} d \nubar ( \xi ) \, \bxiscpx{L}{\tilde{g}_N
        ( P_\xi )}{L} \text{,}
      \end{multline}
      and, as Lebesgue's Dominated Convergence Theorem allows for the
      passage to the limit function under the last integral, we get
      according to \eqref{eq-Exi-approximation} and
      \eqref{eq-Ebullet-approximation}
      \begin{equation}
        \label{eq-open-Delta-disint}
        \bscpx{L}{\EbulletDelta}{L} = \int_{\Xecm} d \nubar ( \xi ) \,
        \bxiscpx{L}{\ExiDelta}{L} \text{.}
      \end{equation}
    \end{subequations}
    This formula, as yet valid only for open Borel sets $\Delta$, is
    readily generalized to closed Borel sets and then, since by
    regularity the spectral measure of an arbitrary Borel set is
    approximated by a sequence in terms of compact subsets of it, to
    any Borel set. By polarization and the fact that the ket vectors
    with entries from $\idealcount$ are dense in $\Hbullet$ and
    $\Hscr_\xi$, respectively, we first conclude with
    \cite[Section~II.2.1, Proposition~1]{dixmier:1981} that all the
    fields $\xi \mapsto \ExiDelta$ are measurable for arbitrary Borel
    sets $\Delta$ and then pass to the aspired formula
    \eqref{eq-spectral-measure-disint} from
    \eqref{eq-open-Delta-disint}.
  \item The unitary operators $V^\bullet ( x )$, $x \in \Rsone$,
    defined in \eqref{eq-unitary-renorm-group} belong to the von
    Neumann algebra $\Mfrak$, according to
    \eqref{eq-condition-max-comm-alg}, and are thus diagonalisable in
    the form
    \begin{subequations}
      \begin{equation}
        \label{eq-Vgroup-disint}
        W \, V^\bullet ( x ) \, W^* = \int_\Xecm^\oplus d \nu ( \xi )
        \; \exp ( i \: p_\xi x ) \, \unit_\xi \text{.}
      \end{equation}
      According to the construction of these operators, we can
      re-express this result in terms of the canonical unitary group
      of \eqref{eq-can-unitary-group}:
      \begin{equation}
        \label{eq-can-group-disint}
        W \, U_c^\bullet ( x ) \, W^* = \int_\Xecm^\oplus d \nu ( \xi
        ) \; \exp ( i \: p_\xi x ) \, U_\xi ( x ) \text{.}
      \end{equation}
    \end{subequations}
    The definition
    \begin{equation}
      \label{eq-can-unitary-group-definition}
      U_\xi^c ( x ) \doteq \exp ( i \: p_\xi x ) \, U_\xi ( x )
      \text{,} \quad x \in \Rsone \text{,} \quad \xi \in \Xecm
      \text{,} 
    \end{equation}
    then provides the asserted canonical choice of a strongly
    continuous unitary group on each Hilbert space $\Hscr_\xi$. Its
    spectral properties are derived from those of the canonical group
    $\bset{U_c^\bullet ( x ) : x \in \Rsone}$ by the methods that have
    already been used repeatedly above. Possibly a further $\nu$-null
    subset of $\Xecm$ gets lost by this procedure. 
    \renewcommand{\qed}{}
  \end{prooflist}
  This finishes the proof of the assertions of
  Theorem~\ref{The-spatial-disintegration}.
\end{proof}
\begin{Rem}
  Theorem~\ref{The-spatial-disintegration} includes the existence of
  a spatial disintegration of the strongly continuous unitary group
  implementing space-time translations in the representation $(
  \pi^\bullet , \Hbullet )$ as well as of the spectral measure
  associated with it. The method used in the demonstration of this
  fact can be generalized to other symmetry groups; however obvious
  a problem of this kind may seem in the present context, it has, to
  our knowledge, not been treated in the literature. Nevertheless, the
  disintegration of unbounded closed operators in Hilbert spaces (the
  self-adjoint generators of strongly continuous unitary groups being
  an example) is the topic of \cite{nussbaum:1964} and also presented
  in \cite[Chapter~12]{schmuedgen:1990}.
\end{Rem}

At present we have no control over the range of energy-momenta $p_\xi$
which enter into the above disintegration theory. It still has to be
investigated if, starting from a physical state of bounded energy in
the constructions of Chapter~\ref{chap-particle-weights} and passing
to the asymptotic limit with respect to a function $h$ that has
support on a small part of the velocity domain, the occurring momenta
are correlated with those defined by the geometric momenta involved in
the limiting procedure. Even tachyonic states cannot be ruled out to
date. These problems might be tackled by introducing a certain
property of `closability' for particle weights, stating that, in case
that a sequence of operators $\set{L_n}_{n \in \Nbb}$ approaches $0$
in a suitable topology and at the same time $\bset{\ket{L_n}}_{n \in
\Nbb}$ is convergent, the limit of the sequence of vectors likewise
vanishes.

Moreover, the spatial disintegration presented above is subject to 
arbitrariness in two respects. There exist different constructions of
the type expounded in Section~\ref{sec-separable-reformulation} and
therefore, according to Theorem~\ref{The-bullet-weight}, one has to
deal with a number of different restricted $\vaccountbar$-particle
weights $( \pi^\bullet , \Hbullet , \Abullet , \alpha^\bullet ,
\Pcount , \vaccountbar)$ derived from the GNS-representa\-tion $(
\pi_w , \Hscr_w )$ pertaining to a given particle weight. As a result,
the object to be disintegrated according to
Theorem~\ref{The-spatial-disintegration} is by no means uniquely
fixed. And even if one has decided to select a system complying with
the requirements of this theorem, there still remains an ambiguity as
to the choice of a maximal abelian von Neumann algebra, with respect
to which the disintegration is to be performed. The same problem is
encountered in the framework of Choquet disintegration theory in its
present status (cf.~the end of Chapter~\ref{chap-choquet}). There a
suitable base in the positive cone of particle weights has to be
chosen with respect to which the disintegration is to be carried
through.

But it should be stressed that these open questions only arise on the
basis of the fact that a disintegration of general particle weights
into pure ones, representing elementary systems, has successfully been
constructed.

\chapter{Phase Space Restrictions and Local Normality}
  \label{chap-local-normality}

A number of criteria have been introduced into the analysis of generic
quantum field theories in order to dismiss those which are not
reasonable form a physical point of view in that they do not allow for
an interpretation in terms of particles. These attempts can be traced
back to the year 1965 when Haag and Swieca proposed a compactness
condition in \cite{haag/swieca:1965}, imposing an effective
restriction to the size of phase space. Subsequently, the notion of
nuclearity has entered the stage, determining maximum values for the
number of local degrees of freedom for physical states of bounded
energy (cf.~the discussions of
\cite{buchholz/wichmann:1986,buchholz/dantoni/longo:1990,haag:1984},
and in addition \cite{buchholz/porrmann:1990} for a treatment of the
interdependence of these various concepts). In the present context we
want to make use of the compactness condition proposed by Fredenhagen
and Hertel \cite{fredenhagen/hertel:1979} to show that, under this
physically motivated presupposition, the arbitrariness in the choice
of a separable $C^*$-subalgebra $\Abullet$ of the quasi-local algebra
$\Afrak$ in Chapter~\ref{chap-disintegration} can be removed.
\begin{Com}[Fredenhagen--Hertel]
  A local quantum field theory, as introduced in
  Chapter~\ref{chap-introduction}, qualifies the Fredenhagen--Hertel
  Compactness Condition if the mappings $\TODprime$, which are defined
  for any bounded Borel set $\Delta' \subseteq \Rsone$ and any bounded
  region $\Oscr$ of Minkowski space through
  \begin{equation*}
    \TODprime : \AO \rightarrow \BH \qquad A \mapsto \TODprime ( A )
    \doteq \EDprime A \EDprime \text{,}
  \end{equation*}
  send bounded subsets of $\AO$ onto precompact subsets of $\BH$ with
  respect to its uniform topology. Precompactness is synonymous with
  totally boundedness and, in the present situation, equivalent to
  relative compactness \cite[Chapter~One, \S\,4,\,5.]{koethe:1983}.
\end{Com}

To be able to demonstrate the main result of this chapter,
Theorem~\ref{The-local-normality}, we have to fall back upon
$\Delta$-bounded particle weights as introduced in
Definition~\ref{Def-weight-classification}. This restriction can be
motivated on physical grounds, as opposed to mere technical needs,
since, according to Lemma~\ref{Lem-Delta-bound}, the asymptotic
functionals in $\CDstarplus$, constructed by use of physical states of
bounded energy in Chapter~\ref{chap-particle-weights}, give rise to
particle weights of this special kind. The corresponding
GNS-representations $( \pi_w , \Hscr_w )$ then meet the
Fredenhagen--Hertel Compactness Condition if the underlying local
quantum field theory does, and the same holds true for the restricted
$\vaccountbar$-particle weights which can be derived from them as
expounded in Chapter~\ref{chap-disintegration}.
\begin{Pro}
  \label{Pro-weight-precompactness}
  Suppose that the given local quantum field theory satisfies the
  Compactness Criterion of Fredenhagen and Hertel.
  \begin{proplist}
  \item If $\scp{~.~}{~.~}$ is a $\Delta$-bounded particle weight on
    $\Lfrak \times \Lfrak$, then the associated GNS-representation $(
    \pi_w , \Hscr_w )$ of the quasi-local algebra $\Afrak$ is subject
    to the compactness condition as well.
  \item The restricted $\vaccountbar$-particle weight $( \pi^\bullet ,
    \Hbullet , \Abullet , \alpha^\bullet , \Pcount , \vaccountbar)$
    derived from the above GNS-representation by virtue of
    Theorem~\ref{The-bullet-weight} likewise inherits the compactness
    property in question.
  \end{proplist}
\end{Pro}
\begin{proof}
  \begin{prooflist}
  \item $\Delta$-boundedness of the particle weight $\scp{~.~}{~.~}$
    means, according to Definition~\ref{Def-weight-classification},
    that to any bounded Borel set $\Delta' \subseteq \Rsone$ there
    exist another such set $\Deltabar$ containing $\Delta + \Delta'$
    and an appropriate positive constant $c$, so that the estimate
    \begin{equation}
      \label{eq-Delta-boundedness-again}
      \norm{\EwDprime \pi_w ( A ) \EwDprime} \leqslant c \cdot
      \norm{\EDbar A \EDbar} \tag{\ref{eq-Delta-boundedness}}
    \end{equation}
    holds for any $A \in \Afrak$. Then a finite cover of $\TODbar
    \bigl( \ArO \bigr) = \EDbar \ArO \EDbar$ by sets of diameter less
    than a given $\delta > 0$ (which exists on account of the
    hypothesis of precompactness) induces a corresponding cover of
    $\EwDprime \pi_w \bigl( \ArO \bigr) \EwDprime$, which is composed
    of sets with a diameter smaller than $c \cdot \delta$ as
    \eqref{eq-Delta-boundedness-again} shows. This establishes totally
    boundedness of the set
    \begin{equation*}
      \EwDprime \pi_w \bigl( \ArO \bigr) \EwDprime \subseteq \BHw
      \text{.}
    \end{equation*}
    By arbitrariness of $\Delta'$ as well as of the bounded region
    $\Oscr$, the representation $( \pi_w , \Hscr_w )$ is thus seen to
    satisfy the Compactness Criterion of Fredenhagen and Hertel in the
    sense that the mappings
    \begin{equation*}
      \TwODprime : \AO \rightarrow \BHw \qquad A \mapsto \TwODprime (
      A ) \doteq \EwDprime \pi_w ( A ) \EwDprime
    \end{equation*}
    are altogether precompact.
  \item According to the construction of $( \pi^\bullet , \Hbullet ,
    \Abullet , \alpha^\bullet , \Pcount , \vaccountbar)$ from $( \pi_w
    , \Hscr_w )$ explained in the proof of
    Theorem~\ref{The-bullet-weight}, both of these representations are
    related by the inequality
    \begin{subequations}
      \begin{equation}
        \label{eq-pibullet-piw-relation}
        \norm{\EbulletDeltaprime \pi^\bullet ( A ) \EbulletDeltaprime}
        \leqslant \norm{\EwDprime \pi_w ( A ) \EwDprime} \text{,}
      \end{equation}
      which holds for any $A \in \Abullet$. Then $\Delta$-boundedness
      of the underlying particle weight again implies the existence of
      a bounded Borel set $\Deltabar \supseteq \Delta + \Delta'$ such
      that
      \begin{equation}
        \label{eq-pibullet-def-rep-relation}
        \norm{\EbulletDeltaprime \pi^\bullet ( A ) \EbulletDeltaprime}
        \leqslant c \cdot \norm{\EDbar A \EDbar} \text{.}
      \end{equation}
    \end{subequations}
    This relation replaces \eqref{eq-Delta-boundedness-again} in the
    proof of the first part, so that we conclude that indeed $(
    \pi^\bullet , \Hbullet )$ inherits the precompactness properties
    of the underlying quantum field theory in the sense that all the
    sets
    \begin{equation*}
      \EbulletDeltaprime \pi^\bullet \bigl( \ArbulletOk \bigr)
      \EbulletDeltaprime \subseteq \BHbullet
    \end{equation*}
    are totally bounded for any $r > 0$ whenever $\Delta'$ is an
    arbitrary bounded Borel set and $\Ok$ is one of the countably many
    localization regions from $\Rcount$. Again that is sufficient to
    establish the fact that the Fredenhagen--Hertel Compactness
    Condition is satisfied in the restricted setting for Local Quantum
    Physics introduced in Section~\ref{sec-separable-reformulation}.
  \end{prooflist}
  \renewcommand{\qed}{}
\end{proof}
Under the presupposition of the Compactness Criterion, a result
corresponding to Proposition~\ref{Pro-weight-precompactness} can be
proved for the irreducible representations $( \pi_\xi , \Hscr_\xi )$
arising in the spatial disintegration of the restricted
$\vaccountbar$-particle weight $( \pi^\bullet , \Hbullet , \Abullet ,
\alpha^\bullet , \Pcount , \vaccountbar)$ by virtue of
Theorem~\ref{The-spatial-disintegration} if the domain of $\xi$ is
further astricted to a $\nu$-measurable non-null subset $\Xecm_0$ of
$\Xecm$.
\begin{Pro}
  \label{Pro-xi-precompactness}
  Let $( \pi_w , \Hscr_w )$ be the GNS-representation of the
  quasi-local algebra $\Afrak$ corresponding to the $\Delta$-bounded
  particle weight $\scp{~.~}{~.~}$, and let $( \pi^\bullet , \Hbullet
  , \Abullet , \alpha^\bullet , \Pcount , \vaccountbar)$ denote the
  restricted $\vaccountbar$-particle weight associated with it
  according to Theorem~\ref{The-bullet-weight}. Under the hypothesis
  that the Compactness Criterion of Fredenhagen and Hertel is in force
  in the underlying quantum field theory, $\nu$-almost all of the
  irreducible representations $( \pi_\xi , \Hscr_\xi )$ occurring in
  the spatial disintegration \eqref{eq-rep-disintegration} of $(
  \pi^\bullet , \Hbullet )$ by course of
  Theorem~\ref{The-spatial-disintegration} comply with this condition
  as well.
\end{Pro}
\begin{proof}
  Select a dense sequence $\set{A_k}_{k \in \Nbb}$ in the
  norm-separable $C^*$-algebra $\Abullet$ and consider the countable
  set of compact balls $\Gamma_N$ of radius $N$ in $\Rsone$. The
  corresponding operators $\EbulletGammaN \pi^\bullet ( A_k )
  \EbulletGammaN \in \BHbullet$ are decomposable according to
  Theorem~\ref{The-spatial-disintegration}:
  \begin{subequations}
    \begin{equation}
      \label{eq-sandwich-decomp}
      W \, \EbulletGammaN \pi^\bullet ( A_k ) \EbulletGammaN \, W^* =
      \int_\Xecm^\oplus d \nu ( \xi ) \; \ExiGammaN \pi_\xi ( A_k )
      \ExiGammaN \text{,}
    \end{equation}
    and \cite[Section~II.2.3, Proposition~2]{dixmier:1981} tells us
    that the respective norms are related in compliance with the
    equation
    \begin{equation}
      \label{eq-ess-sup}
      \norm{\EbulletGammaN \pi^\bullet ( A_k ) \EbulletGammaN} =
      \nusup \bset{\norm{\ExiGammaN \pi_\xi ( A_k ) \ExiGammaN} : \xi
      \in \Xecm} \text{.}
    \end{equation}
  \end{subequations}
  With regard to all possible combinations of operators $A_k$ and
  compact balls $\Gamma_N$ we thus infer that there exists a
  measurable non-null subset $\Xecm_0$ of $\Xecm$ such that for all
  $\xi \in \Xecm_0$ and all indices $k$ and $N$ the estimate
  \begin{equation}
    \label{eq-ess-sup-estimate}
    \norm{\ExiGammaN \pi_\xi ( A_k ) \ExiGammaN} \leqslant
    \norm{\EbulletGammaN \pi^\bullet ( A_k ) \EbulletGammaN}
  \end{equation}
  holds. Now, let $\Delta'$ be an arbitrary bounded Borel set which is
  thus contained in a compact ball $\Gamma_{N_0}$ and note that, by
  continuity of the representations $\pi_\xi$ and $\pi^\bullet$, the
  inequality \eqref{eq-ess-sup-estimate} extends to arbitrary
  operators $A \in \Abullet$. Therefore
  \begin{subequations}
    \begin{equation}
      \label{eq-init-sandwich-estimate}
      \norm{\ExiDeltaprime \pi_\xi ( A ) \ExiDeltaprime} \leqslant
      \norm{\ExiGammaNzero \pi_\xi ( A ) \ExiGammaNzero} \leqslant
      \norm{\EbulletGammaNzero \pi^\bullet ( A ) \EbulletGammaNzero}
      \text{,}
    \end{equation}
    and this implies, according to
    \eqref{eq-pibullet-def-rep-relation}, the existence of a bounded
    Borel set $\Deltabar \supseteq \Delta + \Delta'$ such that
    \begin{equation}
      \label{eq-final-sandwich-estimate}
      \norm{\ExiDeltaprime \pi_\xi ( A ) \ExiDeltaprime} \leqslant c
      \cdot \norm{\EDbar A \EDbar} \text{.}
    \end{equation}
  \end{subequations}
  The arguments given in the proof of
  Proposition~\ref{Pro-weight-precompactness} can then again be
  applied to the present situation to show that for $\xi \in \Xecm_0$
  the irreducible representations $( \pi_\xi , \Hscr_\xi )$ altogether
  meet the requirements of the Fredenhagen--Hertel Compactness
  Condition.
\end{proof}

The central result of the present chapter is the perception that,
under the above assumptions on the structure of phase space, the
representations of the quasi-local $C^*$-algebras $\Afrak$ and
$\Abullet$ which we have come across, i.\,e.~$( \pi_w , \Hscr_w )$ and
$( \pi^\bullet , \Hbullet )$, respectively, as well as $\nu$-almost
all of the irreducible representations $( \pi_\xi , \Hscr_\xi )$
occurring in the direct integral decomposition of the latter, are
locally normal. The representations of $\Abullet$ can thus be
continuously extended to all of $\Afrak$ in such a way that the
formula \eqref{eq-rep-disintegration} describing the disintegration
stays valid for the extended representations when $\Xecm$ is replaced
by the non-null set $\Xecm_0$ occurring in
Proposition~\ref{Pro-xi-precompactness}.
\begin{The}[Local Normality]
  \label{The-local-normality}
  Under the presumptions of Proposition~\ref{Pro-xi-precompactness}
  the following assertions are valid:
  \begin{proplist}
  \item The GNS-representation $( \pi_w , \Hscr_w )$ of the
    quasi-local algebra $\Afrak$ is locally normal, i.\,e.~continuous
    with respect to the relative $\sigma$-weak topologies of both $\AO
    \subseteq \BH$ and $\pi_w \bigl( \AO \bigr) \subseteq \BHw$ for
    arbitrary bounded regions $\Oscr$.
  \item The representation $( \pi^\bullet , \Hbullet )$ of the
    quasi-local algebra $\Abullet$ is locally normal (continuous with
    respect to the relative $\sigma$-weak topologies of both
    $\AbulletOk \subseteq \BH$ and $\pi^\bullet \bigl( \AbulletOk
    \bigr) \subseteq \BHbullet$ for arbitrary bounded regions $\Oscr_k
    \in \Rcount$). The same holds true for the irreducible
    representations $( \pi_\xi , \Hscr_\xi )$ occurring in the spatial
    disintegration of $( \pi^\bullet , \Hbullet )$ when the indices
    $\xi$ are astricted to $\Xecm_0$.
  \item The representation $( \pi^\bullet , \Hbullet )$ as well as the
    irreducible ones $( \pi_\xi , \Hscr_\xi )$ with $\xi \in \Xecm_0$
    allow for unique locally normal extensions to the whole of the
    original quasi-local algebra $\Afrak$ designated $( \pibar^\bullet
    , \Hbullet )$ and $( \pibar_\xi , \Hscr_\xi )$, respectively,
    which are related by
    \begin{equation}
      \label{eq-rep-ext-disintegration}
      ( \pibar^\bullet , \Hbullet ) \simeq \int_{\Xecm_0}^\oplus d \nu
      ( \xi ) \; ( \pibar_\xi , \Hscr_\xi ) \text{,}
    \end{equation}
    where the representations $( \pibar_\xi , \Hscr_\xi )$ are again
    irreducible.
  \end{proplist}
\end{The}
\begin{proof}
  \begin{prooflist}
  \item Let $\Deltabar$ be a bounded Borel set and suppose that $\rho$
    is a normal functional on $\BH$. Then the same applies to the
    functional $\rho_{\Deltabar} (~.~) \doteq \rho \bigl(
    \EDbar~.~\EDbar \bigr)$, and therefore the mapping
    \begin{equation*}
      T_{\Deltabar} : \Afrak \rightarrow \BH \qquad A \mapsto
      T_{\Deltabar} ( A ) \doteq \EDbar A \EDbar
    \end{equation*}
    is continuous with respect to the relative $\sigma$-weak topology
    of $\Afrak$. Now, according to the Compactness Condition,
    $T_{\Deltabar} \restriction \AO = \TODbar$ maps the unit ball
    $\AoneO$ of the local $C^*$-algebra $\AO$ onto the relatively
    compact set $\EDbar \AoneO \EDbar$. The restriction of $\TODbar$
    to $\AoneO$ is now obviously continuous with respect to the
    relative $\sigma$-weak topologies, but this result can be
    tightened up in the following sense: The relative $\sigma$-weak
    topology, being Hausdorff and coarser than the relative norm
    topology, and the relative uniform topology itself coincide on the
    compact norm closure of $\EDbar \AoneO \EDbar$ due to a conclusion
    of general topology \cite[Chapter One,
    \S\,3,\,2.(6)]{koethe:1983}. Therefore $\TODbar$ is still
    continuous on $\AoneO$ when its image is furnished with the norm
    topology instead.

    Now, suppose that $\Delta'$ is an arbitrary bounded Borel set and
    let $\Deltabar \supseteq \Delta + \Delta'$ be another bounded
    Borel set with the property that \eqref{eq-Delta-boundedness} is
    satisfied. Then the linear mapping
    \begin{equation}
      \label{eq-sandwich-mapping}
      \EDbar A \EDbar \mapsto \EwDprime \pi_w ( A ) \EwDprime
    \end{equation}
    is well-defined and continuous with respect to the uniform
    topologies of both domain and image. Therefore, as a consequence
    of the previous paragraph, we infer that the composition of this
    map with the restriction of $T_{\Deltabar}$ to $\AoneO$ is
    continuous, when $\AoneO$ is endowed with the $\sigma$-weak
    topology whereas the range carries the relative norm topology. The
    resulting map is explicitly determined as the restriction to
    $\AoneO$ of
    \begin{equation}
      \label{eq-sandwich-representation}
      \piwDprime : \Afrak \rightarrow \BHw \qquad A \mapsto \piwDprime
      ( A ) \doteq \EwDprime \pi_w ( A ) \EwDprime \text{.}
    \end{equation}
     
    If $\eta$ denotes a $\sigma$-weakly continuous functional on
    $\BHw$, the same is true regarding $\etaDprime (~.~) \doteq \eta
    \bigl( \EwDprime~.~\EwDprime \bigr)$ for any bounded Borel set
    $\Delta' \subseteq \Rsone$, and moreover, due to strong continuity
    of the spectral measure, $\eta$ is the uniform limit of the net of
    functionals $\etaDprime$ for $\Delta' \nearrow \Rsone$. Given a
    $\sigma$-weakly convergent net $\bset{A_\iota : \iota \in J}
    \subseteq \AoneO$ with limit $A \in \AoneO$, we conclude from the
    discussion in the preceding paragraph that
    \begin{equation}
      \label{eq-sigma-weak-Delta-limit}
      \etaDprime \bigl( \pi_w ( A_\iota - A ) \bigr) = \eta \bigl(
      \piwDprime ( A_\iota - A ) \bigr) \xrightarrow[\iota \in J]{} 0
      \text{.}
    \end{equation}
    Therefore, by means of the estimate
    \begin{multline}
      \label{eq-sigma-weak-limit}
      \babs{\eta \circ \pi_w ( A_\iota - A )} \leqslant \babs{\eta
      \bigl( \pi_w ( A_\iota - A ) \bigr) - \etaDprime \bigl( \pi_w (
      A_\iota - A ) \bigr)} + \babs{\etaDprime \bigl( \pi_w ( A_\iota
      - A ) \bigr)} \\
      \leqslant \bnorm{\eta - \etaDprime} \bnorm{\pi_w ( A_\iota - A
      )} + \babs{\etaDprime \bigl( \pi_w ( A_\iota - A ) \bigr)}
      \leqslant 2 \, \bnorm{\eta - \etaDprime} + \babs{\etaDprime
      \bigl( \pi_w ( A_\iota - A ) \bigr)} \text{,}
    \end{multline}
    it is easily seen that, upon selection of a suitable bounded Borel
    set $\Delta'$, the right-hand side can be made smaller than any
    given bound for $\iota \succ \iota_0$ with an appropriate index
    $\iota_0$. This being true for any $\sigma$-weakly continuous
    functional $\eta$ on $\BHw$ and arbitrary nets $\bset{A_\iota :
    \iota \in J}$ in $\AoneO$ converging to $A \in \AoneO$ with
    respect to the $\sigma$-weak topology of $\BH$, we have thus
    established that the restrictions of the representation $\pi_w$ to
    each of the unit balls $\AoneO$ are $\sigma$-weakly
    continuous. According to
    \cite[Lemma~10.1.10]{kadison/ringrose:1986} this assertion extends
    to the entire local $C^*$-algebra $\AO$, so that $\pi_w$ indeed
    turns out to be locally normal.
  \item The arguments given above in the case of $\pi_w$ can be
    transferred literally to the representations $\pi^\bullet$ and
    $\pi_\xi$, $\xi \in \Xecm_0$, in view of the relations
    \eqref{eq-pibullet-def-rep-relation} and
    \eqref{eq-final-sandwich-estimate} established in the proofs of
    Propositions~\ref{Pro-weight-precompactness} and
    \ref{Pro-xi-precompactness}, which substitute
    \eqref{eq-Delta-boundedness} used in the first part. The evident
    modifications to be applied include the restriction to local
    algebras $\AbulletOk$ where $\Oscr_k$ is a member of the countable
    family $\Rcount$.
  \item Complementary to the statements of the second part,
    \cite[Lemma~10.1.10]{kadison/ringrose:1986} exhibits that the
    representations $\pi^\bullet$ and $\pi_\xi$, $\xi \in \Xecm_0$,
    allow for unique $\sigma$-weakly continuous extensions
    $\pibar^\bullet$ and $\pibar_\xi$ onto the weak closures
    $\AbulletOk''$ \cite[Corollary~2.4.15]{bratteli/robinson:1987} of
    the local algebras, which, due to the Bicommutant Theorem
    \cite[Theorem~2.4.11]{bratteli/robinson:1987}, coincide with the
    strong closures and thus, by the very construction of
    $\AbulletOk$, $\Oscr_k \in \Rcount$, expounded in
    Section~\ref{sec-separable-reformulation}, contain the
    corresponding local $C^*$-algebras $\AOk$ of the underlying
    quantum field theory. Now, due to the net structure of $\Oscr_k
    \mapsto \AOk$, the quasi-local $C^*$-algebra $\Afrak$ is its
    $C^*$-inductive limit, i.\,e.~the norm closure of the $^*$-algebra
    $\bigcup_{\Oscr_k \in \Rcount} \AOk$. As the representations
    $\pibar^\bullet$ and $\pibar_\xi$, $\xi \in \Xecm_0$, are
    altogether uniformly continuous on this $^*$-algebra
    \cite[Theorem~1.5.7]{pedersen:1979}, they can in a unique way be
    continuously extended to its completion $\Afrak$ \cite[Chapter
    One, \S\,5,\,4.(4)]{koethe:1983}, and these extensions, again
    denoted $\pibar^\bullet$ and $\pibar_\xi$, respectively, are
    easily seen to be compatible with the algebraic structure of
    $\Afrak$. $( \pibar^\bullet , \Hbullet )$ and $( \pibar_\xi ,
    \Hscr_\xi )$ are thus representations of this quasi-local algebra,
    evidently irreducible in the case of $\pibar_\xi$, and moreover
    locally normal, since, by construction, they are $\sigma$-weakly
    continuous when restricted to local algebras $\AOk$ pertaining to
    the countable subclass of regions $\Oscr_k \in \Rcount$, and an
    arbitrary local algebra $\AO$ is contained in at least one of
    these. The statement on uniqueness of the extensions is then an
    immediate consequence of the fact that they are uniquely
    determined by the property of being $\sigma$-weakly continuous on
    the local $C^*$-algebras $\AOk$.

    Regarding the disintegration of operators $\pibar^\bullet ( A )$
    for arbitrary $A \in \Afrak$, note that any operator $B \in \AOk$ is the
    $\sigma$-weak limit of a \emph{sequence} $\set{B_n}_{n \in \Nbb}$
    in $\ArbulletOk$ with $r = \norm{B}$. For nets in $\ArbulletOk$
    this statement is a consequence of Kaplansky's Density Theorem
    \cite[Theorem~II.4.8]{takesaki:1979} in connection with
    \cite[Lemma~II.2.5]{takesaki:1979} and the various relations
    between the different locally convex topologies on $\BH$. The
    specialization to sequences is justified by
    \cite[Proposition~II.2.7]{takesaki:1979} in view of the
    separability of $\Hscr$. The operators $L \in \idealcount$ define
    fundamental sequences of measurable vector fields $\bset{\xiket{L}
    : \xi \in \Xecm_0}$ (cf.~\cite[Section~II.1.3,
    Definition~1]{dixmier:1981}) and, as the operators $\pi^\bullet (
    B_n )$ are decomposable, all the functions
    \begin{equation*}
      h_n : \Xecm_0 \rightarrow \Cbb \qquad \xi \mapsto h_n ( \xi )
      \doteq \bxiscpx{L_1}{\pi_\xi ( B_n )}{L_2}
    \end{equation*}
    are measurable for arbitrary $L_1 \text{,} L_2 \in \idealcount$.
    The same is valid for the pointwise limit of this sequence
    \cite[II.1.10]{fell/doran:1988a}
    \begin{equation*}
      h : \Xecm_0 \rightarrow \Cbb \qquad \xi \mapsto h ( \xi ) \doteq
      \bxiscpx{L_1}{\pibar_\xi ( B )}{L_2} \text{,}
    \end{equation*}
    and that suffices, according to \cite[Section~II.2.1,
    Proposition~1]{dixmier:1981}, to demonstrate that
    $\bset{\pibar_\xi ( B ) : \xi \in \Xecm_0}$ is a measurable field
    of operators. As the sequence $\bset{\pi^\bullet ( B_n )}_{n \in
    \Nbb}$ converges $\sigma$-weakly to $\pibar^\bullet ( B )$ by
    assumption and since, moreover, $\nu ( \Xecm_0 )$ is finite and
    the family of operators $\bset{\pi_\xi ( B_n ) : \xi \in \Xecm_0}$
    is bounded by $\norm{B}$ for any $n$, we conclude with
    Lebesgue's Dominated Convergence Theorem applied to the
    decompositions of $\pi^\bullet ( B_n )$ with respect to $\Xecm_0$
    (which differs from $\Xecm$ only by a null set), that
    \begin{multline}
      \label{eq-local-operator-disint-limit}
      \bscpx{L_1}{\pi^\bullet ( B_n )}{L_2} = \int_{\Xecm_0} d \nu (
      \xi ) \; \bxiscpx{L_1}{\pi_\xi ( B_n )}{L_2} \\
      \xrightarrow[n \rightarrow \infty]{} \quad \int_{\Xecm_0} d \nu ( \xi
      ) \; \bxiscpx{L_1}{\pibar_\xi ( B )}{L_2} =
      \bscpx{L_1}{\pibar^\bullet ( B )}{L_2} \text{.}
    \end{multline}
    If $W_0$ denotes the isometry which implements the unitary
    equivalence
    \begin{equation*}
      ( \pi^\bullet , \Hbullet ) \simeq \int_{\Xecm_0}^\oplus d \nu (
      \xi ) \; ( \pi_\xi , \Hscr_\xi )
    \end{equation*}
    and has all the properties of the operator $W$ introduced in
    Theorem~\ref{The-spatial-disintegration}, then, by density of the
    set $\bset{\bullket{L} : L \in \idealcount}$ in $\Hbullet$, we
    infer from \eqref{eq-local-operator-disint-limit} that
    \begin{subequations}
      \begin{equation}
        \label{eq-local-operator-disint}
        W_0 \, \pibar^\bullet ( B ) \, W_0^* = \int_{\Xecm_0}^\oplus d
        \nu ( \xi ) \; \pibar_\xi ( B ) \text{.}
      \end{equation}
      This relation has been established under the presupposition that
      $B$ belongs to some local $C^*$-algebra $\AOk$. Now, it is
      possible to reapply the above reasoning in the case of an
      arbitrary element $A$ of the quasi-local algebra $\Afrak$, which
      can be approximated uniformly by a sequence $\set{A_n}_{n \in
      \Nbb}$ from $\bigcup_{\Oscr_k \in \Rcount} \AOk$. In this way,
      \eqref{eq-local-operator-disint} is extended to all of $\Afrak$
      so that we end up with the final equation
      \begin{equation}
        \label{eq-quasilocal-operator-disint}
        W_0 \, \pibar^\bullet ( A ) \, W_0^* = \int_{\Xecm_0}^\oplus d
        \nu ( \xi ) \; \pibar_\xi ( A ) \text{,} \quad A \in \Afrak
        \text{,}
      \end{equation}
    \end{subequations}
    demonstrating that indeed
    \begin{equation*}
      ( \pibar^\bullet , \Hbullet ) \simeq \int_{\Xecm_0}^\oplus d \nu
      ( \xi ) \; ( \pibar_\xi , \Hscr_\xi ) \text{.} \tag*{\qed}
    \end{equation*}
    \renewcommand{\qed}{}
  \end{prooflist}
  \renewcommand{\qed}{}
\end{proof}

Theorem~\ref{The-local-normality} shows that, under the assumption of
sensible phase space restrictions, no information on a physical system
described by a normal state of bounded energy $\omega \in \Sscr (
\Delta )$ gets lost in the entirety of constructions presented in
Chapters~\ref{chap-particle-weights} and \ref{chap-disintegration}. 
These have led us from $\omega$ via an associated particle weight with
representation $( \pi_w , \Hscr_w )$ of the quasi-local algebra
$\Afrak$ to the induced restricted $\vaccountbar$-particle weight $(
\pi^\bullet , \Hbullet , \Abullet , \alpha^\bullet , \Pcount ,
\vaccountbar)$, which comprises a representation $( \pi^\bullet ,
\Hbullet )$ of the algebra $\Abullet$ allowing for a disintegration in
terms of irreducible representations $\bset{( \pi_\xi , \Hscr_\xi ) :
\xi \in \Xecm_0}$. Then, according to the preceding theorem, this
disintegration is extendable in a unique fashion to one in terms of
locally normal representations of the original algebra $\Afrak$ as
expressed by \eqref{eq-rep-ext-disintegration}. Now, due to the
explicit construction in Theorem~\ref{The-bullet-weight} of $(
\pi^\bullet , \Hbullet)$ from $( \pi_w , \Hscr_w)$, the local
normality of both of these representations implies that, actually,
$\pibar^\bullet$ coincides with the restriction of $\pi_w$ to the
subspace $\Hbullet$ of $\Hscr_w$. Thus we arrive at a partial
reconstruction of the GNS-representation $( \pi_w , \Hscr_w )$, which
only depends on the initial choice of a subspace of the Hilbert space
$\Hscr_w$. Moreover, by Theorem~\ref{The-local-normality}, this
entails a spatial disintegration of $\Delta$-bounded particle weights
$\scp{~.~}{~.~}$ according to the following reformulation of
\eqref{eq-rep-ext-disintegration}:
\begin{equation}
  \label{eq-rep-ori-disintegration}
  ( \pi_w , \Hbullet ) \simeq \int_{\Xecm_0}^\oplus d \nu ( \xi ) \; (
  \pibar_\xi , \Hscr_\xi ) \text{.}
\end{equation}

\chapter{Disintegration Revisited: Choquet Theory}
  \label{chap-choquet}

The spatial disintegration as expounded in
Chapter~\ref{chap-disintegration} suffered from a couple of awkward
drawbacks, which, in our belief, are inessential concomitants of this
special approach to a decomposition theory for particle weights and
have no bearing on the physical significance of the concept proper. It
should be noted in this connection that, to be able to apply the
standard disintegration theory for representations made available in
the literature on $C^*$-algebras, we had to fall back upon separable
constructs and countable dense subsets thereof. As a consequence it
had to be accepted, that a theory of disintegration could only be
formulated in terms of restricted $\vaccountbar$-particle
weights. But these technical difficulties seem to be accidental, and
the question obtrudes on us if it is possible to carry through a
disintegration, in the course of which no need arises to leave the
class of particle weights proper, which means that the disintegration
can indeed be formulated in terms of \emph{pure} particle
weights. This is the topic of the present chapter, presenting the
partial results we were able to produce in this direction to date.

As already noted in the remark following
Definition~\ref{Def-particle-weight}, the totality of particle weights
constitutes a positive proper convex cone when supplemented by the
trivial form. This observation opens the way to an application of
another concept of disintegration: the barycentric decomposition in
the special form of a generalization of the well-known Theorem of
Krein--Milman \cite[Theorem~1.4.3]{kadison/ringrose:1983}. This
approach, initiated by Choquet and further developed by Bishop and de
Leeuw \cite{bishop/leeuw:1959,choquet:1956}, is especially
well-adapted to the study of convex sets in infinite-dimensional
spaces. An introduction to this theory can be found in
\cite{alfsen:1971,phelps:1966} and also in
\cite[Section~4.1.2]{bratteli/robinson:1987} where it is applied to
get a decomposition of states on the quasi-local $C^*$-algebra
$\Afrak$ in terms of pure ones. The mathematical structure in this
case is easily accessible from the point of view of Choquet theory:
\begin{indentbulletlist}
\item The positive linear functionals on $\Afrak$ constitute a
  positive convex cone $\Kecm$ in its topological dual $\Afrak^*$. 
\item The quasi-local algebra contains a unit $\unit$, which defines a
  base of this cone when it is considered as a continuous linear
  functional on $\Afrak^*$, thus introducing a convex function which
  intersects $\Kecm$.
\item This convex base of $\Kecm$ coincides with the set of states on
  $\Afrak$ and thus turns out to be compact with respect to the
  weak$^*$ topology \cite[Propositon~2.3.11 and
  Theorem~2.3.15]{bratteli/robinson:1987}.
\end{indentbulletlist}
The situation is much more complicated when particle weights are
considered. These do constitute a positive cone $\Wecm$ in the space
of sesquilinear forms on the left ideal $\Lfrak$ of localizing
operators; but one of the key features in the construction of an
algebra of detectors was the absence of a unit element, the existence
of which would produce infinite values of the asymptotic functionals
$\sigma$ arising in Chapter~\ref{chap-particle-weights}. The obvious
questions to be answered are:
\begin{indentbulletlist}
\item What is an appropriate (metrizable) topology $\Tcal$ to be
  introduced on $\Wecm$ to render relevant subsets compact? 
\item How can a convex base $\Becm_\Wecm$ be fitted into the cone
  $\Wecm$ in a physically meaningful way and such that this base is
  compact with respect to the beforementioned topology?
\end{indentbulletlist}
In our approach the class with respect to which the disintegration is
to be performed will be further restricted
(cf.~Definition~\ref{Def-weights-subclass} below).

The answer to the first of the above problems which we are going to
present in this chapter is based on an effective control of the
dislocalization of almost local operators combined with the
Fredenhagen--Hertel compactness condition. To be more specific, a norm
will be introduced on the space $\Lfrak_0$ of almost local vacuum
annihilation operators which in a way measures their deviation from
being contained in a local algebra. Making use of this norm on the set
$\Lfrak_0 ( \Gamma )$ of those operators with energy-momentum transfer
in a compact and convex subset $\Gamma$ of $\complement \fwcone$, it
can be shown that $\Lfrak_0 ( \Gamma )$ is small in the $\qD$-topology
under the assumption of the Compactness Criterion of Fredenhagen and
Hertel. By its very definition, the notion of almost locality as
introduced in Definition~\ref{Def-almost-locality} imposes a condition
of rapid decrease on the norm difference between almost local
operators and strictly local ones according to the growing extension
of the localization regions. This contrives to introduce the following
norms on $\Lfrak_0$.
\begin{Def}
  \label{Def-disloc-norm}
  Let $m$ be an arbitrary natural number, then the equation
  \begin{equation}
    \label{eq-disloc-norm}
    \Qmx{L_0} \doteq \sup_{r > 0} \, \inf \bset{r^m \norm{L_0 - L_r} :
      L_r \in \AOr} \text{,} \quad L_0 \in \Lfrak_0 \text{,}
  \end{equation}
  defines a norm on the vector space $\Lfrak_0$.
\end{Def}
\begin{Rem}
  The seminorm properties of the mapping $\Qm$ are self-evident, they
  even hold for arbitrary almost local operators as arguments. To
  infer $L_0 = 0$ from the condition $\Qmx{L_0} = 0$ one has to
  restrict attention to vacuum annihilation operators. The latter
  equation means that the operator $L_0$ is contained in the norm
  closure of any local algebra $\AOr$, $r > 0$. As these are
  $C^*$-algebras, hence uniformly closed, $L_0$ itself turns out to be
  a local operator, a property which can be reconciled with it being a
  vacuum annihilation operator only for $L_0 = 0$.
\end{Rem}
The information concerning the localization of the operator $L_0 \in
\Lfrak_0$ embodied by the value of $\Qmx{L_0}$ is highly dependent on
the norm which $L_0$ carries as an element of the $C^*$-algebra
$\Afrak$. Therefore we combine both topologies in the subsequent
definition.
\begin{Def}
  \label{Def-comb-disloc-norm}
  For any natural number $m$ a norm on $\Lfrak_0$ is defined by
  \begin{equation}
    \label{eq-comb-disloc-norm}
    \Norm{L_0}_m \doteq \norm{L_0} + \Qmx{L_0} \text{,} \quad L_0 \in
    \Lfrak_0 \text{.}
  \end{equation}
\end{Def}

As announced above this topology is now to be related to the
$\qD$-seminorms on the subspace $\Lfrak_0 ( \Gamma )$, where $\Gamma$
denotes a compact, convex subset of the complement of the forward
light cone. Although we have the inequality
\eqref{eq-harmonic-analysis-commutator-estimate} at our disposition,
we want to reformulate it here in order to make explicit the
dependence of the integrand on its right-hand side upon the bounded
Borel set $\Delta$ and upon the energy-momentum transfer $\Gamma$. To
this end one has to reapply the arguments given in the Appendix of
\cite{buchholz:1990}.
\begin{Pro}
  Let $\Delta$ be a bounded Borel set and $\Gamma$ a compact and
  convex subset of $\complement \fwcone$. There exists a bounded Borel
  set $\Delta' ( \Delta , \Gamma ) \subseteq \Rsone$, depending on
  $\Delta$ and $\Gamma$ only, such that for any $L_0 \in \Lfrak_0 (
  \Gamma )$ there holds the estimate
  \begin{equation}
    \label{eq-harmonic-analysis-energy-momentum-bound}
    \qDx{L_0}^2 \leqslant N' ( \Delta , \Gamma ) \int_{\Rs} d^s x \;
    \bnorm{\EDG \bcomm{\aibx ( L_0 )}{{L_0}^*} \EDG}
  \end{equation}
  with a suitable constant $N' ( \Delta , \Gamma )$, which is again
  specified by the sets $\Delta$ and $\Gamma$.
\end{Pro}
\begin{proof}
  In a first step it will be shown that, setting
  \begin{equation*}
    Q_\Kib \doteq \int_\Kib d^s x \; \aibx ( {L_0}^* L_0 )
  \end{equation*}
  for any compact subset $\Kib$ of $\Rs$, the following estimate is in
  force for arbitrary bounded Borel sets $\Delta_0$:
  \begin{equation}
    \label{eq-harmonic-analysis-K-estimate}
    \norm{\EDzero Q_\Kib \EDzero} \leqslant N'' \int_{\Kib - \Kib}
    \mspace{-15mu} d^s x \; \bnorm{\EDdprime \bcomm{\aibx ( L_0
    )}{{L_0}^*} \EDdprime}
  \end{equation}
  with a suitable constant $N''$ and an appropriate bounded Borel set
  $\Delta''$. If $\omega_\Psi$ denotes a state on $\BH$ which is
  induced by a vector $\Psi \in \EDzero \Hscr$ we can immediately
  adopt the inequalities of \cite[p.~640]{buchholz:1990} to get
  \begin{subequations}
    \begin{multline}
      \label{eq-harmonic-analysis-adopted}
      \omega_\Psi ( Q_\Kib )^2 \leqslant \omega_\Psi ( Q_\Kib \cdot
      Q_\Kib ) \\
      \shoveleft{\leqslant \omega_\Psi ( Q_\Kib ) \cdot \sup_{\yib
      \in \Rs} \int_{\Kib - \Kib} \mspace{-15mu} d^s x \; \norm{L_0 \,
      U ( - \yib ) \Psi}^{- 1} \bnorm{\bcomm{\aibx ( L_0 )}{{L_0}^*}
      L_0 \, U ( - \yib ) \Psi}} \\
      + \int_\Kib d^s x \; \omega_\Psi \bigl( \aibx ( {L_0}^* ) \,
      Q_\Kib \, \aibx ( L_0 ) \bigr) \text{.}
    \end{multline}
    The integrand of the second term on the right-hand side is subject
    to the relation
    \begin{equation}
      \label{eq-sandwich-energy-bound-estimate}
      \omega_\Psi \bigl( \aibx ( {L_0}^* ) \, Q_\Kib \, \aibx ( L_0 )
      \bigr) \leqslant \norm{\EDbarG Q_\Kib \EDbarG} \cdot \omega_\Psi
      \bigl( \aibx ( {L_0}^* L_0 ) \bigr)
    \end{equation}
    with $\Deltabar_0$ denoting the closure of $\Delta_0$. Upon
    insertion into \eqref{eq-harmonic-analysis-adopted}, removal of
    the resulting common factor $\omega_\Psi ( Q_\Kib )$ on both sides
    and passing to the supremum with respect to all unit vectors $\Psi
    \in \EDzero \Hscr$, we get
    \begin{equation}
      \label{eq-harmonic-analysis-elab}
      \norm{\EDzero Q_\Kib \EDzero} \leqslant \int_{\Kib - \Kib}
      \mspace{-15mu} d^s x \; \bnorm{\bcomm{\aibx ( L_0 )}{{L_0}^*}
      \EDbarG} + \norm{\EDbarG Q_\Kib \EDbarG} \text{,}
    \end{equation}
  \end{subequations}
  where use is made of the fact that all the vectors $L_0 \, U ( -
  \yib ) \Psi$ belong to the subspace $\EDbarG \Hscr$ for arbitrary
  $\yib \in \Rs$ and $\Psi \in \EDzero \Hscr$. The preparatory
  estimate \eqref{eq-harmonic-analysis-K-estimate} is now established
  by complete induction on $n$, where this natural number is defined
  in dependence on the sets $\Delta_0$ and $\Gamma$ through the
  condition $( \Deltabar_0 + \Gamma_n ) \cap \fwcone = \emptyset$
  (cf.~the proof of Proposition~\ref{Pro-harmonic-analysis} on page
  \pageref{page-proof-harmonic-analysis}).

  For $n = 1$ we have, according to the spectrum condition, $\EDbarG =
  0$ so that \eqref{eq-harmonic-analysis-K-estimate} is trivially
  fulfilled since its left-hand side vanishes. Now assume that the
  condition $( \Deltabar_0 + \Gamma_{n + 1} ) \cap \fwcone =
  \emptyset$ is valid, which, stated another way, means that the
  intersection of $( \Deltabar_0 + \Gamma ) + \Gamma_n$ with the
  complement of $\fwcone$ is empty. As $\Deltabar_0 + \Gamma$ is a
  bounded Borel set we can apply the induction hypothesis for $n$,
  i.\,e.~\eqref{eq-harmonic-analysis-K-estimate} with $\Delta_0$
  replaced by $\Deltabar_0 + \Gamma$, to infer that there exists a
  bounded Borel set $\Delta_0''$ which satisfies
  \begin{subequations}
    \begin{equation}
      \label{eq-harmonic-analysis-induction-hypothesis}
      \norm{\EDbarG Q_\Kib \EDbarG} \leqslant N_0'' \int_{\Kib - \Kib}
      \mspace{-15mu} d^s x \; \bnorm{\EDzerodprime \bcomm{\aibx ( L_0
      )}{{L_0}^*}
      \EDzerodprime}
    \end{equation}
    for an appropriate constant $N_0''$. This estimate inserted into
    \eqref{eq-harmonic-analysis-elab} amounts to
    \begin{multline}
      \label{eq-harmonic-analysis-penultimate}
      \norm{\EDzero Q_\Kib \EDzero} \\
      \leqslant \int_{\Kib - \Kib} \mspace{-15mu} d^s x \;
      \bnorm{\bcomm{\aibx ( L_0 )}{{L_0}^*} \EDbarG} + N_0''
      \int_{\Kib - \Kib} \mspace{-15mu} d^s x \; \bnorm{\EDzerodprime
      \bcomm{\aibx ( L_0 )}{{L_0}^*} \EDzerodprime} \text{,}
    \end{multline}
  \end{subequations}
  from which to conclude the validity of
  \eqref{eq-harmonic-analysis-K-estimate} with suitable constant $N''
  = N_0'' + 1$ and proper bounded Borel set $\Delta'' = \Delta_0''
  \cup \bigl( ( \Deltabar_0 + \Gamma ) + \Gamma - \Gamma \bigr)$ is an
  obvious task.

  Now, having established \eqref{eq-harmonic-analysis-K-estimate}, we
  can specialize it to $\Delta_0 \doteq \Delta$ and pass to the limit
  $\Kib \nearrow \Rs$ as in the proof of
  Proposition~\ref{Pro-harmonic-analysis}, noting that
  \begin{equation*}
    Q_{\Delta , \Kib}^{( {L_0}^* L_0 )} = \ED Q_\Kib \ED
  \end{equation*}
  and that, due to almost locality of $L_0$, the integral on the
  right-hand side can be extended over all of $\Rs$. As a result, in
  view of Definition~\ref{Def-seminorms}, one arrives at the desired
  inequality \eqref{eq-harmonic-analysis-energy-momentum-bound}, where
  the formulation chosen makes explicit its dependence on $\Delta$ and
  $\Gamma$.
\end{proof}

The formula \eqref{eq-harmonic-analysis-energy-momentum-bound} just
established is to be applied in the sequel to produce an estimate of
the seminorm $\qDx{L_0}$ for operators $L_0 \in \Lfrak_0 ( \Gamma )$
with compact and convex $\Gamma \subseteq \complement \fwcone$ in
terms of the initial operator norm $\norm{~.~}$ and of the norm
$\Qmx{~.~}$ introduced in Definition~\ref{Def-disloc-norm}. In order
to get a manageable result we specialize to the case $m = 2s$.
\begin{Lem}
  \label{Lem-qD-energy-bound}
  Let $\Delta$ be a bounded Borel set and $\Gamma$ a compact and
  convex subset of the complement of $\fwcone$. Then there exists a
  bounded Borel set $\Delta'' ( \Delta , \Gamma )$, depending on
  $\Delta$ and $\Gamma$, such that for any $L_0 \in \Lfrak_0 ( \Gamma
  )$ the estimate
  \begin{multline}
    \label{eq-qD-energy-bound}
    \qDx{L_0} \leqslant N' ( \Delta , \Gamma )^{1 / 2} \bigl( a ( s )
    \, \norm{L_0}^2 + b ( s ) \, \norm{L_0} + c ( s ) \bigr)^{1 / 2}
    \cdot \\
    {\Qtwosx{L_0}}^{1 / 4} \bnorm{\EDdG L_0 \EDdG}^{1 / 2}
  \end{multline}
  holds with suitable coefficients depending on the spatial dimension
  $s$.
\end{Lem}
\begin{proof}
  We have to calculate the integral on the right-hand side of
  \eqref{eq-harmonic-analysis-energy-momentum-bound} and, to do so, it
  is split into two parts according to $\abs{\xib} > R$ or $\abs{\xib}
  \leqslant R$ with an abitrary radius $R$ which is held fixed for the
  moment. For large $\abs{\xib}$ we use the estimate
  \eqref{eq-commutator-norm-estimate} for the integrand and get in
  terms of the norm $\Qtwos$:
  \begin{subequations}
    \begin{multline}
      \label{eq-integrand-largex-estimate}
      \bnorm{\EDG \bcomm{\aibx ( L_0 )}{{L_0}^*} \EDG} \\
      \mspace{-9mu} \leqslant \bnorm{\bcomm{\aibx ( L_0 )}{{L_0}^*}}
      \leqslant 4 \, \norm{L_0} \, \norm{L_0 - ( L_0 )_{2^{-1}
      \abs{\xib}}} + 2 \, \norm{L_0 - ( L_0 )_{2^{-1} \abs{\xib}}}^2
      \\
      \leqslant 4 \, \norm{L_0} \, 2^{2 s} \, \abs{\xib}^{- 2 s}
      \Qtwosx{L_0} + 2 \, 2^{4 s} \, \abs{\xib}^{- 4 s}
      {\Qtwosx{L_0}}^2 \text{.}
    \end{multline}
    Accordingly, the respective integral is subject to the inequality
    \begin{multline}
      \label{eq-integral-largex-estimate}
      \int_{\abs{\xib} > R} d^s x \; \bnorm{\EDG \bcomm{\aibx ( L_0
      )}{{L_0}^*} \EDG} \\
      \leqslant 2^{2 s + 2} \, \norm{L_0} \, \Qtwosx{L_0}
      \int_{\abs{\xib} > R} d^s x \; \abs{\xib}^{- 2 s} + 2^{4 s + 1}
      \, {\Qtwosx{L_0}}^2 \int_{\abs{\xib} > R} d^s x \; \abs{\xib}^{-
      4 s} \text{.}
    \end{multline}
  \end{subequations}
  The integrand for small $\abs{\xib}$ is evaluated observing the
  spectral projections arising on the right-hand side of
  \eqref{eq-harmonic-analysis-energy-momentum-bound}, which are
  abbreviated as $\Delta' \doteq \Delta' ( \Delta , \Gamma )$ with
  closure $\Deltaprimebar$. This leads to 
  \begin{subequations}
    \begin{multline}
      \label{eq-integrand-smallx-estimate}
      \bnorm{\EDG \bcomm{\aibx ( L_0 )}{{L_0}^*} \EDG} \\
      \mspace{-70mu} \leqslant \bnorm{\EDprime \aibx ( L_0 )
      \EDGprimebarminus {L_0}^* \EDprime} + \bnorm{\EDprime {L_0}^* 
      \EDGprimebarplus \aibx ( L_0 ) \EDprime} \\
      \leqslant \bnorm{\Ecupminus L_0 \Ecupminus}^2 + \bnorm{\Ecupplus
      L_0 \Ecupplus}^2 \\
      \leqslant 2 \, \bnorm{\EDdG L_0 \EDdG}^2
    \end{multline}
    where $\Delta'' ( \Delta , \Gamma ) \doteq \Delta \cup ( \Deltabar
    + \Gamma ) \cup \bigl( \Delta' \cup ( \Deltaprimebar - \Gamma)
    \bigr) \cup \bigl( \Delta' \cup ( \Deltaprimebar + \Gamma)
    \bigr)$---the inclusion of $\Delta \cup ( \Deltabar + \Gamma )$
    into this definition being required at the very end of the present
    argumentation. The corresponding integral satisfies the inequality
    \begin{multline}
      \label{eq-integral-smallx-estimate}
      \int_{\abs{\xib} \leqslant R} d^s x \; \bnorm{\EDG \bcomm{\aibx
      ( L_0 )}{{L_0}^*} \EDG} \\
      \leqslant 2 \, \bnorm{\EDdG L_0 \EDdG}^2 \int_{\abs{\xib}
      \leqslant R} d^s x \text{.}
    \end{multline}
  \end{subequations}
  The integrals remaining in \eqref{eq-integral-largex-estimate} and
  \eqref{eq-integral-smallx-estimate} are known from calculus
  (cf.~\cite[Section~4.11]{courant/john:1974}):
  \begin{subequations}
    \label{eq-s-volumes}
    \begin{align}
      \int_{\abs{\xib} > R} d^s x \; \abs{\xib}^{- 2 s} & = \omega_s
      \int_R^\infty dr \; r^{s - 1} \, r^{- 2 s} = s^{- 1} \, \omega_s
      \, R^{- s} \text{,} \\
      \int_{\abs{\xib} > R} d^s x \; \abs{\xib}^{- 4 s} & = \omega_s
      \int_R^\infty dr \; r^{s - 1} \, r^{- 4 s} = ( 3 \, s )^{- 1} \,
      \omega_s \, R^{- 3 s} \text{,} \\
      \int_{\abs{\xib} \leqslant R} d^s x & = \omega_s \int_0^R dr \;
      r^{s - 1} = s^{- 1} \, \omega_s \, R^s \text{,}
    \end{align}
    where the factor $\omega_s$ is defined via the $\Gamma$-function
    as
    \begin{equation}
      \label{eq-s-volume-factor}
      \omega_s \doteq 2 \; \Gamma ( s / 2 )^{- 1} \sqrt{\pi^s}
      \text{.} 
    \end{equation}
  \end{subequations}

  Collecting the results from \eqref{eq-integral-largex-estimate},
  \eqref{eq-integral-smallx-estimate} and \eqref{eq-s-volumes} one
  gets for the complete integral
  \begin{multline}
    \label{eq-integral-total-estimate}
    \int_{\Rs} d^s x \; \bnorm{\EDG \bcomm{\aibx ( L_0 )}{{L_0}^*}
    \EDG} \\
    \leqslant \omega_s \, R^s \Bigl( a' ( s ) \, {\Qtwosx{L_0}}^2 \,
    R^{- 4 s} + b' ( s ) \, \norm{L_0} \, \Qtwosx{L_0} R^{- 2 s} \\
    + c' ( s ) \bnorm{\EDdG L_0 \EDdG}^2 \Bigr)
  \end{multline}
  with suitable $s$-dependent factors. So far the value of $R$ has
  been left open. To get the concise formula
  \eqref{eq-qD-energy-bound} we deliberately choose
  \begin{equation}
    \label{eq-R-choice}
    R^{2 s} \doteq \bnorm{\EDdG L_0 \EDdG}^{- 2} \, \Qtwosx{L_0}
    \text{,}
  \end{equation}
  so that \eqref{eq-integral-total-estimate} simplifies to
  \begin{multline}
    \label{eq-integral-total-estimate-R-chosen}
    \int_{\Rs} d^s x \; \bnorm{\EDG \bcomm{\aibx ( L_0 )}{{L_0}^*}
    \EDG} \\
    \leqslant \omega_s \, R^s \bnorm{\EDdG L_0 \EDdG}^2 \bigl( a' ( s
    ) \, \norm{L_0}^2 + b' ( s ) \, \norm{L_0} + c' ( s ) \bigr)
    \text{.}
  \end{multline}
  Inserting the square root of \eqref{eq-R-choice} into this estimate
  and carrying the result over to
  \eqref{eq-harmonic-analysis-energy-momentum-bound}, we finally
  arrive at \eqref{eq-qD-energy-bound}, where $\omega_s$ has been
  included in the definition of the coefficients.

  Note, that the above argument is independent of the occurrence of
  $\Qtwosx{L_0} = 0$ or $\bnorm{\EDdG L_0 \EDdG} = 0$, for in this
  case $\qDx{L_0} = 0$, so that \eqref{eq-qD-energy-bound} is
  trivially fulfilled. This consequence is immediate from the norm
  property of $\Qtwos$. As to the second of the above conditions, it
  turns out to be important that we have included $\Delta \cup (
  \Deltabar + \Gamma )$ into the definition of $\Delta'' ( \Delta ,
  \Gamma )$. Thence the named assumption implies $\bnorm{\EDGplus L_0
  \ED} = \bnorm{L_0 \ED} = 0$, and $\qDx{L_0} = 0$ is a result of
  Lemma~\ref{Lem-delta-annihilation}.
\end{proof}

Our next aim is to single out a convex subset in the class of all
particle weights which turns out to be compact in a suitably chosen
topology. To this end it will be assumed from now on that the
underlying quantum field theory satisfies the Fredenhagen--Hertel
Compactness Criterion, under the assumption of which the following
result can be established.
\begin{Pro}
  In a quantum field theory which satisfies the Fredenhagen--Hertel
  Compactness Condition the subsequent mapping, defined for bounded
  Borel subsets $\Delta$ and compact, convex subsets $\Gamma$ of
  $\Rsone$,
  \begin{equation*}
    \SDG : \Lfrak_0 ( \Gamma ) \rightarrow \BH \qquad L_0 \mapsto \SDG (
    L_0 ) \doteq L_0 \ED \text{,}
  \end{equation*}
  sends balls in $\Lfrak_0 ( \Gamma )$ of finite radius with respect
  to the norm $\Norm{~.~}_m$, $m \in \Nbb$, onto precompact subsets of
  $\BH$ in its uniform topology.
\end{Pro}
\begin{proof}
  Let $\Lfrak_{0 , R}^{\; m} ( \Gamma )$ denote the closed $R$-ball,
  $R > 0$, in $\Lfrak_0 ( \Gamma )$ with respect to $\Norm{~.~}_m$. By
  Definition~\ref{Def-comb-disloc-norm}, the condition $\Norm{L_0}_m
  \leqslant R$, $L_0 \in \Lfrak_0 ( \Gamma )$, implies $\Qmx{L_0} <
  R$, stating a property of uniform approximation. This means that,
  given $\varepsilon > 0$, there exists a radius $r_0$, take
  e.\,g.~$r_0 \doteq (2 \, R / \varepsilon)^{- m / 2}$, such that to
  any $L_0 \in \Lfrak_{0 , R}^{\; m} ( \Gamma )$ we can find a local
  operator $( L_0 )_{r_0} \in \Afrak ( \Oscr_{r_0})$ with
  \begin{subequations}
    \begin{equation}
      \label{eq-local-approximation}
      \norm{L_0 - ( L_0 )_{r_0}} < \varepsilon \text{.}
    \end{equation}
    Again according to Definition~\ref{Def-comb-disloc-norm}, one also
    has $\norm{L_0} < R$, so that the collection of local operators
    just introduced belongs to the closed ball of radius $R +
    \varepsilon$ in $\Afrak ( \Oscr_{r_0} )$. Now, the
    Fredenhagen--Hertel Compactness Condition ensues that there exists
    a finite number of operators $L_k$, $k = 1 \text{,} \dots \text{,}
    N ( \varepsilon)$, in this ball such that any $( L_0 )_{r_0}$
    satisfies the condition
    \begin{equation}
      \label{eq-local-precompactness}
      \bnorm{\EDGplus \bigl( ( L_0 )_{r_0} - L_k \bigr) \ED} <
      \varepsilon
    \end{equation}
    for at least one $k$. Combining this with
    \eqref{eq-local-approximation}, we see that for any $L_0 \in
    \Lfrak_{0 , R}^{\; m} ( \Gamma )$ there exists a suitable operator
    $L_k \in \Afrak ( \Oscr_{r_0} )$ with 
    \begin{equation}
      \label{eq-almost-local-precompactness}
      \bnorm{\EDGplus ( L_0 - L_k ) \ED} \leqslant \norm{L_0 - ( L_0
      )_{r_0}} + \bnorm{\EDGplus \bigl( ( L_0 )_{r_0} - L_k \bigr)
      \ED} < 2 \, \varepsilon \text{.}
    \end{equation}
  \end{subequations}
  It is an immediate consequence that finitely many elements from
  $\Lfrak_{0 , R}^{\; m} ( \Gamma )$ can be selected, serving as
  centres of $4 \, \varepsilon$-balls which cover the set
  \begin{equation*}
    \EDGplus \Lfrak_{0 , R}^{\; m} ( \Gamma ) \ED = \Lfrak_{0 , R}^{\;
    m} ( \Gamma ) \ED \text{.}
  \end{equation*}
  By arbitrariness of $\varepsilon$, we have thus established
  precompactness of the mapping $\SDG$ in the sense of the
  Proposition.
\end{proof}

The results presented thus far have only laid down the groundwork for
the topological considerations concerning the set of particle weights
proper. For the moment we return here to the special continuity
properties of the asymptotic functionals resulting from the limiting
procedure expounded in Chapter~\ref{chap-particle-weights}. According
to Proposition~\ref{Pro-positivity-of-limits} in connection with
\eqref{eq-counter-estimate} of Lemma~\ref{Lem-basic-estimate}, one has
for any $L_1 \text{,} L_2 \in \Lfrak$ and any $A \in \Afrak$:
\begin{equation}
  \label{eq-asymptotic-funct-prep-estimate}
  \babs{\sigma ( {L_1}^* A L_2 )} \leqslant \norm{h}_\infty
  \bnorm{\EDGoneplus A \EDGtwoplus} \, \qDx{L_1} \, \qDx{L_2} \text{.}
\end{equation}
Specializing now to operators $A$ from the unit ball of a local
$C^*$-algebra and to vacuum annihilation operators $L_1$ and $L_2$
from the $\Norm{~.~}_{2 s}$-unit balls of $\Lfrak_0 ( \Gamma_1 )$ and
$\Lfrak_0 ( \Gamma_2 )$, respectively, with compact and convex
$\Gamma_k$, we infer from \eqref{eq-asymptotic-funct-prep-estimate} by
use of Lemma~\ref{Lem-qD-energy-bound} in connection with
Definition~\ref{Def-comb-disloc-norm} that there exist constants $C' (
\Delta , \Gamma_1 )$ and $C' ( \Delta , \Gamma_2 )$ such that
\begin{multline}
  \label{eq-asymptotic-funct-final-estimate}
  \babs{\sigma ( {L_1}^* A L_2 )} \leqslant \norm{h}_\infty C'
  ( \Delta , \Gamma_1 ) \; C' ( \Delta , \Gamma_2 ) \cdot \\
  \cdot \bnorm{\EDGoneplus A \EDGtwoplus} \, \bnorm{L_1 \EDdGone}^{1 /
  2} \bnorm{L_2 \EDdGtwo}^{1 / 2}
\end{multline}
with appropriate bounded Borel sets $\Delta'' ( \Delta , \Gamma_1 )$
and $\Delta'' ( \Delta , \Gamma_1 )$, depending on $\Delta$ and both
of the compact sets $\Gamma_1$ and $\Gamma_2$. An inequality of type
\eqref{eq-asymptotic-funct-final-estimate} can likewise be imposed on
the corresponding sesquilinear form, which opens up the way to
distinguish a certain subclass in the space $\Secm$ of \emph{all}
sesquilinear forms on $\Lfrak \times \Lfrak$.
\begin{Def}
  \label{Def-weights-subclass}
  The set $\Secm^b$ of all sesquilinear forms $\Wx{~.~}{~.~}$ on
  $\Lfrak \times \Lfrak$ which are characterized by the existence of
  constants $C' ( \Delta , \Gamma_1 )$, $C' ( \Delta , \Gamma_2 )$ and
  $C''$ such that the following condition
  \eqref{eq-distinguished-sesqui} holds for any operator $A \in
  \AoneO$ as well as for $L_1 \in \Lfrak_{0 , 1}^{2 s} ( \Gamma_1 )$
  and $L_2 \in \Lfrak_{0 , 1}^{2 s} ( \Gamma_2 )$ with compact and
  convex $\Gamma_l$ is a subspace of $\Secm$:
  \begin{multline}
    \label{eq-distinguished-sesqui}
    \babs{\bWx{L_1}{A L_2}} \leqslant C'' \; C' ( \Delta , \Gamma_1 )
    \; C' ( \Delta , \Gamma_2 ) \cdot \\
    \cdot \bnorm{\EDGoneplus A \EDGtwoplus} \, \bnorm{L_1 \EDdGone}^{1
    / 2} \bnorm{L_2 \EDdGtwo}^{1 / 2} \text{.}
  \end{multline}
  Its intersection with the positive cone $\Wecm$ of all particle
  weights according to Definition~\ref{Def-particle-weight} is again a
  positive proper convex cone, denoted $\Wecm^b$, which obviously
  comprises the particle weights induced by asymptotic functionals.
\end{Def}
Due to \eqref{eq-distinguished-sesqui}, the space $\Secm^b$ can be
furnished with various seminorm topologies.
\begin{Def}
  \label{Def-sesqui-seminorms}
  For any combination of bounded regions $\Oscr$ with compact and
  convex $\Gamma_1$ and $\Gamma_2$, a seminorm $\POG$ can be introduced
  on $\Secm^b$ by
  \begin{equation}
    \label{eq-Secm-seminorm}
    \POGx{W} \doteq \sup \Bset{\babs{\bWx{L_1}{A L_2}} : A \in \AoneO
    , L_1 \in \Lfrak_{0 , 1}^{2 s} ( \Gamma_1 ) , L_2 \in \Lfrak_{0 ,
    1}^{2 s} ( \Gamma_2 )} \text{.}
  \end{equation}
\end{Def}
The convex subset of $\Wecm^b$, which is to be used from now on and
will turn out to be compact when furnished with a suitable topology, 
is introduced again in view of \eqref{eq-distinguished-sesqui}.
\begin{Def}
  $\Wecm_c^b$ is the convex set of all particle weights in $\Wecm^b$
  which satisfy the inequality
  \begin{multline}
    \label{eq-def-convex-set-weights}
    \babs{\bWx{L_1}{A L_2}} \leqslant C' ( \Delta , \Gamma_1 ) \; C' (
    \Delta , \Gamma_2 ) \cdot \\
    \cdot \bnorm{\EDGoneplus A \EDGtwoplus} \, \bnorm{L_1 \EDdGone}^{1
    / 2} \bnorm{L_2 \EDdGtwo}^{1 / 2}  
  \end{multline}
  for all bounded regions $\Oscr$ and all compact and convex
  $\Gamma_1$ and $\Gamma_2$. The difference between this condition and
  \eqref{eq-distinguished-sesqui} is that the only $W$-dependent
  constant $C''$ has been omitted.
\end{Def}
\begin{Rem}
  According to \eqref{eq-asymptotic-funct-final-estimate}, all
  particle weights arising from asymptotic functionals with
  $\norm{h}_\infty \leqslant 1$ satisfy
  \eqref{eq-def-convex-set-weights} and are thus contained in
  $\Wecm_c^b$.
\end{Rem}

Now, as a consequence of the Compactness Criterion of Fredenhagen and
Hertel, we know that there exist in each case finitely many operators
in $\AoneO$ as well as in $\Lfrak_{0 , 1}^{2 s} ( \Gamma_l )$, $l = 1
\text{,} 2$, serving as centres of $\delta$-balls to cover the sets
$\Lfrak_{0 , 1}^{2 s} ( \Gamma_l ) \EDdGl$ and $\EDGoneplus \AoneO
\EDGtwoplus$. These operators can be used to span finite-dimensional
subspaces in $\Lfrak_0 ( \Gamma_1 )$ and $\AO \Lfrak_0 ( \Gamma_2
)$. The corresponding space of sesquilinear forms defined on these
domains is again finite-dimensional, so that its unit ball with
respect to the relative $\POG$-topology can be covered by a finite
number of $\varepsilon$-balls ( note that bounded sets in
finite-dimensional vector spaces are relatively compact). The
restriction of any element $W$ of $\Wecm_c^b$ to the named subspaces
of $\Lfrak_0 ( \Gamma_1 )$ and $\AO \Lfrak_0 ( \Gamma_2 )$ is thus
contained in one of these balls. This in turn means, that we can even
select a finite number of elements $W_k \in \Wecm_c^b$, $k = 1
\text{,} \dots \text{,} N ( 2 \, \varepsilon )$, such that any element
of $\Wecm_c^b$ is contained in a $2 \, \varepsilon$-ball around at
least one of these chosen forms with respect to the aforementioned
relative $\POG$-topology. But then it can be shown that the $3 \,
\varepsilon$-balls with respect to the $\POG$-topology proper indeed
cover all of $\Wecm_c^b$. To see this, let $W_K$ be the element
pertaining to $W \in \Wecm_c^b$ and let $L_l \in \Lfrak_{0 , 1}^{2 s}
( \Gamma_l )$, $l = 1 \text{,} 2$, and $A \in \AoneO$ be
arbitrary. Then there exist operators $L_l^\delta \in \Lfrak_{0 ,
1}^{2 s} ( \Gamma_l )$ and $A^\delta \in \AoneO$ which satisfy
\begin{subequations}
  \label{eq-delta-estimates}
  \begin{align}
    \bnorm{( L_l - L_l^\delta ) \EDdGl} & < \delta \text{,} \\
    \bnorm{\EDGoneplus ( A - A^\delta ) \EDGtwoplus} & < \delta \text{.}
  \end{align}
\end{subequations}
Now, making use of condition \eqref{eq-def-convex-set-weights} on
elements of $\Wecm_c^b$ in connection with \eqref{eq-delta-estimates}
as well as of the defining property for $W_K$, we get
\begin{multline}
  \label{eq-conv-weights-precompact}
  \babs{\bWx{L_1}{A L_2} - \bWxK{L_1}{A L_2}} \leqslant
  \babs{\bWx{L_1}{A L_2} - \bWx{L_1^\delta}{A^\delta L_2^\delta}} \\
  + \babs{\bWx{L_1^\delta}{A^\delta L_2^\delta} -
  \bWxK{L_1^\delta}{A^\delta L_2^\delta}} +
  \babs{\bWxK{L_1^\delta}{A^\delta L_2^\delta} - \bWxK{L_1}{A L_2}} \\
  < 2 \, ( 2 \, \delta^{1 / 2} ) + 2 \, \varepsilon \text{.}
\end{multline}
Since we are free to choose $\delta$ appropriately small in dependence
on a given $\varepsilon > 0$, this final relation shows, upon taking
the supremum with respect to the operators appearing on the left-hand
side, that for each $W \in \Wecm_c^b$ there exists at least one $W_K
\in \Wecm_c^b$ such that $\bPOGx{W - W_K} < 3 \, \varepsilon$ in
accordance with our statement.

As in Chapter~\ref{chap-disintegration} we want to pass at this point
to countable families $\set{\Oscr_n}_{n \in \Nbb}$ and
$\set{\Gamma_l}_{l \in \Nbb}$ of bounded regions in space-time and of
compact and convex subsets of $\complement \fwcone$. By
Definition~\ref{Def-sesqui-seminorms} any triple taken from these
sequences defines a seminorm on $\Secm^b$. In this way $\Secm^b$ can
be topologized with a sequence $\set{\Pm}_{m \in \Nbb}$ of
seminorms and thereby becomes a locally convex (Hausdorff) space. This
space is metrizable according to \cite[Chapter~Four,
\S\,18,\,2.(2)]{koethe:1983} and its topology can moreover be derived
from the increasing sequence of seminorms
\begin{equation}
  \label{eq-def-inc-sequence}
  \Rkx{W} \doteq \max_{1 \leqslant m \leqslant k} \Pmx{W} \text{,}
  \quad W \in \Secm^b \text{,} \quad k \in \Nbb \text{.}
\end{equation}
This countable system can then be used to define an (F)-norm
\cite[p.\,163]{koethe:1983} on $\Secm^b$ which furnishes this space
with the same topology. It is given by
\begin{equation}
  \label{eq-F-norm}
  \FNorm{W} \doteq \sum_{k = 1}^\infty \frac{1}{2^k} \frac{\Rkx{W}}{1
    + \Rkx{W}} \text{,} \quad W \in \Secm^b \text{,}
\end{equation}
and generates a translation-invariant metric \cite[Chapter~Four,
\S\,18,\,2.(3)]{koethe:1983}. Given $\varepsilon > 0$ there exists,
according to \eqref{eq-F-norm}, an index $M$ such that for any $W
\text{,} W' \in \Secm^b$
\begin{multline}
  \label{eq-F-norm-metric-estimate}
  \FNorm{W - W'} = \sum_{k = 1}^\infty \frac{1}{2^k} \frac{\Rkx{W -
  W'}}{1 + \Rkx{W - W'}} \\
  < \frac{\varepsilon}{2} + \sum_{k = 1}^M \frac{1}{2^k} \frac{\Rkx{W
  - W'}}{1 + \Rkx{W - W'}} \leqslant \frac{\varepsilon}{2} + \sum_{k =
  1}^M \frac{1}{2^k} \cdot \RMx{W - W'} \text{.}
\end{multline}
A consequence of the preceding paragraph in combination with the
definition \eqref{eq-def-inc-sequence} is the fact that $\Wecm_c^b$
can be covered by a finite number of balls with a given arbitrarily
small radius with respect to the $\Rk$-topologies. It is then an
immediate conclusion from the definitions involved that the
sesquilinear functionals arising as limits with respect to
$\FNorm{~.~}$ of sequences in $\Wecm_c^b$ are again elements of
$\Wecm_c^b$. This convex subset thus turns out to be closed. Since it
has been seen above to be precompact, it is indeed compact in the
$\FNorm{~.~}$-topology.
\begin{Pro}
  The convex set $\Wecm_c^b$ in the class of all particle weights is
  compact with respect to the metric derived from the (F)-norm
  $\FNorm{~.~}$.
\end{Pro}
The above work has laid the foundation for an application of Choquet's
Theorem \cite[Corollary~I.4.9]{alfsen:1971} which tells us that any
particle weight $\scp{~.~}{~.~}$ in the metrizable compact convex set
$\Wecm_c^b$ can be represented by a positive and normalized measure
vanishing off its extreme boundary $\partial_e \Wecm_c^b$:
\begin{equation}
  \label{eq-choquet-prep-integral}
    \scp{~.~}{~.~} = \int_{\partial_e \Wecm_c^b} d \upsilon ( \zeta ) \;
    \scp{~.~}{~.~}_\zeta \text{.}
\end{equation}

The above result represents the present status of the Choquet approach
to a disintegration theory for particle weights. The problem to be
tackled at this point is the open question of how a base can be fitted
into the cone $\Wecm^b$ which is completely contained in $\Wecm_c^b$.
This would allow for a disintegration of a particle weight on this
base in terms of extremal points, defining extremal rays of the cone
$\Wecm^b$ and representing pure particle weights. On this foundation a
complete theory in parallel to that developed for states on a
$C^*$-algebra $\Afrak$ in \cite[Sections~4.1 and
4.2]{bratteli/robinson:1987} still awaits its completion. The
advantage of this approach in comparison to the spatial disintegration
presented in Chapter~\ref{chap-disintegration} is that, apart from the
somewhat intricate topological considerations, it is more direct and
the resulting pure particle weights are no longer subject to the
restrictive Definition~\ref{Def-restr-particle-weight}. On the other
hand, the mathematical problems concerning convex sets in
infinite-dimensional spaces are far from being trivial. Therefore, a
lot of work remains to be done until eventually the particle content
of a quantum field theory is seen to be encoded in the geometrical
structure (the set of extreme rays) of a positive cone of particle
weights.

\chapter{Summary and Outlook}
  \label{chap-summary}

The present work is based on the general point of view that the
concept of `particles' is asymptotic in nature and simultaneously has
to be founded by making appropriate use of the notion of locality.
This reflects our conviction that the long-standing problem of
`asymptotic completeness' of quantum field theory, i.\,e., the
question if a quantum field theoretic model can be interpreted
completely in terms of particles, has to be tackled by the aid of
further restrictions on the general structure, which essentially are
of a local character. The question is, what the local structure of a
theory should be in order that it governs scattering processes in such
a way that asymptotically the physical states appear to clot in terms
of certain entities named particles. The compactness and nuclearity
conditions discussed in \cite{buchholz/porrmann:1990} and the
references therein are examples of this kind of approach. We do not
claim that they already give a complete answer, but believe that they
indicate the right direction.

In this thesis we have constructed asymptotic functionals on a
certain algebra of detectors giving rise to particle weights which can
be interpreted as mixtures of particle states. A disintegration theory
has been developed for restricted particle weights by means of a
highly technical procedure in Chapter~\ref{chap-disintegration}. This
constitutes the basis for the definition of mass and spin even in the
case of charged states \cite{buchholz/porrmann}. We are convinced that
the technicalities involved can be dissolved by future research. In
this connection the analysis of concrete models may be helpful. Such
investigations are already under way. They concern the Schwinger model
\cite{gruening:1999} and an application of our formalism to quantum
electrodynamics \cite{fredenhagen/freund}. It is expected that some
insight may be gained with respect to the open questions mentioned in
the various chapters. E.\,g., the convergence problem in connection
with Theorem~\ref{The-singular-limits} can perhaps be solved with
additional information at hand, and the direct integral decomposition
of Chapter~\ref{chap-disintegration} might get more manageable,
unfolding the connection between the intrinsic energy-momenta
pertaining to the irreducible representations and the geometrical
energy-momenta (velocities) that stem from the asymptotic functionals.

So far we have considered \emph{single} particle weights. Another
field of future research is the inspection of coincidence arrangements
of detectors as in \cite{araki/haag:1967}. In this respect, too, the
analysis of concrete models is helpful.

As indicated by Chapter~\ref{chap-local-normality} and in view of the
partial results presented in Chapter~\ref{chap-choquet} phase space
restrictions seem to be a key ingredient in the general analysis, in
particular of the Choquet approach to disintegration. This theory is
still in its initial stage. But, difficult as the mathematical
problems concerning convex sets in infinite-dimensional spaces are, it
deserves further efforts. Presumably, both the spatial disintegration
and the Choquet decomposition will eventually turn out to be
essentially equivalent, revealing relations similar to those
encountered in the disintegration theory of states on
$C^*$-algebras \cite[Chapter~4]{bratteli/robinson:1987}. Further
studies have to disclose the geometrical structure of the positive
cone of particle weights, as the particle content of a theory seems to
be encoded in this kind of information.

\appendix

\chapter{Concepts of Differentiability}
  \label{chap-differentiability}

Various notions of differentiability have to be used in this work and
some of them take on a somewhat unusual shape. So it seems right to
collect in this appendix a number of definitions and propositions,
both to assign a precise meaning to the concepts proper and to their
consequences as well as to fix the notation.

\section{Differentiation in Locally Convex Spaces}

\begin{Def}
  \label{Def-diff-lqs}
  Let $\Xfrak$ be a (real or complex) normed space and let $\Vfrak$ be
  a locally convex space over the same field whose topology is defined
  by the family $\bset{\ql : \lambda \in L}$ of seminorms which
  separate the points of $\Vfrak$. Suppose further that we are given
  an open subset $\Gfrak$ of $\Xfrak$.
  \begin{deflist}
  \item A mapping $F : \Gfrak \rightarrow \Vfrak$ is called
    differentiable at the point $\xfrak \in \Gfrak$ if there exists a
    continuous linear mapping $T : \Xfrak \rightarrow \Vfrak$ such
    that for any vector $\hfrak$ in a certain
    $\zerofrak$-neighbourhood $\Ufrak \subseteq \Xfrak$ the increment
    $F ( \xfrak + \hfrak ) - F ( \xfrak )$ allows for the linearized
    approximation 
    \begin{subequations}
      \begin{equation}
        \label{eq-diff-lqs}
        F ( \xfrak + \hfrak ) - F ( \xfrak ) = T \hfrak + R [ F ,
        \xfrak] ( \hfrak ) \text{,}
      \end{equation}
      where $R [ F , \xfrak ]$ is a mapping on $\Ufrak$ to $\Vfrak$
      subject to the condition
      \begin{equation}
        \label{eq-diff-lqs-residual}
        \lim_{\hfrak \rightarrow \zerofrak} \norm{\hfrak}^{-1} \bqlx{R
        [ F , \xfrak ] ( \hfrak )} = 0
      \end{equation}
    \end{subequations}
    for any seminorm $\ql$, $\lambda \in L$. The linear operator $T$
    occurring in \eqref{eq-diff-lqs} is signified as $\Dfrak F (
    \xfrak )$ and called the derivative of $F$ at $\xfrak$.
  \item The mapping $F : \Gfrak \rightarrow \Vfrak$ is called
    differentiable if it is differentiable at any $\xfrak \in
    \Gfrak$.
  \item The differentiable mapping $F : \Gfrak \rightarrow \Vfrak$ is
    called continuously differentiable if the mapping $\Gfrak \ni
    \xfrak \mapsto \Dfrak F ( \xfrak ) \hfrak \in \Vfrak$, which
    exists by assumption, is continuous with respect to the locally
    convex topology of $\Vfrak$ for any given $\hfrak \in \Xfrak$.
  \end{deflist}
\end{Def}
\begin{Rem}
  The definition of the continuous linear operator $\Dfrak F ( \xfrak
  )$ requires uniqueness of the corresponding $T$ in
  \eqref{eq-diff-lqs}, but this is easily established. Assume the
  existence of another $\zerofrak$-neighbourhood $\Ufrak'$, a
  continuous linear operator $T' : \Xfrak \rightarrow \Vfrak$ and a
  mapping $R' [ F , \xfrak ] : \Ufrak' \rightarrow \Vfrak$ which, upon
  insertion into \eqref{eq-diff-lqs}, represent the increment $F (
  \xfrak + \hfrak ) -  F ( \xfrak )$ such that $R' [ F , \xfrak ]$
  fulfills a condition analogous to \eqref{eq-diff-lqs-residual}. Then
  \begin{equation*}
    T \hfrak - T' \hfrak = R' [ F , \xfrak ] ( \hfrak ) - R [ F ,
    \xfrak ] ( \hfrak ) \text{,} \quad \hfrak \in \Ufrak \cap \Ufrak'
    \text{.}
  \end{equation*}
  Let $\yfrak \in \Xfrak$, $\yfrak \not= \zerofrak$, be arbitrary but
  fixed, then for $\alpha \in \Cbb \setminus \set{0}$ small enough we
  infer from the above equation due to the linearity of both $T$ and
  $T'$
  \begin{multline*}
    \bqlx{T \yfrak - T' \yfrak} = \bqlx{\alpha^{-1} \bigl( R' [ F ,
    \xfrak ] ( \alpha \yfrak ) - R [ F , \xfrak ] ( \alpha \yfrak )
    \bigr)} \\
    = \norm{\yfrak} \, \norm{\alpha \yfrak}^{-1} \bqlx{R' [ F , \xfrak
    ] ( \alpha \yfrak ) - R [ F , \xfrak ] ( \alpha \yfrak )} \text{,}
  \end{multline*}
  where the right-hand side vanishes in the limit $\alpha \rightarrow
  0$ for any seminorm $\ql$, according to
  \eqref{eq-diff-lqs-residual}. This yields $\bqlx{T \yfrak} = 
  \bqlx{T' \yfrak}$, valid also for $\yfrak = \zerofrak$, and as a
  consequence $T \yfrak = T' \yfrak$ for any $\yfrak \in \Xfrak$ since
  the seminorms $\ql$ separate the points in $\Vfrak$.
\end{Rem}
An immediate consequence of the presumed continuity of the linear
operators $\Dfrak F ( \xfrak )$, entering as derivatives the
representation \eqref{eq-diff-lqs} of the increment of $F$ at
$\xfrak$, is the fact that differentiability implies continuity.
\begin{Cor}
  \label{Cor-diff-continuity}
  Let $\Xfrak$ be a normed space and let $\Vfrak$ be a locally convex
  space. If the mapping $F : \Gfrak \rightarrow \Vfrak$, $\Gfrak
  \subseteq \Xfrak$ open, is differentiable at the point $\xfrak \in
  \Gfrak$ then it is also continuous in $\xfrak$.
\end{Cor}
The methods used in the standard theory of differentiable functions
yield the following propositions when applied to the concept laid open
in Definition~\ref{Def-diff-lqs}, the main modification being the
occurrence of seminorms $\ql$ on $\Vfrak$ in
\eqref{eq-diff-lqs-residual}. 
\begin{Pro}[Product Rule for Derivatives]
  \label{Pro-product-rule}
  Let $\Xfrak$ be a normed space and $\Gfrak$ an open subset of
  $\Xfrak$.
  \begin{proplist}
  \item Suppose that $\Vfrak$ is a locally convex space and that the
    mappings $F : \Gfrak \rightarrow \Vfrak$ and $f : \Gfrak
    \rightarrow \Kbb$, $\Kbb$ the scalar field of both $\Xfrak$ and
    $\Vfrak$, are differentiable at $\xfrak \in \Gfrak$. Then their 
    product $f F$ is differentiable at this point, too, and the
    derivative at $\xfrak$ is given by
    \begin{equation*}
      \Dfrak ( f F ) ( \xfrak ) \hfrak = \Dfrak f ( \xfrak ) \hfrak \,
      F ( \xfrak ) + f ( \xfrak ) \, \Dfrak F ( \xfrak ) \hfrak
      \text{,} \quad \hfrak \in \Xfrak \text{.}
    \end{equation*}
  \item Let $\Yfrak$ be a normed algebra and assume that the mappings
    $F : \Gfrak \rightarrow \Yfrak$ and $G : \Gfrak \rightarrow
    \Yfrak$ are differentiable at $\xfrak \in \Gfrak$. Then their 
    product $F G$ is differentiable at $\xfrak$, too, and the
    derivative is
    \begin{equation*}
      \Dfrak ( F G ) ( \xfrak ) \hfrak = \Dfrak F ( \xfrak ) \hfrak \,
      G ( \xfrak ) + F ( \xfrak ) \, \Dfrak G ( \xfrak ) \hfrak
      \text{,} \quad \hfrak \in \Xfrak \text{.}
    \end{equation*}
  \end{proplist}
\end{Pro}
\begin{Pro}[Chain Rule for Derivatives]
  \label{Pro-chain-rule}
  Let $\Xfrak$ and $\Yfrak$ be normed spaces and let $\Vfrak$ be a
  locally convex space. Assume further that the mapping $G : \Gfrak_1
  \rightarrow \Yfrak$ is differentiable at $\xfrak \in \Gfrak_1$ and
  that the mapping $F : \Gfrak_2 \rightarrow \Vfrak$ is differentiable
  at $G ( \xfrak )$, where $\Gfrak_1$ and $\Gfrak_2$ are open subsets
  of $\Xfrak$ and $\Yfrak$, respectively, and $G ( \Gfrak_1 )
  \subseteq \Gfrak_2$. Then the composition of $F$ and $G$: $F \circ G
  : \Gfrak_1 \rightarrow \Vfrak$, exists and is differentiable at
  $\xfrak$ with a derivative connected to those of $F$ and $G$ through
  \begin{equation*}
    \Dfrak ( F \circ G ) ( \xfrak ) = \Dfrak G \bigl( F ( \xfrak )
    \bigr) \circ \Dfrak F ( \xfrak ) \text{.}
  \end{equation*}
\end{Pro}

The fundamental Mean Value Theorem which has to be formulated in the
setting of Definition~\ref{Def-diff-lqs} is based on the following two
lemmas. Their proof as well as that of the theorem proper is an
adaptation of the reasoning to be found in \cite[Kapitel~XX,
Abschnitt~175]{heuser:1993b}.
\begin{Lem}
  \label{Lem-lqs-diff-0-const-relation}
  Let $F : [ a , b ] \rightarrow \Vfrak$ be a continuous mapping on
  the compact interval $[ a , b ] \subseteq \Rbb$ to the locally
  convex space $\Vfrak$ and suppose that it is differentiable on the
  interior of this set with $\Dfrak F ( x ) = 0$ for any $x \in \: ] a
  , b [$. Then $F$ is constant on $[ a , b ]$.
\end{Lem}
\begin{proof}
  Let $s$ and $t$ be arbitrary distinct points in $] a , b [$. We
  shall assume $s < t$ and want to show that $F ( s ) = F ( t )$.
  Define $u \doteq 2^{-1} ( t - s )$ and consider one of the seminorms
  $\ql$ topologizing $\Vfrak$. There are two possibilities:
  \begin{subequations}
    \begin{align}
      \label{eq-a}
      \bqlx{F ( u ) - F ( s )} & \geqslant \bqlx{F ( t ) - F ( u )}
      \text{,} \\
      \label{eq-b}
      \bqlx{F ( t ) - F ( u )} & > \bqlx{F ( u ) - F ( s )} \text{.}
    \end{align}
  \end{subequations}
  Depending on the actual situation we define an interval $] s_1 , t_1
  [ \: \subseteq [ a , b ]$, choosing $s_1 \doteq s$, $t_1 \doteq u$
  in case \eqref{eq-a} and $s_1 \doteq u$, $t_1 \doteq t$ in case
  \eqref{eq-b}. Independent of this selection is the estimate
  \begin{equation}
    \label{eq-indep-estimate}
    \bqlx{F ( t ) - F ( s )} \leqslant \bqlx{F ( t ) - F ( u )} +
    \bqlx{F ( u ) - F ( s )} \leqslant 2 \, \bqlx{F ( t_1 ) - F ( s_1
    )} \text{.}
  \end{equation}
  The same procedure can then be applied to the interval $] s_1 , t_1
  [$, to the resulting interval $] s_2 , t_2 [$ and so on. In this way
  a sequence of intervals $] s_n , t_n [$ is constructed, which is
  decreasing with respect to the inclusion relation: $] s_{n+1} ,
  t_{n+1} [ \: \subseteq \: ] s_n , t_n [$. Furthermore the lengths
  are explicitly known as $t_n - s_n =  2^{-n} ( t - s )$ and the
  estimate \eqref{eq-indep-estimate} can be generalized to
  \begin{equation}
    \label{eq-indep-gen-estimate}
    \bqlx{F ( t ) - F ( s )} \leqslant 2^n \bqlx{F ( t_n ) - F ( s_n
    )} \text{.}
  \end{equation}
  There exists exactly one point $u_0 \in \: ] a , b [$ belonging to
  all intervals of this sequence and by assumption $\Dfrak F ( u_0 ) =
  0$, so that for $h$ in a small 0-neighbourhood $\Uscr \subseteq
  \Rbb$ the increment of $F$ at $u_0$ is represented by
  \begin{subequations}
    \begin{equation}
      \label{eq-increment-rep}
      F ( u_0 + h ) - F ( u_0 ) =  h \, R ( h )
    \end{equation}
    with a mapping $R : \Uscr \rightarrow \Vfrak$ satisfying
    \begin{equation}
      \label{eq-residual-rep}
      \lim_{h \rightarrow 0} \bqlx{R ( h )} = 0 \text{.}
    \end{equation}
  \end{subequations}
  Hence, given $\epsilon > 0$, there exists $N \in \Nbb$ such that
  $\bqlx{R ( u_0 - s_n)}$ and $\bqlx{R ( t_n - u_0)}$ are majorized by
  $( t - s )^{-1} \epsilon$ for $n > N$. According to
  \eqref{eq-increment-rep} this implies
  \begin{multline*}
    \bqlx{F ( t_n ) - F ( s_n )} \leqslant \bqlx{F ( t_n ) - F ( u_0
    )} + \bqlx{F ( s_n ) - F ( u_0 )} \\ 
    \mspace{120mu} \leqslant | t_n - u_0 | \, \bqlx{R ( t_n - u_0 )} +
    | u_0 - s_n | \, \bqlx{R ( u_0 - s_n )} \\ 
    \leqslant ( t_n - u_0 ) \frac{\epsilon}{t - s} + ( u_0 - s_n )
    \frac{\epsilon}{t - s} = ( t_n - s_n ) \frac{\epsilon}{t - s} =
    \frac{\epsilon}{2^n} \text{,}
  \end{multline*}
  where we made use of the length formula for the interval $] s_n ,
  t_n [$. From \eqref{eq-indep-gen-estimate} one then infers
  \begin{equation*}
    \bqlx{F ( t ) - F ( s )} \leqslant 2^n \frac{\epsilon}{2^n} =
    \epsilon \text{,}
  \end{equation*}
  so that, by arbitrariness of $\epsilon$ and $\ql$ together with the
  separation property of the seminorms, we see that $F ( t ) = F ( s
  ) = \vfrak_0 \in \Vfrak$. This relation holds for any $s$, $t \in
  \: ] a , b [$ and extends by the supposed continuity of $F$ to all
  of $[ a , b ]$, establishing $ F \equiv \vfrak_0$ as stated.
\end{proof}
\begin{Lem}
  \label{Lem-lqs-diff-int-relation}
  Let $F : [ a , b ] \rightarrow \Vfrak$ be a continuous mapping on
  the compact interval $[ a , b ] \subseteq \Rbb$ to the locally
  convex space $\Vfrak$ and define $G : [ a , b ] \rightarrow
  \Vfrakbar$, $\Vfrakbar$ the completion of $\Vfrak$, through the
  integral
  \begin{equation*}
    G ( x ) \doteq \int_a^x d \vartheta \; F ( \vartheta ) \text{,}
    \quad x \in [ a , b ] \text{.}
  \end{equation*}
  Then the mapping $G$ is differentiable for any $x_0 \in ] a , b [$
  and the action of the derivative $\Dfrak G ( x_0 )$ as a linear
  operator on $\Rbb$ is given by
  \begin{equation}
    \label{eq-diff-int-derivative}
    \Dfrak G ( x_0 ) h = h \, F ( x_0 ) \text{,} \quad h \in \Rbb
    \text{.}
  \end{equation}
\end{Lem}
\begin{proof}
  By \cite[II.6.2]{fell/doran:1988a} $G$ is a well-defined
  $\Vfrakbar$-valued mapping on the compact interval $[ a , b ]$. For
  $x_0 \in \: ] a , b [$ and $h \in \Rbb$ satisfying $ x_0 + h \in [ a
  , b ]$ we have
  \begin{equation*}
    G ( x_0 + h ) - G ( x_0 ) = \int_{x_0}^{x_0 + h} d \vartheta \; F
    ( \vartheta ) \text{,} 
  \end{equation*}
  hence
  \begin{subequations}
    \begin{equation}
      \label{eq-diff-int-representation}
      G ( x_0 + h ) - G ( x_0 ) - h \, F ( x_0 ) = \int_{x_0}^{x_0 +
      h} d \vartheta \; \bigl( F ( \vartheta ) - F ( x_0 ) \bigr)
      \doteq \rho ( h ) \text{.} 
    \end{equation}
    Now by assumption, $\vartheta \mapsto \bqlx{F ( \vartheta ) - F (
    x_0 )}$ is continuous on the compact interval $I_h$ of integration
    for any of the defining seminorms $\ql$ of $\Vfrak$, and,
    according to \cite[II.6.2 in connection II.5.4]{fell/doran:1988a},
    one has for any $h$ the estimate  
    \begin{equation}
      \label{eq-diff-int-residual-estimate}
      \abs{h}^{-1} \bqlx{\rho ( h )} \leqslant \abs{h}^{-1}
      \Babs{\int_{x_0}^{x_0 + h} d \vartheta \; \bqlx{F ( \vartheta )
      - F ( x_0 )}} \leqslant \max_{\vartheta \in I_h} \bqlx{F (
      \vartheta ) - F ( x_0 )} \text{,}
    \end{equation}
  \end{subequations}
  where the right-hand side vanishes in the limit $h \rightarrow 0$.
  Thus \eqref{eq-diff-int-representation} corresponds to the
  representation \eqref{eq-diff-lqs} of
  Definition~\ref{Def-diff-lqs} in terms of the increment $G ( x_0 + h
  ) - G (  x_0 )$ with a residual term $\rho ( h )$ satisfying
  \eqref{eq-diff-lqs-residual}. This proves differentiability of $G$
  on $] a , b [$ along with relation \eqref{eq-diff-int-derivative}.
\end{proof}
\begin{The}[Mean Value Theorem]
  \label{The-mean-value}
  Let $\Xfrak$ be a normed space and $\Vfrak$ be a locally convex
  space. Let furthermore $F : \Gfrak \rightarrow \Vfrak$, $\Gfrak
  \subseteq \Xfrak$ open, be a continuously differentiable mapping (in
  the sense of Definition~\ref{Def-diff-lqs}) and consider $\xfrak_0
  \in \Gfrak$ and $\hfrak \in \Xfrak$ small enough so that $\xfrak_0 +
  \vartheta \hfrak \in \Gfrak$ for $0 \leqslant \vartheta \leqslant
  1$. Then
  \begin{equation}
    \label{eq-mean-value-theorem}
    F ( \xfrak_0 + \hfrak ) - F ( \xfrak_0 ) = \int_0^1 d \vartheta \;
    \Dfrak F ( \xfrak_0 + \vartheta \hfrak ) \, \hfrak \text{.}
  \end{equation}
\end{The}
\begin{proof}
  Given $\xfrak_0 \in \Gfrak$ and $\hfrak \in \Xfrak$ as above we
  define two mappings $F_1$ and $F_2$ on the compact interval $[ 0 , 1
  ]$ to $\Vfrak$ respectively $\Vfrakbar$ through
  \begin{subequations}
    \label{eq-mean-value-def}
    \begin{align}
      s & \mapsto F_1 ( s ) \doteq F ( \xfrak_0 + s \hfrak ) \text{,}
      \\ 
      s & \mapsto F_2 ( s ) \doteq \int_0^s d \vartheta \; \Dfrak F (
      \xfrak_0 + \vartheta \hfrak ) \, \hfrak \text{.}
    \end{align}
  \end{subequations}
  From Lemma~\ref{Lem-lqs-diff-int-relation} and
  Proposition~\ref{Pro-chain-rule} we infer $\Dfrak F_2 ( s ) = \Dfrak
  F ( \xfrak_0 + s \hfrak ) \hfrak = \Dfrak F_1 ( s )$ for any $s \in
  \: ] 0 , 1 [$. This implies, according to
  Lemma~\ref{Lem-lqs-diff-0-const-relation}, that the mapping $F_1 -
  F_2$ is constant on the interval $[ 0 , 1 ]$ (Note, that $F_1$ as
  well as $F_2$ are continuous.). Hence
  \begin{equation*}
    F ( \xfrak_0 ) = F_1 ( 0 ) - F_2 ( 0 ) = F_1 ( 1 ) - F_2 ( 1 ) = F
    ( \xfrak_0 + \hfrak ) - \int_0^1 d \vartheta \; \Dfrak F (
    \xfrak_0 + \vartheta \hfrak ) \hfrak \text{,}
  \end{equation*}
  which is just equation \eqref{eq-mean-value-theorem} re-written.
\end{proof}

\section{Differentiation on Analytic Manifolds}

Being of a local nature, the concept of differentiability set out in
Definition~\ref{Def-diff-lqs} can be generalized to $\Vfrak$-valued
mappings on analytic manifolds in the following way.
\begin{Def}
  \label{Def-M-differentiability}
  Let $\Mscr$ be a (real or complex) analytic manifold of dimension
  $d$ and let $\Vfrak$ be a locally convex space over the same
  field. Let furthermore $( \Uscr , \phi )$ denote a local chart on
  $\Mscr$, which means that $\phi ( \Uscr ) \subseteq \Kbb^d$, $\Kbb =
  \Rbb$ or $\Kbb = \Cbb$.
  \begin{deflist}
  \item The mapping $F : \Uscr \rightarrow \Vfrak$ is called
    differentiable (with respect to $\phi$) at $m_0 \in \Uscr$ if $F
    \circ \phi^{-1} : \phi ( \Uscr ) \rightarrow \Vfrak$ is
    differentiable at $\phi ( m_0 )$ in the sense of
    Definition~\ref{Def-diff-lqs}. The derivative is denoted
    $\Dfrak_\phi F ( m_0 ) \doteq \Dfrak \bigl( F \circ \phi^{-1}
    \bigr) \bigl( \phi ( m_0 ) \bigr)$. 
  \item $F : \Uscr \rightarrow \Vfrak$ is called (continuously)
    differentiable if $F \circ \phi^{-1}$ is (continuously)
    differentiable in the sense of Definition~\ref{Def-diff-lqs}.
  \item The mapping $F : \Mscr \rightarrow \Vfrak$ is called
    (continuously) differentiable if to any $m_0 \in \Mscr$ there
    exists a local chart $( \Uscr , \phi )$ containing $m_0$ such that
    $F \restriction \Uscr$ is (continuously) differentiable with
    respect to $\phi$. 
  \item Let $\set{\eib_i : i = 1 \text{,} \dots \text{,}d}$ be the
    canonical orthonormal basis of $\Kbb^d$. Then $F : \Uscr
    \rightarrow \Vfrak$ is said to have continuous partial derivatives
    if there exist $d$ continuous mappings $F_\phi^i : \Uscr
    \rightarrow \Vfrak$, such that the increment of $F$ in direction
    $\eib_i$ at any $m_0 = \phi^{-1} ( \tib_0 ) \in \Uscr$ allows for
    the representation
    \begin{subequations} 
      \begin{equation}
        \label{eq-diff-partial}
        F \circ \phi^{-1} ( \tib_0 + h \, \eib_i ) - F \circ \phi^{-1}
        ( \tib_0 ) = h \, F_\phi^i ( m_0 ) + R \bigl[ F \circ
        \phi^{-1} , \tib_0 \bigr] ( h )
      \end{equation}
      if $h \in \Kbb$ is small enough, where the residual term
      satisfies 
      \begin{equation}
        \label{eq-diff-partial-residual}
        \lim_{h \rightarrow 0} \abs{h}^{-1} \bqlx{R \bigl[ F \circ
        \phi^{-1} , \tib_0 \bigr] ( h )} = 0 
      \end{equation}
    \end{subequations}
    for any seminorm $\ql$, $\lambda \in L$.
  \item Higher derivatives of the mapping $F : \Uscr \rightarrow
    \Vfrak$ are defined recursively in terms of partial derivatives of
    the mappings $F_\phi^i$, $i = 1 \text{,} \dots \text{,} d$, and,
    if they happen to exist, are denoted $F_\phi^\kappa$ for
    multi-indices $\kappa \doteq ( k_1 , \dots , k_d ) \in \Nbb_0^d$
    in an obvious fashion (for given $i$ let $F_\phi^{\kappa_i} \doteq
    F_\phi^i$ where all entries in $\kappa_i$ apart from $k_i = 1$
    vanish). $F : \Uscr \rightarrow \Vfrak$ is called $N$-fold (or
    infinitely often) continuously differentiable if the mappings
    $F_\phi^\kappa$ exist and are continuous for any $\abs{\kappa}
    \doteq \sum_i k_i \leqslant N$ (or $\abs{\kappa} < \infty$). These
    concepts apply equally to mappings $F$ defined on all of $\Mscr$.
  \end{deflist}
\end{Def}
\begin{Rem}
  If $F$ is differentiable at $m_0 \in \Uscr$ with respect to the
  local chart $( \Uscr , \phi )$ it is also differentiable with
  respect to any other local chart $( \Vscr , \psi )$ containing
  $m_0$, and according to Proposition~\ref{Pro-chain-rule} one has
  \begin{equation}
    \label{eq-diff-different-charts}
    \Dfrak_\psi F ( m_0 ) = \Dfrak_\phi F ( m_0 ) \circ \bigl( \phi
    \circ \psi^{-1} \bigr)' \bigl( \psi ( m_0 ) \bigr) \text{,}
  \end{equation}
  where $\bigl( \phi \circ \psi^{-1} \bigr)'$ denotes the first
  derivative (Jacobi matrix) of the analytic function $ \phi \circ 
  \psi^{-1} : \psi ( \Uscr \cap \Vscr ) \rightarrow \phi ( \Uscr \cap
  \Vscr )$. 
\end{Rem}
Strictly speaking, the definition of and notation for higher
derivatives of a mapping $F : \Uscr \rightarrow \Vfrak$ is justified
only after the following two results are established.
\begin{Pro}
  $F : \Uscr \rightarrow \Vfrak$ is continuously differentiable if and
  only if it has continuous  partial derivatives in all directions
  $\eib_i$, $i = 1 \text{,} \dots \text{,} d$.
\end{Pro}
\begin{proof}
  \begin{prooflist}
  \item If $F$ is continuously differentiable the mappings 
    \begin{equation}
      \label{eq-diff-partial-derivatives}
      \Uscr \ni m_0 \mapsto F_\phi^i ( m_0 ) \doteq \Dfrak_\phi F (
      m_0) \eib_i 
    \end{equation}
    are continuous for any $i$; furthermore \eqref{eq-diff-partial}
    and \eqref{eq-diff-partial-residual} correspond for each $i$
    exactly to \eqref{eq-diff-lqs} and \eqref{eq-diff-lqs-residual} of
    Definition~\ref{Def-diff-lqs} setting $\hfrak = h \, \eib_i$, so
    that the first part of the statement is almost trivial. 
  \item Let all the partial derivatives of $F$ exist as continuous
    mappings $F_\phi^i : \Uscr \rightarrow \Vfrak$, then, for small
    $\hib = \sum_i h_i \, \eib_i \in \Kbb^d$, we have through an
    application of the Mean Value Theorm~\ref{The-mean-value} for any
    $m_0 = \phi^{-1} ( \tib_0 ) \in \Uscr$  
    \begin{multline}
      \label{eq-first-appl-mean-value}
      F \circ \phi^{-1} ( \tib_0 + \hib ) - F \circ \phi^{-1} ( \tib_0
      ) \\ 
      \mspace{-120mu} = \sum_{i = 1}^d \Bigl[ F \circ \phi^{-1}
      \Bigl( \tib_0 + \sum_{j = 1}^i h_j \, \eib_j \Bigr) - F \circ
      \phi^{-1} \Bigl( \tib_0 + \sum_{j = 1}^{i - 1} h_j \, \eib_j
      \Bigr) \Bigr] \\ 
      = \sum_{i = 1}^d h_i \, F_\phi^i ( m_0 ) + \sum_{i = 1}^d
      \int_0^1 d \vartheta \; h_i \, \Bigl[ F_\phi^i \circ \phi^{-1}
      \Bigl( \tib_0 + \sum_{j = 1}^{i - 1} h_j \, \eib_j + \vartheta
      \, h_i \, \eib_i \Bigr) - F_\phi^i ( m_0 ) \Bigr] \text{.}
    \end{multline}
    Due to continuity of the mappings $F_\phi^i$, the second term on
    the right-hand side multiplied with $\abs{\hib}^{-1}$ can be
    estimated by
    \begin{multline}
      \label{eq-first-appl-mean-value-residual}
      \abs{\hib}^{-1} \bgqlx{\sum_{i = 1}^d \int_0^1 d \vartheta \;
      h_i \, \Bigl[ F_\phi^i \circ \phi^{-1} \Bigl( \tib_0 + \sum_{j =
      1}^{i - 1} h_j \, \eib_j + \vartheta \, h_i \, \eib_i \Bigr) -
      F_\phi^i ( m_0 ) \Bigr]} \\
      \leqslant \abs{\hib}^{-1} \sum_{i = 1}^d \abs{h_i} \max_{0
      \leqslant \vartheta \leqslant 1} \Bqlx{F_\phi^i \circ \phi^{-1}
      \Bigl( \tib_0 + \sum_{j = 1}^{i - 1} h_j \, \eib_j + \vartheta
      \, h_i \, \eib_i \Bigr) - F_\phi^i ( m_0 )} \\
      \leqslant \sum_{i = 1}^d \max_{0 \leqslant \vartheta \leqslant
      1} \Bqlx{F_\phi^i \circ \phi^{-1} \Bigl( \tib_0 + \sum_{j =
      1}^{i - 1} h_j \, \eib_j + \vartheta \, h_i \, \eib_i \Bigr) -
      F_\phi^i ( m_0 )} \text{,}  
    \end{multline}
    where the last expression  of the above inequality is seen to
    vanish in the limit $\hib \rightarrow \zeroib$ by assumption. Thus
    \eqref{eq-first-appl-mean-value} in connection with
    \eqref{eq-first-appl-mean-value-residual} establishes continuous
    differentiability of the mapping $F : \Uscr \rightarrow \Vfrak$
    with 
    \begin{equation*}
      \Dfrak_\phi F ( m_0 ) \hib = \sum_{i = 1}^d h_i \, F_\phi^i (
      m_0 ) \text{,} \quad \hib \in \Kbb^d \text{.} \tag*{\qed}
    \end{equation*}
    \renewcommand{\qed}{}
  \end{prooflist}
  \renewcommand{\qed}{}
\end{proof}
\begin{Pro}
  Assume that the mixed derivatives $F_\phi^{i j} \doteq \bigl(
  F_\phi^i \bigr)_\phi^j$ and $F_\phi^{j i} \doteq \bigl( F_\phi^j
  \bigr)_\phi^i$, $i \text{,} j \in \{ 1 , \dots , d\}$ of the mapping
  $F : \Uscr \rightarrow \Vfrak$ exist and are continuous on
  $\Uscr$. Then they coincide: 
  \begin{equation*}
    F_\phi^{i j} ( m_0 ) = F_\phi^{j i} ( m_0 ) \text{,} \quad m_0 \in
    \Uscr \text{.}
  \end{equation*}
\end{Pro}
\begin{proof}
  For $m_0 = \phi^{-1} ( \tib_0 ) \in \Uscr$ and sufficiently small $h
  \text{,} k \in \Kbb$ consider the following expression which
  involves two increments of $F \circ \phi^{-1}$:
  \begin{equation*}
    F \circ \phi^{-1} ( \tib_0 + h \, \eib_j + k \, \eib_i ) - F \circ
    \phi^{-1} ( \tib_0 + h \, \eib_j ) - F \circ \phi^{-1} ( \tib_0 +
    k \, \eib_i ) + F \circ \phi^{-1} ( \tib_0 ) \text{.}  
  \end{equation*}
  By assumption on the existence and continuity of the mixed
  derivatives we can apply the Mean Value Theorem~\ref{The-mean-value}
  twice to the above expression: One can consider the increments with
  respect to $\eib_i$ and apply the Mean Value Theorem to them first
  and afterwards to the resulting integrand which takes on the form of
  an increment with respect to $\eib_j$, or one carries out the same 
  procedure with the roles of $\eib_i$ and $\eib_j$ interchanged. Upon
  division by $h \, k \not= 0$ this yields the integrals 
  \begin{multline*}
    \int_0^1 d \vartheta \int_0^1 d \vartheta' \; F_\phi^{i j} \circ
    \phi^{-1} ( \tib_0 + \vartheta \, h \, \eib_j + \vartheta' \, k \,
    \eib_i ) \text{,} \\
    = \int_0^1 d \vartheta \int_0^1 d \vartheta' \; F_\phi^{j i} \circ
    \phi^{-1} ( \tib_0 + \vartheta \, h \, \eib_j + \vartheta' \, k \,
    \eib_i ) \text{,}
  \end{multline*}
  for any $h \text{,} k \in \Kbb \setminus \set{0}$. Specializing to
  sequences $\set{h_n}_{n \in \Nbb}$ and $\set{k_n}_{n \in \Nbb}$ in
  this set which converge to $0$, it is a consequence of Lebesgue's
  Dominated Convergence Theorem (cf.~\cite[II.5.6 and
  II.6.2]{fell/doran:1988a}) that for $n \rightarrow \infty$ the
  left-hand side converges to $F_\phi^{i j} ( m_0 )$ whereas the
  right-hand side approaches $F_\phi^{j i} ( m_0 )$ in the locally
  convex topology of $\Vfrak$. Since this topology separates the
  elements of $\Vfrak$, we conclude that these limits coincide and get
  the assertion by arbitrariness of $m_0 \in \Uscr$.
\end{proof}

\section{Differentiation on Automorphism Lie Groups}

The concepts developed thus far can now be applied to the case where
the underlying analytic manifold is a (real or complex) Lie group
$\Gscr$ acting via a strongly continuous group of automorphisms
$\bset{\alpha_g : g \in \Gscr} \subseteq \Aut \Bfrak$ on the
$C^*$-algebra $\Bfrak$. These automorphisms $\alpha_g$, when applied
to a given element $B \in \Bfrak$, define a $\Bfrak$-valued mapping on
$\Gscr$, for which statements can be proved that go beyond the above
results. In doing so we shall be concerned with the canonical
coordinates $( \Uscr_0 , \phi_0 )$ of the first kind around the
neutral element $\neutral$ of $\Gscr$ where $\neutral = \phi_0^{-1} (
\zeroib )$ (cf.~\cite[Section~2.10]{varadarajan:1984}). Note also,
that, for given $g \in \Gscr$, the left and right translations $l_g$
and $r_g$ on $\Gscr$ as well as their composition $i_g = l_g \circ
r_{g^{-1}}$ are analytic diffeomorphisms, so that their application to
$( \Uscr_0 , \phi_0 )$ yields local charts around $g$ and $\neutral$,
respectively (cf.~\cite[Section~2.1]{varadarajan:1984}).
\begin{Pro}
  \label{Pro-G-differentiability}
  Let $\Gscr$ be a $d$-dimensional real or complex Lie group and let
  $\Bfrak$ be a $C^*$-algebra. For given $B \in \Bfrak$ define the
  mapping 
  \begin{equation*}
    \Xi_B : \Gscr \rightarrow \Bfrak \quad g \mapsto \Xi_B ( g )
    \doteq \alpha_g ( B ) \text{.}
  \end{equation*}
  \begin{proplist}
  \item $\Xi_B$ is continuously differentiable on $\Gscr$ if and only
    if it is differentiable at $\neutral \in \Gscr$.
  \item If $\Xi_B$ is differentiable at $\neutral \in \Gscr$, then
    $\Xi_{\alpha_{g'} ( B )}$ is differentiable for any $g' \in \Gscr$
    and the mapping
    \begin{equation*}
      \Gscr \times \Uscr \ni ( g' , g ) \mapsto \Dfrak_\phi
      \Xi_{\alpha_{g'} ( B )} ( g ) \hib
    \end{equation*}
    is jointly continuous in $g'$ and $g$ for any local chart $( \Uscr
    , \phi )$ around $g$ and any $\hib \in \Kbb^d$.
  \end{proplist}
\end{Pro}
\begin{proof}
  \begin{prooflist}
  \item To prove the non-trivial part, suppose that $g \in \Gscr$ is
    arbitrary but fixed. Then $( g \Uscr_0 , \phi_g )$, $\phi_g \doteq
    \phi_0 \circ l_{g^{-1}}$, is a local chart around $g$ with
    $\phi_g^{-1} = l_g \circ \phi_0^{-1}$. According to the definition
    of $\Xi_B$ we have 
    \begin{equation*}
      \Xi_B \circ \phi_g^{-1} = \Xi_B \circ l_g \circ \phi_0^{-1} =
      \alpha_g \circ \Xi_B \circ \phi_0^{-1}
    \end{equation*}
    and, since the automorphisms are norm-preserving, the assumed
    differentiability of the mapping $\Xi_B \circ \phi_0^{-1}$ at
    $\zeroib$ carries over to $\Xi_B \circ \phi_g^{-1}$ which by
    Definition~\ref{Def-M-differentiability} means that $\Xi_B$ is
    differentiable at $g = \phi_g^{-1} ( \zeroib )$:
    \begin{equation*}
      \Dfrak_{\phi_g} \Xi_B ( g ) = \Dfrak \bigl( \Xi_B \circ
      \phi_g^{-1} \bigr) ( \zeroib ) = \alpha_g \circ \Dfrak \bigl(
      \Xi_B \circ \phi_0^{-1} \bigr) ( \zeroib ) = \alpha_g \circ
      \Dfrak_{\phi_0} \Xi_B ( \neutral ) \text{.}
    \end{equation*}
    In view of \eqref{eq-diff-different-charts} this relation can be
    re-written with respect to an arbitrary local chart $( \Uscr ,
    \phi )$ on $\Gscr$ containing $g$:
    \begin{equation}
      \label{eq-general-G-derivatives}
      \Dfrak_\phi \Xi_B ( g ) = \Dfrak_{\phi_g} \Xi_B ( g ) \circ
      \bigl( \phi_g \circ \phi^{-1} \bigr)' \bigl( \phi ( g ) \bigr) =
      \alpha_g \circ \Dfrak_{\phi_0} \Xi_B ( \neutral ) \circ
      \Mbf^\phi ( g ) \text{,}
    \end{equation}
    where the matrix elements of $\Mbf^\phi ( g ) \doteq \bigl( \phi_g
    \circ \phi^{-1} \bigr)' \bigl( \phi ( g ) \bigr)$ are analytic in
    $g \in \Uscr$. Since the automorphisms are norm-preserving and act
    stongly continuous on $\Bfrak$, it is evident that application of
    the above operator to any vector $\hib \in \Kbb^d$ yields a
    continuous mapping on $\Kbb^d$ to $\Bfrak$, thus establishing
    continuous differentiability of $\Xi_B$ on $\Gscr$ as stated.
  \item Let $g' \in \Gscr$ be arbitrary and consider the local chart
    $( \Uscr_0 g' , \psi_{g'} )$, $\psi_{g'} \doteq \phi_0 \circ
    r_{{g'}^{-1}}$, around $g'$ with inverse $\psi_{g'}^{-1} = r_{g'}
    \circ \phi_0^{-1}$. Then
    \begin{equation*}
      \Xi_{\alpha_{g'} ( B )} \circ \phi_0^{-1} = \Xi_B \circ r_{g'}
      \circ \phi_0^{-1} = \Xi_B \circ \psi_{g'}^{-1} \text{,}
    \end{equation*}
    so that the assumed differentiablity of $\Xi_B$ at $\neutral$ and
    thus, according to the first part, at $g'$ with respect to the
    local chart $( \Uscr_0 g' , \psi_{g'} )$ implies differentiability
    of $\Xi_{\alpha_{g'} ( B )}$ at $\neutral$. By an application of
    \eqref{eq-general-G-derivatives} we have
    \begin{equation}
      \label{eq-transformed-G-derivatives}
      \Dfrak_{\phi_0} \Xi_{\alpha_{g'} ( B )} ( \neutral ) =
      \Dfrak_{\psi_{g'}} \Xi_B ( g' ) = \alpha_{g'} \circ
      \Dfrak_{\phi_0} \Xi_B ( \neutral ) \circ \Nbf ( g') \text{,}
    \end{equation}
    where the matrix elements of $\Nbf ( g') \doteq \bigl( \phi_g
    \circ \psi_{g'}^{-1} \bigr)' ( \zeroib )$ are analytic in
    $g'$. This in turn can, again by use of
    \eqref{eq-general-G-derivatives}, be generalized to any $g \in
    \Gscr$ lying in the local chart $( \Uscr , \phi )$:
    \begin{equation}
      \label{eq-generalized-transformed-G-derivatives}
      \Dfrak_\phi \Xi_{\alpha_{g'} ( B )} ( g ) = \alpha_g \circ
      \Dfrak_{\phi_0} \Xi_{\alpha_{g'} ( B )} ( \neutral ) \circ
      \Mbf^\phi ( g ) = \alpha_{g g'} \circ \Dfrak_{\phi_0} \Xi_B (
      \neutral ) \circ \Nbf ( g') \circ \Mbf^\phi ( g ) \text{,}
    \end{equation}
    an expression which is obviously continuous in both variables $g'$
    and $g$ when applied to an arbitrary element $\hib$ of $\Kbb^d$. 
  \end{prooflist}
  \renewcommand{\qed}{}
\end{proof}
\begin{Rem}
  Note, that in the case of differentiability of $\Xi_B$ the mapping
  $g \mapsto \Dfrak_\phi \Xi_B ( g )$ need not be continuous in the
  operator-norm topology of the Banach space of linear operators on
  $\Kbb^d$ to $\Bfrak$, since the automorphism group $\bset{\alpha_g :
  g \in \Gscr} \subseteq \Aut \Bfrak$ is only supposed to be strongly
  continuous.
\end{Rem}

Consider those operators $B \in \Bfrak$ for which the mapping $\Xi_B$
is continuously differentiable on $\Gscr$. According to
Proposition~\ref{Pro-G-differentiability} this is equivalent to
differentiability at $\neutral$ with respect to the canonical 
coordinates $( \Uscr_0 , \phi_0 )$. Therefore one can define mappings
$\delta^i$ corresponding to the partial derivatives of $\Xi_B$ at
$\neutral$ (cf.~\eqref{eq-diff-partial-derivatives}) by
\begin{equation*}
  \delta^i ( B ) \doteq \Dfrak_{\phi_0} \Xi_B ( \neutral ) \eib_i
  \text{,} \quad i = 1 \text{,} \dots \text{,} d \text{.}
\end{equation*}
Since $\Xi_B$ depends linearly on $B$, it is easily seen that 
\begin{subequations}
  \begin{align}
    \label{eq-derivation-sum}
    \delta^i ( B_1 + B_2 ) & = \delta^i ( B_1 ) + \delta^i ( B_2 )
    \text{,} \\ 
    \label{eq-derivation-scalmult}
    \delta^i ( \lambda \, B_1 ) & = \lambda \, \delta^i ( B_1 )
    \text{,} 
  \end{align}
  for any $\lambda \in \Kbb$ and $B_1$, $B_2$ in $\Bfrak$ subject to
  Proposition~\ref{Pro-G-differentiability}. Moreover, $\Xi_{B_1 B_2}
  = \Xi_{B_1} \Xi_{B_2}$, so that Proposition~\ref{Pro-product-rule}
  yields
  \begin{equation}
    \label{eq-derivation-product}
    \delta^i ( B_1 B_2 ) = \delta^i ( B_1 ) \, B_2 + B_1 \, \delta^i (
    B_2 ) \text{.}
  \end{equation}
\end{subequations}
Equations \eqref{eq-derivation-sum} through
\eqref{eq-derivation-product} show that the mappings $\delta^i$ act as
derivations of the $C^*$-algebra $\Bfrak$
(cf.~\cite[Chapter~III.9]{dixmier:1981} and
\cite[Section~8.6]{pedersen:1979}). Their domains are certain
subalgebras which are invariant under transformations from the
automorphism group $\bset{\alpha_g : g \in \Gscr}$, since by
\eqref{eq-transformed-G-derivatives} for any $g' \in \Gscr$ and any $B
\in \Bfrak$ with differentiable $\Xi_B$ one has
\begin{equation}
  \label{eq-derivations-transformed}
  \delta^i \bigl( \alpha_{g'} ( B ) \bigr) = \Dfrak_{\phi_0}
  \Xi_{\alpha_{g'} ( B )} ( \neutral ) \eib_i = \alpha_{g'} \bigl(
  \Dfrak_{\phi_0} \Xi_B ( \neutral ) \Nbf ( g' ) \eib_i \bigr) =
  \sum_{j = 1}^d \Nbf_{j i} ( g' ) \alpha_{g'} \bigl( \delta^j ( B )
  \bigr) \text{.}
\end{equation}

Let $\iota_M$ denote the $M$-tuple $( i_1 , \dots , i_M )$ with
integer entries $1 \leqslant i_l \leqslant d$, then the corresponding
products of derivations $\delta^{\iota_M} \doteq \delta^{i_M} \dots
\delta^{i_1}$ act as linear operators on certain subspaces of $\Bfrak$
which are again invariant with respect to $\bset{\alpha_g : g \in
\Gscr}$, possibly the trivial space $\set{0}$ (note, that in general
the derivations will not commute). Making use of the concepts of
differentiability introduced above together with the fact that left
and right translations act as analytic diffeomorphisms on the group
$\Gscr$, it is a matter of elementary considerations to establish the
following connection between products $\delta^{\iota_M}$ of the above
kind and the partial derivatives of the mapping $\Xi_B$ indexed by
multi-indices $\kappa$:
\begin{subequations}
  \begin{align}
    \label{eq-xi-delta}
    \Xi^\kappa_{B , \phi} ( g ) & = \sum_{\iota_M , M \leqslant
    \abs{\kappa}} C^\phi_{\kappa , \iota_M} ( g ) \, \alpha_g \bigl(
    \delta^{\iota_M} ( B ) \bigr) \text{,} \\
    \label{eq-delta-xi}
    \delta^{\iota_M} ( B ) & = \sum_{\kappa , \abs{\kappa} \leqslant
    M} D^{\phi_0}_{\iota_M , \kappa} ( \neutral ) \, \Xi^\kappa_{B ,
    \phi_0} ( \neutral ) \text{.}
  \end{align}
\end{subequations}
Here the real or complex functions $C^\phi_{\kappa , \iota_M}$ and
$D^{\phi_0}_{\iota_M , \kappa}$ are analytic on the respective charts
$( \Uscr , \phi )$ and $( \Uscr_0 , \phi_0 )$, containing $g$ and
$\neutral$ respectively. Implicit in \eqref{eq-xi-delta} and
\eqref{eq-delta-xi} is the perception that the mapping $\Xi_B$ is
$N$-fold (or infinitely often) continuously differentiable if and only
if the operator $B$ belongs to the domain of all $\delta^{\iota_M}$
for $M \leqslant N$ (or any $M < \infty$).

We formulate these results in the following definition and subsequent
proposition.
\begin{Def}
  \label{Def-partial-derivations}
  Let $\delta^i$, $i = 1 \text{,} \dots \text{,} d$, denote the
  partial derivations pertaining to the mappings $\Gscr \ni g \mapsto
  \Xi_B ( g ) = \alpha_g ( B ) \in \Bfrak$ for certain $B \in \Bfrak$
  via
  \begin{equation}
    \label{eq-def-partial-derivations}
    \delta^i ( B ) \doteq \Dfrak_{\phi_0} \Xi_B ( \neutral ) \eib_i
    \text{.} 
  \end{equation}
  For given $N \in \Nbb$ the domain of arbitrary $N$-fold products
  $\delta^{\iota_N}$ of these derivations is an invariant subspace of
  $\Bfrak$ with respect to the automorphism group $\bset{\alpha_g : g
  \in \Gscr}$ and denoted $\Dscr^{( N )} ( \Bfrak )$: the space of
  $N$-fold differentiable operators. The elements of the space
  $\Dscr^{( \infty )} ( \Bfrak ) \doteq \bigcap_{N \in \Nbb} \Dscr^{(
  N )} ( \Bfrak )$ in turn are called infinitely often differentiable
  with respect to $\bset{\alpha_g : g \in \Gscr}$. Accordingly, the
  resulting operators $\delta^{\iota_N} ( B )$ are designated as the
  derivatives of $B \in \Bfrak$, if this element happens to lie in
  their domain.
\end{Def}
\begin{Pro}
  \label{Pro-partial-derivations}
  For given $B \in \Bfrak$ the mapping
  \begin{equation*}
    \Xi_B : \Gscr \rightarrow \Bfrak \quad g \mapsto \Xi_B ( g )
    \doteq \alpha_g ( B )
  \end{equation*}
  is $N$-fold or infinitely often continuously differentiable if and
  only if the operator $B$ belongs to $\Dscr^{( N )} ( \Bfrak )$
  respectively $\Dscr^{( \infty )} ( \Bfrak )$.
\end{Pro}

\section{Differentiable Linear Mappings}

In this section a special notion of differentiability for linear
mappings on a locally convex space $\Vfrak$ is introduced, which is
motivated by the following result that is valid under the assumption
of continuity.
\begin{Pro}
  \label{Pro-differentiable-lin-mappings}
  Let $\Xfrak$ be a (real or complex) normed space, $\Gfrak \subseteq
  \Xfrak$ open, and let $\Vfrak$ and $\Wfrak$ be locally convex spaces
  over the same field $\Kbb \doteq \Rbb$ or $\Kbb \doteq \Cbb$ with
  topologies defined by the families $\bset{\ql : \lambda \in L}$  and
  $\bset{\qprim : \mu \in M}$ of seminorms separating the points of
  $\Vfrak$ and $\Wfrak$, respectively. If $F : \Gfrak \rightarrow
  \Vfrak$ is differentiable at the point $\xfrak \in \Gfrak$ and $\Psi
  : \Vfrak \rightarrow \Wfrak$ is a continuous linear mapping then the
  composition
  \begin{equation*}
    \Psi \circ F : \Gfrak \rightarrow \Wfrak
  \end{equation*}
  is differentiable at $\xfrak$, too, and its derivative is given by
  \begin{equation}
    \label{eq-diff-lin-mapping-derivative}
    \Dfrak ( \Psi \circ F ) ( \xfrak ) = \Psi \circ \Dfrak F ( \xfrak
    ) \text{.}
  \end{equation}
  If $F$ is differentiable on all of $\Gfrak$ the same holds true for
  $\Psi \circ F$ and \eqref{eq-diff-lin-mapping-derivative} is valid
  for any $\xfrak \in \Gfrak$.
\end{Pro}
\begin{proof}
  By assumption on $F$ (relations \eqref{eq-diff-lqs} and
  \eqref{eq-diff-lqs-residual}) in connection with linearity of
  $\Psi$, the increment of $\Psi \circ F$ at $\xfrak$ allows for the
  representation
  \begin{equation}
    \label{eq-diff-lin-mapping}
    ( \Psi \circ F ) ( \xfrak + \hfrak ) - ( \Psi \circ F ) ( \xfrak )
    = \Psi \circ \Dfrak F ( \xfrak ) \hfrak + \Psi \bigl( R [ F ,
    \xfrak] ( \hfrak ) \bigr) \text{,}
  \end{equation}
  where
  \begin{equation*}
    \lim_{\hfrak \rightarrow \zerofrak} \norm{\hfrak}^{-1} \bqlx{R [ F
    , \xfrak ] ( \hfrak )} = 0
  \end{equation*}
  for any seminorm $\ql$, $\lambda \in L$. But, due to continuity of
  $\Psi$, there exist to any seminorm $\qprim$ on $\Wfrak$ a finite
  number of seminorms $q_{\lambda_i}$ on $\Vfrak$, $i = 1 \text{,}
  \dots \text{,} N$, and a positive constant $C_\mu$ such that
  for any $\vfrak \in \Vfrak$
  \begin{equation*}
    \bqprimx{\Psi ( \vfrak )} \leqslant C_\mu \max_{1 \leqslant i
    \leqslant N} q_{\lambda_i} ( \vfrak ) \text{,}
  \end{equation*}
  and therefore
  \begin{equation*}
    0 \leqslant \norm{\hfrak}^{-1} \bqprimx{\Psi \bigl( R [ F ,
    \xfrak] ( \hfrak ) \bigr)} \leqslant C_\mu \max_{1 \leqslant i
    \leqslant N} \Bigl( \norm{\hfrak}^{-1} q_{\lambda_i} \bigl( R [ F
    , \xfrak] ( \hfrak ) \bigr) \Bigr) \xrightarrow[\hfrak \rightarrow
    \zerofrak]{} 0 \text{.}
  \end{equation*}
  This is just the formulation of \eqref{eq-diff-lqs-residual} for
  $\Psi \circ F$ and thus proves, according to
  \eqref{eq-diff-lin-mapping}, differentiability of this mapping at
  $\xfrak$ together with \eqref{eq-diff-lin-mapping-derivative}. The
  remainder of the assertion is a trivial consequence.
\end{proof}
The above results can easily be generalized to $\Vfrak$-valued
mappings on an analytic manifold $\Mscr$.
\begin{Cor}
  \label{Cor-differentiable-M-lin-mappings}
  Let $\Mscr$ be a (real or complex) analytic manifold of dimension
  $d$ and let $\Vfrak$ and $\Wfrak$ be locally convex spaces over the
  same field. If $F : \Uscr \rightarrow \Vfrak$ is differentiable at
  the point $m_0 \in \Uscr$, $( \Uscr , \phi )$ a local chart on
  $\Mscr$, and $\Psi : \Vfrak \rightarrow \Wfrak$ is a continuous
  linear mapping then
  \begin{equation*}
    \Psi \circ F : \Uscr \rightarrow \Wfrak
  \end{equation*}
  is differentiable at $m_0$, and its derivative is given by
  \begin{equation}
    \label{eq-diff-lin-mapping-M-derivative}
    \Dfrak_\phi ( \Psi \circ F ) ( m_0 ) = \Psi \circ \Dfrak_\phi F (
    m_0 ) \text{.}
  \end{equation}
  Accordingly, $\Psi \circ F$ is differentiable on all of $\Mscr$ in
  case that $F$ is. 
\end{Cor}

Proposition~\ref{Pro-differentiable-lin-mappings} motivates the
following definition which does no longer depend on the assumption of
continuity.
\begin{Def}
  \label{Def-generalized-diff-lin-mappings}
  \begin{deflist}
  \item Let $\Xfrak$ be a normed space and $\Fscr$ a family of
    differentiable mappings on $\Xfrak$ with values in a locally
    convex space $\Vfrak$. A linear mapping $\Psi$ on $\Vfrak$ to the
    locally convex space $\Wfrak$ is called $\Fscr$-differentiable if
    and only if $\Psi \circ F : \Xfrak \rightarrow \Wfrak$ is
    differentiable on $\Xfrak$ for any $F \in \Fscr$ with 
    \begin{equation*}
      \Dfrak ( \Psi \circ F ) ( \xfrak ) = \Psi \circ \Dfrak F (
      \xfrak ) \text{,} \quad \xfrak \in \Xfrak \text{.}
    \end{equation*}
  \item Let $\Mscr$ be an analytic manifold and let $\Vfrak$, $\Wfrak$
    and $\Psi$ be as above. Assume furthermore that $\Fscr$ is a
    family of differentiable $\Vfrak$-valued mappings on $\Mscr$. Then
    $\Psi$ is called $\Fscr$-differentiable if and only if $\Psi \circ
    F : \Mscr \rightarrow \Wfrak$ is differentiable on $\Mscr$ for any
    $F \in \Fscr$  and 
    \begin{equation*}
      \Dfrak_\phi ( \Psi \circ F ) ( m_0 ) = \Psi \circ \Dfrak_\phi F
      ( m_0 )
    \end{equation*}
    for any chart $( \Uscr , \phi )$ around the arbitrary element $m_0
    \in \Mscr$.
  \end{deflist}
\end{Def}

\chapter[A Lemma on Norm-Separable $\mathib{C^*}$-Algebras]{A Lemma
  on Norm-Separable $\mathib{C^*}$-Algebras}
  \label{chap-separable-algebras}

The following result is an adaptation of
\cite[Lemma~14.1.17]{kadison/ringrose:1986} to our needs.
\begin{Lem}
  Let $\Afrak$ be a unital $C^*$-subalgebra of $\BH$, where the
  Hilbert space $\Hscr$ is separable. There exists a norm-separable
  $C^*$-algebra $\Afrak^0$, containing the unit element $\unit$, that
  lies strongly dense in $\Afrak$.
\end{Lem}
\begin{proof}
  Let $\bset{\phi_n}_{n \in \Nbb}$ be a dense sequence of non-zero
  vectors in $\Hscr$ and let $\Mfrak \doteq \Afrak''$ denote the von
  Neumann algebra generated by $\Afrak$. According to von Neumann's
  Density Theorem, $\Mfrak$ coincides with the strong closure
  $\Afrak^-$ of the algebra $\Afrak$, which by assumption acts
  non-degenerately on $\Hscr$ (cf.~\cite[Section~I.3.4]{dixmier:1981},
  \cite[Corollary~2.4.15]{bratteli/robinson:1987}).

  First we assume the existence of a separating vector for $\Mfrak$,
  which is thus cyclic for $\Mfrak'$
  \cite[Section~I.1.4]{dixmier:1981}. Then any normal functional on
  $\Mfrak$ is of the form $\omega_{\psi , \psi'} \restriction \Mfrak$
  with $\psi \text{,} \psi' \in \Hscr$
  \cite[Theorem~V.3.15]{takesaki:1979}. Choose operators $A_{j , k}
  \in \Afrak_1$ satisfying
  \begin{equation}
    \label{eq-first-estimate}
    \omega_{\phi_j , \phi_k} ( A_{j , k} ) \geqslant
    \norm{\omega_{\phi_j , \phi_k} \restriction \Mfrak} - 2^{-1}
    \text{,}
  \end{equation}
  which is possible due to Kaplansky's Density Theorem
  \cite[Theorem~2.3.3]{pedersen:1979}. Let $\Afrak^0$ denote the
  norm-separable $C^*$-algebra generated by the unit element $\unit$
  together with all the operators $A_{j , k}$, $j \text{,} k \in
  \Nbb$, and select a normal functional $\omega_{\xi , \theta}$ on
  $\Mfrak$ with the properties $\norm{\omega_{\xi , \theta}
  \restriction \Afrak^0} = 0$ and $\norm{\omega_{\xi , \theta}
  \restriction \Mfrak} > 0$. Without loss of generality we can assume
  $\norm{\omega_{\xi , \theta} \restriction \Mfrak} = 1$. To any
  $\epsilon > 0$ there exist vectors $\phi_j \text{,} \phi_k$ from the
  dense sequence in $\Hscr$ rendering $\norm{\phi_j - \xi}$ and
  $\norm{\phi_k - \theta}$ small enough so that
  \begin{equation}
    \label{eq-second-estimate}
    \norm{( \omega_{\xi , \theta} - \omega_{\phi_j , \phi_k} )
    \restriction \Mfrak} < \epsilon \text{.}
  \end{equation}
  Making use of \eqref{eq-first-estimate} we then get the estimate
  \begin{multline*}
    \epsilon > \norm{( \omega_{\xi , \theta} - \omega_{\phi_j ,
    \phi_k} ) \restriction \Mfrak} \geqslant \norm{( \omega_{\xi ,
    \theta} - \omega_{\phi_j , \phi_k} ) ( A_{j , k} )} \\
    = \norm{\omega_{\phi_j , \phi_k} ( A_{j , k} )} \geqslant
    \norm{\omega_{\phi_j , \phi_k} \restriction \Mfrak} - 2^{-1}
    \text{,} 
  \end{multline*}
  which in connection with \eqref{eq-second-estimate} implies
  \begin{equation*}
    \norm{\omega_{\xi , \theta} \restriction \Mfrak} \leqslant \norm{(
    \omega_{\xi , \theta} - \omega_{\phi_j , \phi_k} ) \restriction
    \Mfrak} + \norm{\omega_{\phi_j ,\phi_k} \restriction \Mfrak} < 2
    \epsilon + 2^{-1} \text{.}
  \end{equation*}
  By arbitraryness of $\epsilon$ we infer $\norm{\omega_{\xi , \theta}
  \restriction \Mfrak} \leqslant 2^{-1}$ in contradiction to the
  assumption that $\omega_{\xi , \theta} \restriction \Mfrak$ be
  normalized. Thus, $\omega_{\xi , \theta} \restriction \Afrak^0 = 0$
  implies $\omega_{\xi , \theta} \restriction \Mfrak = 0$, i.\,e.~any
  normal functional on $\Mfrak$ annihilating $\Afrak^0$ annihilates
  $\Mfrak$ as well. Now, since the $C^*$-algebra $\Afrak^0$ acts
  non-degenerately on $\Hscr$, von Neumann's Density Theorem tells us
  that its strong and $\sigma$-weak closures coincide with
  ${\Afrak^0}'' = {\Afrak^0}^-$, and this in turn is equal to the von
  Neumann algebra $\Mfrak$; for the existence of an element $A \in
  \Mfrak$ not contained in ${\Afrak^0}^-$ would, by the
  Hahn-Banach-Theorem, imply a $\sigma$-weakly continuous (normal)
  functional that vanishes on $\Afrak^0$ but not on $A \in \Mfrak
  \setminus {\Afrak^0}^-$ in contradiction to the above result.

  Now suppose that there does not exist a separating vector for the
  von Neumann algebra $\Mfrak = \Afrak^-$. Then the sequence 
  \begin{equation*}
    \Bigl( ( n \norm{\phi_n} )^{-1} \phi_n \Bigr)_{n \in \Nbb}
    \subseteq \underline{\Hscr} \doteq \bigoplus_{n=1}^\infty \Hscr
  \end{equation*}
  is such a vector for the von Neumann algebra $\underline{\Mfrak}
  \doteq \bigl( \bigoplus_{n=1}^\infty \iota \bigr) ( \Mfrak )$, where
  $\iota$ denotes the identity representation of $\Mfrak$ in
  $\Hscr$. The result of the preceding paragraph thus applies to the
  $C^*$-algebra $\underline{\Afrak} \doteq \bigl(
  \bigoplus_{n=1}^\infty \iota \bigr) ( \Afrak )$ of operators on the
  separable Hilbert space $\underline{\Hscr}$ which is weakly dense in
  $\underline{\Mfrak}$: $\underline{\Afrak}^- =
  \underline{\Mfrak}$. We infer that there exists a norm-separable
  $C^*$-subalgebra $\underline{\Afrak}^0$ of $\underline{\Afrak}$
  including its unit $\underline{\unit} \doteq ( \unit )_{n \in
  \Nbb}$, which is strongly dense in $\underline{\Afrak}$. Now,
  $\underline{\iota} \doteq \bigoplus_{n=1}^\infty \iota$ is a
  faithful representation of $\Afrak$ on $\underline{\Hscr}$ and its
  inverse $\underline{\iota}^{-1} : \underline{\Afrak} \rightarrow
  \Afrak$ is a faithful representation of $\underline{\Afrak}$ on
  $\Hscr$ which is continuous with respect to the strong topologies of
  $\underline{\Afrak}$ and $\Afrak$. Therefore $\Afrak^0 \doteq
  \underline{\iota}^{-1} \bigl( \underline{\Afrak}^0 \bigr)$ is a
  norm-separable $C^*$-subalgebra of $\Afrak$, containing the unit
  element $\unit$ and lying strongly dense in $\Afrak$.
\end{proof}

\cleardoublepage

\addcontentsline{toc}{chapter}{\protect\numberline{}%
                               {Bibliography}}

% \bibliographystyle{acm}
% \bibliography{preamble,articles,books,unpublished}

\providecommand{\SortNoop}[1]{}

\cleardoublepage

\addcontentsline{toc}{chapter}{\protect\numberline{}%
                               {Acknowledgements}}

\chapter*{Acknowledgements}
  \label{chap-acknowledgements}

My sincere gratitude is due to Prof.~Dr.~Detlev Buchholz for his
support and forbearance. I not only learned a lot from his views on
theoretical physics and its relationship to mathematics, but his ideas
also constituted the basis on which I could erect my contributions to
the topic presented in this thesis. \\[1.5ex]
I should furthermore like to thank Prof.~Dr.~Klaus Fredenhagen for his
immediate readiness to write the additional report on this
work. \\[1.5ex]
Financial support by the Deutsche Forschungsgemeinschaft is gratefully
acknowledged which I obtained from the Graduiertenkolleg at the
II. Institut f\"{u}r Theoretische Physik of the University of
Hamburg. \\[2ex]
I want to thank AnnA for all her support and encouragement, and for
unrepiningly sharing the burden of my strain in the final stages of
this project. \\[1.5ex]
Eventually, I want to express my deepest gratitude to my parents for
their patience and confidence. I am afraid that I shall be unable to
pass back even only part of what they have done for me.

\clearpage

\lhead[\fancyplain{}{\bfseries\thepage}]{}
\rhead[]{\fancyplain{}{\bfseries\thepage}}

\vspace*{\fill}

\begin{otherlanguage}{german}
  \label{cit-prokrustes-transl}
  \noindent
  {\bfseries
   {\large German translation of the quotation on page
     \pageref{cit-prokrustes}} \\[0.5cm]
   {\Large\textsc {Diodoros}: Griechische Weltgeschichte IV, 59 (5)}
     \\[2mm] 
  }
  (nach der \"{U}bersetzung von Otto Veh) \\[1cm]
  Theseus beseitigte auch bei Eleusis den Kerkyon, der die Passanten
  zum Ringkampf veranla\ss{}te und den, der unterlag,
  umbrachte. Sodann t\"{o}tete er auch den Prokrustes, wie er
  hie\ss{}, der am sogenannten Korydallos in Attika hauste. Der
  n\"{o}tigte die vor\"{u}berziehenden Wanderer, sich auf ein Bett
  niederzulegen und war einer zu lang, dann schlug er ihm die
  herausragenden K\"{o}rperteile ab; denen aber, die kleiner waren,
  zog er die F\"{u}\ss{}e in die L\"{a}nge, weshalb er den Namen
  Prokrustes erhielt. 
\end{otherlanguage}

\vspace*{\fill}

\vspace*{\fill}

\end{document}